\documentclass[11pt]{article}
\usepackage{jheppub}

\pdfoutput=1

\usepackage{amssymb,amsmath,amsthm,graphics,graphicx}
\usepackage{hyperref} 
\usepackage{latexsym}
\usepackage{bm}
\usepackage{float}
\usepackage{hyperref}

\usepackage{mathabx}
\usepackage{mathtools}
\usepackage{pdfsync}
\usepackage{slashed}
\usepackage[normalem]{ulem}
\usepackage{upgreek}
\usepackage{url}
\usepackage{booktabs}
\allowdisplaybreaks

\def\be{\begin{equation}}
\def\ee{\end{equation}}
\def\beq{\begin{eqnarray}}
\def\eeq{\end{eqnarray}}

\newcommand{\eg}{{\it e.g.,}\ }
\newcommand{\ie}{{\it i.e.\,}\ }

\usepackage{color}
    \definecolor{darkgreen}{rgb}{0,0.5,0}
    \definecolor{darkred}{rgb}{0.5,0,0}
    \definecolor{darkblue}{rgb}{0,0,0.6}
    \definecolor{purple}{rgb}{0.4,.2,0.7}

\usepackage{placeins}

\begin{document}

\def\lsim{\mathrel{\rlap{\lower4pt\hbox{\hskip1pt$\sim$}}
    \raise1pt\hbox{$<$}}}
\def\gsim{\mathrel{\rlap{\lower4pt\hbox{\hskip1pt$\sim$}}
    \raise1pt\hbox{$>$}}}
\def\be{\begin{equation}}
\def\ee{\end{equation}}
\def\bea{\begin{eqnarray}}
\def\eea{\end{eqnarray}}
\newcommand{\dd}{\mathrm{d}}
\newcommand{\LL}{\mathcal{L}}
\newcommand{\DD}{\mathcal{D}}

\title{Scalar QNM spectra of Kerr and Reissner-Nordstr\"om revealed by eigenvalue repulsions in Kerr-Newman}

\author{Alex Davey$^{a}$, \'Oscar J.~C.~Dias$^{a}$, Jorge E.~Santos$^{b}$}
\affiliation{$\,$\\$^{a}$STAG research centre and Mathematical Sciences, University of Southampton, Southampton SO17~1BJ, UK, $\,$\\$^{b}$DAMTP, Centre for Mathematical Sciences,
    University of Cambridge, Wilberforce Road, Cambridge CB3~0WA, United Kingdom}
\emailAdd{amd1g13@soton.ac.uk, ojcd1r13@soton.ac.uk, jss55@cam.ac.uk}

\abstract{Recent studies of the gravito-electromagnetic frequency spectra of Kerr-Newman (KN) black holes have revealed two families of quasinormal modes (QNMs), namely \emph{photon sphere} modes and \emph{near-horizon} modes. However, they can only be unambiguously distinguished in the Reissner-Nordstr\"om (RN) limit, due to a phenomenon called \emph{eigenvalue repulsion} (also known as  \emph{level repulsion},  \emph{avoided crossing} or the \emph{Wigner-Teller effect}), whereby the two families can interact strongly near extremality. We find that these features are also present in the QNM spectra of a scalar field in KN, where the perturbation modes are described by ODEs and thus easier to explore. Starting from the RN limit, we study how the scalar QNM spectra of KN dramatically changes as we vary the ratio of charge to angular momentum, all the way until the Kerr limit, while staying at a fixed distance from extremality. This scalar field case  clarifies the (so far puzzling) relationship between the QNM spectra of RN and Kerr black holes and the nature of the eigenvalue repulsions in KN, that ultimately settle the fate of the QNM spectra in Kerr. We study not just the slowest-decaying QNMs (both for $\ell=m=0$ and $\ell=m=2$), but several sub-dominant overtones as well, as these turn out to play a crucial role understanding the KN QNM spectra. We also give a new high-order WKB expansion of KN QNMs that typically describes the photon sphere modes beyond the eikonal limit, and use a matched asymptotic expansion to get a very good approximation of the near-horizon modes near extremality.}

\maketitle


\section{Introduction}

The Kerr-Newman (KN) metric~\cite{Adamo:2014baa,Newman:1965my} is the most general stationary, axisymmetric and asymptotically flat electro-vacuum solution of the Einstein-Maxwell equations\footnote{Alternatively, one can drop the axisymmetry condition, and consider instead real analytic black hole spacetimes.}~\cite{Robinson:2004zz,Chrusciel:2012jk,Chrusciel:2023onh}. Parameterized by mass $M$, charge $Q$ and angular momentum $J \equiv M a$, it encompasses the Schwarzschild ($a = Q = 0$)~\cite{schwarzschildGravitationalFieldMass1916}, Kerr ($Q = 0$)~\cite{kerrGravitationalFieldSpinning1963} and Reissner-Nordstr\"om ($a = 0$)~\cite{reissnerUberEigengravitationElektrischen1916,nordstromEnergyGravitationField1918} black holes as limiting cases. In those special cases the quasinormal mode (QNM) spectrum has been known for a long time~\cite{Regge:1957td,Zerilli:1974ai,Moncrief:1974am,Chandrasekhar:1975zza,Moncrief:1974gw,Moncrief:1974ng,Newman:1961qr,Geroch:1973am,Teukolsky:1972my,Detweiler:1980gk,Chandra:1983,Leaver:1985ax,Whiting:1988vc,Onozawa:1996ux,Glampedakis:2003dn,Berti:2003jh,Berti:2005eb,Yang:2012pj}, since the perturbation equations reduce to a system of two coupled ordinary differential equations~\cite{Regge:1957td,Zerilli:1974ai,Moncrief:1974am,Teukolsky:1972my}. However, while the theoretical basis for a computation of the QNM spectrum of Kerr-Newman was set out in the 1980s, when Chandrasekhar reduced the equations governing gravito-electromagnetic perturbations to a system of two coupled partial differential equations~\cite{Chandra:1983} (see also \cite{Dias:2015wqa,Dias:2022oqm} for a gauge invariant derivation of this system), a numerical computation of the QNMs remained an open problem for several decades, due to a lack of further separability.

The gravito-electromagnetic QNM spectrum of Kerr-Newman was finally computed in~\cite{Dias:2015wqa}, but it focused the attention in the near-extremal part of the parameter space most relevant to search for linear mode instabilities (which are not present~\cite{Dias:2015wqa})\footnote{Recent advances towards a proof of stability of the full linear problem, with the assumption of linear mode stability, have been made in \cite{Giorgi:2020ujd}.}. Recently, efficiency improvements have made a full KN parameter space search feasible, and these results have been used to construct templates that model existing gravitational wave data, to constrain the range of remnant charge in a binary merger~\cite{Dias:2021yju,Carullo:2021oxn,Dias:2022oqm}. In these later studies of KN~\cite{Dias:2021yju,Dias:2022oqm}, a surprising phenomenon $-$ called \emph{eigenvalue repulsion} $-$ was observed. This phenomenon is common in some eigenvalue problems of quantum mechanical systems where it is also known as  \emph{level repulsion},  \emph{avoided crossing} or \emph{Wigner-Teller effect} \cite{Landau1981Quantum,Cohen-Tannoudji:1977}. 
Typically, two different QNM families of a black hole can have eigenfrequencies that may simply cross in the real or imaginary plane (but not in both), but they do not interact in any way. However, in KN an intricate interaction between the gravito-electromagnetic $\ell=m=2$ modes was observed, where in certain parts of the parameter space the frequencies of two QNM families approach in the complex plane very closely, without crossing (i.e. without matching in frequency), before repelling violently and moving apart again. These repulsions are very strongly dependent on the black hole parameters $(M, a, Q)$ $-$ a relatively minor change of the black hole parameters can cause the repulsion to be absent $-$ and hence are crucial to understanding the structure of the QNM spectrum in KN and how the latter bridges the Reissner-Nordst\"om and Kerr cases to solve some puzzling properties of the QNM spectra of the latter two. This full understanding will be completed only after the present study since scalar modes behave qualitatively similarly to the gravito-electromagnetic modes but are much easier to explore. Eigenvalue repulsions have also recently been observed in charged~\cite{Dias:2020ncd} and rotating de Sitter black holes in higher dimensions~\cite{Davey:2022vyx}, but not in studies of (four-dimensional) Schwarzschild, RN or Kerr~\cite{Regge:1957td,Zerilli:1974ai,Moncrief:1974am,Chandrasekhar:1975zza,Moncrief:1974gw,Moncrief:1974ng,Newman:1961qr,Geroch:1973am,Teukolsky:1972my,Detweiler:1980gk,Chandra:1983,Leaver:1985ax,Whiting:1988vc,Onozawa:1996ux,Berti:2003jh,Berti:2005eb,Yang:2012pj}. In section~\ref{sec:EigenvalueRepulsionsA}, we review a first-principles argument that explains why eigenvalue repulsions have a better chance of occurring in black hole families with two or more dimensionless parameters~\cite{Dias:2022oqm}.

Despite recent technical advancements, the computation of gravito-electromagnetic perturbations remains numerically costly. Although gravito-electromagnetic perturbations on KN do not separate, scalar (and Dirac) perturbations are separable, and reduce to a pair of ODEs for the radial and angular components. Thus, in this manuscript, we study the scalar QNM spectra for the full parameter space of KN in fine detail to better understand the phenomena of eigenvalue repulsions. Typically, it is the dominant (\ie slowest decaying) quasinormal mode that is of interest, however we take advantage of the reduced numerical complexity to also compute several sub-dominant radial overtones (often denoted by integer $n\geq 0$). This is necessary for a complete understanding of the QNM spectra, as one consequence of eigenvalue repulsions is that modes trade dominance in a highly non-trivial way as we move around the parameter space near extremality.
We focus our study on scalar QNMs with harmonic numbers $m=\ell=2$ because these the dominant modes in the gravito-electromagnetic sector and we want to use the scalar field as a proxy for the latter case, but we also study the ground state scalar QNMs with $m=\ell=0$ because they are the dominant scalar modes, and provide an example where eigenvalue repulsions are {\it not} present (here $\ell$ is the wave quantum number that is related to the number of nodes in the angular eigenfunction along the polar direction and $m$ is the azimuthal quantum number).

To provide context for this study, we first describe the two families of QNMs that are present in Reissner-Nordstr\"om, namely the \emph{photon sphere} (PS) modes and \emph{near-horizon} (NH) modes. The photon sphere modes (also denoted as  {\it damped modes} in \cite{Yang:2012pj,Yang:2013uba,Zimmerman:2015trm}) are typically  described by the well-known eikonal approximation of the quasinormal mode spectrum, where the angular momentum quantum numbers $m$ and $\ell$ are taken to be large and we thus have a null particle limit \cite{Goebel:1972,Ferrari:1984zz,Ferrari:1984ozr,Mashhoon:1985cya,Schutz:1985km,Bombelli:1991eg,Cornish:2003ig,Cardoso:2008bp,Dolan:2010wr,Yang:2012he,Zimmerman:2015trm,Dias:2022oqm}. It provides a geometric interpretation, first presented by Goebel \cite{Goebel:1972} and Ferrari and Mashhoon~\cite{Ferrari:1984zz,Mashhoon:1985cya}, in terms of the dynamics of null geodesics in the equatorial plane: the real part is proportional to the Keplerian velocity of the photon orbit while the imaginary part is proportional to the Lyapunov exponent, which characterises the instability timescale of the null geodesic. However, the eikonal approximation is strictly valid in the limit $|m| = \ell \to \infty$. Thus, it coincides with a leading order WKB analysis of the QNM problem, as first discussed by Schutz and Will \cite{Schutz:1985km} and completed for Schwarzschild, RN and Kerr in
\cite{Iyer:1986vv,Iyer:1986nq,Kokkotas:1988fm,Seidel:1989bp}. In section~\ref{sec:AnalyticalPS-NH} we perform a WKB expansion that extends the WKB approximation of Schutz and Will beyond the eikonal result, by a further three orders in $\mathcal{O}(m)$, significantly improving the accuracy for the small values of $m$ and $\ell$ that one typically considers. To the best of our knowledge, this extension has never been done for the QNMs of KN although it is a standard higher-order WKB analysis first discussed in the context of QNMs by Will and Guinn \cite{Will:1988zz} and reduces to the higher-order WKB results of \cite{Iyer:1986vv,Iyer:1986nq,Kokkotas:1988fm,Seidel:1989bp} in the Schwarzschild, RN and Kerr limits. 

The second family of QNMs are known as \emph{near-horizon} modes (a.k.a \emph{near-extremal} or \emph{zero-damped} modes)~\cite{Teukolsky:1974yv,Detweiler:1980gk,Sasaki:1989ca,Andersson:1999wj,Glampedakis:2001js,Hod:2008zz,Yang:2012pj,Yang:2013uba,Hod:2014uqa,Zimmerman:2015trm,Hod:2015xlh,Dias:2021yju,Dias:2022oqm}. They are characterised by a frequency with a vanishing imaginary part in the extremal limit (with the real part saturating the superradiant bound exactly at extremality), and a wavefunction that is very localised near the event horizon (at least near-extremality), and are related to the near-horizon geometry of the (extremal) black hole. We capture these modes by performing a matched asymptotic expansion (MAE) that is similar to the one performed in \cite{Teukolsky:1974yv,Detweiler:1980gk,Yang:2012pj,Yang:2013uba,Zimmerman:2015trm}. 
We first solve for the eigenfunction near the horizon, then for the eigenfunction far from the horizon, before matching the two in the overlap region were both eigenfunctions overlap. As pointed out in \cite{Zimmerman:2015trm},
the RN PS modes (a.k.a. damped modes in \cite{Zimmerman:2015trm}) are very well known in the literature, starting with the WKB analysis of \cite{Kokkotas:1988fm}. However, the existence of the RN NH modes (a.k.a. zero-damped modes in \cite{Zimmerman:2015trm}) seem to have been missed till the work of \cite{Zimmerman:2015trm} in spite of the seminal work of Teukolsky and Press \cite{Teukolsky:1974yv} already suggesting that such a family might or should be present in any black hole with an extremal configuration. To the best of our knowledge, the scalar field NH QNMs of RN are first computed exactly (within numerical accuracy) in the present manuscript (see Fig.~\ref{Fig:RN}); the gravito-electromagnetic NH QNMs of RN were computed in \cite{Zimmerman:2015trm}.

Given a QNM in any part of the parameter space of RN, we {\it can} uniquely classify it (but would not need to) as either a photon sphere mode or a near-horizon mode by tracing it to the extremal limit, where it agrees with either the WKB or MAE approximation of the PS or NH modes of RN. As we add angular momentum to the system (while simultaneously decreasing the charge, so that the black hole remains subextremal) and move along the KN space to approach the Kerr configuration, the situation is more complicated. Depending on the harmonic wave numbers $m$ and $\ell$, the two QNM families may interact in some part of the parameter space. Namely, the curves describing the imaginary part of the two frequencies can merge and bifurcate again (but in a different way) to form QNM families that can no longer be clearly identified as PS or NH modes, since they are often well approximated by both the WKB and the near-horizon matched asymptotic expansion (possibly with a higher radial overtone). Whether this is the case or not depends on the angular momentum quantum numbers $m$ and $\ell$ of the perturbation. The precise set of values of $m$ and $\ell$ for which the NH modes are an independent family of the PS modes have been partially studied previously in~\cite{Yang:2013uba,Yang:2012he,Yang:2012pj}. To complement this analysis and get a deeper understanding of this transition, we will do a first-principles analysis (first identified in \cite{Davey:2022vyx}) that finds that the boundary between the two behaviours can be approximately determined by finding the case where modes start violating the effective AdS$_2$ Breitenl\"ohner-Freedman mass bound that characterizes scalar perturbations of the near-horizon geometry (which is the product of AdS$_2$ times a compact space) of the (1-parameter) extremal KN geometry (here, AdS$_2$ stands for 2-dimensional Anti-de Sitter spacetime). We then focus our detailed discussion on two important representatives cases: $m = \ell = 2$, for which the PS and NH modes get entangled and lose their original independence (that they had in the RN limit), and $m = \ell = 0$ for which both families remain independent and clearly distinguishable as we span the KN parameter space. The $m = \ell = 2$ case (as a representative element of its class) is particularly interesting since the two families of QNM that exist in the RN limit, unlike in the $\ell=m=0$ case, become a {\it single} one in the Kerr limit (that we can denote as a combined PS-NH family, with its radial overtones); see Fig.~\ref{Fig:Kerr}. Thus the KN spectra and its eigenvalue repulsions will help us  understand a long puzzling fact.  Namely, for example for $\ell=m=2$, how can it be that we start with {\it two distinct} QNM families of damped and zero-damped modes (and their tower of overtones) in RN (Fig.~\ref{Fig:RN}) and end up with a single QNM family of zero damped modes (and its tower of overtones) in Kerr (Fig.~\ref{Fig:Kerr})? (This is to be contrasted, e.g. with the $m=\ell=0$ case where we start with two QNM families in RN and end up with the same two QNM families in Kerr; see Fig.~\ref{Fig:NH-PS-full-m0KerrRN}.)

In more detail, the aim of this manuscript is to connect the RN and Kerr QNM spectra explicitly, while navigating through the eigenvalue repulsions that settle in the pathway. Starting with the two QNM families in RN, we track how they change as we increase the rotation and/or vary the charge, until we reach either the Kerr limit or the extremal limit of Kerr-Newman. While the PS and NH modes are unambiguously discernible in the RN limit, as stated above, these distinctions are often blurred by eigenvalue repulsions as we turn on angular momentum. We will study the KN QNM spectra with a focus on the $m = \ell = 2$ modes which are most closely analogous to the gravito-electromagnetic modes in~\cite{Dias:2021yju,Dias:2022oqm} (most likely because $m=\ell=2$ then equals the spin of gravitational perturbations) but also the $m=\ell=0$ modes since these are the dominant spin-0 modes. Overall, our study is complementary to \cite{Yang:2012pj,Yang:2013uba,Zimmerman:2015trm} and it offers a fresh perspective of the RN/KN/Kerr QNM spectra, identifies and studies the features of eigenvalue repulsions and helps understanding the Kerr QNM spectra when coming from the RN QNM spectra. 

The plan of the manuscript is as follows. In the next section we introduce a novel ``polar parameterisation'' of the Kerr-Newman parameter space, which has the advantage of allowing us to smoothly transition between the RN and Kerr limits while always keeping at the same `distance' from extremality. We also formulate the scalar QNM eigenvalue problem including its boundary conditions. In section~\ref{sec:AnalyticalPS-NH}, we first review the well-known eikonal limit of the QNM problem in Kerr-Newman (section~\ref{sec:PSeikonal}), which is universal to any perturbation spin and defines the photon sphere family of modes. We then proceed beyond this leading WKB order and derive a high-order WKB expansion of the scalar field QNM spectrum (section~\ref{sec:PSwkbHighOrders}), which is novel and necessarily more accurate for finite $m$ (in a wide neighbourhood about the Reissner-Nordstr\"om and Kerr solutions) than the eikonal approximation. We also perform a near-horizon matched asymptotic expansion that identifies the near-horizon family of modes (section~\ref{sec:NHanalytics}). In section~\ref{sec:EigenvalueRepulsions}, after reviewing a first-principles analysis of the phenomenon of eigenvalue repulsion (section~\ref{sec:EigenvalueRepulsionsA}), we compute the exact $m = \ell = 2$ spectra of KN QNMs (using numerical methods), focusing on the near-extremal region where eigenvalue repulsions often blur the distinction between the photon sphere and near-horizon QNM families (section~\ref{sec:EigenvalueRepulsionsB}). We then discuss the full KN spectrum, for both $m = \ell = 2$ and $m = \ell = 0$ modes, in section~\ref{sec:QNMspectra}.

\section{Klein-Gordon equation in the Kerr-Newman background}\label{sec:Pert}

\subsection{KN black hole and its polar parameterization}\label{sec:coupled}

The uniqueness theorems \cite{Robinson:2004zz,Chrusciel:2012jk} state that the Kerr-Newman (KN) black hole (BH) is the unique, most general family of stationary, axisymmetric and asymptotically flat BHs of Einstein-Maxwell theory. It is characterised by 3 dimensionful parameters: mass $M$, angular momentum $J\equiv M a$ and charge $Q$. The Kerr, Reissner-Nordstr\"om (RN) and Schwarzschild BHs constitute limiting cases: ${Q=0}$, $a=0$ and $Q=a=0$, respectively.

The KN BH solution is most commonly expressed in standard Boyer-Lindquist coordinates $\{t,r,\theta,\phi\}$ (time, radial, polar, azimuthal coordinates)~\cite{Adamo:2014baa}, in which the metric and Maxwell potential take the form
\begin{eqnarray}\label{KNsoln}
{\rm d}s^2&=&-\frac{\Delta}{\Sigma} \left(\dd t-a \sin^2\theta \dd \phi  \right)^2+\frac{\Sigma }{\Delta }\,\dd r^2 + \Sigma \,\dd \theta^2 
+ \frac{\sin ^2\theta}{\Sigma }\left[\left(r^2+a^2\right)\dd \phi -a \dd t \right]^2, \nonumber\\
A&=& \frac{Q \,r}{\Sigma}\left(\dd t-a \sin^2\theta \dd \phi \right),
\end{eqnarray}
with $\Delta = r^2 -2Mr+a^2+Q^2$ and $\Sigma=r^2+a^2 \cos^2\theta$. We will find it convenient to work with the angular coordinate $x=\cos\theta$ with range $x\in[-1,1]$.

The roots of the function $\Delta(r)$ correspond to the inner $(r_-)$ and outer $(r_+)$ horizons, $r_- \le r_+$. We can solve $\Delta(r_+)=0$ with respect to $M$ to find that
\begin{equation}\label{Mr+}
M=\frac{r_+^2+a^2+Q^2}{2 r_+}.
\end{equation}
Moreover, the system has a scaling symmetry that allows one to write all physical quantities in units of $r_+ $ (or $M$).\footnote{The scaling symmetry is 
$\{t,r,\theta,\phi\}\to \{\lambda t,\lambda r,\theta,\phi\}$ and $\{r_+,a,Q\}\to \left\{\lambda \, r_+, \lambda\, a, \lambda Q \right\}$ which rescales the metric and Maxwell potential as $g_{ab}\to \lambda^2 \, g_{ab}$ and $A_{a}\to \lambda \, A_{a}$ but  leaves the equations of motion invariant (since the affine connection $\Gamma^c_{\phantom{c}ab}$, and the Riemann ($R^a_{\phantom{a}bcd}$), Ricci ($R_{ab}$) and energy-momentum ($T_{ab}$) tensors are left invariant).} Thus the KN black hole is effectively a 2-parameter family of solutions that we can parametrize by the dimensionless quantities
\begin{equation}
\alpha=\frac{a}{r_+},  \qquad\qquad \tilde{Q}=\frac{Q}{r_+}.
\end{equation}
The outer event horizon ($r=r_+$) is a Killing horizon generated by the Killing vector
$ K=\partial_t +\Omega_H \partial_\phi$,
with angular velocity $\Omega_H$ and temperature $T_H$ given by  
\begin{eqnarray}\label{KN:thermo}
&&\hspace{-0.4cm} \tilde{\Omega}_H \equiv \Omega_H r_+= \frac{\alpha}{1+\alpha^2} \,, \qquad 
 \tilde{T}_H \equiv T_H r_+=\frac{1}{4 \pi}\frac{1-\alpha^2-\tilde{Q}^2}{1+\alpha^2}. 
\end{eqnarray}
If $r_-=r_+$, \ie  $\alpha=\alpha_{\hbox{\footnotesize ext}}=\sqrt{1-\tilde{Q}^2}$, the KN BH has a regular extremal (``ext") configuration with $T_H^{\hbox{\footnotesize ext}} =0$, and maximum angular velocity $\tilde{\Omega}_H^{\hbox{\footnotesize ext}} =\alpha_{\hbox{\footnotesize ext}}/(1+\alpha_{\hbox{\footnotesize ext}}^2)$.

Finally, for our purposes, we will find it very enlightening to parametrize the KN black hole by ``polar" parameters. For that, we first define the parameter $\mathcal{R}(\sigma)$ as:
\begin{equation}\label{def:sigma}
 \sigma=1-\frac{r_-}{r_+}=1-\alpha^2-\tilde{Q}^2\,,\qquad \mathcal{R}=\sqrt{1-\sigma}\,,
\end{equation}  
and we then introduce the polar parametrization
\begin{equation}\label{PolarParametrization}
\alpha= \mathcal{R}\, \sin \Theta\,, \qquad \tilde{Q}=\mathcal{R}\, \cos \Theta\,.
\end{equation}
This parametrization $(\mathcal{R},\Theta)$ has the property that $\mathcal{R}$ is an off-extremality radial measure (since it vanishes in the Schwarzschild limit and attains its maximum value of $\mathcal{R}=1$ at extremality), while the polar parameter $\Theta$ ranges between the Reissner-Nordstr\"om solution (where $\Theta=0$ and thus $\alpha=0$) and the Kerr solution (where $\Theta=\pi/2$ and thus $\tilde{Q}=0$). So, with this parametrization we will be able to follow a family of KN black holes that starts at the Reissner-Nordstr\"om solution ($\Theta=0$) and evolves in $\Theta$ towards the Kerr solution ($\Theta=\pi/2$) while staying always at fixed distance from extremality (\ie at fixed $\mathcal{R}$).  

\subsection{Klein-Gordon equation and boundary conditions of the problem}\label{sec:BCs}

We are interested in studying massless scalar field perturbations in the KN background which are described by the Klein-Gordon equation $\Box \, \Phi=0$. Since $\partial_t$ and $\partial_\phi$ are Killing vectors of the KN background we can perform a Fourier decomposition of the modes along these directions, which introduces the frequency $\omega$ and azimuthal quantum number $m\in\mathbb{Z}$. Moreover, we can look into perturbations that admit a separation {\it ansatz}. Altogether, we look for scalar perturbations of the form
 \begin{equation}
 \Phi=e^{-i \omega t}e^{i m \phi} R(r)S(x)\,.
\end{equation}
In this case the Klein-Gordon equation separates into a set of radial and angular ODEs:
\begin{subequations}
\begin{align}
&\frac{d}{dr}\left( \Delta\, \frac{d R}{dr} \right)+\left[\frac{\left[\left(r^2+a^2\right)\omega -m\,a \right]^2}{\Delta }  + 2m a \omega -a^2\omega^2-\lambda \right] R=0, \label{KG:radial}\\
& \frac{d}{dx}\left( (1-x^2)\, \frac{d S}{dx} \right)+\left( -a^2\omega^2(1- x^2) -\frac{m^2}{1-x^2}+a^2\omega^2+\lambda \right) S =0, \label{KG:ang}
\end{align}
\end{subequations}
where $\lambda$ is the separation constant of the problem.

The angular equation \eqref{KG:ang} is a standard oblate spheroidal harmonic equation, namely
$\left[\left(1-x^2\right) S'\right]'+\left[ \gamma ^2 \left(1-x^2\right)-\frac{m^2}{1-x^2}+\Lambda \right]S=0$ with $\gamma\equiv i\,a\,\omega$ and $\Lambda\equiv \lambda+a^2\omega^2$,  whose regular solutions are given by the oblate spheroidal harmonics $S=S_\ell^m(\gamma;x)$. Here,  $\ell$ is a non-negative integer that essentially gives the number of zeros of the eigenfunction along the polar angle and  regularity at the North and South poles ($x=\pm 1$) requires that $m$ is an integer that obeys the constraint $|m|\leq \ell$. 

Consider now the radial equation \eqref{KG:radial}.
To have a well-posed boundary value problem we must supplement this ODE with appropriate (physical) boundary conditions. 
At spatial infinity we require only outgoing waves, and at the future event horizon we only keep modes that are regular in ingoing Eddington-Finkelstein coordinates.

In detail, recall that $\omega$ and $m$ are the frequency and azimuthal quantum number of the linear mode perturbations, respectively. The $t -\phi$ symmetry of the KN BH allows us to consider only modes with  Re$(\omega)\geq 0$, as long as we study both signs of $m$.
Then, to solve the coupled ODEs~\eqref{KG:radial}-\eqref{KG:ang}, we need to impose physical boundary conditions (BCs).
At spatial infinity, a Frobenius analysis of \eqref{KG:radial} yields the two independent asymptotic solutions:
\begin{equation}\label{BC:inf}
R {\bigl |}_\infty \simeq A_{out} e^{i \,\omega \,r} \left(\frac{r_+}{r}\right)^{1-i\,\frac{ \omega}{r_+}\left(r_+^2+a^2+Q^2\right)}\left(1+\cdots\right)
+A_{in} e^{-i \,\omega \,r} \left(\frac{r_+}{r}\right)^{1+i\,\frac{ \omega}{r_+}\left(r_+^2+a^2+Q^2\right)}\left(1+\cdots\right),
\end{equation}
where $A_{out}$ and $A_{in}$ are two arbitrary amplitudes.
For the QNM problem we require that only outgoing waves are permitted and thus we set the boundary condition $A_{in} \equiv 0$.

At the event horizon, a Frobenius analysis yields the expansion 
\begin{equation}\label{BC:H}
R{\bigl |}_H \!\simeq\! B_{in}\left(r-r_+\right)^{-i\,\frac{\omega -m\Omega_H}{4 \pi T_H}}\left(1+\cdots\right)
+B_{out}\left(r-r_+\right)^{i\,\frac{\omega -m\Omega_H}{4 \pi T_H}}\left(1+\cdots\right), \nonumber
\end{equation}
where $B_{in}$ and $B_{out}$ are two arbitrary amplitudes, and $\Omega_H, T_H$ are defined in \eqref{KN:thermo}. We want to keep only the solution that is regular in ingoing Eddington-Finkelstein coordinates, \ie that excludes outgoing waves across the future event horizon, and this requires that we set $B_{out}\equiv 0$.\footnote{The ingoing Eddington-Finkelstein coordinates $\{v,r,x,\widetilde{\phi} \}$, which extend the solution through the horizon, are defined via
$t=v-\int \frac{r^2+a^2}{\Delta}\,\mathrm{d}r\,,\: \phi=\widetilde{\phi}-\int \frac{a}{\Delta}\,\mathrm{d}r$.
\label{footnote:ingoingEF}} These two boundary conditions can be automatically imposed if we redefine the radial function as 
\begin{equation}\label{eigenfunction_redef}
R= e^{i \,\omega \,r} \left(\frac{r_+}{r}\right)^{1-i\,\frac{ \omega}{r_+}\left(r_+^2+a^2+Q^2\right)} \left(1-\frac{r_+}{r}\right)^{-i\,\frac{\omega -m\Omega_H}{4 \pi T_H}} \chi(r)
\end{equation}
and search for eigenfunctions $\chi(r)$ that are smooth everywhere in the outer domain of communications. 

Similarly, we impose regularity on the angular eigenfunction solutions to~\eqref{KG:ang} at the North and South poles $(x = \pm 1)$ by the redefinition
\begin{equation}\label{eigenfunction_redef_angular}
  S = (1-x^{2})^{\frac{|m|}{2}} Y(x)
\end{equation}
and then solve for smooth eigenfunctions $Y(x)$. After inserting the field redefinitions~\eqref{eigenfunction_redef} and~\eqref{eigenfunction_redef_angular} into~\eqref{KG:radial}-\eqref{KG:ang}, we have a coupled eigenvalue problem that is quadratic in $\omega$ and linear in $\lambda$. To solve this numerically, we first discretise the system using pseudospectral collocation methods. Then, starting with an appropriate seed, we use a Newton-Raphson algorithm to march QNMs from one part of the parameter space to another. See ~\cite{Dias:2015nua} for a review and ~\cite{Dias:2010gk,Dias:2014eua,Dias:2011jg,Dias:2009iu,Dias:2010eu,Dias:2010ma,Dias:2011tj,Cardoso:2013pza,Dias:2018etb} for details and examples of this method. Our numerical results are accurate to at least the eighth decimal place.

\section{Two families of QNM: photon sphere and near-horizon modes} \label{sec:AnalyticalPS-NH}

In subsection~\ref{sec:PSeikonal}, we review the well-known eikonal limit of the QNM problem in Kerr-Newman, which is independent of the perturbation spin and defines the photon sphere family of modes. Then, in subsection~\ref{sec:PSwkbHighOrders} we go beyond this leading WKB order and derive a high-order WKB expansion of the scalar field QNM spectrum, which is novel and necessarily more accurate for finite $m$ (in a wide neighbourhood about the RN and Kerr solutions) than the eikonal approximation. Finally, in subsection~\ref{sec:NHanalytics}, we introduce a near-horizon matched asymptotic expansion that identifies the near-horizon family of modes.
\subsection{WKB expansion of photon sphere modes} \label{sec:PSwkb}

\subsubsection{Photon sphere modes in the eikonal limit (the leading WKB result)} \label{sec:PSeikonal}

In the eikonal or geometric optics limit, whereby we consider the WKB limit $\ell\sim |m| \gg 1$, there are QNM  frequencies $-$ known as ``photon sphere"(PS) QNMs $-$  that are closely related to the properties of the unstable circular photon orbits in the equatorial plane of the KN black hole. Namely, the real part of the PS frequency is proportional to the Keplerian frequency $\Omega_c$ of the circular null orbit and the imaginary part of the PS frequency is proportional to the Lyapunov exponent $\lambda_L$ of the orbit \cite{Goebel:1972,Ferrari:1984zz,Ferrari:1984ozr,Mashhoon:1985cya,Bombelli:1991eg,Cornish:2003ig,Cardoso:2008bp,Dolan:2010wr,Yang:2012he,Zimmerman:2015trm,Dias:2022oqm}. The latter describes how quickly a null geodesic congruence on the unstable circular orbit increases its cross-section under infinitesimal radial deformations. 

We will study the eikonal limit of modes with  $\ell=m$ or $\ell=-m$ and we denote the associated frequency by $\omega^{\hbox{\tiny eikn}}_{\hbox{\tiny PS}}$.  The final analytical formula for these frequencies is strictly valid in the WKB limit $\ell= |m| \to \infty$. That is, it only captures the leading behaviour of a WKB expansion in $1/m$ with $\ell=|m|\to \infty$. This analysis is independent of the spin of the perturbation and was already performed previously in several references in the literature, either for the KN background or its limiting solutions (Kerr, Reissner-Nordstr\"om or Schwarzschild), see \eg \cite{Goebel:1972,Ferrari:1984zz,Ferrari:1984ozr,Mashhoon:1985cya,Bombelli:1991eg,Cornish:2003ig,Cardoso:2008bp,Dolan:2010wr,Yang:2012he,Zimmerman:2015trm,Dias:2022oqm}. We review it here because we want to compare our numerical results with the eikonal limit (to identify the nature of some of the modes), and more importantly, in the next subsection we will extend this WKB expansion to higher-orders in a $1/m$ expansion, so it is good to have a self-contained analysis of the leading term at hand.

The geodesic equation that describes the motion of pointlike particles around a KN BH leads to a set of quadratures. This may be an unexpected result given that KN only possesses two Killing fields  $K=\partial/\partial_t$ and $\xi=\partial/\partial_\phi$, seemingly one short of leading to an integrable system. There is however another conserved quantity $-$ the Carter constant $-$ associated to a Killing tensor $K_{ab}$, which rescues the day \cite{Chandra:1983}.

The Hamilton-Jacobi equation \cite{Chandra:1983} provides a quick way to identify the integrable structure of the system:
\be\label{HJeq}
\frac{\partial S}{\partial x^\mu}\frac{\partial S}{\partial x^\nu}g^{\mu\nu}=0\,,
\ee
with $S$ being denoted as the principal function.  The motion of null particles is obtained noting that, according to Hamilton-Jacobi's theory, the particle momenta can be obtained from the principal function as
\be \label{HJ:defP}
\frac{\partial S}{\partial x^\mu}\equiv p_\mu\quad \text{and}\quad p^\mu =\frac{\mathrm{d}x^\mu}{\mathrm{d}\tau}\,,
\ee
where $\tau$ is an affine parameter. To proceed, we take a separation \emph{ansatz} of the form (using $x=\cos\theta$ where $\theta$ is the polar angle)
\be \label{HJ:ansatz}
S=-e\,t+j\,\phi+R(r)+X(x)\,,
\ee
with the constants $e$ and $j$ being the conserved charges associated with the Killing fields $K$ and $\xi$ \footnote{For massive particles, these coincide with the energy and angular momentum of the particle, but for massless particles $e$ and $j$ have no physical meaning since they can be rescaled. The ratio $j/e$, however, is invariant under such rescalings.} through
\be
e\equiv - K_\mu \dot{x}^\mu\qquad \text{and}\qquad j\equiv \xi_\mu \dot{x}^\mu\,,
\label{eq:conserved}
\ee
where the dot ( $\dot{}$ ) describes the derivative w.r.t. the affine parameter $\tau$. With \eqref{HJ:ansatz}, the Hamilton-Jacobi equation \eqref{HJeq} for null geodesics yields coupled ODEs for $R(r)$ and $X(x)$ (the prime $(^{\prime})$ describes a derivative w.r.t. the argument, $r$ or $x$, respectively)
\begin{gather} \label{HJ:Rad}
\Delta ^2 R'^2 -\left[e \left(r^2+a^2\right)-a j\right]^2+\Delta  \left[\mathcal{Q}+(j-a e)^2\right]=0\,,
\\ \label{HJ:Ang}
X'^2-\frac{(j-a e)^2+\mathcal{Q}}{1-x^2 }+\frac{\left[ a e \left(1-x^2\right)-j\right]^2}{ \left(1-x^2\right)^2}=0\,,
\end{gather}
where $\mathcal{Q}$ is Carter's separation constant. Additionally, from \eqref{HJ:defP}, \ie $\dot{x}^{\mu}=g^{\mu \nu }\frac{\partial S}{\partial x^{\mu }}$, one has
\begin{eqnarray} \label{HJ:tphi}
&& \hspace{-0.4cm}\dot{t}=\frac{\left(r^2+a^2\right) \left[e \left(r^2+a^2\right)-a j \right]+a \Delta  \left[ j-a e \left(1-x^2\right)\right]}{\Delta  \left(r^2+a^2 x^2\right)},
\nonumber \\
&& \hspace{-0.4cm} \dot{\phi }=\frac{\left(1-x^2\right) a\left[e \left(r^2+a^2\right)-a j\right]+\Delta  \left[j-a e \left(1-x^2\right)\right]}{\Delta  \left(1-x^2\right) \left(r^2+a^2 x^2\right)}.
\end{eqnarray}

We want null geodesics whose behaviour matches that of large $\ell=|m|$ QNMs, \ie geodesics confined to the equatorial plane $x=0$. It follows from \eqref{HJ:Ang} that such geodesics exist only if at $\tau=0$ one has $X(0)=\dot{X}(0)=0$ and $\mathcal{Q}=0$. Introducing the geodesic impact parameter  
\be
b\equiv\frac{j}{e},
\ee
the equation \eqref{HJ:Rad} governing the radial motion can be rewritten as
\be
\label{HJ:geodesic}
\dot{r}^2=V(r;b)\,,
\ee
where the potential is
\be \label{HJ:pot}
V(r;b)=\frac{j^2}{b^2}\left(1+\frac{a^2-b^2}{r^2}+\frac{2 M (b-a)^2}{r^3}-\frac{Q^2 (b-a)^2}{r^4}\right).
\ee

\begin{figure}[t]
\centering
\includegraphics[width=.5\textwidth]{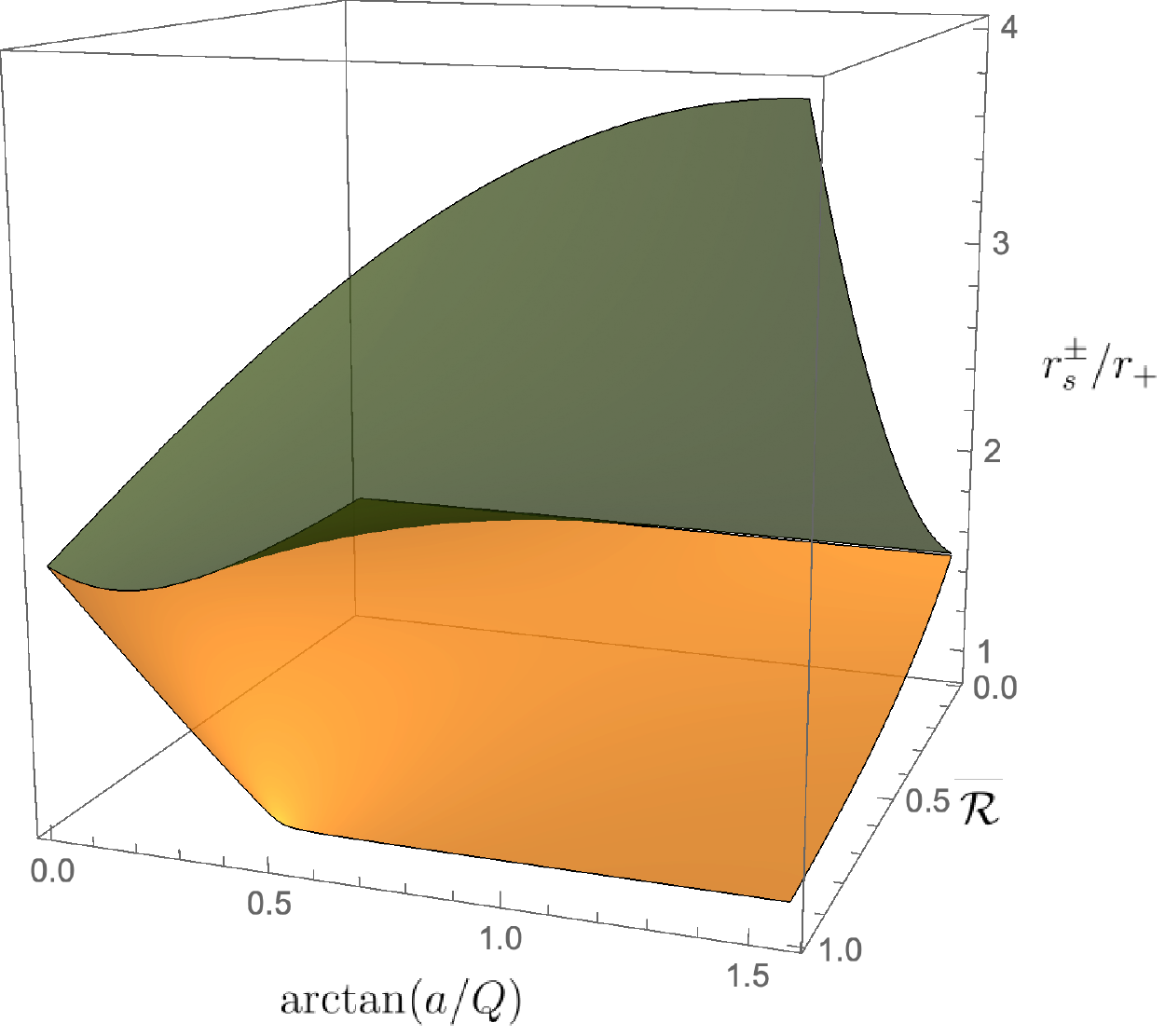}
\caption{The radii $r_s^{\pm}$ (with $r_s^{+}\geq r_s^{-}\geq r_+$) of the two unstable circular orbits in the equatorial plane of the KN black hole that ultimately yield the co-rotating $m=\ell$ (in the $r_s^{-}$ case) and the counter-rotating $m=-\ell$  (in the $r_s^{+}$ case) PS QNM frequencies in the eikonal limit.}
\label{Fig:PSradii}
\end{figure}

We now want to find the photon sphere (the region where null particles are trapped on unstable circular orbits), \ie the values of $r=r_s$ and $b=b_s$, such that
\be\label{HJ:CircOrb}
V(r_s;b_s)=0\quad\text{and}\quad \left.\partial_r V(r;b)\right|_{r=r_s,b=b_s}=0.
\ee
The first equation allows to find
\be\label{PS:bs}
b_s(r_s)=\frac{r_s^2 \sqrt{\Delta(r_s)}+a \left(Q^2-2 M r_s\right)}{r_s^2-2 M r_s+Q^2}\,,
\ee
which is then inserted in the second equation of  \eqref{HJ:CircOrb} to get (after algebraic manipulations) a fourth order polynomial equation for $r_s$:
\be \label{PS:4thPolyn}
4\left[r_s^2+2 a \left(\sqrt{\Delta(r_s)}+a\right)\right]^2-\left(3 M r_s+\sqrt{9 M^2 r_s^2-8 Q^2\left[r_s^2+2 a \left(\sqrt{\Delta(r_s)}+a\right)\right]}\right)^2=0\,,
\ee
where $\Delta(r)$ is defined  below \eqref{KNsoln} and we are interested in solutions with $r_s>r_+$.
Alternatively, \eqref{HJ:CircOrb} can be solved to get the black hole parameters $M$ and $Q$ that have circular orbits with radius $r_s$ and impact parameter $b_s$, namely
\be\label{PS:MQrsbs}
M=\frac{r_s \left(b_s^2-a^2-2 r_s^2\right)}{\left(b_s-a\right)^2}\,, \qquad Q=\frac{r_s\sqrt{b_s^2-a^2-3 r_s^2}}{\sqrt{\left(b_s-a\right){}^2}}.
\ee
This system has two real roots $r_s$ larger than $r_+$, in correspondence with the two PS modes: the {\it co-rotating} one (with $m=\ell$) which is in correspondence with the eikonal orbit with radius $r_s=r_s^-$ and $b_s>0$ (and that has the lowest $|\mathrm{Im}\,\tilde{\omega}|$, as we will see) and the {\it counter-rotating} mode with $m=-\ell$  that maps to  the orbit with radius $r_s=r_s^+$ and $b_s<0$, with $r_s^+\geq r_s^- \geq r_+$.  The two real roots $r_s^{\pm}$ larger than $r_+$ are displayed in
Fig.~\ref{Fig:PSradii}.
In the RN $\Theta=\arctan{(a/Q)}=0$ or Schwarzschild $\mathcal{R}=0$ (\ie $a=0=Q$) limits, one has $r_s^{+}=r_s^{-}$, and at extremality ($\mathcal{R}=1$, \ie $\alpha=\alpha_{\rm ext}=\sqrt{1-\tilde{Q}^2})$ the co-rotating orbit radius equals the event horizon radius, $r_s^{-}=r_+$, when $\Theta \ge \pi/6 \simeq 0.52$.

Finally, we can compute the orbital angular velocity (a.k.a. Kepler frequency) of the null circular photon orbit, which is simply given by
\be \label{PS:Omega_c}
\Omega_c \equiv \frac{\dot{\phi}}{\dot{t}}=\frac{1}{b_s}\,,
\ee
where we used \eqref{HJ:tphi} evaluated at $r=r_s$ and $b=b_s$.
Moreover, we can also compute the largest Lyapunov exponent $\lambda_L$, measured in units of $t$, associated with infinitesimal fluctuations around photon orbits with $r(\tau)=r_s$. This can be  obtained by perturbing the geodesic equation (\ref{HJ:geodesic}) with the potential \eqref{HJ:pot} evaluated on an orbit with impact parameter $b=b_s$ and setting $r(\tau)=r_s+\delta r(\tau)$. We find that small deviations from the orbit decay exponentially in time as $\delta r \sim e^{-\lambda_L t}$ with Lyapunov exponent given  by
\begin{eqnarray}
\lambda_L &=& \sqrt{\frac{1}{2}\frac{V''(r;b)}{\dot{t}(\tau )^2}}\bigg|_{r=r_s,b=b_s} \nonumber\\
&=& \frac{1}{b_s r_s^2}\sqrt{\frac{\left(r_s^2+a^2-a b_s\right)^2 \left(6 r_s^2+a^2-b_s^2\right)}{\left(b_s-a\right)^2}}\,.
\end{eqnarray}
where a prime $(')$ denotes a derivative with respect to $r$. We finally obtain the approximate spectrum of the photon sphere family of QNMs in the leading WKB limit $\ell=|m|\to \infty$ using the correspondence \cite{Goebel:1972,Ferrari:1984zz,Ferrari:1984ozr,Mashhoon:1985cya,Schutz:1985km,Bombelli:1991eg,Cornish:2003ig,Cardoso:2008bp,Dolan:2010wr,Yang:2012he,Zimmerman:2015trm, Dias:2022oqm}:
\begin{eqnarray} \label{PS:eikonal}
\omega^{\hbox{\tiny eikn}}_{\hbox{\tiny PS}} &\simeq& m\,\Omega_c-i\left(n+\frac{1}{2}\right)\lambda_L \nonumber \\
&\simeq& \frac{m}{b_s}  -i\,\frac{n+1/2}{b_s r_s^2}\frac{\left | r_s^2+a^2-a b_s\right |}{\left | b_s-a\right |} \sqrt{6 r_s^2+a^2-b_s^2}\,,  
\end{eqnarray}
where $n=0,1,2,\ldots$ is the radial overtone. The frequency $\omega^{\hbox{\tiny eikn}}_{\hbox{\tiny PS}}$ describes the eikonal approximation for the PS modes. This expression is blind to the spin of the perturbation, \ie it is the same for scalar and gravito-electromagnetic perturbations (at higher order in the $1/m$ expansion, the result does depend on the spin; see the next subsection for the scalar field case in KN and the WKB spin dependence for Kerr in \cite{Seidel:1989bp}).

Strictly speaking, \eqref{PS:eikonal} is valid only in the geometric optics limit $\ell=|m|\to \infty$, with corrections to $\mathrm{Im}\,\tilde{\omega}$ and $\mathrm{Re}\,\tilde{\omega}$ being of order $\mathcal{O}\left(|m|\right)$ and  $\mathcal{O}\left(1\right)$, respectively. However, in practice we find that it is already a good approximation for $\ell=m=2$ in a wide window of the KN parameter space centred around the Kerr and Reissner-Nordstr\"om limiting solutions. In the next subsection we do a higher-order WKB expansion that finds the corrections to the leading eikonal result \eqref{PS:eikonal}.

\subsubsection{Photon sphere modes in a WKB expansion: beyond the eikonal limit} \label{sec:PSwkbHighOrders}

The eikonal limit of the previous subsection was first studied by Goebel \cite{Goebel:1972} and Ferrari and Mashhoon \cite{Ferrari:1984zz,Ferrari:1984ozr,Mashhoon:1985cya}.
Naturally, this eikonal limit is the leading order result of a WKB expansion in $1/m$ in the limit $m\gg 1$ initiated by Schutz and Will \cite{Schutz:1985km} and completed for Schwarzschild, RN and Kerr in
\cite{Iyer:1986vv,Iyer:1986nq,Kokkotas:1988fm,Seidel:1989bp}. In this subsection, we extend the WKB expansion of KN QNMs (which is also valid for the sub-families of this black hole) to higher orders to capture the next-to-leading order WKB contributions to the photon sphere QNM frequencies. We will only consider modes with large $\ell=m>0$.
To the best of our knowledge, this extension has never been done for the QNMs of KN although it is a standard higher-order WKB analysis first discussed in the context of QNMs by Will and Guinn \cite{Will:1988zz}, and reduces to the higher-order WKB results of \cite{Iyer:1986vv,Iyer:1986nq,Kokkotas:1988fm,Seidel:1989bp} in the Schwarzschild, RN and Kerr limits. Unlike the leading eikonal WKB result, the next-to-leading order WKB corrections depend on the spin of the perturbation. Our analysis is valid for spin-0 perturbations.

Consider first the angular equation \eqref{KG:ang} for the oblate spheroidal harmonics $S(x)$.
To leading order in $1/m$ (and for modes with $\ell=m$) the leading WKB solution that is regular at $x=\pm 1$ is given by $S(x)=(1-x^2)^{m/2}$ with eigenvalue $\lambda=\ell(\ell+1)$. At higher WKB order, the angular eigenfunction and the angular eigenvalue receive $(1/m)^k$ corrections (with integer $k>0$). To find them we assume the WKB ansatz for these quantities,
\begin{subequations}\label{Ang:WKBansatz1}
\begin{align}
& S(x)=(1-x^2)^{\frac{m}{2}}\left(1+\frac{\mathcal{S}_1(x)}{m}+\frac{\mathcal{S}_2(x)}{m^2}+\frac{\mathcal{S}_3(x)}{m^3} \right)+\mathcal{O}\left(1/m^4\right), \\
& \lambda=\ell(\ell+1)+\frac{\lambda_1}{m}+\frac{\lambda_2}{m^2}+\mathcal{O}\left(1/m^3\right), 
\label{Ang:WKBansatz1b}
\end{align}
\end{subequations}
and we solve the angular equation \eqref{KG:ang} order by order in a small $1/m$ series expansion to find the WKB coefficients:
\begin{subequations}\label{Ang:WKBansatz2}
\begin{align}
&  \mathcal{S}_1(x) =\frac{1}{4} a^2 \omega^2 x^2 ,\quad 
\mathcal{S}_2(x) =\frac{1}{32} a^2 \omega^2 x^2  \left(a^2 \omega^2 x^2 -12\right), \nonumber \\
& \qquad
\mathcal{S}_3(x) =\frac{1}{384} a^2 \omega^2 x^2  \left(a^4 \omega^4 x^4 +24 a^2 \omega^2 \left(1-2 x^2\right)+216\right);  \\
&\lambda_1 =-\frac{1}{2} a^2 \omega ^2, \quad  \lambda_2 =\frac{3}{4}a^2 \omega ^2\,;\label{Ang:WKBansatz2b}
\end{align}
\end{subequations}
where the coefficients $\lambda_1$ and $\lambda_2$ were chosen to be such that $\mathcal{S}_2(x)$ and $\mathcal{S}_3(x)$ are everywhere regular (\ie the choices made eliminate $\ln(x)$ divergences at $x=0$ at each order).

Next, we want to solve the radial equation \eqref{KG:radial} also in a $1/m$ expansion to obtain the higher-order corrections to the leading order solution obtained in the previous subsection.
For that, we insert \eqref{PS:MQrsbs} and the expansion for $\lambda$ of \eqref{Ang:WKBansatz1b} and \eqref{Ang:WKBansatz2b} into \eqref{KG:radial} and we assume the WKB ansatz\"e for the radial eigenfunction and eigenfrequency
\begin{eqnarray}\label{Rad:WKBansatz1}
R(r)&=&e^{-m\chi(r)}\left(Q_0(r)+\frac{Q_1(r)}{m}+\frac{Q_2(r)}{m^2} +\mathcal{O}\left(1/m^3\right)\right),\nonumber \\
\omega&=&\frac{m}{b_s}+\omega_0 + \frac{\omega_1}{m}+\frac{\omega_2}{m^2}+\mathcal{O}\left(1/m^3\right), 
\end{eqnarray}
where the leading order contribution ($m/b_s$) to the frequency is the one we already determined in the eikonal limit \eqref{PS:eikonal}. Inserting these WKB ansatz\"e into the radial equation \eqref{KG:radial} we can solve the latter order by order in a small $1/m$ series expansion.
At each order, the requirement that the radial equation must be valid, in particular at $r=r_s$, yields a condition that allows one to determine the eigenfrequency correction $\omega_{0,1,2}$.
Then, before proceeding to the next order, we just need to find the equation of motion for the eigenfunction's correction $\chi$ and $Q_{0,1,2}$ (but not the solution itself) to use it at next order. At the end of the day, the WKB frequency coefficients of \eqref{Rad:WKBansatz1} are given as a function of $(a,r_s,b_s)$ by: 
\begin{subequations}\label{Rad:WKBansatz2}
\begin{align} 
\omega_0&=-\frac{r_s^2+a^2-a b_s}{4 b_s^2 r_s^2 (b_s-a)}\left(a^2-2b_s^2 +2 i b_s \sqrt{6 r_s^2+a^2-b_s^2} \right), \label{Rad:WKBansatz2w0} 
\\ 
\omega_1&=\frac{r_s^2+a^2-a b_s}{32 b_s^3 r_s^4 (a-b_s)^3 \left(6 r_s^2+a^2-b_s^2\right)^2}
\bigg[
2 a^{11}-10 a^{10} b_s+a^9 \left(8 b_s^2+26 r_s^2\right)
\nonumber \\
&+3 a^8 \left(20 b_s^3-37 b_s r_s^2\right) +a^7 \left(92 b_s^2 r_s^2-110 b_s^4+96 r_s^4\right)+a^6 \left(486 b_s^3 r_s^2-50 b_s^5-396 b_s r_s^4\right)
\nonumber \\
&
+2 a^5 \left(-443 b_s^4 r_s^2+204 b_s^2 r_s^4+98 b_s^6+36 r_s^6\right)
 -3 a^4 \left(29 b_s^5 r_s^2-356 b_s^3 r_s^4+16 b_s^7+180 b_s r_s^6\right)
\nonumber \\
&+a^3 \left(864 b_s^6 r_s^2-1976 b_s^4 r_s^4+864 b_s^2 r_s^6-96 b_s^8\right)
+4 a^2 \left(-84 b_s^7 r_s^2+112 b_s^5 r_s^4+77 b_s^3 r_s^6+12 b_s^9\right)
\nonumber \\
&
-8 a \left(6 b_s^8 r_s^2-41 b_s^6 r_s^4+74 b_s^4 r_s^6\right)
 +4 \left(50 b_s^3 r_s^8-37 b_s^5 r_s^6+6 b_s^7 r_s^4\right)
\nonumber \\
&
+4 i a b_s  \sqrt{6 r_s^2+a^2-b_s^2}\bigg(a^8-8 a^7 b_s+a^6 \left(12 b_s^2+13 r_s^2\right)+a^5 \left(14 b_s^3-61 b_s r_s^2\right)
\nonumber \\
&
+a^4 \left(77 b_s^2 r_s^2-37 b_s^4+48 r_s^4\right)
+a^3 \left(49 b_s^3 r_s^2+6 b_s^5-150 b_s r_s^4\right)+6 a^2 \big(-23 b_s^4 r_s^2+29 b_s^2 r_s^4+4 b_s^6\nonumber \\
&+6 r_s^6\big)-12 a \left(-4 b_s^5 r_s^2+b_s^3 r_s^4+b_s^7+6 b_s r_s^6\right)+12 b_s^2 r_s^2 \left(-5 b_s^2 r_s^2+b_s^4+6 r_s^4\right)\bigg)
\bigg], \label{Rad:WKBansatz2w1} 
\\ 
\omega_2&=\frac{r_s^2+a^2-a b_s}{128 b_s^4 r_s^6 (a-b_s)^5 \left(6 r_s^2+a^2-b_s^2\right)^5}
\bigg[ 
2 a^{22}-20 b_s a^{21}+\left(74 b_s^2+64 r_s^2\right) a^{20}
\nonumber \\
& -2 \left(32 b_s^3+301 r_s^2 b_s\right) a^{19} 
+\left(-670 b_s^4+2128 r_s^2 b_s^2+842 r_s^4\right) a^{18}
\nonumber \\
&
+\left(2428 b_s^5-2028 r_s^2 b_s^3-7530 r_s^4 b_s\right) a^{17}
+\left(-654 b_s^6-15920 r_s^2 b_s^4+25383 r_s^4 b_s^2+5820 r_s^6\right) a^{16}\nonumber \\
&
-2 \left(4616 b_s^7-30185 r_s^2 b_s^5+12927 r_s^4 b_s^3+25290 r_s^6 b_s\right) a^{15}
\nonumber \\
&
+\left(11710 b_s^8-31432 r_s^2 b_s^6-147595 r_s^4 b_s^4+164382 r_s^6 b_s^2+22320 r_s^8\right) a^{14}
\nonumber \\
&
+4 \left(2857 b_s^9-44084 r_s^2 b_s^7+146495 r_s^4 b_s^5-44532 r_s^6 b_s^3-49140 r_s^8 b_s\right) a^{13}
\nonumber \\
&
+\left(-28802 b_s^{10}+266496 r_s^2 b_s^8-443136 r_s^4 b_s^6-680476 r_s^6 b_s^4+631512 r_s^8 b_s^2+45792 r_s^{10}\right) a^{12}
\nonumber \\
&
+2 \left(736 b_s^{11}+58709 r_s^2 b_s^9-599046 r_s^4 b_s^7+1440568 r_s^6 b_s^5-368712 r_s^8 b_s^3-220320 r_s^{10} b_s\right) a^{11}
\nonumber \\
&
+2 \big(15131 b_s^{12}-233552 r_s^2 b_s^{10}+1092082 r_s^4 b_s^8-1409730 r_s^6 b_s^6-799010 r_s^8 b_s^4+719064 r_s^{10} b_s^2
\nonumber \\
&
+22032 r_s^{12}\big) a^{10}
-2 \big(7590 b_s^{13}-73778 r_s^2 b_s^{11}-30543 r_s^4 b_s^9+1715704 r_s^6 b_s^7-3849468 r_s^8 b_s^5
\nonumber \\
&
+907200 r_s^{10} b_s^3+260496 r_s^{12} b_s\big) a^9
+\big(-13338 b_s^{14}+307472 r_s^2 b_s^{12}-2463177 r_s^4 b_s^{10}\nonumber \\
&
+8094764 r_s^6 b_s^8-9116984 r_s^8 b_s^6-1532896 r_s^{10} b_s^4+1714608 r_s^{12} b_s^2+15552 r_s^{14}\big) a^8
\nonumber \\
&
+2 \big(6216 b_s^{15}-114257 r_s^2 b_s^{13}+673253 r_s^4 b_s^{11}-1173570 r_s^6 b_s^9
-1560404 r_s^8 b_s^7+5369000 r_s^{10} b_s^5
\nonumber \\
&
-1108080 r_s^{12} b_s^3-116640 r_s^{14} b_s\big) a^7
+b_s^2 \big(552 b_s^{14}-42248 r_s^2 b_s^{12}
+713245 r_s^4 b_s^{10}-4681810 r_s^6 b_s^8
\nonumber \\
&
+13463880 r_s^8 b_s^6
-14621120 r_s^{10} b_s^4+208752 r_s^{12} b_s^2+692064 r_s^{14}\big) a^6
-8 b_s^3 \big(408 b_s^{14}-10395 r_s^2 b_s^{12}
\nonumber \\
&
+99510 r_s^4 b_s^{10}-438984 r_s^6 b_s^8
+844913 r_s^8 b_s^6-278048 r_s^{10} b_s^4-823092 r_s^{12} b_s^2
+93312 r_s^{14}\big) a^5
\nonumber \\
&
+2 b_s^4 \big(432 b_s^{14}-9776 r_s^2 b_s^{12}+66617 r_s^4 b_s^{10}-64822 r_s^6 b_s^8
-985532 r_s^8 b_s^6+4000144 r_s^{10} b_s^4
\nonumber \\
&
-4892832 r_s^{12} b_s^2+362016 r_s^{14}\big) a^4
-8 b_s^5 r_s^2 \big(128 b_s^{12}-4260 r_s^2 b_s^{10}+45727 r_s^4 b_s^8-217699 r_s^6 b_s^6
\nonumber \\
&
+473398 r_s^8 b_s^4-318564 r_s^{10} b_s^2-203688 r_s^{12}\big) a^3
+4 b_s^4 r_s^2 \big(24 b_s^{14}-740 r_s^2 b_s^{12}+5834 r_s^4 b_s^{10}
\nonumber \\
&
-5685 r_s^6 b_s^8
-107628 r_s^8 b_s^6+462432 r_s^{10} b_s^4-612144 r_s^{12} b_s^2+82512 r_s^{14}\big) a^2
\nonumber \\
&
-32 b_s^5 r_s^4 \big(3 b_s^{12}-97 r_s^2 b_s^{10}+1068 r_s^4 b_s^8-5289 r_s^6 b_s^6
+11814 r_s^8 b_s^4-7884 r_s^{10} b_s^2-5400 r_s^{12}\big) a
\nonumber \\
&
+8 b_s^6 r_s^6 \left(b_s^2-6 r_s^2\right)^3 \left(6 b_s^4-37 r_s^2 b_s^2+50 r_s^4\right)
\nonumber \\
&+ 2 i b_s \sqrt{6 r_s^2+a^2-b_s^2}\bigg(
2 a^{20}-26 b_s a^{19}+\left(153 b_s^2+64 r_s^2\right) a^{18}-\left(362 b_s^3+727 r_s^2 b_s\right) a^{17}
\nonumber \\
&
-\left(271 b_s^4-3772 r_s^2 b_s^2-842 r_s^4\right) a^{16}+\left(2556 b_s^5-8769 r_s^2 b_s^3-8399 r_s^4 b_s\right) a^{15}-\big(2426 b_s^6
\nonumber \\
&
+3010 r_s^2 b_s^4-37394 r_s^4 b_s^2-5820 r_s^6\big) a^{14}
-2 \left(2366 b_s^7-24551 r_s^2 b_s^5+41584 r_s^4 b_s^3+25845 r_s^6 b_s\right) a^{13}
\nonumber \\
&
+2 \big(4566 b_s^8-28560 r_s^2 b_s^6+1424 r_s^4 b_s^4
+98757 r_s^6 b_s^2+11160 r_s^8\big) a^{12}
\nonumber \\
&
+2 \big(619 b_s^9-28977 r_s^2 b_s^7+171002 r_s^4 b_s^5
-207154 r_s^6 b_s^3-90900 r_s^8 b_s\big) a^{11}
\nonumber \\
&
-\big(12035 b_s^{10}-147852 r_s^2 b_s^8+472290 r_s^4 b_s^6-157566 r_s^6 b_s^4-616644 r_s^8 b_s^2-45792 r_s^{10}\big) a^{10}
\nonumber \\
&
+\big(5094 b_s^{11}-28091 r_s^2 b_s^9-173950 r_s^4 b_s^7+1099064 r_s^6 b_s^5-1224580 r_s^8 b_s^3-362016 r_s^{10} b_s\big) a^9
\nonumber \\
&
+\big(6369 b_s^{12}-123188 r_s^2 b_s^{10}+773952 r_s^4 b_s^8-1781082 r_s^6 b_s^6+819180 r_s^8 b_s^4+1149336 r_s^{10} b_s^2
\nonumber \\
&
+44064 r_s^{12}\big) a^8
+\big(-5400 b_s^{13}+82355 r_s^2 b_s^{11}-350365 r_s^4 b_s^9+158096 r_s^6 b_s^7+1637020 r_s^8 b_s^5
\nonumber \\
&
-2178848 r_s^{10} b_s^3-371952 r_s^{12} b_s\big) a^7
-2 \big(246 b_s^{14}-11611 r_s^2 b_s^{12}+150842 r_s^4 b_s^{10}-760087 r_s^6 b_s^8
\nonumber \\
&
+1586638 r_s^8 b_s^6-921032 r_s^{10} b_s^4-581256 r_s^{12} b_s^2-7776 r_s^{14}\big) a^6
+2 \big(816 b_s^{15}-18214 r_s^2 b_s^{13}
\nonumber \\
&
+144727 r_s^4 b_s^{11}-492003 r_s^6 b_s^9+583154 r_s^8 b_s^7+416552 r_s^{10} b_s^5-1057824 r_s^{12} b_s^3-73872 r_s^{14} b_s\big) a^5
\nonumber \\
&
-\big(432 b_s^{16}-8456 r_s^2 b_s^{14}+42434 r_s^4 b_s^{12}+38464 r_s^6 b_s^{10}-855632 r_s^8 b_s^8+2340904 r_s^{10} b_s^6
\nonumber \\
&
-1895226 r_s^{12} b_s^4-435456 r_s^{14} b_s^2\big) a^4
+4 b_s^3 r_s^2 \big(128 b_s^{12}-3906 r_s^2 b_s^{10}+35243 r_s^4 b_s^8-136975 r_s^6 b_s^6
\nonumber \\
&
+224308 r_s^8 b_s^4-44484 r_s^{10} b_s^2-188352 r_s^{12}\big) a^3
-4 b_s^4 r_s^2 \big(12 b_s^{12}-343 r_s^2 b_s^{10}+1990 r_s^4 b_s^8
\nonumber \\
&
+1678 r_s^6 b_s^6-48290 r_s^8 b_s^4+153429 r_s^{10} b_s^2-159738 r_s^{12}\big) a^2
\nonumber \\
&
+8 b_s^3 r_s^4 \big(6 b_s^{12}-212 r_s^2 b_s^{10}+2109 r_s^4 b_s^8-9354 r_s^6 b_s^6+19398 r_s^8 b_s^4-13632 r_s^{10} b_s^2-5400 r_s^{12}\big) a
\nonumber \\
&
+2 b_s^4 r_s^8 \big(146 b_s^8-2044 r_s^2 b_s^6+11133 r_s^4 b_s^4-28212 r_s^6 b_s^2+28212 r_s^8\big)
\bigg)
\bigg].\label{Rad:WKBansatz2w2}\
\end{align}
\end{subequations}
We can immediately compare the WKB result~\eqref{Rad:WKBansatz1}-\eqref{Rad:WKBansatz2}, which is obtained by directly solving the radial Klein-Gordon equation and is valid for the first radial overtone $n=0$, with the eikonal
result~\eqref{PS:eikonal}, which is obtained solving the geodesic equation for a point particle. We see that the leading WKB frequency (which is $\mathcal{O}(m)/\mathcal{O}(1)$ for the real/imaginary part of the frequency) indeed agrees with the eikonal frequency, thus confirming the validity of the latter. On the other hand, the WKB result~\eqref{Rad:WKBansatz1}-\eqref{Rad:WKBansatz2} now finds the next-to-leading order corrections in the frequency up to  $\mathcal{O}(1/m^2)$.

 The WKB result~\eqref{Rad:WKBansatz1}-\eqref{Rad:WKBansatz2} describes the first radial overtone, $n=0$. In the simplest scenario, one expects that the overtone dependence ~\eqref{PS:eikonal} of the eikonal result \cite{Schutz:1985km} extends to the higher-order WKB corrections. This expectation is supported by the comparison with our numerical data: we find that the WKB frequencies of higher overtones are well approximated by
 \begin{equation}\label{WKBfrequency}
\omega_{\hbox{\tiny WKB}}=\frac{m}{b_s} + \operatorname{Re}(\omega_0) + (2n+1) \left(i \operatorname{Im}(\omega_0) + \frac{\omega_1}{m}+\frac{\omega_2}{m^2}+\mathcal{O}\left(1/m^3\right)\right), 
\end{equation} 
with $\omega_{0,1,2}$ given by \eqref{Rad:WKBansatz2}, and radial overtone $n=0,1,2,\ldots$.
We have split the $\omega_0$ correction into real and imaginary parts, which is consistent with the eikonal result~\eqref{PS:eikonal}, since the real part $\operatorname{Re}(\omega_0)$ is a sub-leading correction not present in~\eqref{PS:eikonal}. To use this formula, recall that given a KN black hole with parameters $(M,Q,a)$ we can find $r_s$ solving \eqref{PS:4thPolyn} and then $b_s$ is given by \eqref{PS:bs}. Further recall that we can use the polar parametrization~\eqref{PolarParametrization} to express the rotation and charge of the KN black hole in terms of $(\mathcal{R},\Theta)$.

\begin{figure}[t]
\centering
\includegraphics[width=.49\textwidth]{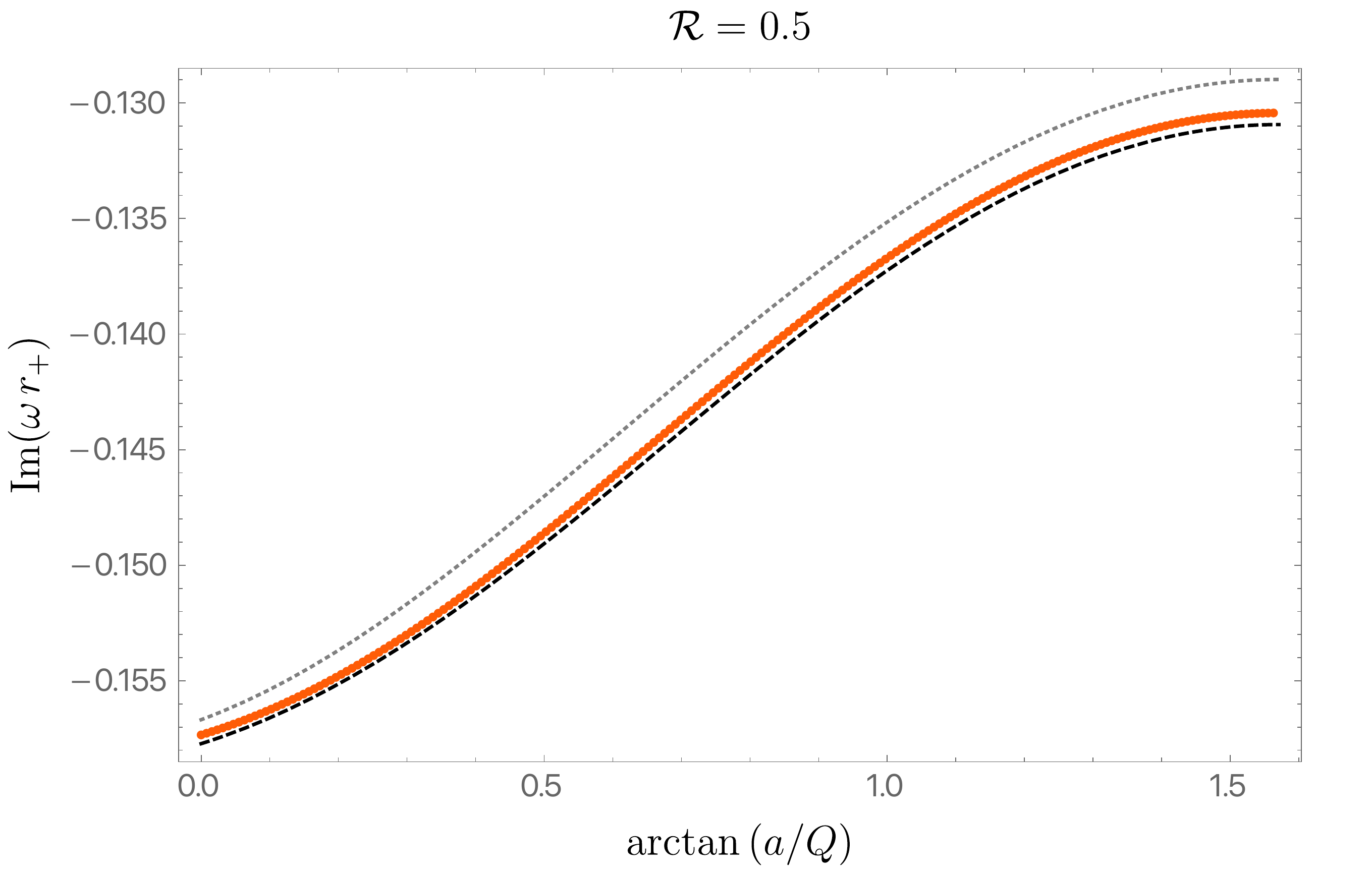}
\hspace{0.0cm}
\includegraphics[width=.49\textwidth]{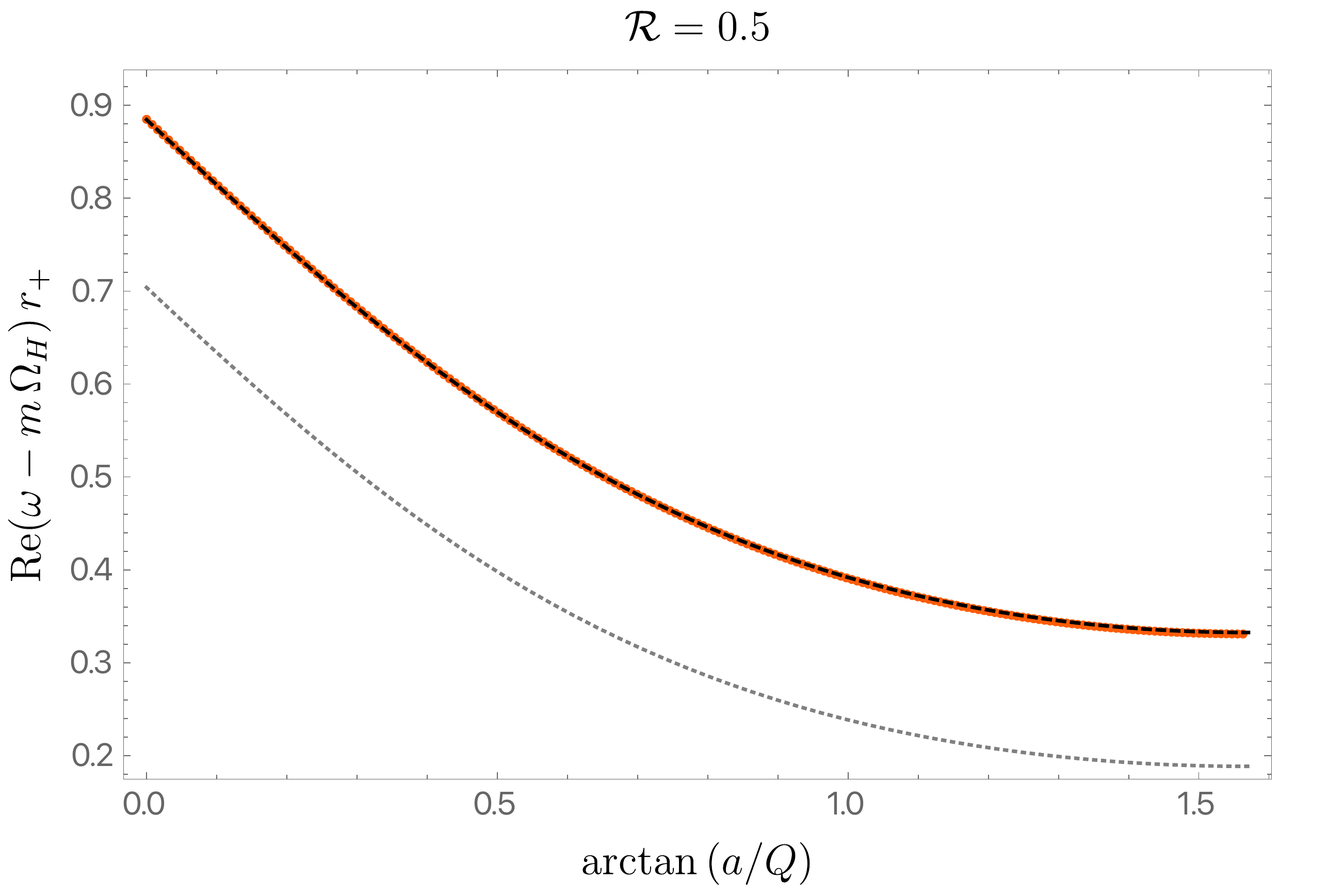}
\caption{Comparing the eikonal prediction $\omega^{\hbox{\tiny eikn}}_{\hbox{\tiny PS}}$ (dotted gray line) with the WKB result $\omega_{\hbox{\tiny WKB}}$ (dashed black line) and with the actual numerical frequencies (orange points) for co-rotating PS modes with $m=\ell=2, n=0$ in a KN black hole family with fixed $\mathcal{R}=0.5$. {\bf Left panel:} Imaginary part of the dimensionless frequency as a function of $\Theta$. {\bf Right panel:} Real part of the frequency measured with respect to the superradiant bound $m\Omega_H$ as a function of $\Theta$.}
\label{Fig:WKB-Eik_r05}
\end{figure}  

\begin{figure}[t]
\centering
\includegraphics[width=.49\textwidth]{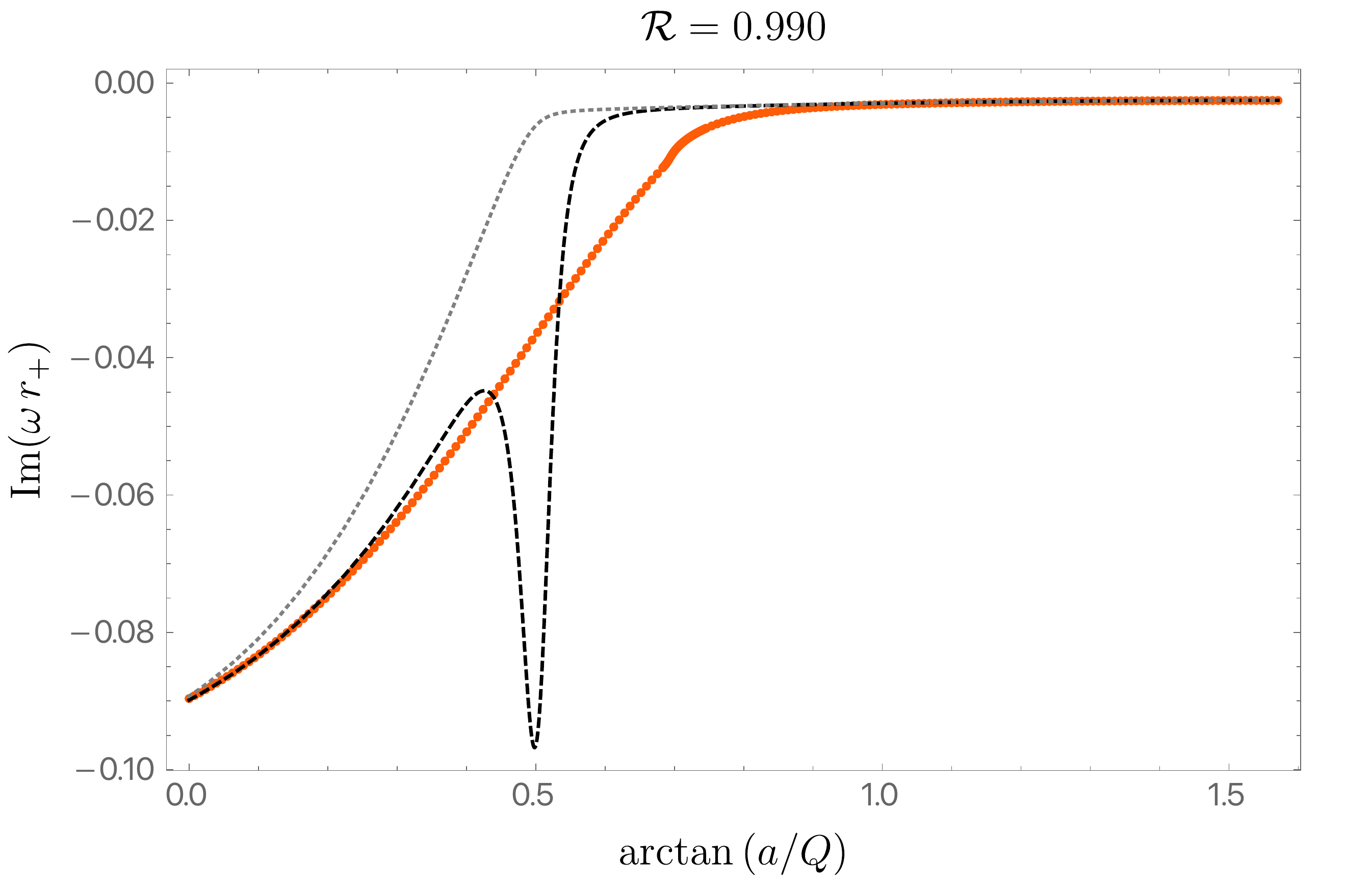}
\hspace{0.0cm}
\includegraphics[width=.49\textwidth]{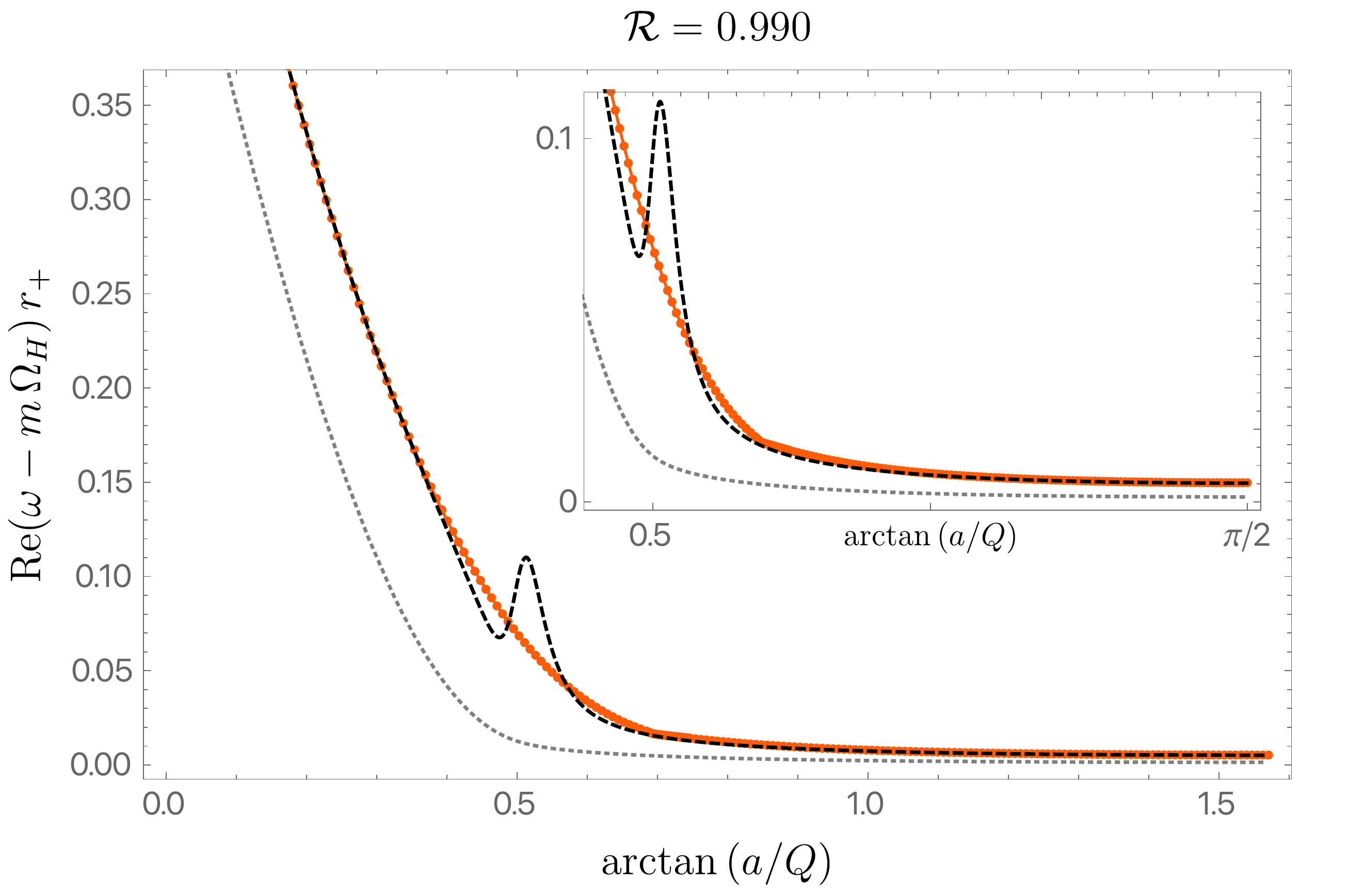}
\caption{Comparing the eikonal prediction $\omega^{\hbox{\tiny eikn}}_{\hbox{\tiny PS}}$ (dotted gray line) with the WKB result $\omega_{\hbox{\tiny WKB}}$ (dashed black line) and with the actual numerical frequencies (orange points) for co-rotating PS modes with $m=\ell=2, n=0$ in a KN black hole family with fixed $\mathcal{R}=0.99$. {\bf Left panel:} Imaginary part of the dimensionless frequency as a function of $\Theta$. {\bf Right panel:} Real part of the frequency measured with respect to the superradiant bound $m\Omega_H$ as a function of $\Theta$.}
\label{Fig:WKB-Eik_r99}
\end{figure}  

Naturally, the WKB result $\omega_{\hbox{\tiny WKB}}$  in  \eqref{Rad:WKBansatz1}-\eqref{Rad:WKBansatz2} with corrections up to  $\mathcal{O}(1/m^2)$ represents a considerable improvement over the (leading order WKB) eikonal approximation $\omega^{\hbox{\tiny eikn}}_{\hbox{\tiny PS}}$ in \eqref{PS:eikonal}. This is best illustrated in Figs.~\ref{Fig:WKB-Eik_r05} and~\ref{Fig:WKB-Eik_r99}. Here we plot the numerical PS frequency (orange points)  for a KN family with $\mathcal{R}=0.5$ (Fig.~\ref{Fig:WKB-Eik_r05})  and $\mathcal{R}=0.99$ (Fig.~\ref{Fig:WKB-Eik_r99}) as $\Theta=\arctan{\frac{a}{Q}}$ ranges from the Reissner-Nordstr\"om black hole ($\Theta=0$) to the Kerr solution ($\Theta=\pi/2$). We compare this exact numerical result with the WKB result $\omega_{\hbox{\tiny WKB}}$  in~\eqref{Rad:WKBansatz1}-\eqref{Rad:WKBansatz2} (dashed black line) and with the eikonal approximation $\omega^{\hbox{\tiny eikn}}_{\hbox{\tiny PS}}$ in \eqref{PS:eikonal} (dotted gray line). We see that the WKB prediction is an excellent approximation for any $\Theta$ for values of $\mathcal{R}$ that are not too close to extremality (Fig.~\ref{Fig:WKB-Eik_r05}). Moreover, even for values of $\mathcal{R}$ close to extremality (Fig.~\ref{Fig:WKB-Eik_r99}), the WKB prediction is still an excellent approximation 
in a wide neighbourhood around $\Theta=0$ and again in a large vicinity around $\Theta=\pi/2$. On the other hand, in both plots, one sees that the eikonal approximation is a less good approximation (as expected, since it is strictly valid only in the limit $\ell=m\to \infty$).

In the extremal limit $\mathcal{R} \to 1$, the imaginary part of the (co-rotating) eikonal approximation $\omega_{\hbox{\tiny PS}}^{\hbox{\tiny eikn}}$ vanishes when $\Theta\ge \Theta_{\star}^{\footnotesize \rm eik}$ with $ \Theta_{\star}^{\footnotesize \rm eik} \equiv \pi/6 \simeq 0.52$, since the orbit radius reaches the event horizon $r_{s}^{-} \to r_{+}$ (see Fig.~\ref{Fig:PSradii}). This occurs because the peak of the eikonal effective Schr\"oedinger potential reaches the event horizon $r_{s} = r_{+}$, as previously mentioned in~\cite{Zimmerman:2015trm}. However, going beyond the eikonal approximation, this is not the case for the numerical frequencies computed at finite $m$, and this is reflected by the higher-order WKB corrections~\eqref{WKBfrequency} which have a non-zero imaginary part when $\Theta \gtrsim \pi/6$ in the extremal limit (accordingly, from Fig.~\ref{Fig:WKB-Eik_r99} it is also clear that the dashed black and dotted gray lines are distinct near $\Theta \sim \pi/6$). However, for sufficiently large $\Theta$, the higher-order WKB approximation still predicts that PS modes have vanishing imaginary part and $\mathrm{Re}\, \omega\to m\Omega_H^{\hbox{\footnotesize{ext}}}$ at extremality. This happens for $\Theta\ge \Theta_{\star}^{\footnotesize \rm WKB}$ with $ \Theta_{\star}^{\footnotesize \rm WKB} > \Theta_{\star}^{\footnotesize \rm eik} \equiv \pi/6$. However, later $-$ see in particular the discussion of Figs.~\ref{Fig:WKBn0-NHn0-Im}-\ref{Fig:Star} $-$ we will find that it is not entirely clear whether the PS modes do approach $\mathrm{Im}\, \omega\to 0$ and $\mathrm{Re}\, \omega\to m\Omega_H^{\hbox{\footnotesize{ext}}}$ for large $\Theta$ at extremality (although it is probably the case that they indeed do so). On the other hand, we will find that the near-horizon QNM family is very well described by \eqref{WKBfrequency} close to extremality.

\subsection{Near-extremal QNM frequencies: a matched asymptotic expansion} \label{sec:NHanalytics}

Near extremality, the scalar wavefunctions of relevant classes of modes about KN (this is the case \eg for $m=\ell=2$ modes when $\Theta$ is small, but not for the $m=\ell=0$ modes at any $\Theta$; see Fig.~\ref{fig:eigenfunctions_ml0} for the latter case)  are very localized near the horizon and quickly decay away from it. This suggests that we might be able to analytically study the problem within perturbation theory, with the expansion parameter being the off-extremality quantity $\sigma$ introduced in \eqref{def:sigma} 
\cite{Teukolsky:1974yv,Detweiler:1980gk,Sasaki:1989ca,Andersson:1999wj,Glampedakis:2001js,Hod:2008zz,Yang:2012pj,Yang:2013uba,Hod:2014uqa,Zimmerman:2015trm,Hod:2015xlh,Dias:2021yju,Dias:2022oqm}. 
This turns out to be indeed possible if we resort to a matched asymptotic expansion (MAE) whereby we split the spacetime into a near-region (where the wavefunction is mainly localized) and a far region  (where the wavefunction is considerably smaller).
The near-region is defined as  $\frac{r}{r_+}-1\ll 1$ and the wavefunction must be regular in this region. In particular it must be regular in ingoing Eddington-Finkelstein coordinates at the event horizon $r=r_+$. On the other hand, the far-region covers the region $\frac{r}{r_+}-1\gg \sigma$ and the associated wavefunction must satisfy the outgoing boundary condition at $r\to +\infty$. The two solutions must then be simultaneously valid $-$ and thus the free parameters of the two regions must be matched $-$ in the matching region $\sigma \ll \frac{r}{r_+}-1\ll 1$. The latter is guaranteed to exist since the expansion parameter is small, $\sigma\ll 1$. In each of these regions we find that the radial Klein-Gordon simplifies considerably and can be solved analytically.

We can now formulate and perform the matched asymptotic expansion in detail.
As stated above, the expansion parameter of our perturbation theory is the dimensionless off-extremality quantity  $\sigma=1 - \frac{r_{-}}{r_{+}}$ defined in \eqref{def:sigma}. 
At extremality ($\sigma=0$),  we numerically find that the modes with slowest decay rate {\it always} approach $\mathrm{Im}\,\omega=0$ and $\mathrm{Re}\,\omega=m \Omega_H^{\hbox{\footnotesize{ext}}}$.\footnote{In several studies of perturbations of RN, Kerr, KN \cite{Teukolsky:1974yv,Detweiler:1980gk,Sasaki:1989ca,Andersson:1999wj,Glampedakis:2001js,Hod:2008zz,Yang:2012pj,Yang:2013uba,Hod:2014uqa,Zimmerman:2015trm,Hod:2015xlh,Dias:2021yju,Dias:2022oqm} and even de Sitter black holes \cite{Cardoso:2017soq,Dias:2018etb,Dias:2018ufh}, it was also found that there are near-horizon modes that saturate the superradiant bound $\omega=m\Omega_H$ at extremality. This happens for bosonic perturbations with spin $s$, {\it not} only for $s=0$. Our analysis here is very similar to the one presented in \cite{Zimmerman:2015trm} for $s=0$ (the MAE analysis of NH gravito-electromagnetic modes is considerably more elaborated than the $s=0$ case \cite{Dias:2022oqm}).}
Therefore, onwards we assume that the eigenfrequency we search for has an expansion in $\sigma$ about this superradiant bound: 
\be\label{freqExp}
\tilde{\omega} =  m \tilde{\Omega}_H^{\hbox{\footnotesize ext}} + \sigma \,\delta\tilde{\omega} +\mathcal{O}(\sigma^2)\,,
\ee 
where $\tilde{\omega}=\omega\, r_+$, $\delta\tilde{\omega}=\delta\omega\,r_+$, and our task is to find $\delta\tilde{\omega}$. In \eqref{freqExp} and onwards, $\tilde{\Omega}_H$ and  $\alpha$ always refer to their extremal values although we will drop the super/subscripts `ext' (present in \eqref{freqExp}) for notational simplicity.
By now, it is clear that it's useful to parameterize the background using the inner and event horizon locations, $r_\pm$. For that, recall that $\Delta = r^2 -2Mr+a^2+Q^2$ which can be equivalently written as $\Delta=(r-r_-)(r-r_+)$. Equating these two expressions and their derivatives we can express $M$ and $Q$ as a function of $(r_-,r_+,a)$:
\be\label{MQexpressions}
M=\frac{1}{2} \left(r_-+r_+\right)\,,\qquad Q=\sqrt{r_- r_+-a^2}\,.
\ee
We will insert these relations in the radial Klein-Gordon equation.

Consider first the far-region, $\frac{r}{r_+}-1\gg \sigma$. It is then natural to redefine the radial coordinate as
\be\label{FR:radialCoord}
y = \frac{r}{r_+}-1\,,
\ee
in which case the far-region is simply defined as the region $y\gg \sigma$.
Inserting \eqref{FR:radialCoord}, \eqref{MQexpressions} and \eqref{freqExp} into the radial equation \eqref{KG:radial} one finds that, to leading order in a $\sigma$ expansion about extremality, it reduces to
\begin{equation}\label{KGrad:farRegion}
y^2 R''(y)+2 y R'(y)+ \left[m^2 \tilde{\Omega}_H ^2 \left(6+\alpha ^2+4y+y^2 \right)-\lambda \right]R(y)\simeq 0.
\end{equation}
Introducing a new radial coordinate $\bar{y}$ and a redefinition of the radial function,
\begin{equation}
\bar{y}=2\,i\,m \tilde{\Omega}_H\,  y\,, \qquad R=e^{-i\, m \tilde{\Omega}_H y}y^{-\frac{1}{2}+i \,\delta} K
\end{equation}
where
\begin{equation}\label{def:delta} 
\delta=\sqrt{(6+\alpha^2)m^2 \tilde{\Omega}_H^2 -\frac{1}{4}-\lambda}\,,
\end{equation}
we find that \eqref{KGrad:farRegion} is a standard Kummer equation, $\bar{y} K''(\bar{y})+(\mathfrak{b}-\bar{y})K'(\bar{y})-\mathfrak{a} K(\bar{y})=0$
with 
\begin{equation}
\mathfrak{a}=\frac{1}{2}+i \left( 2 m \tilde{\Omega}_H+\delta \right) \,, \qquad
\mathfrak{b}=1+2 i \delta \,.
\end{equation}
Its most general solution is a sum of two independent functions,
$_1F_1(\mathfrak{a};\mathfrak{b};\bar{y})$ and $\bar{y}^{1-\mathfrak{b}}\, _1F_1(\mathfrak{a}-\mathfrak{b}+1;2-\mathfrak{b};\bar{y})$ where $\, _1F_1$ is the Kummer confluent hypergeometric function (a.k.a. of the first kind) \cite{AbramowitzStegun64}\footnote{There is a third solution $U(\mathfrak{a},\mathfrak{b}, y)$, known as the confluent hypergeometric function of the second kind, that is sometimes also used as one of the two independent solutions. It can be written as a linear combination of the two independent solutions that we use as $U(\mathfrak{a},\mathfrak{b},y)=\frac{\pi}{ \sin(\pi  \mathfrak{b}) }  \left(\frac{\, _1F_1(\mathfrak{a};\mathfrak{b};y)}{(\mathfrak{b}-1)! (\mathfrak{a}-\mathfrak{b})!}-\frac{ \, _1F_1(\mathfrak{a}-\mathfrak{b}+1;2-\mathfrak{b};y)}{(\mathfrak{a}-1)! (1-\mathfrak{b})!}\,y^{1-\mathfrak{b}}\right)$ \cite{AbramowitzStegun64}.}.
Thus, the most general solution of the far-region equation \eqref{KGrad:farRegion} is
\begin{eqnarray}\label{FR:finalSol}
R&=&A_1 e^{-i\, m \tilde{\Omega}_H y}y^{-\frac{1}{2}+i \,\delta}  \, _1F_1\left(\mathfrak{a};\mathfrak{b}; 2 i \,m \tilde{\Omega}_H \, y\right)
\nonumber \\
&&
+A_2  e^{-i\, m \tilde{\Omega}_H y}y^{-\frac{1}{2}-i \,\delta} 
 \, _1F_1\left(\mathfrak{a}-\mathfrak{b}+1;2-\mathfrak{b}; 2 i \, m \tilde{\Omega}_H \, y\right),
\end{eqnarray}
for arbitrary integration constants $A_1$ and $A_2$.
Asymptotically, this solution behaves as  
\begin{eqnarray}\label{FR:asympSol}
R\big|_{y\to\infty}&\simeq& 
e^{-i\, m \tilde{\Omega}_H y}y^{-1-2 i\, m \tilde{\Omega}_H}  
 \bigg( 
A_1 \frac{(-2 i m \tilde{\Omega}_H )^{-\frac{1}{2}-i (\delta +2 m \tilde{\Omega}_H )} \Gamma (2 i \delta +1)}{\Gamma \left(\frac{1}{2}-i (2 m \tilde{\Omega}_H -\delta )\right)}
\nonumber \\
&& \hspace{3.5cm} 
+A_2 \frac{(-2 i m \tilde{\Omega}_H )^{-\frac{1}{2}-i (2 m \tilde{\Omega}_H -\delta )} \Gamma (1-2 i \delta )}{\Gamma \left(\frac{1}{2}-i (\delta +2 m \tilde{\Omega}_H ) \right)}
\bigg) \nonumber \\
&& +e^{i\, m \tilde{\Omega}_H y}y^{-1+2 i\, m \tilde{\Omega}_H} 
\bigg(
A_1 \frac{(2 i m \tilde{\Omega}_H )^{-\frac{1}{2}+i (2 m \tilde{\Omega}_H -\delta )} \Gamma (2 i \delta +1)}{\Gamma \left(\frac{1}{2}+i (\delta +2 m \tilde{\Omega}_H )\right)}
\nonumber \\
&& \hspace{3.5cm} 
+A_2 \frac{(2 i m \tilde{\Omega}_H )^{-\frac{1}{2}+i (\delta +2 m \tilde{\Omega}_H )} \Gamma (1-2 i \delta )}{\Gamma \left(\frac{1}{2}+i (2 m \tilde{\Omega}_H -\delta )\right)}
\bigg).
\end{eqnarray}
The first contribution (proportional to $e^{-i\, m \tilde{\Omega}_H y}$) represents an ingoing wave while the second (proportional to $e^{+i\, m \tilde{\Omega}_H y}$)  describes an outgoing wave. For the QNM problem we want to impose boundary conditions in the asymptotic region that keep only the outgoing waves. This fixes $A_1$ to be
\begin{equation}\label{FR:outgoingBC}
A_1=A_2 e^{\delta  \left[\pi +2 i \ln (2 m \tilde{\Omega}_H)\right]}\frac{ \Gamma (-2 i \delta ) \Gamma \left(\frac{1}{2}-i (2 m \tilde{\Omega}_H-\delta )\right)}{\Gamma (2 i \delta ) \Gamma \left(\frac{1}{2}-i (2 m \tilde{\Omega}_H+\delta )\right)}.
\end{equation}
For the matching with the near-region, we will need the small $y$ behaviour of the far-region solution \eqref{FR:asympSol} with the boundary condition \eqref{FR:outgoingBC}: 
\begin{eqnarray}\label{FR:smallRadius}
R_{\rm far}\big|_{y\ll 1} \simeq
A_2\left( y^{-\frac{1}{2}-i\,\delta} + y^{-\frac{1}{2}+i\,\delta}e^{\pi  \delta } (2 m \tilde{\Omega}_H ) ^{2 i \delta }
\frac{ \Gamma (-2 i \delta ) \Gamma \left(\frac{1}{2}-i (2 m \tilde{\Omega}_H-\delta )\right)}{\Gamma (2 i \delta ) \Gamma \left(\frac{1}{2}-i (2 m \tilde{\Omega}_H+\delta )\right)}
\right).
\end{eqnarray}

Consider now the near-region $0 \leq y\ll 1$. 
This time we should proceed cautiously when doing the perturbative expansion in $\sigma \ll 1$ since this small expansion parameter can now be of similar order as the radial coordinate $y$. This is closely connected with the fact that the far-region solution breaks down when $y/\sigma \sim \mathcal{O}(1)$. This suggests that to proceed with the near-region analysis we should define a new radial coordinate as\footnote{At the heart of the matching expansion procedure, note that a factor of $\sigma$ (the expansion parameter!) is absorbed in the new near-region radial coordinate.}
\be\label{NR:radialCoord}
z =\frac{y}{\sigma} = \frac{r-r_+}{r_+-r_-}\,,
\ee
The near-region now corresponds to $0 \leq  z\ll \sigma^{-1}$. 
So,  in the near-region we simultaneously zoom in around the horizon and approach extremality. 
Inserting \eqref{NR:radialCoord}, \eqref{MQexpressions} and \eqref{freqExp} into the radial equation  \eqref{KG:radial} one finds that (again to leading order in a $\sigma$ expansion about extremality) it reduces to
\begin{equation}\label{KGrad:nearRegion}
z (1+z) R''(z)+(1+2 z) R'(z)+ \left(\frac{\left(2  m \tilde{\Omega}_H z+\left(1+\alpha ^2\right) \delta\tilde{\omega}\right)^2}{z (1+z)}
+\left(2+\alpha ^2\right) m^2 \tilde{\Omega}_H^2-\lambda \right)R(z) \simeq 0.
\end{equation}
Introducing a new radial coordinate $\bar{z}$ and a redefinition of the radial function,
\begin{equation}
\bar{z}=-z\,, \qquad R=\bar{z}^{-i\,(1+\alpha^2)\delta\tilde{\omega}} (1-\bar{z})^{i\,[2 m\tilde{\Omega}_H-(1+\alpha^2)\delta\tilde{\omega}]} F,
\end{equation}
we find that \eqref{KGrad:nearRegion} reduces to a standard hypergeometric equation, 
$(1-\bar{z}) \bar{z} F''(\bar{z})+[c-\bar{z} (a_+ + a_- +1)] F'(\bar{z})-a_+ a_- F(\bar{z})=0$
with 
\begin{equation}\label{NR:delta}
a_\pm=\frac{1}{2}+i \left[2 m \tilde{\Omega}_H -2(1+\alpha^2)\delta\tilde{\omega}\pm \delta \right]\,, \qquad
c=1-2\,i (1+\alpha^2)\delta\tilde{\omega}\,,
\end{equation}
and $\delta$ defined in \eqref{def:delta}.
Its most general solution is a sum of two hypergeometric functions,
$F=\bar{C}_1 \, _2F_1(a_+,a_-;c;\bar{z})+ \bar{C}_2 \bar{z}^{1-c} \, _2F_1(a_+ -c+1,a_--c+1;2-c;\bar{z})$ \cite{AbramowitzStegun64} (for an arbitrary integration constants $\bar{C}_1,\bar{C}_2$), but regularity at the horizon in Eddington-Finkelstein coordinates requires that we eliminate the solution that is outgoing at the event horizon. So we set $\bar{C}_2\equiv 0$.  Thus, the solution of \eqref{KGrad:nearRegion} that describes ingoing waves at the event horizon is
\begin{equation}\label{NR:finalSol}
 R=C_1 z^{-i\,(1+\alpha^2)\delta\tilde{\omega}} (1+z)^{i\,[2 m\tilde{\Omega}_H-(1+\alpha^2)\delta\tilde{\omega}]} \, _2F_1(a_+,a_-;c;-z)
\end{equation}
for arbitrary constant $C_1$.
Later, to match the near-region solution \eqref{NR:finalSol} with the far-region one, we will need the large $z=\frac{y}{\sigma}$ behaviour of  \eqref{NR:finalSol} which is given by
\begin{eqnarray}\label{NR:largeRadius}
\hspace{-1.5cm} R_{\rm near}\big|_{z\gg 1}&\simeq& C_1 y^{-\frac{1}{2}-i \delta }
\frac{\sigma ^{\frac{1}{2}+i \delta } \Gamma (-2 i \delta ) \Gamma \Big(1-2 i (1+\alpha^2) \delta\tilde{\omega} \Big)}{\Gamma \left(\frac{1}{2} -2 i m \tilde{\Omega}_H -i \delta \right) \Gamma \left(\frac{1}{2}-2 i \left(1+\alpha^2\right) \delta\tilde{\omega} +2 i m \tilde{\Omega}_H -i \delta \right)}\nonumber\\
\hspace{-1.5cm} &&+C_1 y^{-\frac{1}{2}+i \delta }
\frac{\sigma ^{\frac{1}{2}-i \delta } \Gamma (2 i \delta ) \Gamma \left(1-2 i \left(1+\alpha^2\right) \delta\tilde{\omega} \right)}{\Gamma \left(\frac{1}{2} -2 i m \tilde{\Omega}_H +i \delta \right) \Gamma \left(\frac{1}{2} -2 i \left(1+\alpha^2\right) \delta\tilde{\omega} +2 i m \tilde{\Omega}_H +i \delta\right)}.
\end{eqnarray}

In the near-region we have used the horizon boundary condition to fix one of the two amplitudes of the most general solution. Similarly, in the far-region we have used the asymptotic boundary condition to fix one of the two amplitudes of the most general far-region solution. In each of the regions we are left with a free integration constant, $C_1$ in the far-region and $A_2$ in the near-region. These are now fixed in the matching region $\sigma \ll \frac{r}{r_+}-1\ll 1$ by requiring that the small radius expansion \eqref{FR:smallRadius} of the far-region matches with the large radius expansion \eqref{NR:largeRadius} of the near-region. 
Concretely, matching first the coefficients of $y^{-\frac{1}{2}-i \delta }$ of \eqref{FR:smallRadius} and \eqref{NR:largeRadius} requires that
\begin{equation}\label{Match1}
A_2=C_1\, \frac{\sigma ^{\frac{1}{2}+i \delta } \Gamma (-2 i \delta ) \Gamma \left(1-2 i \left(1+\alpha ^2\right) \delta\tilde{\omega}\right)}{\Gamma \left(\frac{1}{2}-i (2 m \tilde{\Omega}_H +\delta)\right) \Gamma \left(\frac{1}{2}+i [2 m \tilde{\Omega}_H -\delta -2 \left(1+\alpha ^2\right)\delta\tilde{\omega} ]\right)}\,.
\end{equation}
We can now insert this relation into  \eqref{FR:smallRadius} and the final matching, this time of  the coefficients of $y^{-\frac{1}{2}+i \delta }$ of \eqref{FR:smallRadius} and \eqref{NR:largeRadius},
 requires that the following condition holds
\begin{eqnarray}\label{Match2}
&& \beta  \,\Gamma \left(\frac{1}{2}+\frac{2 i\, m \alpha}{1+\alpha^2}-2 i \left(1+\alpha^2\right)\delta\tilde{\omega}-i \delta \right)-1=0\\ 
&&\hbox{with}\quad  \beta=\sigma ^{-2 i \delta } \frac{e^{-\pi  \delta } \left(\frac{2 \alpha  m}{1+\alpha ^2}\right)^{-2 i \delta }  \Gamma (2 i \delta )^2 \Gamma \left(\frac{1}{2}-\frac{2 i m \alpha }{1+\alpha^2}-i \delta \right)^2}{\Gamma (-2 i \delta )^2 \Gamma \left(\frac{1}{2}-\frac{2 i m \alpha }{1+\alpha^2}+i \delta \right)^2 \Gamma \left(\frac{1}{2}+\frac{2 i m \alpha }{1+\alpha^2}-2 i \left(1+\alpha^2\right) \delta\tilde{\omega}+i \delta \right)}. \nonumber
\end{eqnarray}
Here, we make the important observation that $\sigma\ll 1$   and thus  $\beta\,\propto\, \sigma ^{-2 i \delta } $ is generally a very small number,  $\beta\ll 1$. In these conditions,  \eqref{Match2} can be obeyed is if the gamma function multiplying $\beta$ is very large \ie if the frequency correction $\delta\tilde{\omega}$ is such that the argument of the gamma function is near one of its poles.
Recalling that $\Gamma(-p)\to +\infty$ when $p$ is a non-negative integer, the matching condition \eqref{Match2} quantizes the frequency correction as
\begin{equation}\label{Match3}
\delta\tilde{\omega}\simeq \frac{m \alpha}{\left(1+\alpha^2\right)^2}
-\frac{\delta}{2 \left(1+\alpha^2\right)}
-\frac{i \left(p+\frac{1}{2}\right)}{2 \left(1+\alpha^2\right)}\,,\qquad p=0,1,2,\cdots
\end{equation}
Inserting this quantization into \eqref{freqExp}, we conclude that QNMs that approach $\tilde{\omega} =  m \tilde{\Omega}_H^{\hbox{\footnotesize ext}} $ at extremality should have a frequency that is well approximated near extremality by\footnote{Note that, as discussed below \eqref{KG:ang}, the eigenvalue $\lambda$ is related to the eigenvalue $\Lambda$ of the standard oblate spheroidal equation by $\lambda=\Lambda-\alpha^2\tilde{\omega}^2$ and, in the near-horizon analysis of this subsection, one has $\tilde{\omega}=m\tilde{\Omega}_H^{\hbox{\footnotesize ext}}=m\,\alpha/(1+\alpha^2)$.}
\be\label{NEfreq}
\tilde{\omega}_{\hbox{\tiny MAE}} =  \frac{m \alpha}{1+\alpha^2} + \sigma \!\left(
\frac{m \alpha}{\left(1+\alpha^2\right)^2}
-\frac{1}{2 \left(1+\alpha^2\right)}
\sqrt{\frac{m^2 \alpha^2(6+\alpha^2)}{\left(1+\alpha^2\right)^2} -\frac{1}{4}-\lambda}
-\frac{i \left(p+\frac{1}{2}\right)}{2 \left(1+\alpha^2\right)}\right)+\mathcal{O}\left(\sigma^2\right)\!,
\ee   
where $p=0,1,2,\cdots$ is the radial overtone of the mode. To compare with our numerical results generated using the polar parametrization  \eqref{def:sigma}-\eqref{PolarParametrization}, we should now replace $\sigma = 1- \mathcal{R}^2$ and $\alpha= \mathcal{R}\, \sin \Theta$ in \eqref{NEfreq}. This approximation should be good for $\mathcal{R}\simeq 1$ and any $0 \leq \Theta\leq \frac{\pi}{2}$. 
    
Note that if we wish, we can convert \eqref{NEfreq} into units of $M$ by multiplying \eqref{NEfreq} by $M/r_+$ (since $\omega M=\tilde{\omega}M/r_+$) and expanding it in terms of $\sigma$ while keeping terms only up to $\mathcal{O}(\sigma)$ (since all our analysis is valid only up to this order). The near-horizon modes (a.k.a. zero damped \cite{Yang:2012pj,Yang:2013uba,Zimmerman:2015trm} or near-extremal modes) and a matched asymptotic analysis of such modes similar to the one above that leads to their frequencies near-extremality have already been considered for RN, Kerr and KN for several bosonic fields in \cite{Teukolsky:1974yv,Detweiler:1980gk,Sasaki:1989ca,Andersson:1999wj,Glampedakis:2001js,Hod:2008zz,Yang:2012pj,Yang:2013uba,Hod:2014uqa,Zimmerman:2015trm,Hod:2015xlh,Dias:2021yju,Dias:2022oqm}. In the appropriate limits, our frequency \eqref{NEfreq} reduces to the expressions presented in this literature.

\begin{figure}[t]
\centering
\includegraphics[width=.49\textwidth]{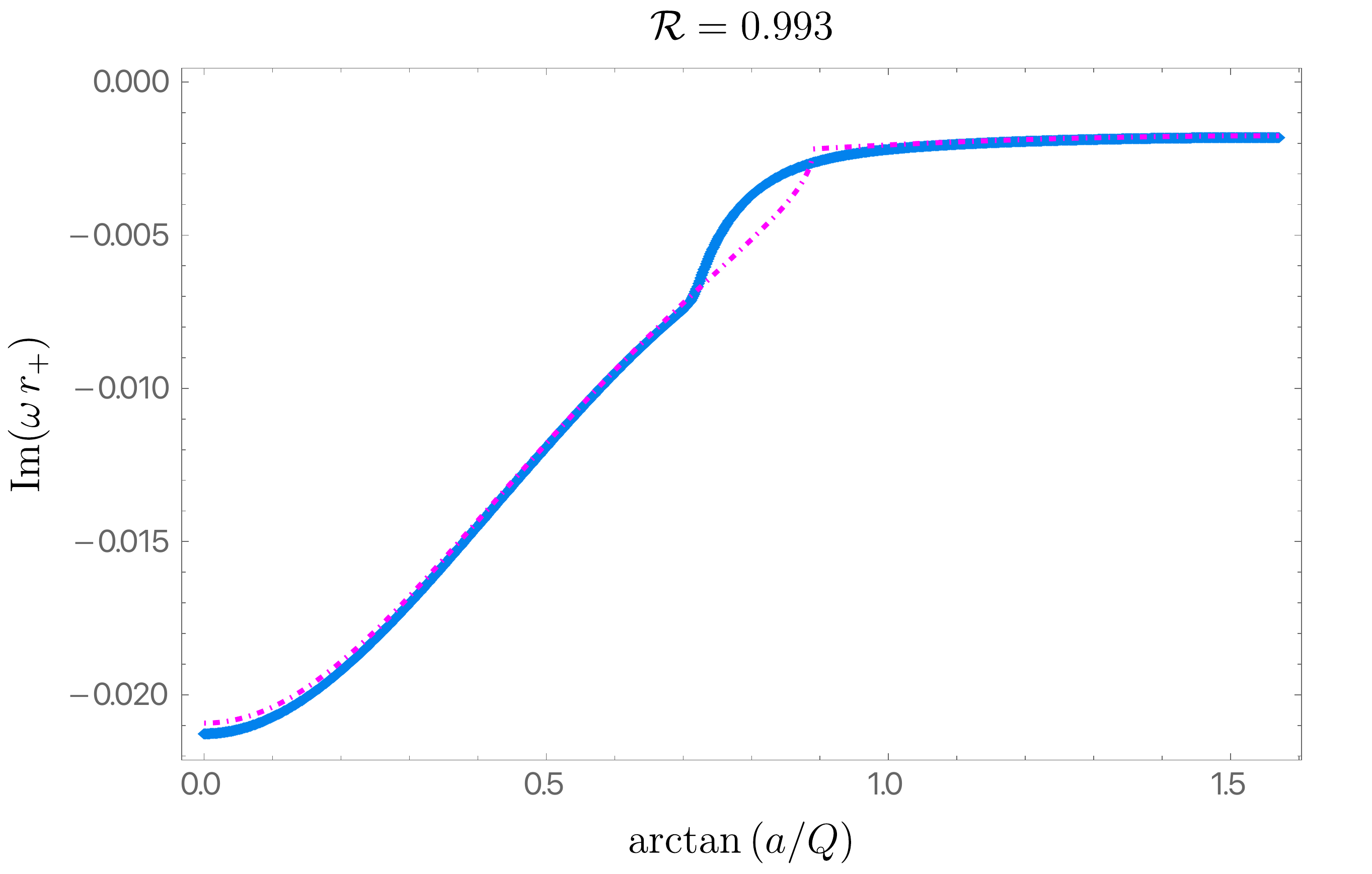}
\hspace{0.0cm}
\includegraphics[width=.49\textwidth]{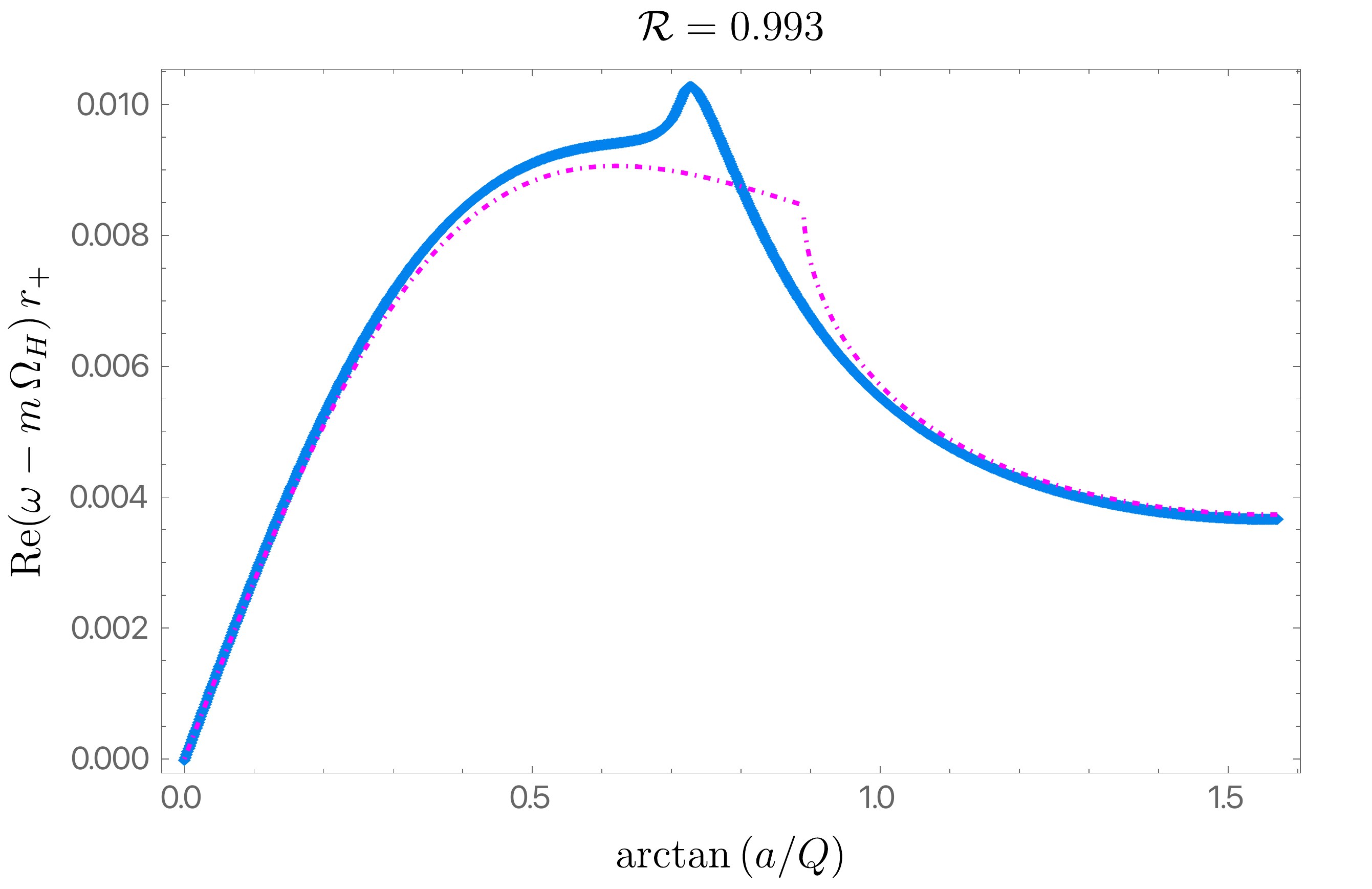}
\caption{The  NH (blue diamonds) family of QNMs with  $m=\ell=2, n=0$ for a KN family with $\mathcal{R}=0.993$.
We also display the near-extremal frequency $\tilde{\omega}_{\hbox{\tiny MAE}} $ for $p=0$ (dot-dashed magenta line). {\bf Left panel:} Imaginary part of the dimensionless frequency as a function of $\Theta$. {\bf Right panel:} Real part of the frequency measured with respect to the superradiant bound $m\Omega_H$ as a function of $\Theta$.}
\label{Fig:NH-R0.93}
\end{figure}

In Fig.~\ref{Fig:NH-R0.93} we compare the matched asymptotic expansion $\tilde{\omega}_{\hbox{\tiny MAE}}$ (dot-dashed magenta curve) to the exact NH modes (blue diamond curve) of KN for $\mathcal{R} = 0.993$ as we go from RN ($a = 0$, \ie $\Theta=0$) to Kerr ($Q = 0$, \ie $\Theta=\pi/2$). For most values of $\Theta \equiv \arctan(a/Q)$, $\tilde{\omega}_{\hbox{\tiny MAE}}$ is an excellent approximation. There is a sharp change in the behaviour of the matched asymptotic expansion at $\Theta \sim 0.881$, which turns out to be an important feature of this approximation, and is only present for certain values of the azimuthal quantum numbers $m$ and $\ell$, as already noted in \cite{Yang:2012pj,Yang:2013uba,Zimmerman:2015trm,Dias:2021yju,Dias:2022oqm}. 

\begin{figure}[t]
\centering
\includegraphics[width=.49\textwidth]{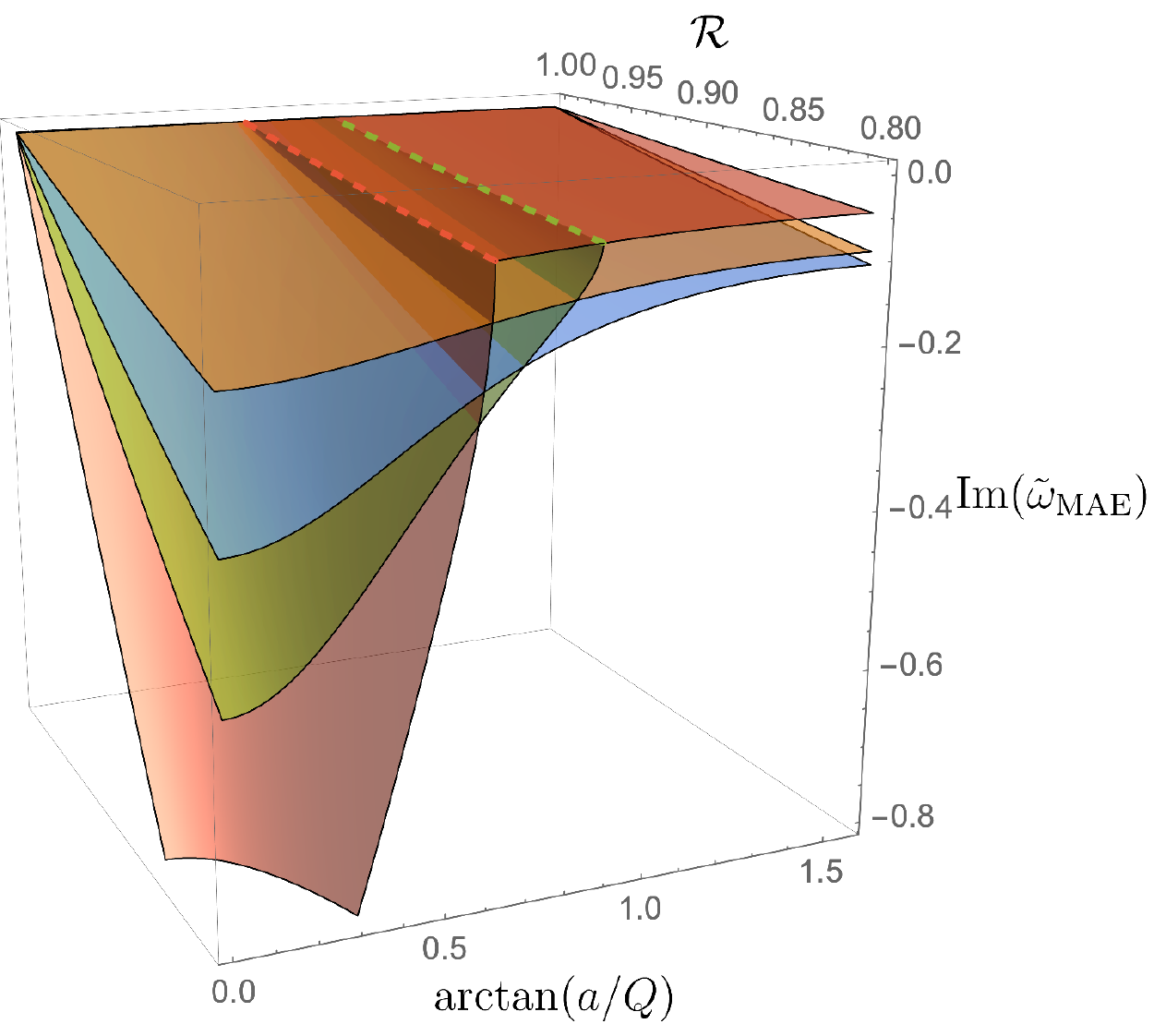}
\includegraphics[width=.49\textwidth]{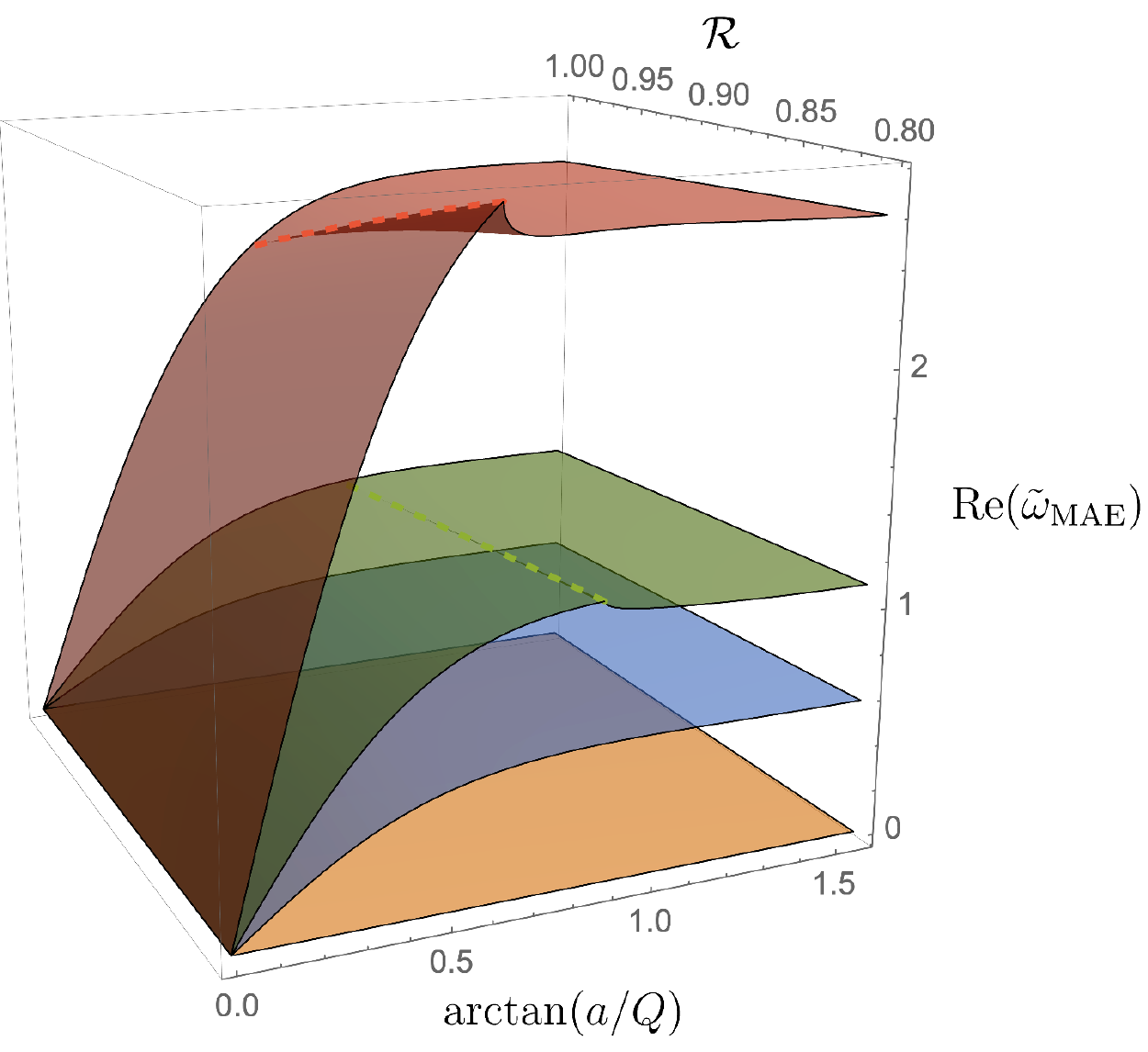}
\caption{Imaginary (left) and real (right) parts of the matched asymptotic expansion $\tilde{\omega}_{\hbox{\tiny MAE}}$ given by~\eqref{NEfreq}, for KN families with $\mathcal{R} \ge 0.8$, with $m = \ell = 0, 1$  (yellow, blue surfaces) and $m = \ell =2, 5$ (green, red surfaces). In the latter two cases, which are representative of the $m=\ell\geq 2$ cases, there are two regions with distinct behaviours demarcated by a critical value $\Theta_{\rm c}$ indicated by a dashed line.}
\label{Fig:NH-MAE_all_m}
\end{figure}

To understand this feature of the matched asymptotic expansion better, in Fig.~\ref{Fig:NH-MAE_all_m} we plot the imaginary (left panel) and real (right panel) parts of the matched asymptotic expansion $\tilde{\omega}_{\hbox{\tiny MAE}}$, for several values of $m = \ell = 0, 1, 2, 5$ (yellow, blue, green, red). For $m=\ell\geq 2$, as the cases $m = \ell = 2$ and $m = \ell = 5$ in Fig.~\ref{Fig:NH-MAE_all_m} illustrate, this sharp change in behaviour is present, but not for $m = \ell = 0$ or $m = \ell = 1$. This change occurs at a critical value $\Theta_{\rm c}$, indicated by dashed lines, which is the value of $\Theta$ at which the argument of the square root in~\eqref{NEfreq} changes sign, or equivalently when $\delta^{2}$ changes sign, where $\delta$ is defined by~\eqref{def:delta}: $\delta^{2}$ is negative when $\Theta < \Theta_{\rm c}$ and positive when $\Theta > \Theta_{\rm c}$.


It is important to note that the matched asymptotic expansion we performed captures any modes that approach the superradiant bound $\omega  \to m \,\Omega_H$ (with vanishing imaginary part) in the extremal limit, regardless of whether they are associated to `NH' or `PS' modes in the RN limit (or any other mode classification we choose). In particular, we will find `PS modes' that approach the superradiant bound, and when this occurs the matched asymptotic expansion frequency $\tilde{\omega}_{\hbox{\tiny MAE}}$ provides an excellent approximation near extremality, even for small values of $m$ and $\ell$ where the WKB expansion~\eqref{WKBfrequency} might fail to give a good approximation (see the last three plots of the later Fig.~\ref{Fig:WKBn0-NHn0-Im}).

In the eikonal limit ($m\to \infty$) and at extremality, one finds that $\Theta_{\rm c}|_{m\to\infty} \to \pi/6\simeq 0.524$. Interestingly, we can see that this value is in agreement with the eikonal expectation that the PS modes have vanishing imaginary part in the near-extremal limit when $\Theta \ge \Theta_{\star}^{\footnotesize \rm eik}$ (where $ \Theta_{\star}^{\footnotesize \rm eik} \equiv \pi/6$ was introduced in the last paragraph of section~\ref{sec:PSwkbHighOrders} and in the discussion of Fig.~\ref{Fig:PSradii}). However, as discussed in the same paragraph, this is no longer the case once we include the higher-order WKB corrections~\eqref{WKBfrequency} since this frequency approaches $\omega=m\Omega_H$ at extremality only for $\Theta\ge \Theta_{\star}^{\footnotesize \rm WKB}$ with $ \Theta_{\star}^{\footnotesize \rm WKB} > \Theta_{\star}^{\footnotesize \rm eik}$. 
The expectation is that if we extend the large $m$ WKB analysis of section~\ref{sec:PSwkbHighOrders} beyond third order $\mathcal{O}(m^{-3})$ to increasingly higher orders (so that it progressively more accurately describes small $m$ modes), one would observe $\Theta_{\star}^{\footnotesize \rm WKB}$ approaching $\Theta_{\rm c}\sim 0.881$ from below. Conversely, we have observed that as $m=\ell$ increases, the associated $\Theta_{\rm c}$ approaches $\Theta_{\star}^{\footnotesize \rm WKB}$.

For gravito-electromagnetic perturbations in KN~\cite{Dias:2022oqm}, there is a separation constant $\lambda_{2}$ which plays the role of $\delta^{2}$, and there it was shown that the vanishing of $\lambda_{2}$ 
provides a very good indication of the point where PS modes want to reach vanishing imaginary part in the extremal limit~\cite{Dias:2022oqm}.
 For the scalar field case, the sign of $\delta^{2}$ has been shown to at least roughly correspond to whether there are one or two families~\cite{Zimmerman:2015trm} in the extremal limit, however, establishing whether the sign of $\delta^{2}$ is a sharp criteria for KN in the scalar field case was not previously studied in detail. Later, when discussing Figs.~\ref{Fig:WKBn0-NHn0-Im}$-$\ref{Fig:WKBn1-NHn4-ImB}, we will find that 1) 
PS modes do attempt to approach $\mathrm{Im}\, \omega\to 0$ and $\mathrm{Re}\, \omega\to m\Omega_H^{\hbox{\footnotesize{ext}}}$ for large $\Theta$ at extremality, and 2) NH modes always start at $\mathrm{Im}\, \omega\to 0$ and $\mathrm{Re}\, \omega\to m\Omega_H^{\hbox{\footnotesize{ext}}}$ at extremality for any value of $\Theta$. Proceeding with caution, the critical values $\Theta_{\star}^{\footnotesize \rm eik}$ and $\Theta_{\star}^{\footnotesize \rm WKB}$ that emerge from the eikonal/WKB analysis and $\Theta_{\rm c}$ that emerges from the near-horizon analysis are to be seen only as rough reference values signalling where one expects some change in the qualitative behaviour of the QNM spectra. They are rough references because these quantities emerge from WKB or near-horizon analytical considerations that are just approximation analyses, but also because 
these expectations can be subverted by eigenvalue repulsions in KN (which are not present in the RN or Kerr limits), as we discuss next.

\section{Eigenvalue repulsions} \label{sec:EigenvalueRepulsions}

\subsection{Eigenvalue or level repulsion, avoided crossing or Wigner-Teller effect}\label{sec:EigenvalueRepulsionsA}

Eigenvalue repulsions are ubiquitous in eigenvalue problems, for both classical and quantum mechanical systems, where it goes by the name \emph{level repulsion}, \emph{avoided crossing} or the\emph{Wigner-Teller effect}~\cite{Landau1981Quantum,Cohen-Tannoudji:1977}. For example, in solid state physics eigenvalue repulsion is responsible for the energy gap between different energy bands of simple lattice models~\cite{Kittel:2004}. However, this phenomenon has only recently been observed or, at least,  correctly understood/identified as such in the QNM spectra of black holes. In this section, we give a brief discussion of this phenomenon using the analogy of a two-level system, and explain why one only expects eigenvalue repulsions to occur in the QNM spectra of black hole families with two or more dimensionless parameters (\eg in Kerr-Newman) but not in black holes parametrized by a single dimensionless parameter (\eg RN or Kerr). We ask the reader to see section 4.1 of~\cite{Dias:2022oqm} for a more thorough treatment of the argument sketched here.

As an example, let us consider $\mathcal{L}_{0}$ to be an operator schematically representing the eigenvalue problem $\mathcal{L}_{0} \psi = \omega \, \psi$ given by~\eqref{KG:radial}-\eqref{KG:ang} subject to QNM boundary conditions, for some fixed value of $\Theta$, \eg RN $(\Theta = 0)$. We select two eigenfunctions ${\psi_{1}, \psi_{2}}$ whose associated eigenvalues $\omega_{1}$ and $\omega_{2}$  are distinct but very close in the complex plane. For example, $\omega_{1}$ could be the dominant PS eigenvalue and $\omega_{2}$ the dominant NH eigenvalue for some specific RN BH, which never coincide in the complex plane for any value of $\mathcal{R}$ (as discussed later in Fig.~\ref{Fig:RN}).

Now, we perturb the operator, $\mathcal{L} = \mathcal{L}_{0} + \mathcal{K}$, by turning on angular momentum, $\Theta > 0$, and ask how the eigenvalues change. We make the zeroth-order approximation that the perturbed eigenfunctions $\bar{\psi}$ are a linear combination of the unperturbed basis, $\bar{\psi} = c_{1} \psi_{1} + c_{2} \psi_{2}$, which leads to the perturbed eigenvalue problem $(\mathcal{L}_{0} + \mathcal{K}) \bar{\psi} = \bar{\omega} \,\bar{\psi}$. The matrix representing this eigenvalue problem can then be written as
\begin{equation}
  \left[\begin{matrix}
    \omega_{1} + \mathcal{K}_{11} & \mathcal{K}_{12} \\
    \mathcal{K}_{21}              & \omega_{2} + \mathcal{K}_{22}
  \end{matrix}\right]
\end{equation}
where $\mathcal{K}_{ij}$ are the matrix elements of $\mathcal{K}$ in the $\{\psi_{1}, \psi_{2}\}$ basis, defined with respect to some suitable inner product, and the eigenvalues $\bar{\omega}_{\pm}$ of this perturbed equation are given by
\begin{equation}
  \bar{\omega}_{\pm} = \frac{\tilde{\omega}_{1} + \tilde{\omega}_{2}}{2} \pm \sqrt{\frac{(\tilde{\omega}_{1} - \tilde{\omega}_{2})^{2}}{4} + \mathcal{K}_{12}\mathcal{K}_{21}}\,,
\end{equation}
where $\tilde{\omega}_{i} \equiv \omega_{i} - \mathcal{K}_{ii}$ (no Einstein summation over $i$). The perturbed eigenvalues $\bar{\omega}_{\pm}$ can only cross if the argument of the square root vanishes, \ie if and only if
\begin{equation}\label{ERcond}
  \frac{(\tilde{\omega}_{1} - \tilde{\omega}_{2})^{2}}{4} + \mathcal{K}_{12} \mathcal{K}_{21} = 0.
\end{equation}
This complex condition gives rise to two real conditions, which both need to be satisfied for an eigenvalue crossing. In general, the matrix elements of the perturbed operator $\mathcal{K}_{ij}$ will depend on the $N$ real black hole parameters. With the exception of some symmetry that reduces the number of conditions required, we thus expect that eigenvalue crossing can only occur on an $N-2$ dimensional subspace. KN is parameterized by two dimensionless parameters $\{\mathcal{R}, \Theta\}$, and hence eigenvalue crossing (in the complex plane) can only occur at isolated points in the parameter space. This simple argument might explain why eigenvalue repulsions have been observed in Kerr-Newman~\cite{Dias:2021yju,Dias:2022oqm}, Reissner-Nordstr\"om-de Sitter~\cite{Dias:2020ncd} and Myers-Perry-de Sitter~\cite{Davey:2022vyx}, but not in RN, Kerr or Schwarzschild~\cite{Regge:1957td,Zerilli:1974ai,Moncrief:1974am,Chandrasekhar:1975zza,Moncrief:1974gw,Moncrief:1974ng,Newman:1961qr,Geroch:1973am,Teukolsky:1972my,Detweiler:1980gk,Chandra:1983,Leaver:1985ax,Whiting:1988vc,Onozawa:1996ux,Berti:2003jh,Berti:2005eb,Yang:2012pj}.
It is also important to note that the above analysis leaves room for the following scenario. If the background system has a parameter space with a boundary (e.g. the 1-dimensional extremal boundary in the KN black hole case or the 0-dimensional extremal endpoint in the Kerr and RN cases), two eigenvalue families might be able to meet in the complex plane at this extremal boundary (or at a portion of it if 1-dimensional). This is not  an eigenvalue crossing  in the complex plane (since it occurs at a boundary) and thus is not  ruled out by the above analysis; instead it is a special case where two eigenvalue families meet and {\it terminate} at a {\it boundary} of the parameter space. 
 
Having understood (in section~\ref{sec:AnalyticalPS-NH}) that the QNM spectra of KN has two families of modes (PS and NH) and that KN is a 2-dimensional parameter family of black holes, we might now expect the existence of one point (or, at most, a few isolated points) in the KN parameter space where we might see the PS and NH modes trying to approach each other in the frequency complex plane.
What might this point be? 
Well, from the matched asymptotic analysis of section~\ref{sec:NHanalytics} we know that NH modes always start at $\mathrm{Im}\, \omega\to 0$ and $\mathrm{Re}\, \omega\to m\Omega_H^{\hbox{\footnotesize{ext}}}$ at extremality for any value of $\Theta$ and there is an associated critical point $\{\mathcal{R},\Theta_{\rm c}\}\sim\{1, 0.881\}$.\footnote{\label{foot:ThetaStar}Recall that the quantity  $\Theta_{\rm c}$ was introduced in the last paragraph of section~\ref{sec:NHanalytics} (when discussing the cusps in Figs.~\ref{Fig:NH-R0.93} and~\ref{Fig:NH-MAE_all_m}), and that $\Theta_{\star}^{\footnotesize \rm eik}$ and $\Theta_{\star}^{\footnotesize \rm WKB} \ge \Theta_{\star}^{\footnotesize \rm eik}$ were introduced in the discussion of Fig.~\ref{Fig:PSradii} in the last paragraph of section~\ref{sec:PSwkbHighOrders}.} 
On the other hand, the eikonal and WKB analyses of section~\ref{sec:PSwkb}, suggest that PS modes want to approach $\mathrm{Im}\, \omega\to 0$ and $\mathrm{Re}\, \omega\to m\Omega_H^{\hbox{\footnotesize{ext}}}$ at extremality  for $\Theta\geq \Theta_{\star}^{\footnotesize \rm WKB}$ with  $\Theta_{\star}^{\footnotesize \rm WKB}> \Theta_{\star}^{\footnotesize \rm eik}\equiv\pi/6\simeq 0.524$ 
which singles out the special point  $\{\mathcal{R},\Theta_{\star}^{\footnotesize \rm WKB} \}$. The expectation is that if we extend the large $m$ WKB analysis of section~\ref{sec:PSwkb} to increasingly higher orders so that it progressively describes the small $m$ modes more accurately, one would observe $\Theta_{\star}^{\footnotesize \rm WKB}$ approaching $\Theta_{\rm c}$ from below. Onwards, for simplicity, let us thus denote this point simply as $\Theta_\star \sim 0.881$. Given the restrictions on eigenvalue crossings argued previously, and the special point $\Theta_\star$ given by our analytic predictions, there are thus three possibilities for the Kerr-Newman QNM spectra:
\begin{enumerate}
    \item In one of the simplest scenarios, the PS and NH modes have the same frequency at a single point. If so, the MAE and WKB results suggest that this point should be at $\mathcal{R}=1$ and around $\Theta=\Theta_\star\sim 0.881$.
    \item However, since $\mathcal{R} = 1$ happens to be at the 1-dimensional extremal boundary of the KN parameter space, there is also room to actually have both the PS and NH eigenfrequencies meeting and terminating with $\mathrm{Im}\, \omega\to 0$ and $\mathrm{Re}\, \omega\to m\Omega_H^{\hbox{\footnotesize{ext}}}$, not only at the single point $\{\mathcal{R},\Theta_{\star}\}$ but actually along the portion of the extremal boundary parametrized by $\mathcal{R}=1$ and $\Theta_{\star}\leq \Theta \leq \pi/2$. In fact, we will see that this situation certainly occurs for modes within the same family of QNM: all the overtones of the NH modes (the exact numerical frequencies) meet with $\mathrm{Re}\,\omega=m\Omega_H$ and $\mathrm{Im}\,\omega =0$ at extremality for RN, Kerr and KN. Therefore, it seems reasonable that two distinct families of QNMs (namely, the PS and NH modes) might also meet and terminate along a 1-parameter portion of the extremal KN boundary ($\mathcal{R}=1$ and $\Theta_{\star}\leq \Theta \leq \pi/2$), perhaps with the appearance of eigenvalue repulsions near-extremality when they do or attempt to do so.
    \item The final scenario, that cannot be excluded, is that the PS and NH eigenvalues never coincide, not even at $\{\mathcal{R},\Theta_{\star}\}$.
\end{enumerate}
 What ends up happening in the KN QNM spectra? This question will be addressed in the next section. We will do a detailed numerical search of the $m=\ell=2$ PS and NH frequencies, some of which will be displayed in Figs.~\ref{Fig:RN}$-$\ref{Fig:Kerr}. From this analysis, we will conclude that: 1) NH modes indeed exist and always approach $\mathrm{Im}\, \omega\to 0$ and $\mathrm{Re}\, \omega\to m\Omega_H^{\hbox{\footnotesize{ext}}}$ at extremality, and 2) PS modes indeed seem to be strongly willing to approach $\mathrm{Im}\, \omega\to 0$ and $\mathrm{Re}\, \omega\to m\Omega_H^{\hbox{\footnotesize{ext}}}$ at extremality for $\Theta>\Theta_{\star}$.  However, intricate eigenvalue repulsions will typically kick in close to extremality and for $\Theta \gtrsim \Theta_\star$ (as a rough indication) which will break the monotony of the system that was present for smaller values of $\mathcal{R}$ and/or $\Theta$.

\subsection{Eigenvalue repulsions in the scalar field spectra of KN}\label{sec:EigenvalueRepulsionsB}

The quasinormal mode spectra of the Kerr-Newman black hole has two distinct families of modes.
In the Reissner-Nordstr\"om (RN) limit (\ie $a=0$ or $\Theta=0$) we can undoubtedly associate one of these families to the photon sphere (PS) modes and the other to the near-horizon (NH) modes. This is because when $\Theta=0$, the PS family is well approximated by $\tilde{\omega}_{\hbox{\tiny WKB}}$ in~\eqref{WKBfrequency}, while the NH family is well described by $\tilde{\omega}_{\hbox{\tiny MAE}}$ in~\eqref{NEfreq}. 

\begin{figure}[t]
\centering
\includegraphics[width=.49\textwidth]{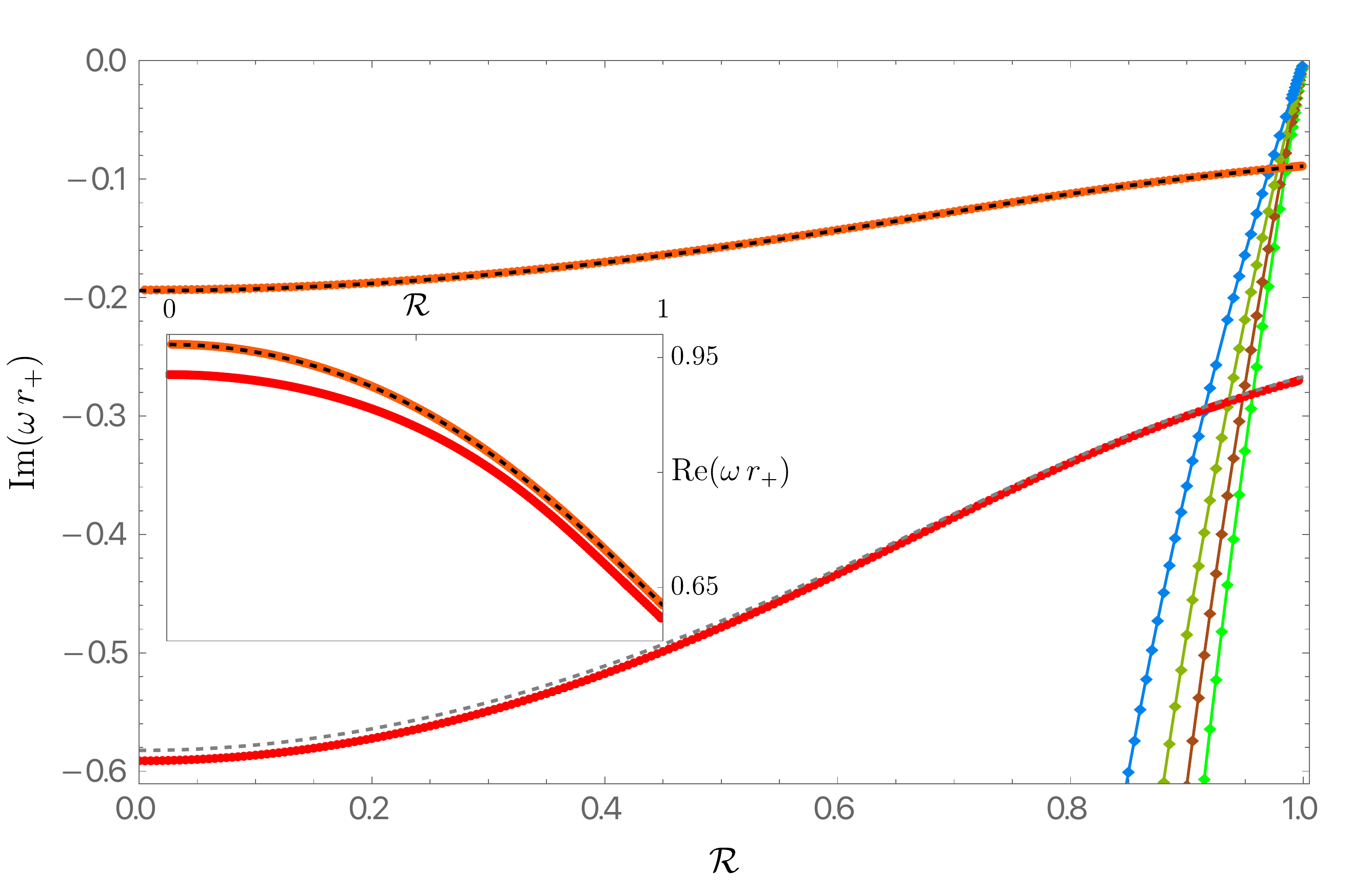}
\hspace{0.0cm}
\includegraphics[width=.49\textwidth]{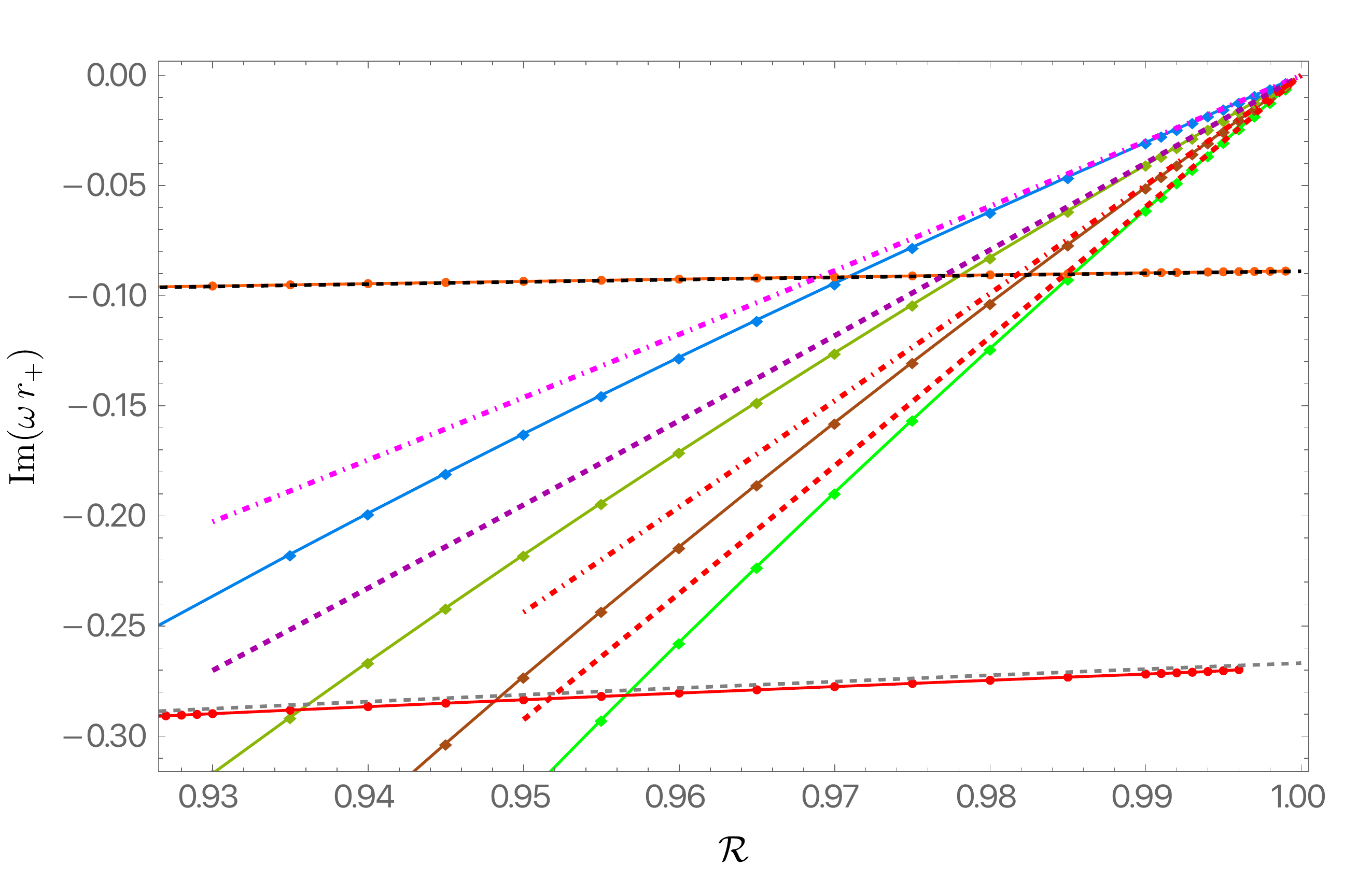}
\caption{The first few overtone PS and  NH families of QNMs with  $m=\ell=2$ for the Reissner-Nordstr\"om black hole (\ie the KN black hole with $\Theta=0$ and thus $Q/r_+=\mathcal{R}, a=0$).
The orange (red) disk curve is the $n=0$ ($n=1$) PS family, while the blue, dark-green, brown and green diamond curves are the NH families with $n=0$, $n=1$, $n=2$ and $n=3$, respectively.
 {\bf Left panel:} the main plot displays the imaginary part of the dimensionless frequency as a function of $\mathcal{R}$. On the other hand the inset plot displays the real part of the frequency of the $n=0$ and $n=1$ PS families (all the NH families have $\hbox{Re}\,\omega =0$ in the RN limit). The black (gray) dashed line that is almost on the top of the $n=0$ ($n=1$) PS numerical curve is the analytical WKB approximation
$\tilde{\omega}_{\hbox{\tiny WKB}}$ given by \eqref{WKBfrequency}.
 {\bf Right panel:} Zoom of the left panel in the near-extremal region (\ie around $\mathcal{R}\sim 1$ where the NH families approach $\hbox{Im}\,\tilde{\omega} \to 0$ as $\mathcal{R}\to 1$). This time we also show, as dot-dashed or dashed lines, the near-extremal approximation 
$\tilde{\omega}_{\hbox{\tiny MAE}}$ as from~\eqref{NEfreq} for $p=0,1,2,3$. We see that the latter approximate the NH frequencies very well when we are close to extremality $\mathcal{R}\sim 1$, as expected. (The counterpart of this figure for the Kerr case is displayed in Fig.~\ref{Fig:Kerr}).}
\label{Fig:RN}
\end{figure}  

This is illustrated in Fig.~\ref{Fig:RN}, where we plot the $n=0$ (orange disks) and $n=1$ (red disks) PS QNM frequencies as well as $\tilde{\omega}_{\hbox{\tiny WKB}}$ given by the black ($n=0$) and gray ($n=1$) dashed lines. We see that the latter higher-order WKB curves are on top of the numerical PS curves, indicating that \eqref{WKBfrequency} provides an excellent approximation for the PS family of RN QNMs and its overtones, and allows one to identify them in the RN limit.
Additionally, in Fig.~\ref{Fig:RN} we also plot the $n=0,1,2,3$ (blue, dark-green, brown and green diamonds) NH QNM frequencies and  $\tilde{\omega}_{\hbox{\tiny MAE}}$ (magenta and red dot-dashed and dashed lines). We see that the latter matched asymptotic expansions approximate the NH frequencies of RN very well when we are close to extremality (\ie as $\mathcal{R}\to 1$). This clearly identifies the NH family of QNMs and their overtones in the RN limit.
As pointed out in \cite{Zimmerman:2015trm},
the PS modes (a.k.a. damped modes in \cite{Zimmerman:2015trm}) of RN are very well known in the literature, starting with the WKB analysis of \cite{Kokkotas:1988fm}. However, the existence of the NH modes (a.k.a. zero-damped modes in \cite{Zimmerman:2015trm}) in RN seems to have been missed until the work of \cite{Zimmerman:2015trm}, in spite of the seminal work of Teukolsky and Press \cite{Teukolsky:1974yv} and Detweiler \cite{Detweiler:1980gk} already suggesting that such family might be present in any black hole with an extremal configuration. Our PS frequencies in Fig.~\ref{Fig:RN} agree with those first computed in \cite{Kokkotas:1988fm,Leaver:1990zz,Andersson:1993,Onozawa:1995vu,Andersson:1996xw}. 
On the other hand, the NH QNM spectrum in Fig.~\ref{Fig:RN} agrees with the data obtained in \cite{Zimmerman:2015trm} (see its figures 9 and 10). 

One final property of the RN QNM spectra that is worth observing in the context of the eigenvalue repulsions discussed in section~\ref{sec:EigenvalueRepulsionsA} is the fact that the several NH overtone frequencies do meet and terminate  with $\omega=0$ at the extremal RN point $\mathcal{R}=1=Q/r_+$. So we clearly can have different modes meeting and terminating at the boundary of the RN parameter space.

We will {\it lock} the color code of Fig.~\ref{Fig:RN} for the rest of the figures of our manuscript, since this settles a nomenclature to frame our discussions (this rule will not be respected only in Fig.~\ref{Fig:Star}). That is to say,
in all our figures (except Fig.~\ref{Fig:Star}) we will always use orange and red disks to represent the KN QNM families that continuously connect to the  RN $n=0$ and $n=1$ PS families of Fig.~\ref{Fig:RN}, respectively, when $\Theta\to 0$. Similarly,  in all our figures  we will always use the blue, dark-green, brown and green diamonds to represent the KN QNM families that continuously connect to the RN $n=0,1,2,3$ NH families of Fig.~\ref{Fig:RN}, respectively, when $\Theta\to 0$. Moreover, to keep the discussion fluid (but, unfortunately, often misleadingly), we will keep denoting these modes as the PS and NH families. 
However, we will find that, generically, this sharp distinction between the PS and NH families only holds in (a neighbourhood of) the Reissner-Nordstr\"om limit (\ie for small $a/r_+$ or small $\Theta$) and is often lost as $\Theta$ grows and approaches the Kerr limit ($Q=0$ \ie $\Theta=\pi/2$). So much that at a certain point it will be more appropriate to denote the different KN QNM families as `PS-NH' families and their overtones, rather than separate PS or NH families.

\begin{figure}[th]
\centering
\includegraphics[width=.49\textwidth]{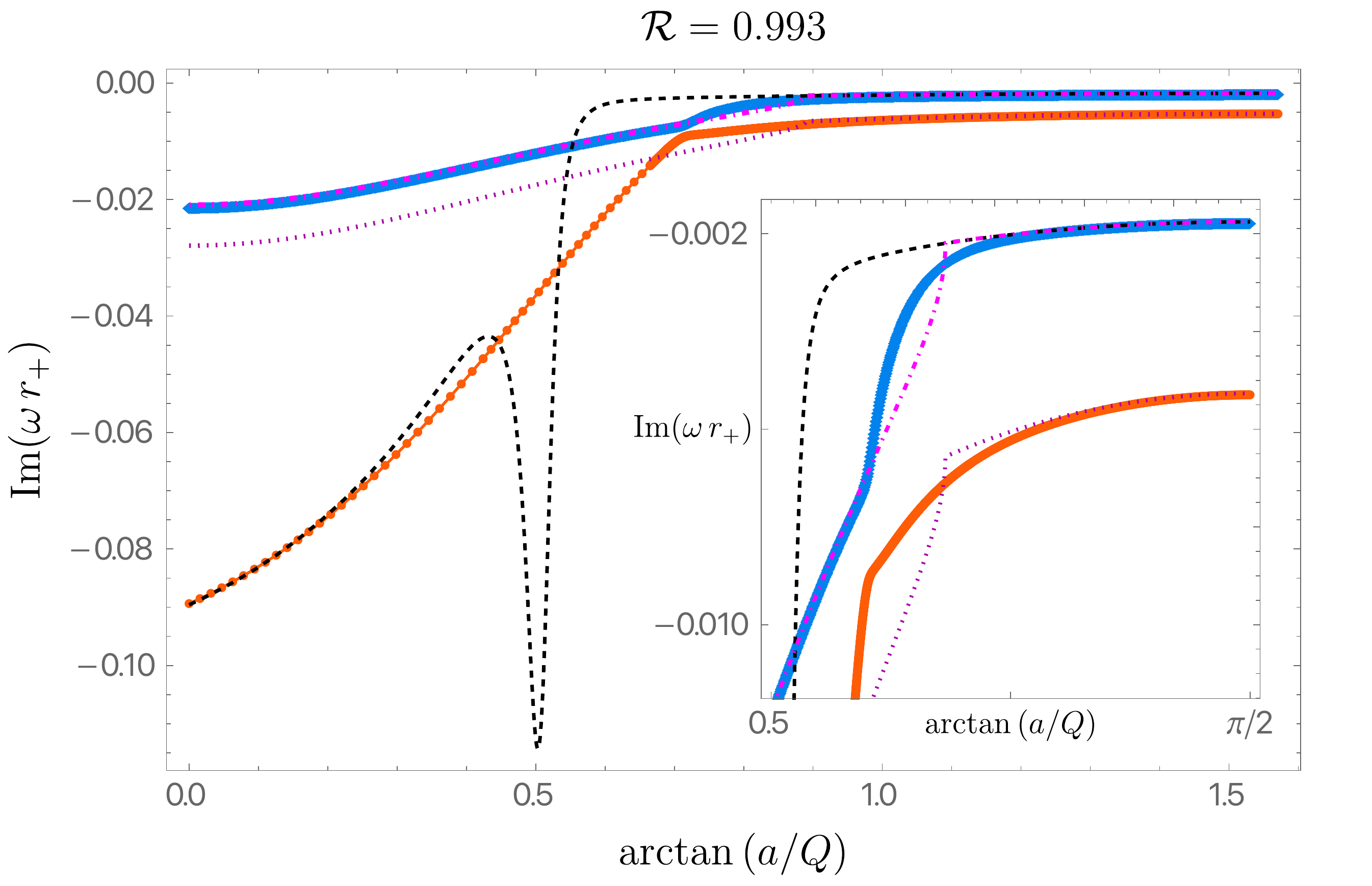}
\hspace{0.0cm}
\includegraphics[width=.49\textwidth]{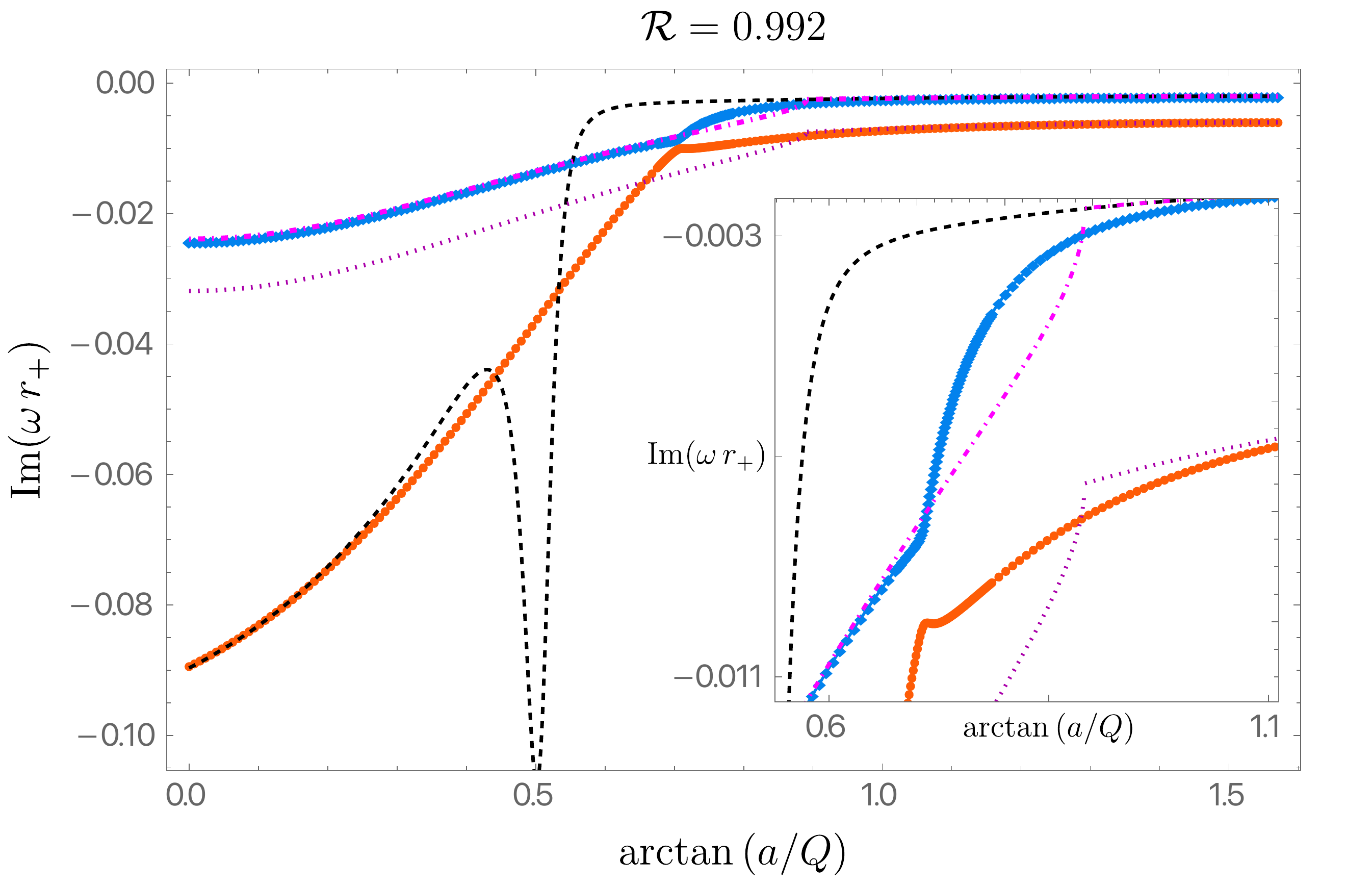}
\vskip 0.2cm
\includegraphics[width=.49\textwidth]{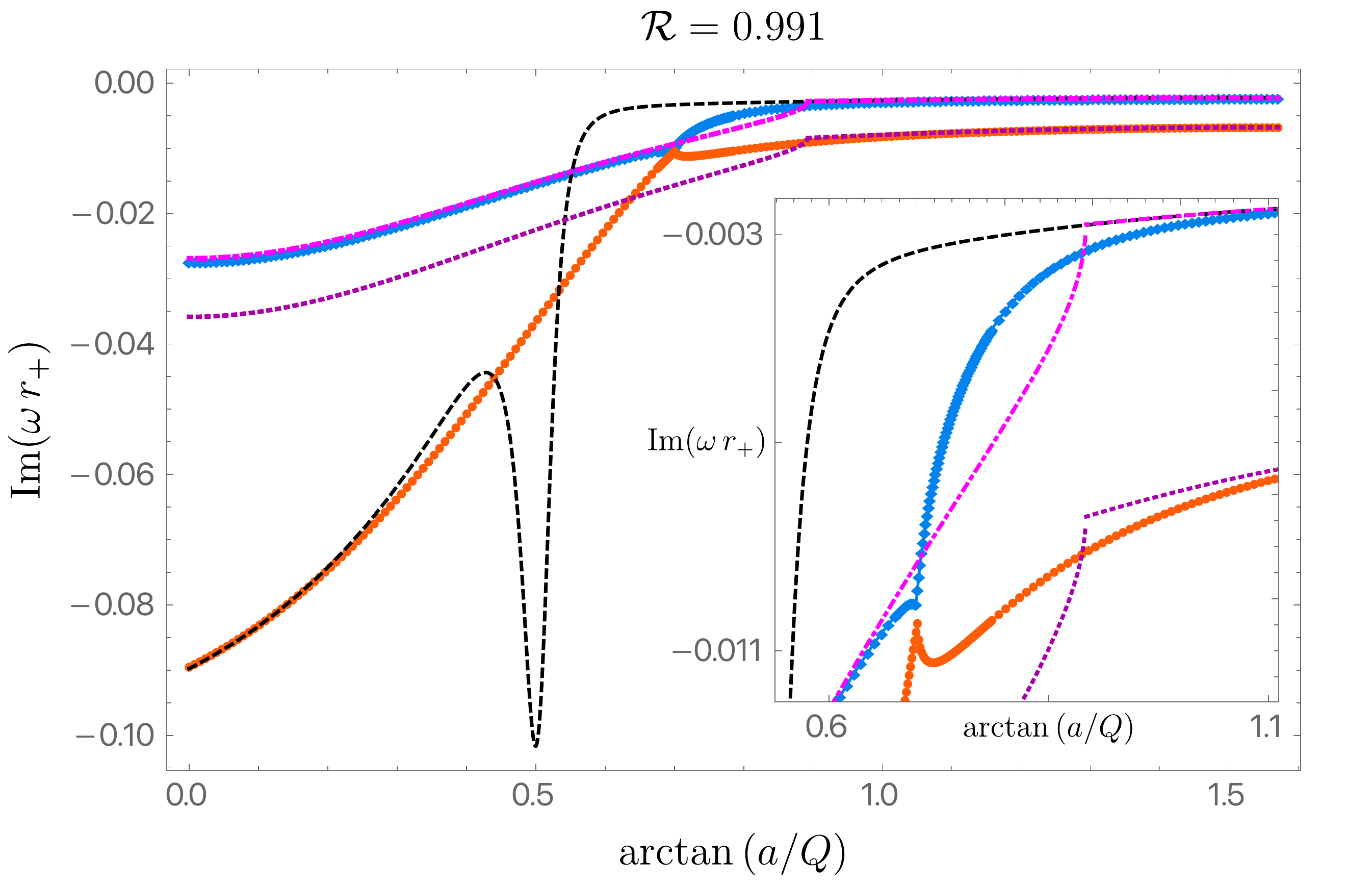}
\hspace{0.0cm}
\includegraphics[width=.49\textwidth]{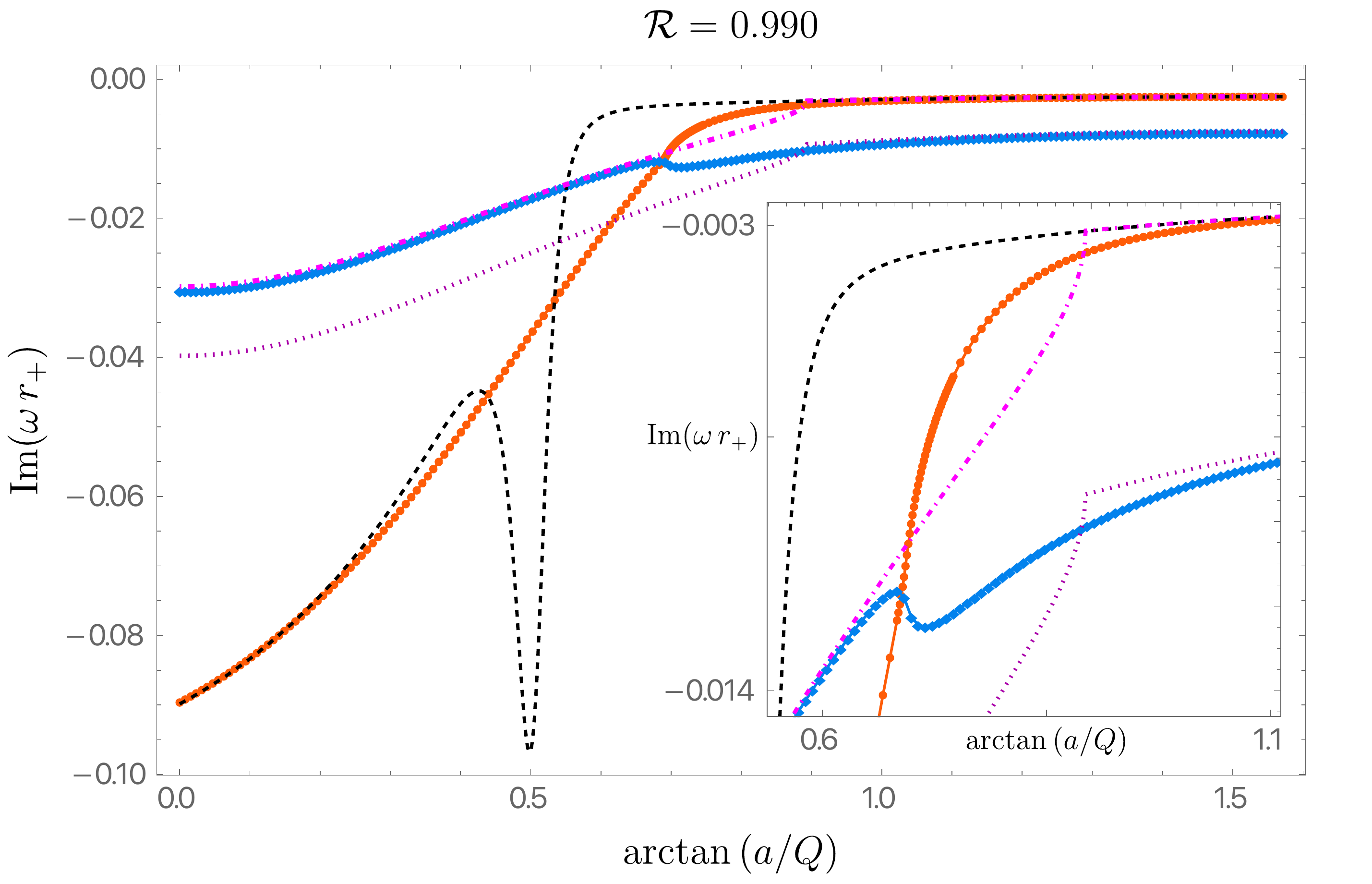}
\vskip 0.2cm
\includegraphics[width=.49\textwidth]{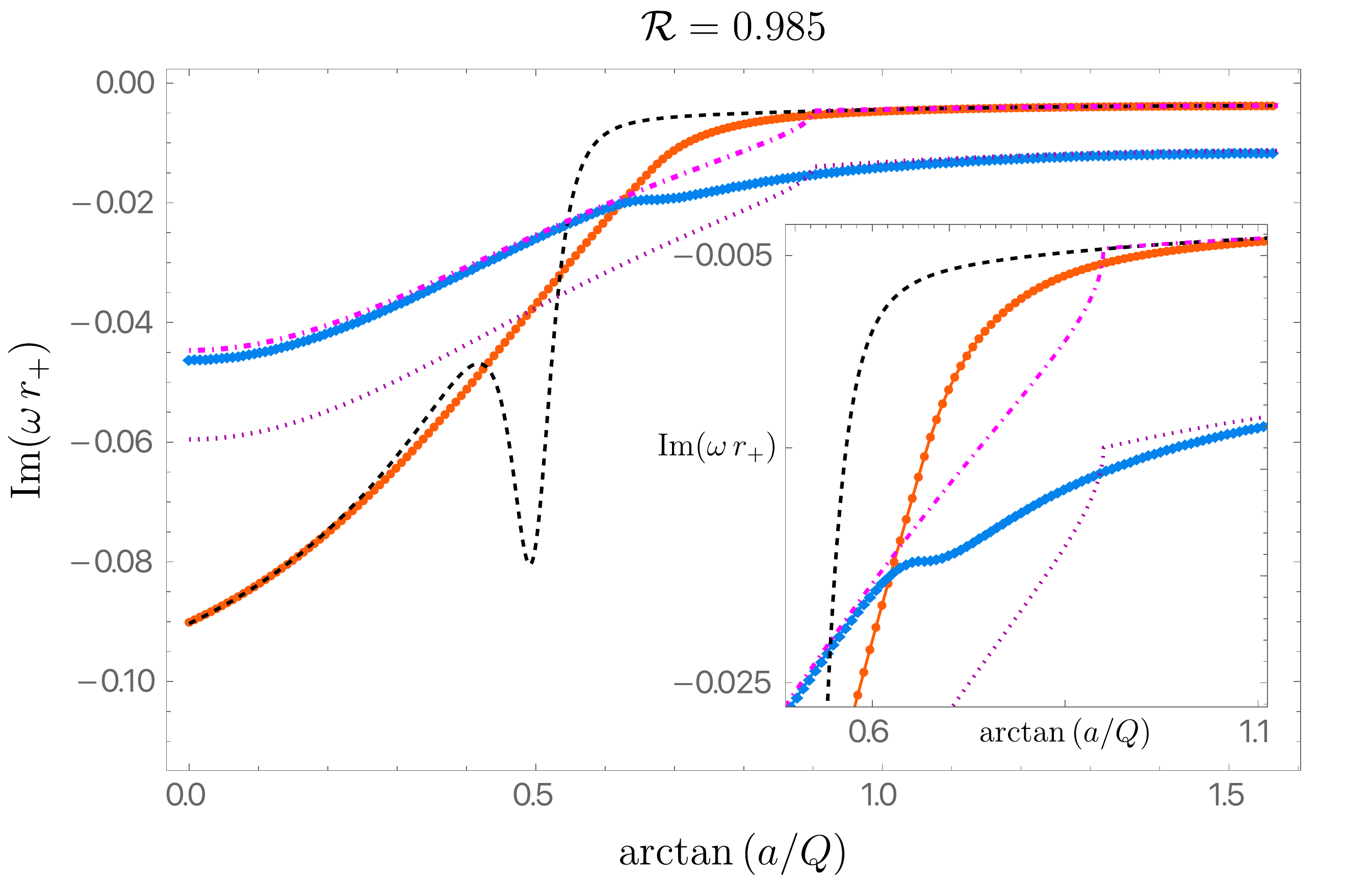}
\hspace{0.0cm}
\includegraphics[width=.49\textwidth]{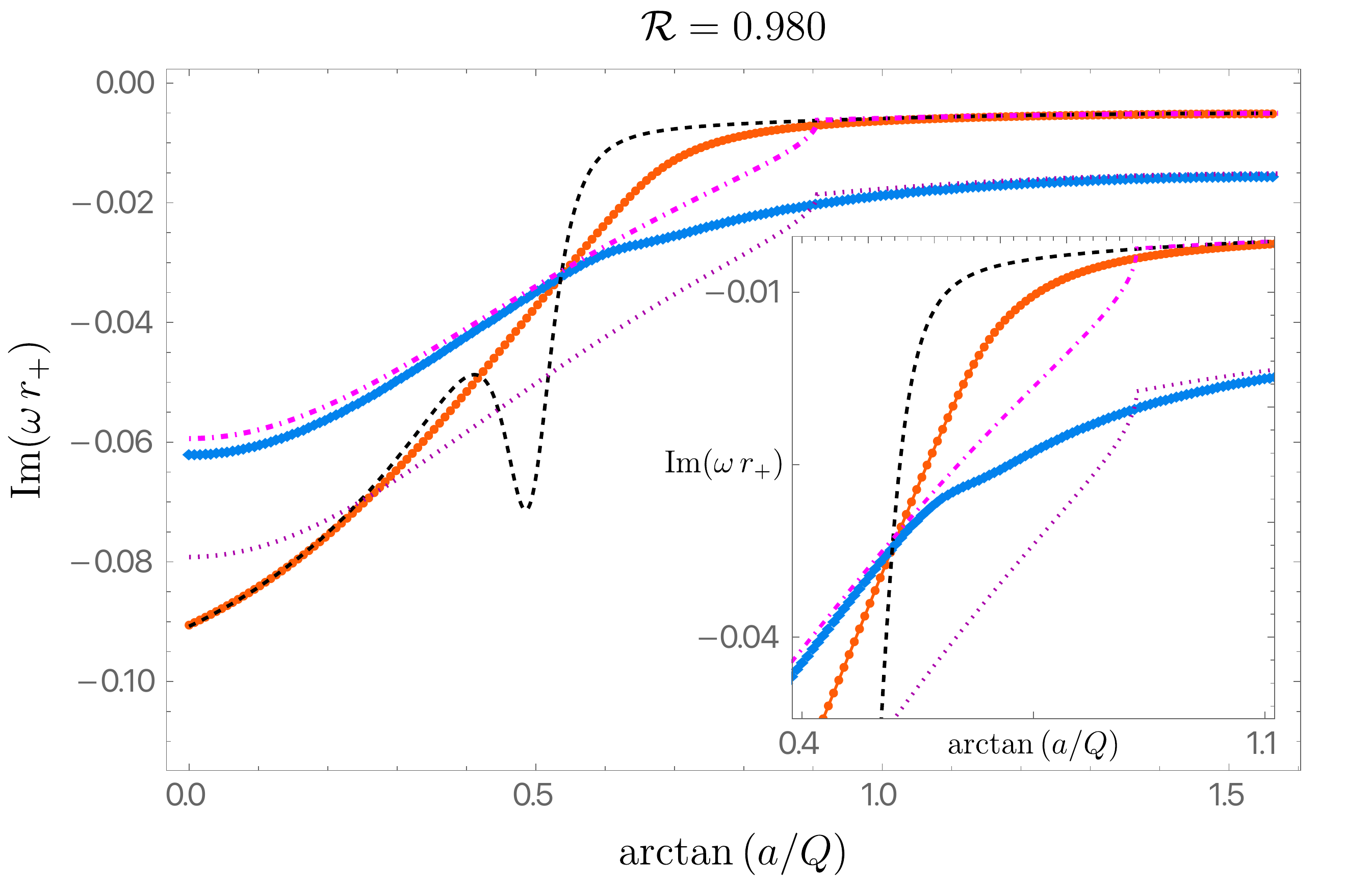}
\caption{Imaginary part of the frequency as a function of $\Theta=\arctan\left( a/Q\right) $ for the PS (orange disks) and the NH (blue diamonds) families of QNMs with  $m=\ell=2, n=0$ for a KN family with, following the lexicographic order, $\mathcal{R}=0.993$, $\mathcal{R}=0.992$, $\mathcal{R}=0.991$, $\mathcal{R}=0.990$,  $\mathcal{R}=0.985$ and $\mathcal{R}=0.980$. We also display the WKB result $\tilde{\omega}_{\hbox{\tiny WKB}}$ (dashed black line) and the near-extremal frequency $\tilde{\omega}_{\hbox{\tiny MAE}} $ for $p=0$ (dot-dashed magenta line) and $p=1$ (dotted dark magenta  line).}
\label{Fig:WKBn0-NHn0-Im}
\end{figure}

\begin{figure}[th]
\centering
\includegraphics[width=.49\textwidth]{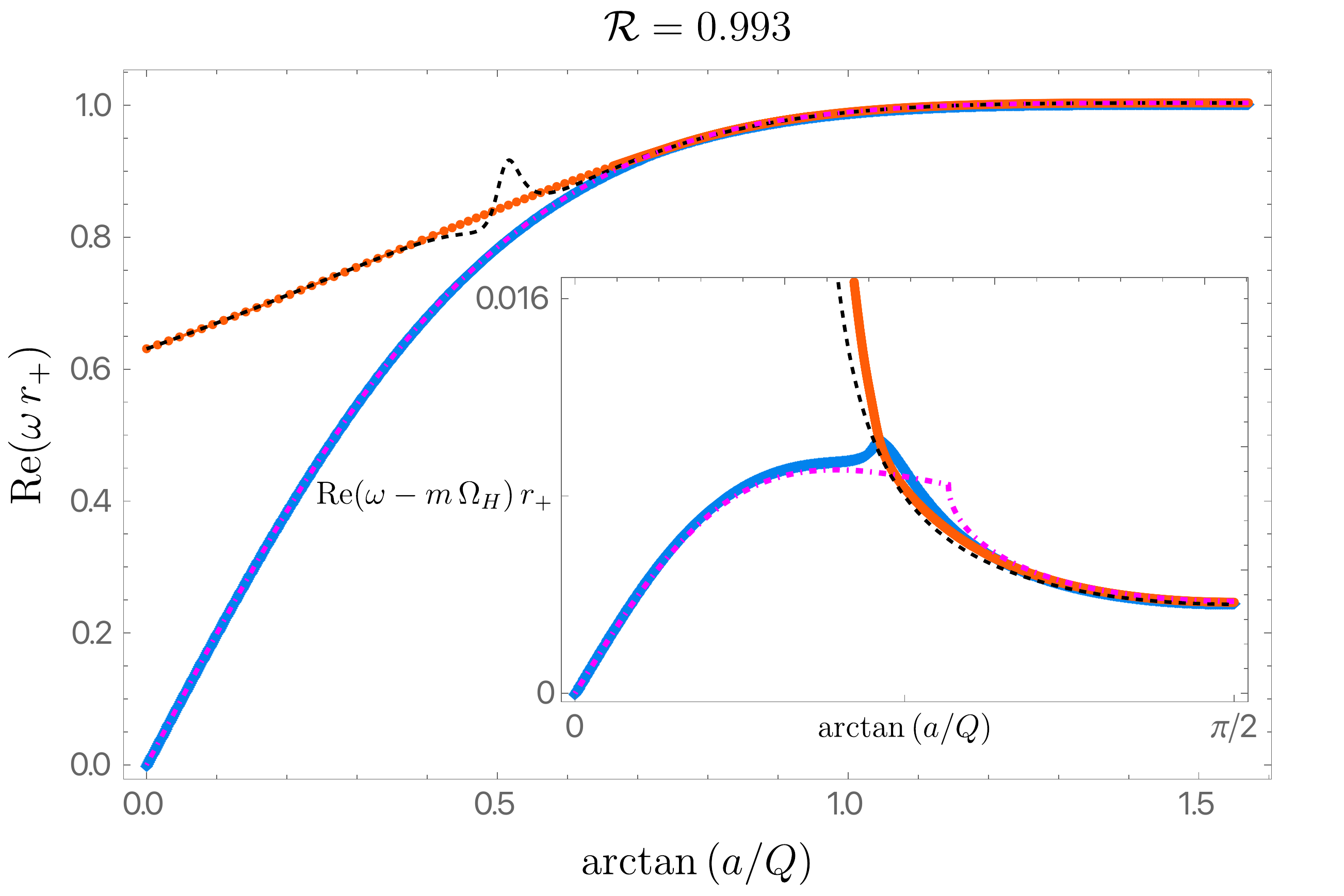}
\hspace{0.0cm}
\includegraphics[width=.49\textwidth]{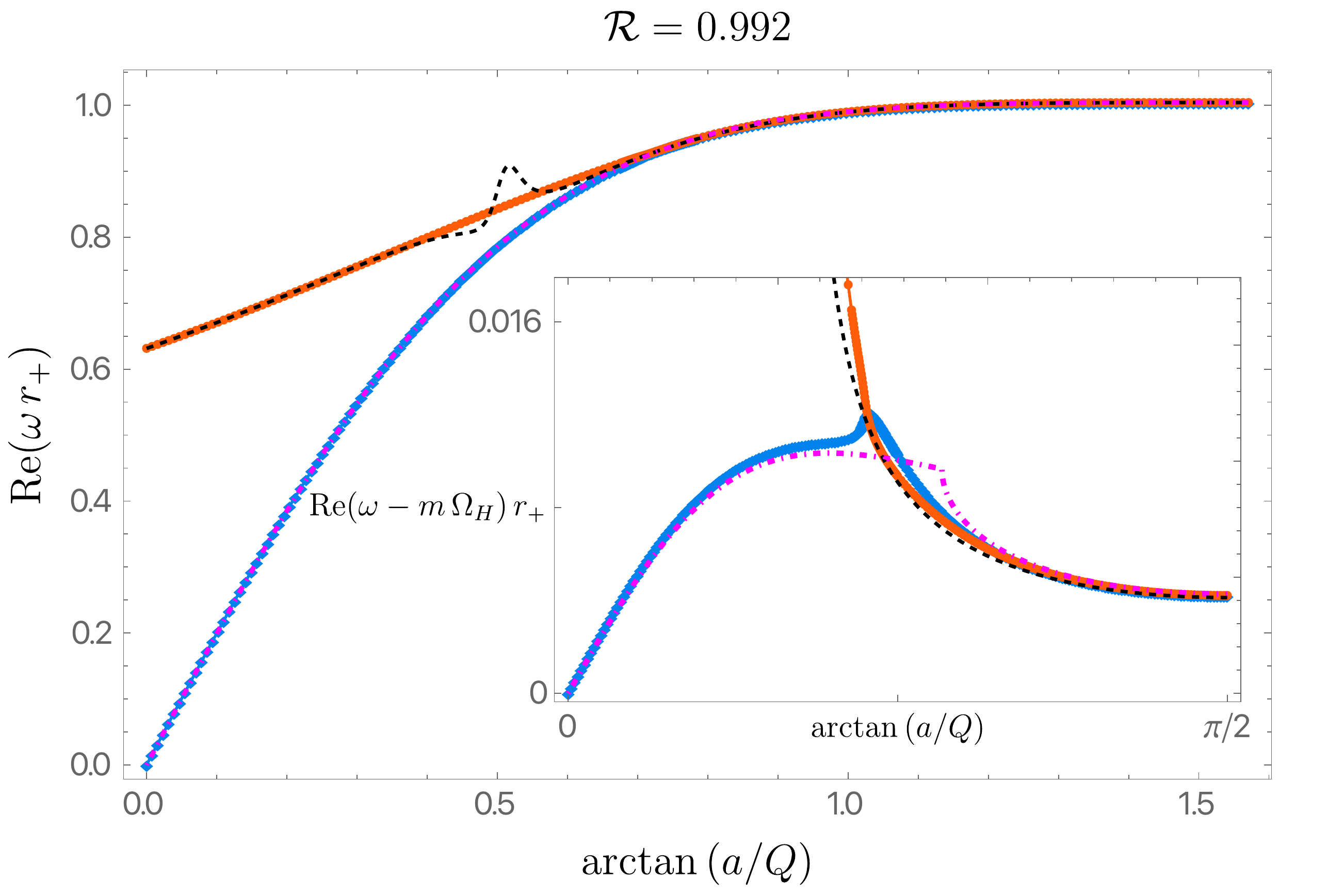}
\vskip 0.2cm
\includegraphics[width=.49\textwidth]{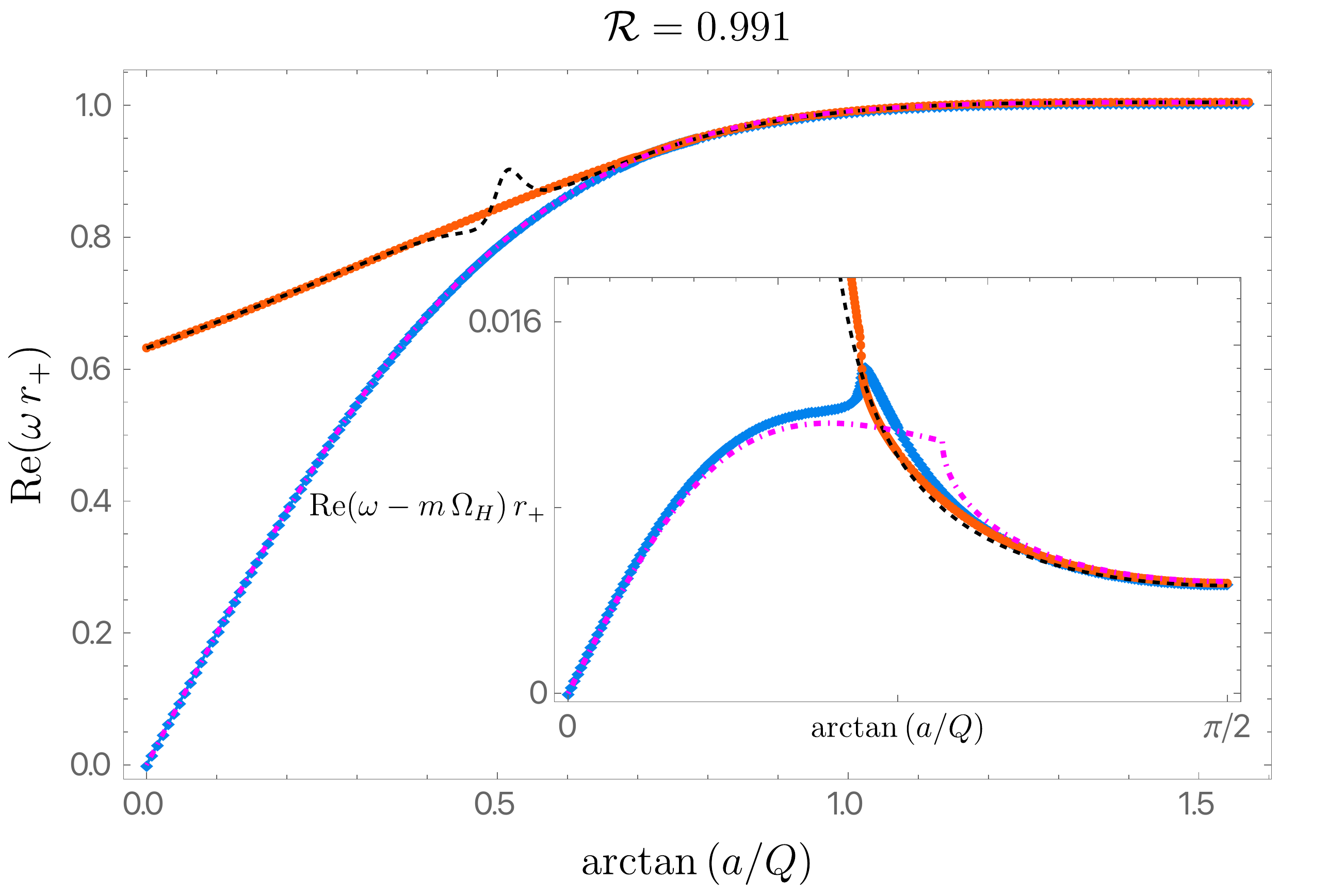}
\hspace{0.0cm}
\includegraphics[width=.49\textwidth]{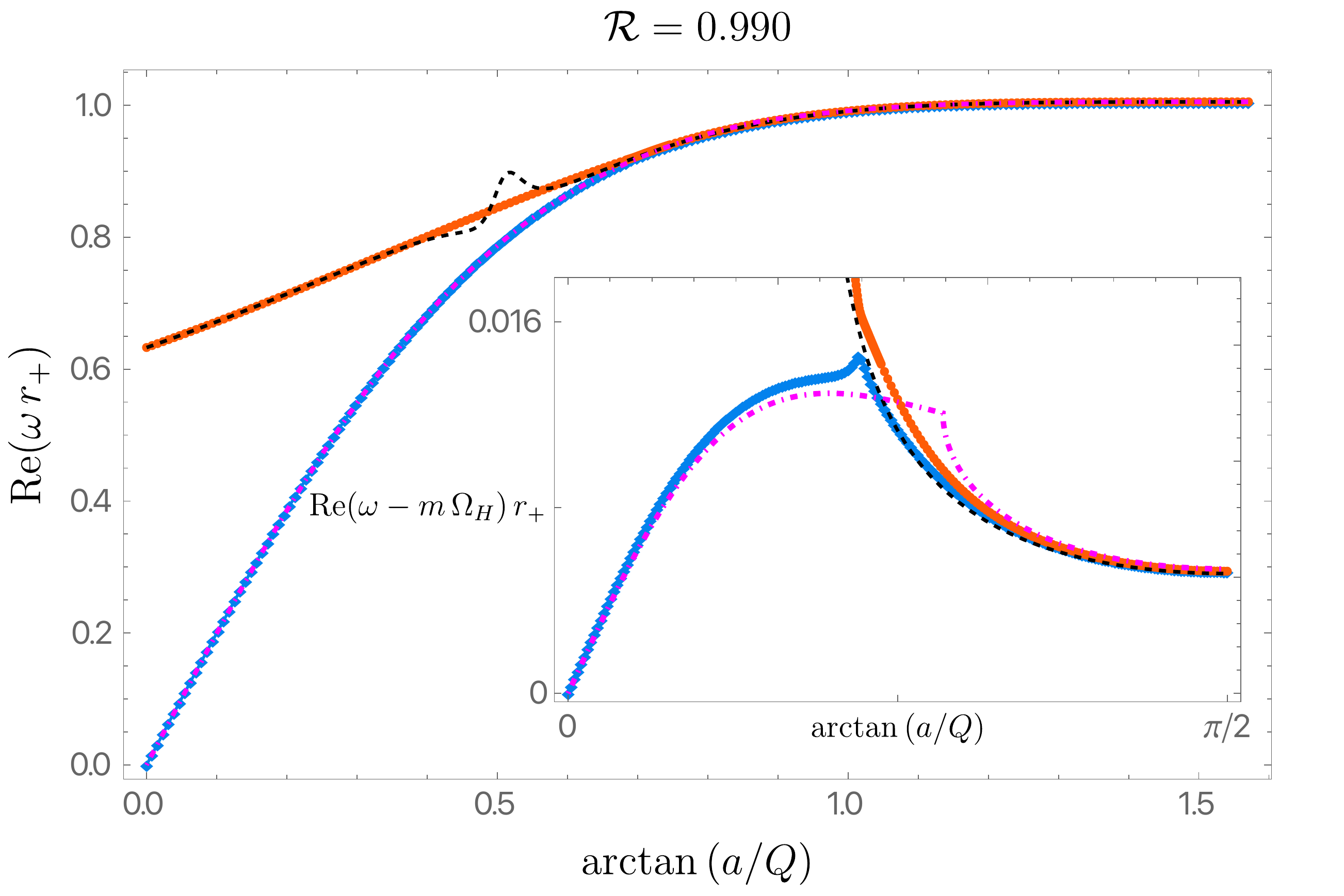}
\vskip 0.2cm
\includegraphics[width=.49\textwidth]{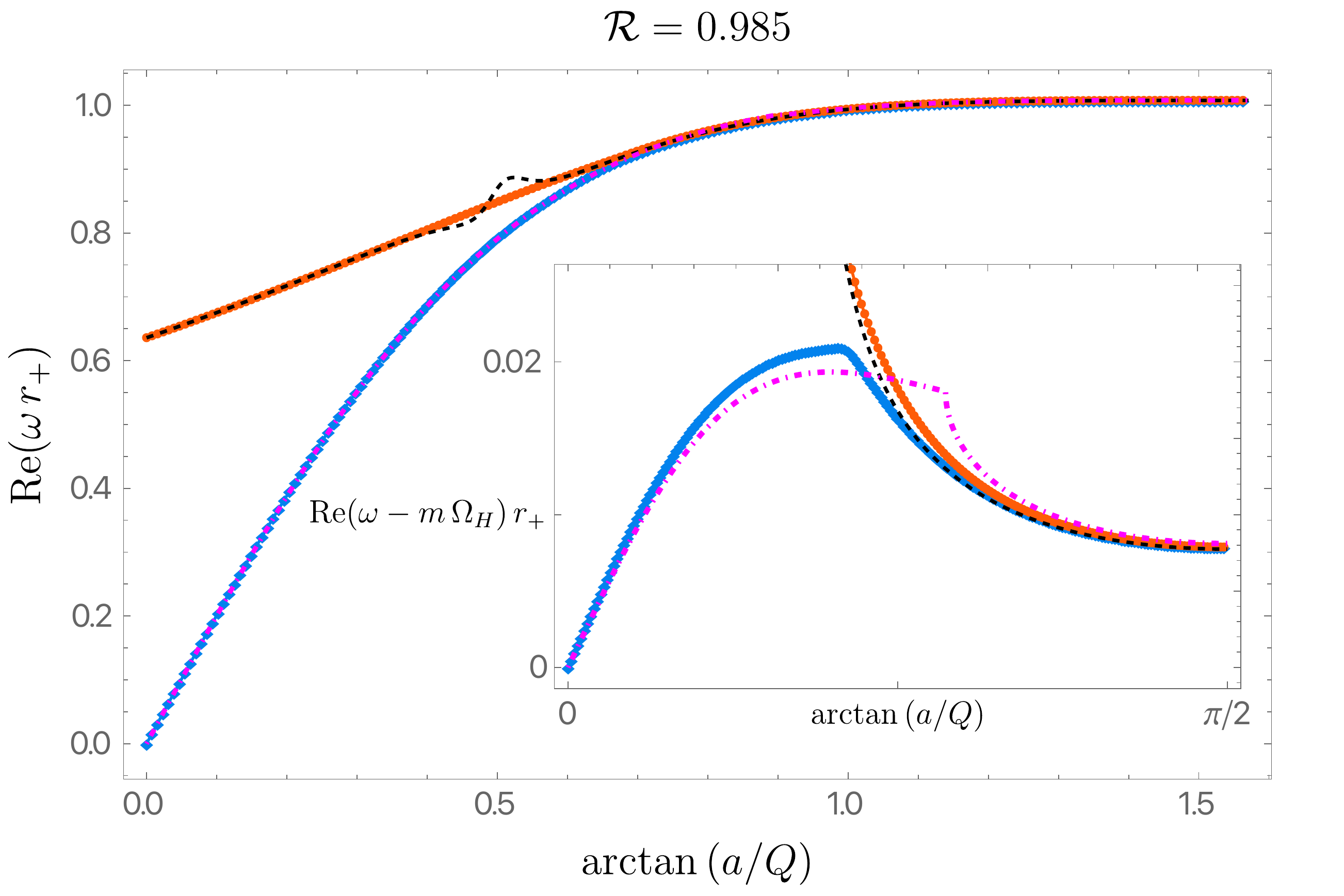}
\hspace{0.0cm}
\includegraphics[width=.49\textwidth]{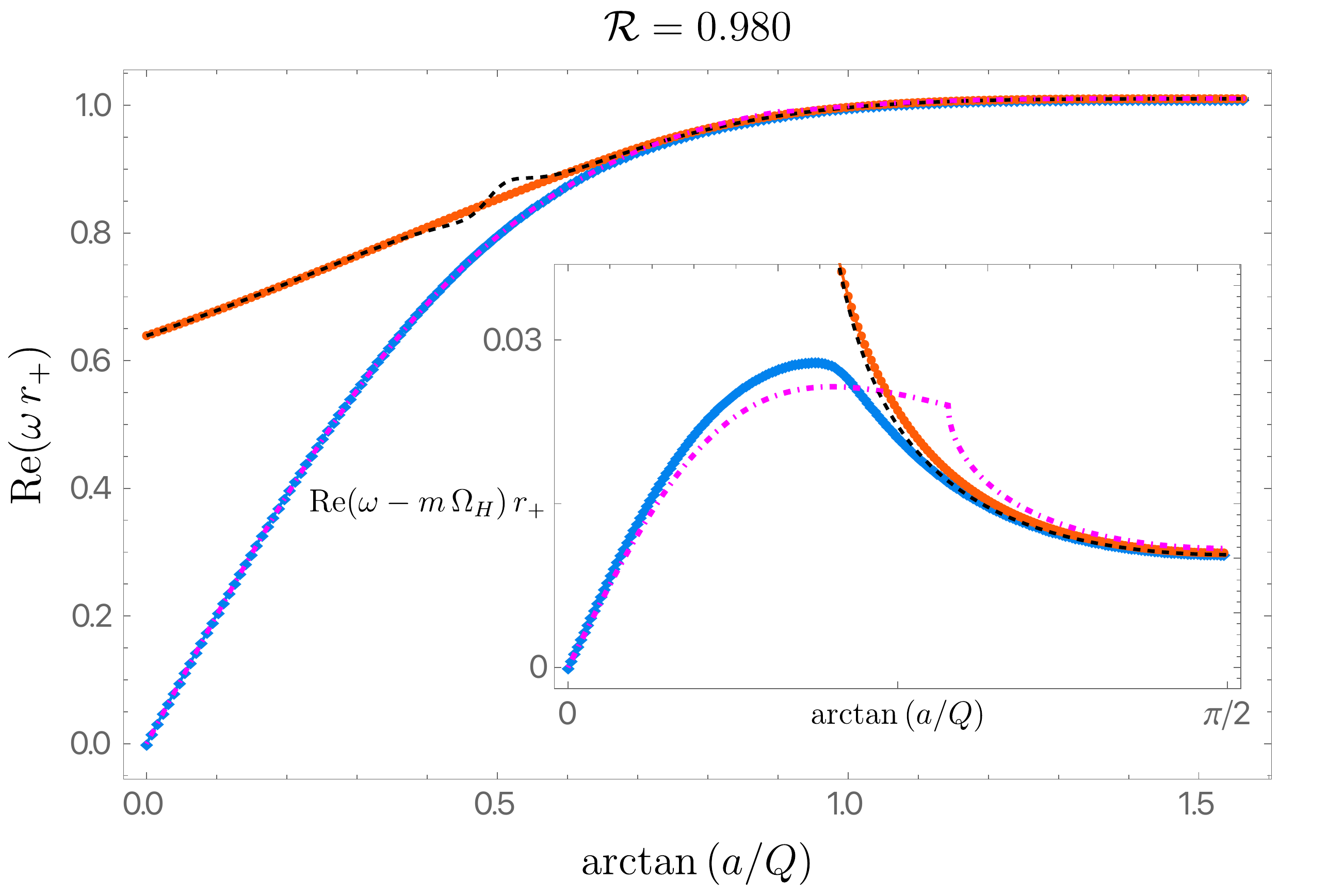}
\caption{This series of figures is for the same cases reported in Fig.~\ref{Fig:WKBn0-NHn0-Im} but this time we plot the real part of the frequency as a function of $\Theta$ for the PS (orange disks) and the NH (blue diamonds) families of QNMs with $m=\ell=2, n=0$ for a KN family with, following the lexicographic order, $\mathcal{R}=0.993$, $\mathcal{R}=0.992$, $\mathcal{R}=0.991$, $\mathcal{R}=0.990$, $\mathcal{R}=0.985$ and $\mathcal{R}=0.980$. We also display the WKB result $\tilde{\omega}_{\hbox{\tiny WKB}}$ (dashed black line) and the near-extremal frequency $\tilde{\omega}_{\hbox{\tiny MAE}} $ for $p=0$ (dot-dashed magenta line) and $p=1$ (dotted dark magenta line).
In the inset plots, we still show the real part of the frequency but this time measured with respect to the superradiant bound $m\Omega_H$.}
\label{Fig:WKBn0-NHn0-Re}
\end{figure}

To illustrate how the PS and NH families of RN QNM evolve when we extend them to the KN case, we plot the imaginary part (Fig.~\ref{Fig:WKBn0-NHn0-Im}) and real part (Fig.~\ref{Fig:WKBn0-NHn0-Re}) of the KN QNM frequencies as a function of $\Theta=\arctan\left( a/Q\right) $ for the PS (orange disks) and the NH (blue diamonds) families of QNM with  $m=\ell=2$, $n=0$ for a series of KN families with fixed $\mathcal{R}$. Recall $-$ see \eqref{def:sigma}-\eqref{PolarParametrization} $-$ that $\mathcal{R}$ is a `radial' parameter that effectively measures the distance away from extremality, with the extremal KN family being described by $\mathcal{R}=1$ (and the Schwarzschild solution by $\mathcal{R}=0$). In Figs.~\ref{Fig:WKBn0-NHn0-Im}$-$\ref{Fig:WKBn0-NHn0-Re} we have selected six KN families at constant $\mathcal{R}$ that illustrate a key feature of the QNM spectra as we progressively move away from extremality. These values are $\mathcal{R}=0.993$, $\mathcal{R}=0.992$, $\mathcal{R}=0.991$, $\mathcal{R}=0.990$,  $\mathcal{R}=0.985$ and $\mathcal{R}=0.980$ (please find this value on the top of each plot). For each plot, we increase $\Theta$ to follow the $m=\ell=2, n=0$ PS and NH  QNM families from their RN limit ($\Theta=0$), shown in Fig.~\ref{Fig:RN},  all the way up to their Kerr limit  ($\Theta=\pi/2$), which will be shown in later Fig.~\ref{Fig:Kerr}.

The most relevant property of the system is found in Fig.~\ref{Fig:WKBn0-NHn0-Im}, which plots the imaginary part. We see that for $\mathcal{R}\lesssim 0.991$ the blue diamond NH family has smaller $|{\rm Im}\,\tilde{\omega}|$ than the orange disk PS family for all values of $\Theta$. However, at $\mathcal{R}\sim 0.991$ (middle-left plot; see in particular the zoom in the inset plot) we see that both the NH and PS curves develop a cusp around $\Theta\sim 0.7$ where the two families approach arbitrarily close. (Hereafter, we denote the piece of the curve to the left/right of this cusp as the `\emph{old left/right branches}' of the PS or NH family). For slightly smaller values of $\mathcal{R}$ the `old left/right' branches of the NH curve break, and the same happens for the `old left/right branches' of the PS curve. In particular, for $\mathcal{R}\sim 0.990$ (middle-right plot; see in particular the zoom in the inset plot) we see that the `old NH left branch' is now smoothly merged with the `old PS right branch', and a similar trade-off occurs with the other two branches, \ie the `old PS left branch' is now smoothly merged with the `old NH right branch'. That is to say, in the small window $0.991 \gtrsim \mathcal{R} \gtrsim  0.990$ the identification of the PS and NH families is no longer clean but fuzzy. Up to the point where it becomes more appropriate to consider the two families of QNMs displayed in Fig.~\ref{Fig:WKBn0-NHn0-Im} as two `merged PS-NH' mode families that intersect each other by simple crossovers (the $\mathrm{Im}\,\tilde{\omega}$ curves but not the real part) for values $\mathcal{R}< 0.990$, as illustrated in the two bottom plots of Fig.~\ref{Fig:WKBn0-NHn0-Im} for $\mathcal{R}= 0.985$ (bottom-left panel) and for $\mathcal{R}= 0.980$ (bottom-right panel). The non-trivial interaction between the imaginary part of the $m=\ell=2, n=0$ NH and PS QNM families is a consequence of the eigenvalue repulsion phenomenon reviewed in the previous section. Eigenvalue repulsions were previously identified in~\cite{Dias:2022oqm} for gravito-electromagnetic perturbations of KN, and we see here that they are also present in the scalar field QNM spectra. The polar parametrization adopted here for the KN black hole is particularly useful to study this phenomenon.

Three important observations are still in order. First, note that for small values of $\Theta$, and for any value of $\mathcal{R}$, it is true that the two families of QNM can be unequivocally traced back to the PS and NH QNM families of the RN black hole when $\Theta\to 0$. It is only at intermediate values of $\Theta$ (roughly, $\Theta\gtrsim 0.5$, say) that the two curves approximate and develop cusps (for $0.992 \gtrsim \mathcal{R} \gtrsim  0.990$) and finally break/merge to form the two `PS-NH' curves (for $\mathcal{R} \lesssim  0.990$). 

Secondly, notice that the formation of cusps and the associated  breakup/merge process between two branches of the old PS and NH families (described above) only occurs
at the level of the imaginary part of the frequencies. Indeed, in Fig.~\ref{Fig:WKBn0-NHn0-Re} we analyse the evolution of the real part of the frequency as a function of $\Theta$ for the same fixed values of $\mathcal{R}$ as those displayed in Fig.~\ref{Fig:WKBn0-NHn0-Im} and we conclude that nothing special occurs to the real part of the frequency as $\mathcal{R}$ decreases. In particular there is no formation of cusps or breakups/mergers in the range $0.992 \gtrsim \mathcal{R} \gtrsim  0.990$. As discussed in section~\ref{sec:EigenvalueRepulsionsA}, this is consistent with the fact that for a 2-parameter family of black holes (KN in our case), two eigenvalues can coincide only at a isolated point (or a discrete set of isolated points) in the parameter space (where the analogue of the two conditions~\eqref{ERcond} are satisfied).  Elsewhere, except possibly at a portion of the extremal KN boundary where the modes terminate, the real and imaginary part of the frequency of one family of QNMs cannot be simultaneously the same as those of another QNM family. If we are to summarize our key findings in a single sentence, our numerical data strongly suggests that PS and NH modes can meet and terminate at the portion of the extremal boundary described by $\mathcal{R}=1$ and $\Theta_\star \lesssim \Theta\leq \pi/2$ (with $\Theta_\star\sim 0.881$), but we find no eigenvalue crossing anywhere else away from extremality.
Interestingly, the repulsions around the portion of the extremal boundary where the PS and NH modes do (or do attempt to) meet and terminate creates ripple effects relatively far away that can produce intricate interactions between the imaginary part of the frequency of modes of two QNM families like those observed in Fig.~\ref{Fig:WKBn0-NHn0-Im} that could not be anticipated before doing the actual computation. (We will analyse key aspects of this discussion in more detail later when we discuss Fig.~\ref{Fig:Star}).

There is a third observation that emerges from 
Figs.~\ref{Fig:WKBn0-NHn0-Im}$-$\ref{Fig:WKBn0-NHn0-Re} which is crucial to interpret the nature of the QNM families. In the plots of these figures we also show the higher-order WKB frequency $\tilde{\omega}_{\hbox{\tiny WKB}}$ (dashed black line) as given by~\eqref{WKBfrequency} with $p=0$,  and the near-horizon frequency $\tilde{\omega}_{\hbox{\tiny MAE}} $ of~\eqref{NEfreq} for $p=0$ (dot-dashed magenta line) and $p=1$ (dotted dark magenta  line).
For small $\Theta$, and independently of the value of $\mathcal{R}$, $\tilde{\omega}_{\hbox{\tiny WKB}}$ approximates the orange disk PS curve well; moreover, \eqref{NEfreq} with $p=0$ is an excellent approximation for the blue diamond NH curve. This is what we expect from the discussions of sections~\ref{sec:PSwkb} and ~\ref{sec:NHanalytics}. In particular, we used these criteria to unambiguously identify the PS and NH families of modes in the RN limit ($\Theta\to 0$). The situation is however much more intricate in the opposite Kerr limit ($\Theta=\pi/2$). To start with, for $\mathcal{R}\gtrsim 0.991$ (see \eg the first three plots of Figs.~\ref{Fig:WKBn0-NHn0-Im}$-$\ref{Fig:WKBn0-NHn0-Re} for $\mathcal{R}= 0.993, 0.992, 0.991$), the black-dashed $\tilde{\omega}_{\hbox{\tiny WKB}}$ describes the orange PS modes well for small $\Theta$, but (and this comes as a surprise) it fails to do so at large $\Theta$! Instead, at large $\Theta$,  $\tilde{\omega}_{\hbox{\tiny WKB}}$ describes the blue NH family very well (in particular, in the Kerr limit $\Theta\to \pi/2$)! This is a first indication that the clean criterion used to classify and distinguish PS and NH families in the RN limit becomes absolutely misleading as we approach the Kerr limit. 
On the other hand, without surprise, in this range $\mathcal{R}\gtrsim 0.991$ the dot-dashed magenta $\tilde{\omega}_{\hbox{\tiny MAE}}$ (with $p=0$) is an excellent approximation to the blue NH family for all $0 \leq \Theta \leq \pi/2$. 
However, $\tilde{\omega}_{\hbox{\tiny MAE}}$ coincides with $\tilde{\omega}_{\hbox{\tiny WKB}}$ for large $\Theta$! This is a second indication that the RN criteria for the PS/NH distinction does not extend to the Kerr limit.
Still in the range $\mathcal{R}\gtrsim 0.991$ we have yet another surprise: for large $\Theta$, the orange PS family, that is not well approximated by $\tilde{\omega}_{\hbox{\tiny WKB}}$, is instead well described by 
$\tilde{\omega}_{\hbox{\tiny MAE}}\,$\dots with $p=1$ (dotted dark magenta  line)!! So not only is the orange PS curve not well described by the eikonal/WKB approximation, but it is instead well described by a {\it higher overtone} MAE frequency: this orange disk family starts at $\Theta=0$ as a $n=0$  family but terminates at $\Theta=\pi/2$ with radial overtone $n=1$!\footnote{Note that here we are not including discussions of the $p=1$ $\tilde{\omega}_{\hbox{\tiny WKB}}$. This will be discussed in Figs.~\ref{Fig:WKBn1-NHn4-Im}$-$\ref {Fig:WKBn1-NHn4-ImB}.} Summarizing, for $\mathcal{R}\gtrsim 0.991$, the mode we naively called the ground state NH family (in the RN limit) is simultaneously described by $p=0$ $\tilde{\omega}_{\hbox{\tiny WKB}}$ and $p=0$ $\tilde{\omega}_{\hbox{\tiny MAE}}$ at large $\Theta$, while the mode we naively called the ground state PS family (in the RN limit) is described by $p=1$ $\tilde{\omega}_{\hbox{\tiny MAE}}$ (and $p = 1$ $\tilde{\omega}_{\hbox{\tiny WKB}}$) at large $\Theta$.
The three conclusive facts above confirm that the criteria used to classify and distinguish QNM families in the RN limit cannot be extended without contradictions/inconsistencies to high values of $\Theta$ and, in particular, to the Kerr limit. The PS/NH classification at the RN limit remains valid for small values of $\Theta$ but gets absolutely misleading at high values of $\Theta$. Up to the point that it should be dropped because it simply cannot be formulated in equal terms in the Kerr limit.  

\begin{figure}[th]
\centering
\includegraphics[width=.49\textwidth]{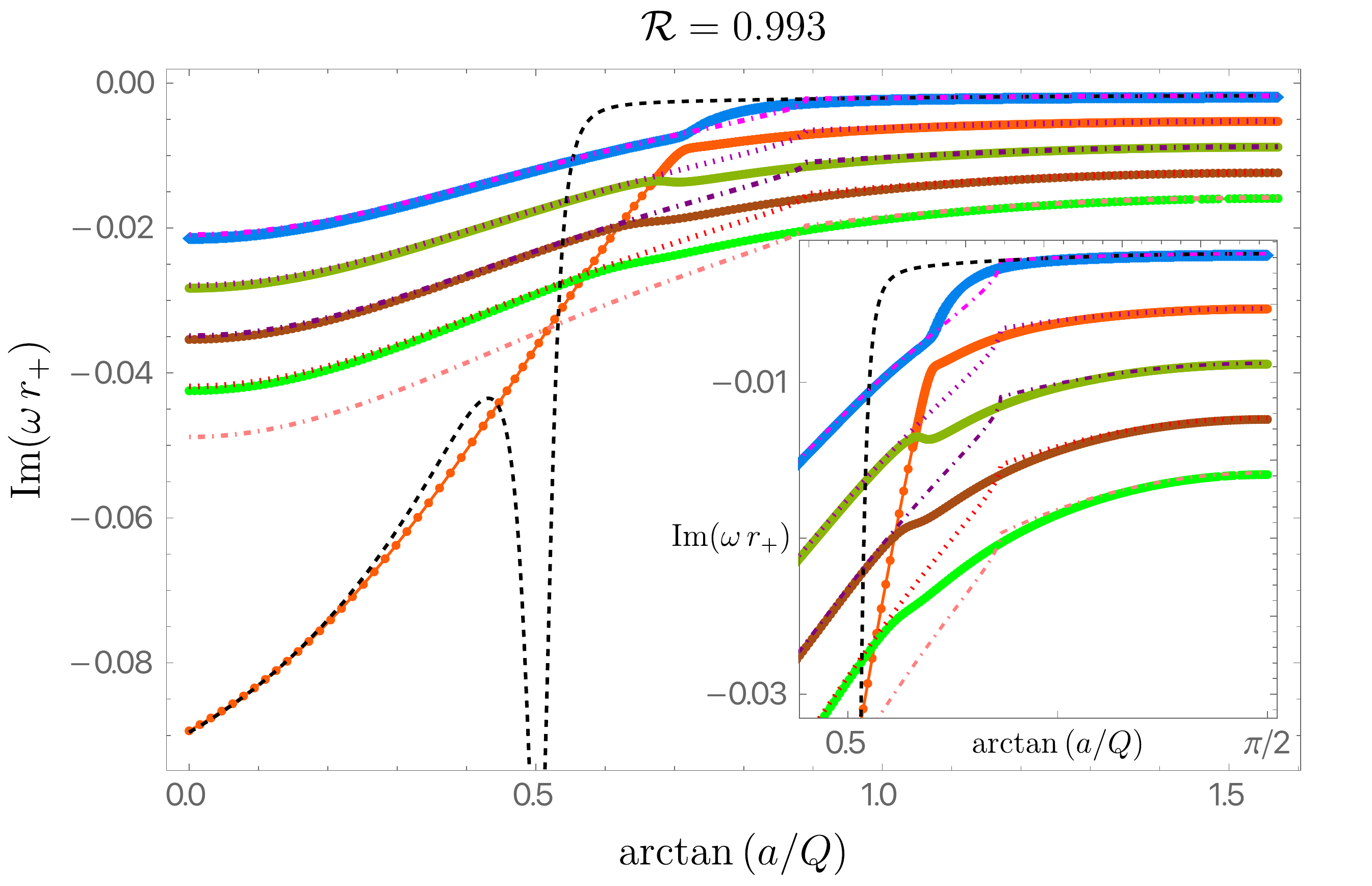}
\hspace{0.0cm}
\includegraphics[width=.49\textwidth]{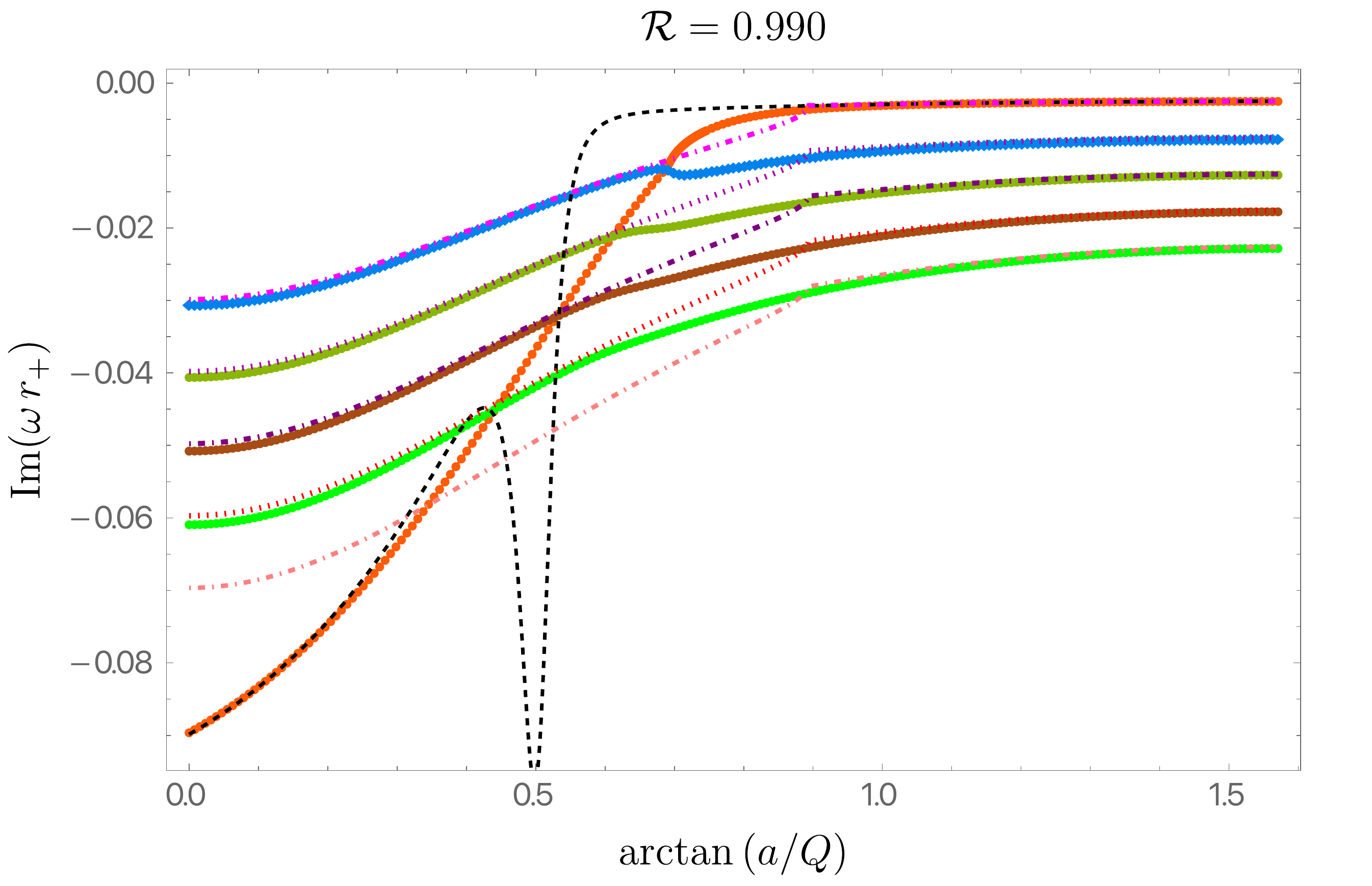}
\vskip 0.2cm
\includegraphics[width=.49\textwidth]{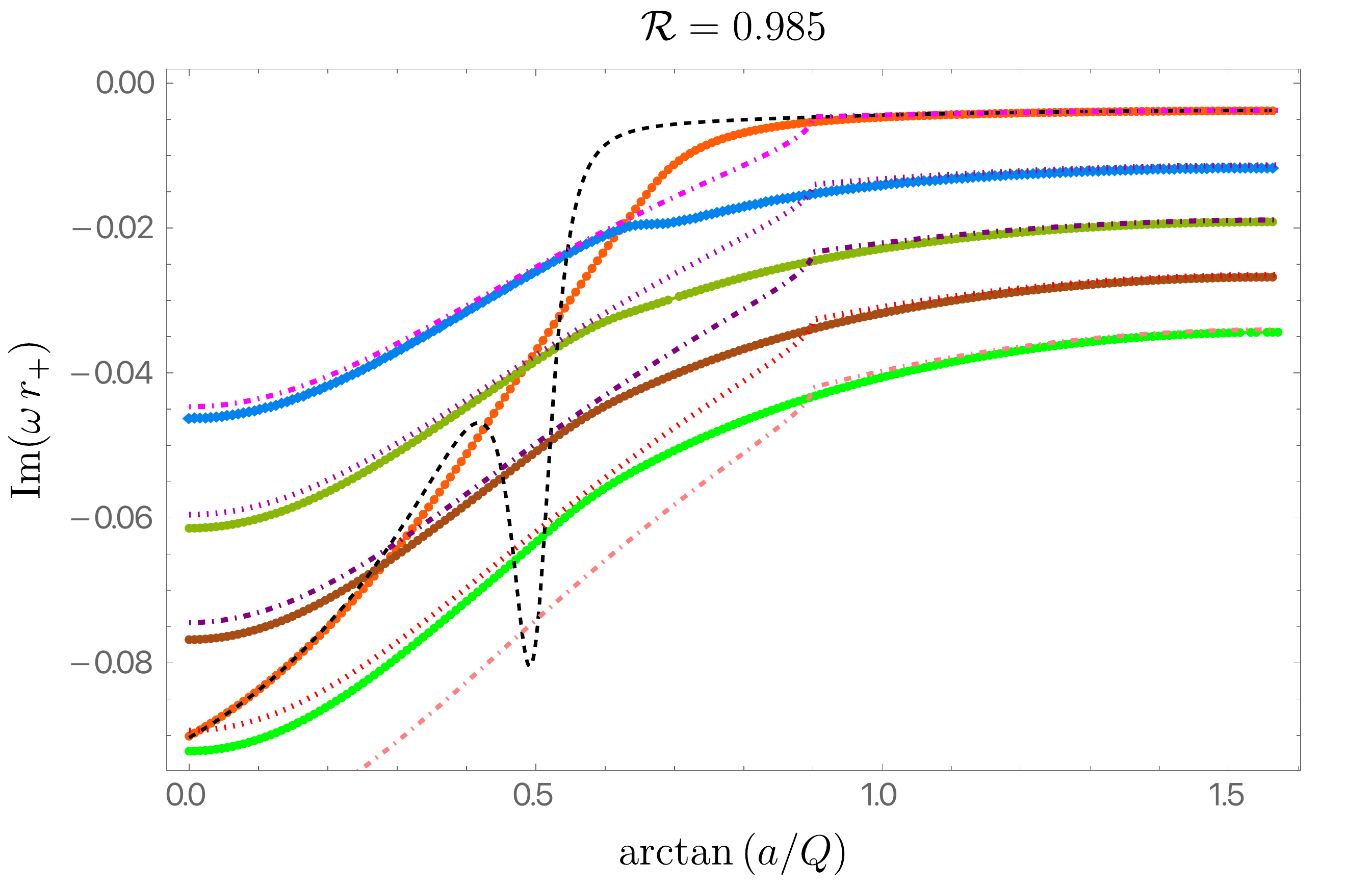}
\hspace{0.0cm}
\includegraphics[width=.49\textwidth]{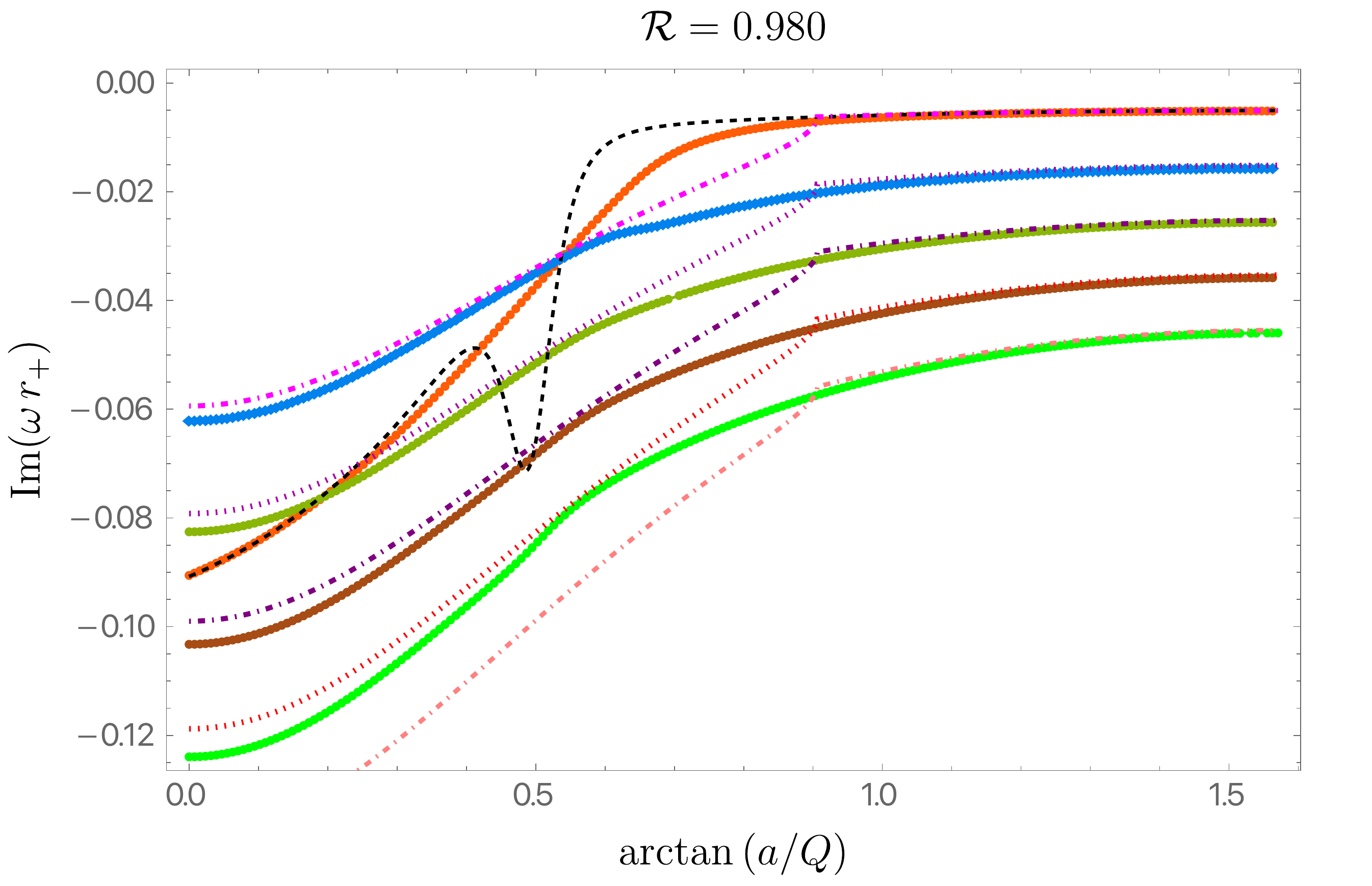}
\vskip 0.2cm
\includegraphics[width=.49\textwidth]{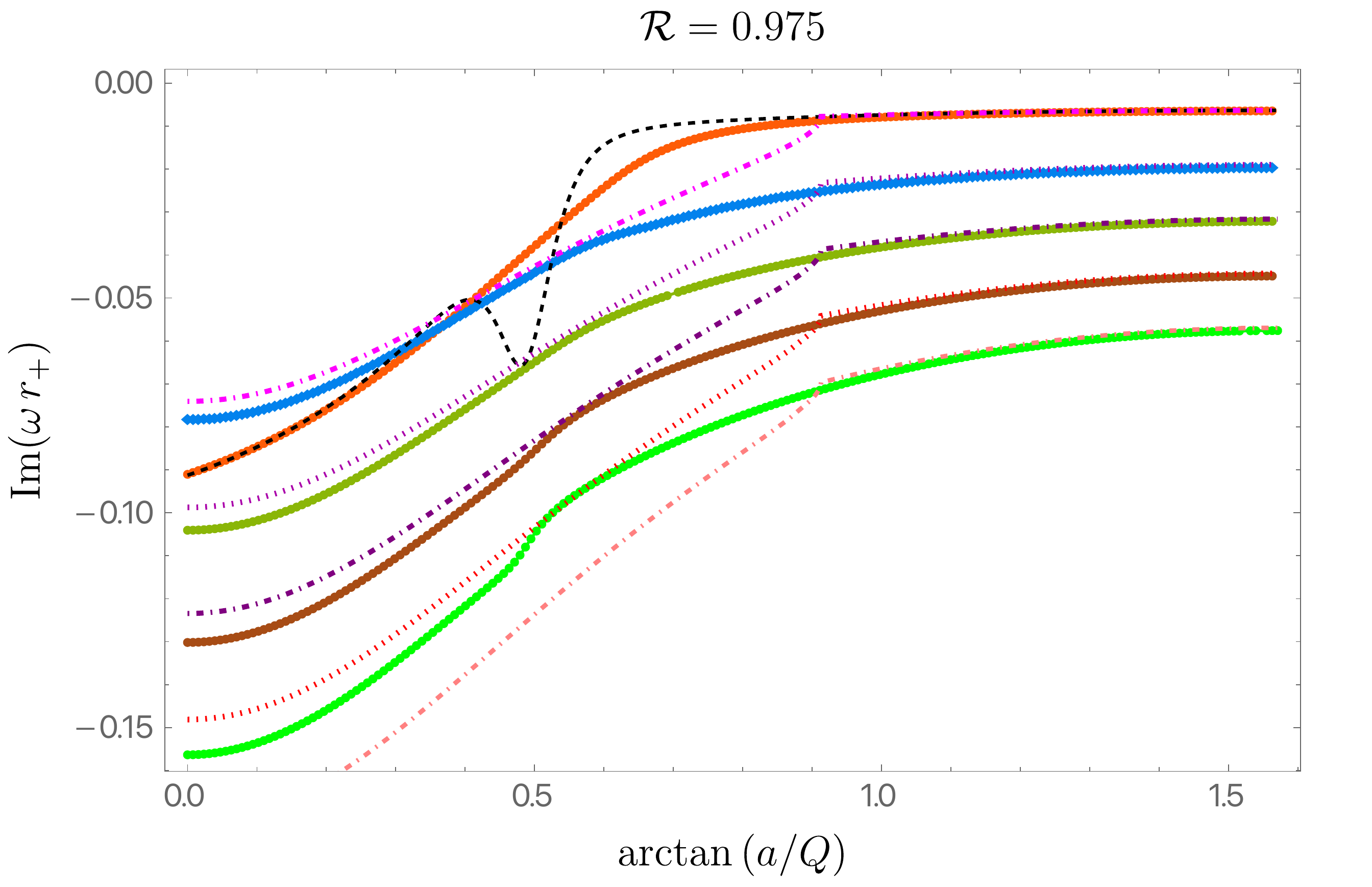}
\hspace{0.0cm}
\includegraphics[width=.49\textwidth]{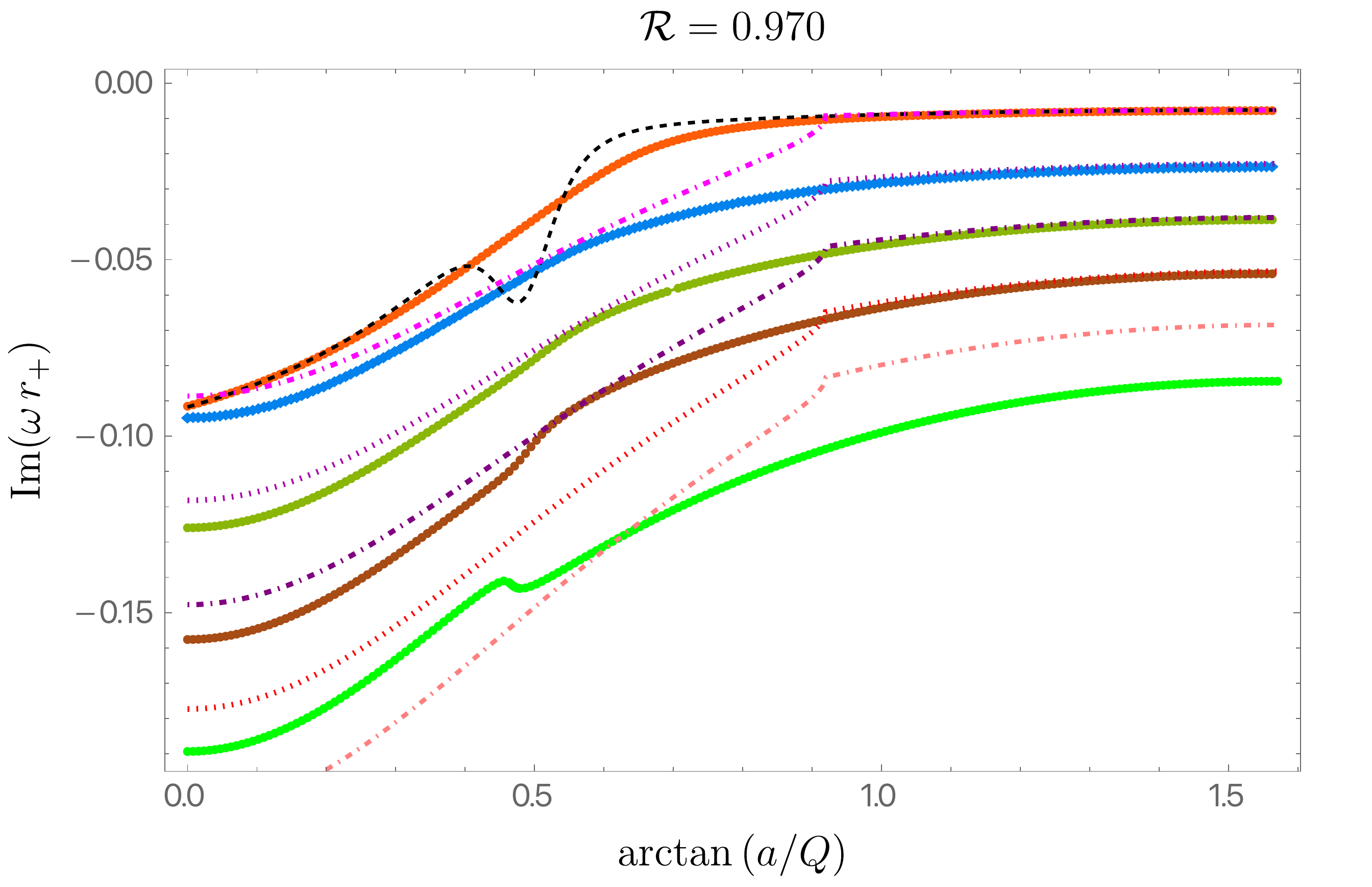}
\caption{Imaginary part of the frequency as a function of $\Theta=\arctan\left( a/Q\right) $  for the PS (orange disks) and the $n=0,1,2,3$ NH (blue, dark-green, brown, green diamonds) families of QNM with  $m=\ell=2$ for a KN family with $\mathcal{R}=0.993$, $\mathcal{R}=0.990$, $\mathcal{R}=0.985$, $\mathcal{R}=0.980$,  $\mathcal{R}=0.975$ and $\mathcal{R}=0.970$ (following the lexicographic order). We also display the WKB result $\tilde{\omega}_{\hbox{\tiny WKB}}$ (dashed black line) and the near-extremal frequency $\tilde{\omega}_{\hbox{\tiny MAE}} $ for $p=0,1,2,3$ (dot-dashed magenta, dotted dark magenta, dot-dashed purple, dotted pink, dot-dashed pink  lines, respectively).}
\label{Fig:WKBn0-NHn4-Im}
\end{figure}  

If not already sufficiently intricate, another level of complexity is added when we move to the region $0.991 \gtrsim \mathcal{R} \gtrsim  0.990$ where the phenomenon of eigenvalue repulsion occurs, as already described in detail previously. Moving further away from extremality, for $\mathcal{R} \leq  0.990$, the modes that previously repelled now simply cross (i.e. the imaginary part of the frequency crosses but not the real part) as we increase $\Theta$. One now finds that it is the orange disk mode $-$ that we initially (in the RN limit) called the PS family $-$ that is simultaneously described by $p=0$ $\tilde{\omega}_{\hbox{\tiny WKB}}$ and $p=0$ $\tilde{\omega}_{\hbox{\tiny MAE}}$ at large $\Theta$ (this is possible because the original PS modes approach ${\rm Im}\,\tilde{\omega}=0$ at extremality for large $\Theta$)! And the blue diamond curve that was initially (\ie at the RN limit) denoted as the ground state NH family is the one that is now well approximated  by $\tilde{\omega}_{\hbox{\tiny MAE}}$ with $p=1$ (not $p=0$) at large $\Theta$! Altogether, and with hindsight, it would have been more appropriate to denote all the QNM families simply as an entangled `PS-NH' family and its radial overtones, with the photon sphere and near-horizon nature of the modes unequivocally disentangling only for small values of $\Theta$ as one approaches the Reissner-Nordstr\"om limit.

In Figs.~\ref{Fig:WKBn0-NHn0-Im}$-$\ref{Fig:WKBn0-NHn0-Re} we have only displayed the ground state modes, \ie the first overtone ($n=0$) of the PS family and the first overtone ($n=0$) of the NH family (as we unambiguously classify them in the RN limit).
To further explore the properties of our system, in Fig.~\ref{Fig:WKBn0-NHn4-Im}
we display the $n=0$ PS (orange disks) and $n=0$ NH (blue diamonds) curves that were already presented in Fig.~\ref{Fig:WKBn0-NHn0-Im} but, this time, we additionally display the $n=1,2,3$ NH (dark-green, brown, green diamonds) families of QNM with  $m=\ell=2$ for a KN family with $\mathcal{R}=0.993$, $\mathcal{R}=0.990$, $\mathcal{R}=0.985$, $\mathcal{R}=0.980$,  $\mathcal{R}=0.975$ and $\mathcal{R}=0.970$ (following the lexicographic order). Moreover, we also display the $p=0$ WKB result $\tilde{\omega}_{\hbox{\tiny WKB}}$ (dashed black line) and the near-extremal frequency $\tilde{\omega}_{\hbox{\tiny MAE}} $ for $p=0,1,2,3$ (dot-dashed magenta, dotted dark magenta, dot-dashed purple, dotted pink, dot-dashed pink  lines, respectively). As pointed out above when discussing  Fig.~\ref{Fig:WKBn0-NHn0-Im}, we see that for $\mathcal{R}\leq 0.990$, the blue diamond curve is well approximated by  $\tilde{\omega}_{\hbox{\tiny MAE}}$ with $p=0$ for small $\Theta$ and then by $\tilde{\omega}_{\hbox{\tiny MAE}}$ with $p=1$ for large $\Theta$. A similar behavior is found in the higher overtone NH families. Indeed, near extremality, \ie for large $\mathcal{R}$, the $n=1,2,3$ NH curves are well described by $\tilde{\omega}_{\hbox{\tiny MAE}}$ with $p=1,2,3$ for small $\Theta$ but, for large $\Theta$, then are instead well described by $\tilde{\omega}_{\hbox{\tiny MAE}}$ with $p=2,3,4$. That is to say,  the family that at RN is described by the MAE result with overtone $p$ turns out to become, at large $\Theta$, well approximated by the MAE result with overtone $p+1$!  

\begin{figure}[t]
\centering
\includegraphics[width=.49\textwidth]{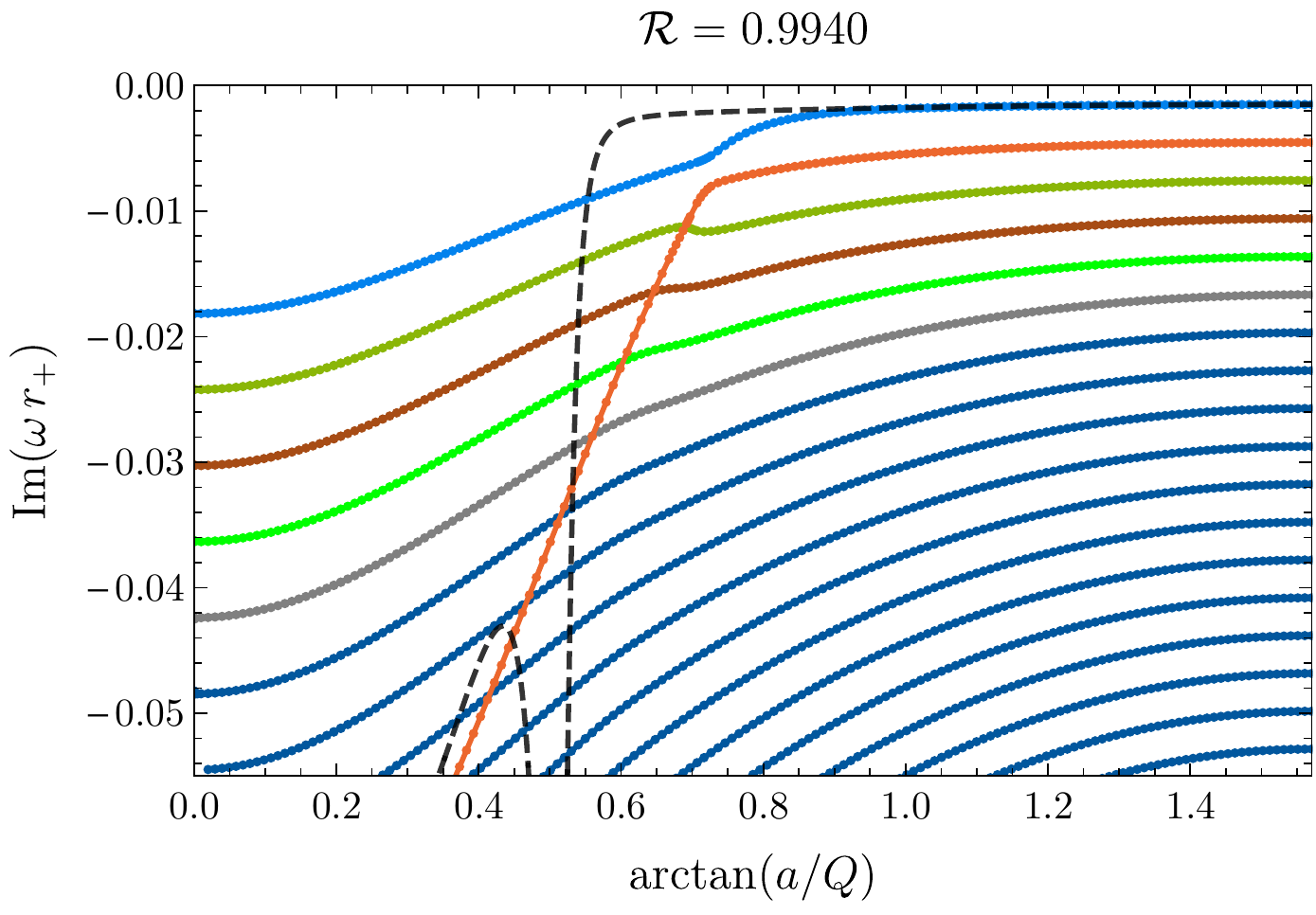}
\hspace{0.0cm}
\includegraphics[width=.49\textwidth]{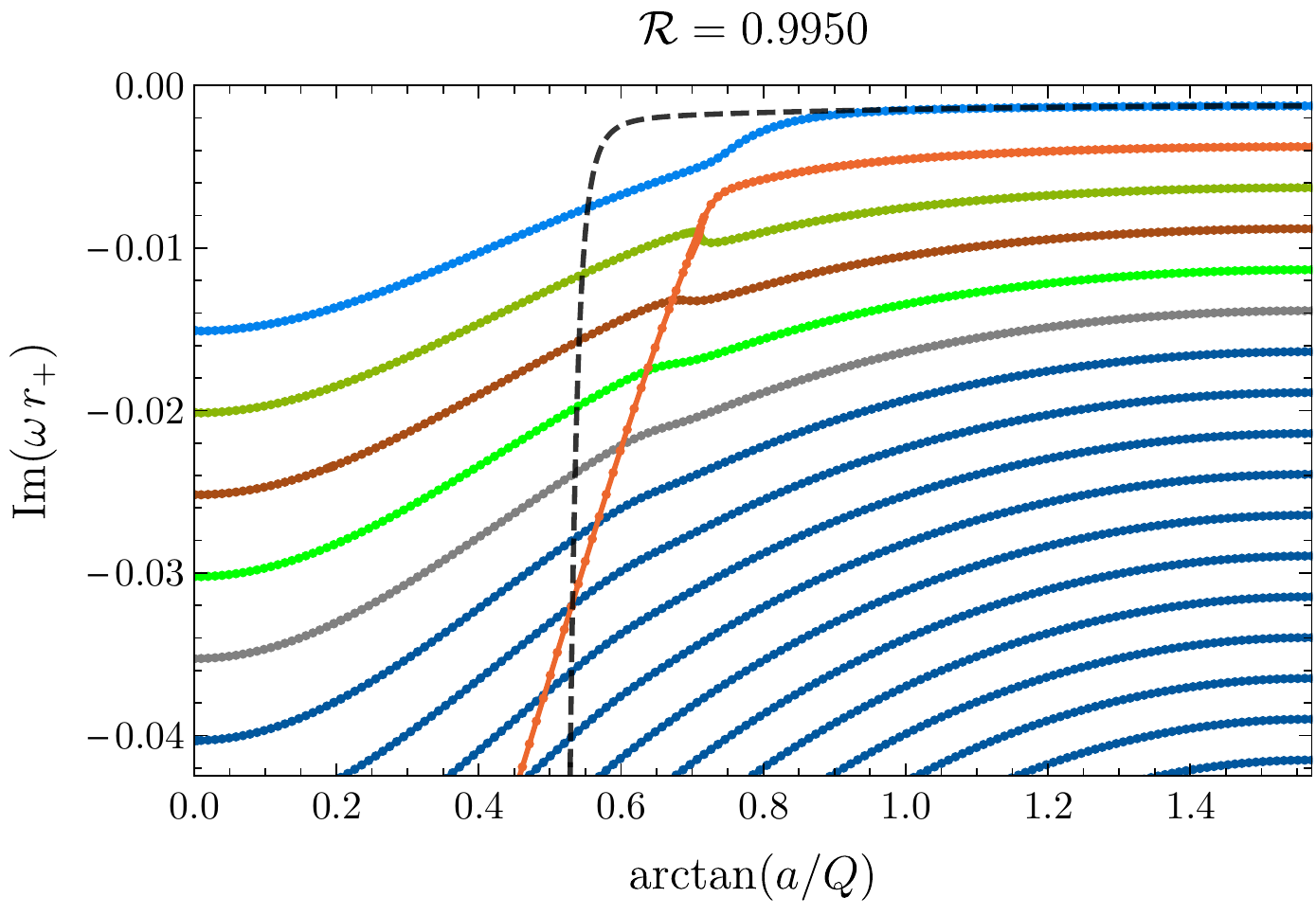}
\vskip 0.2cm
\includegraphics[width=.49\textwidth]{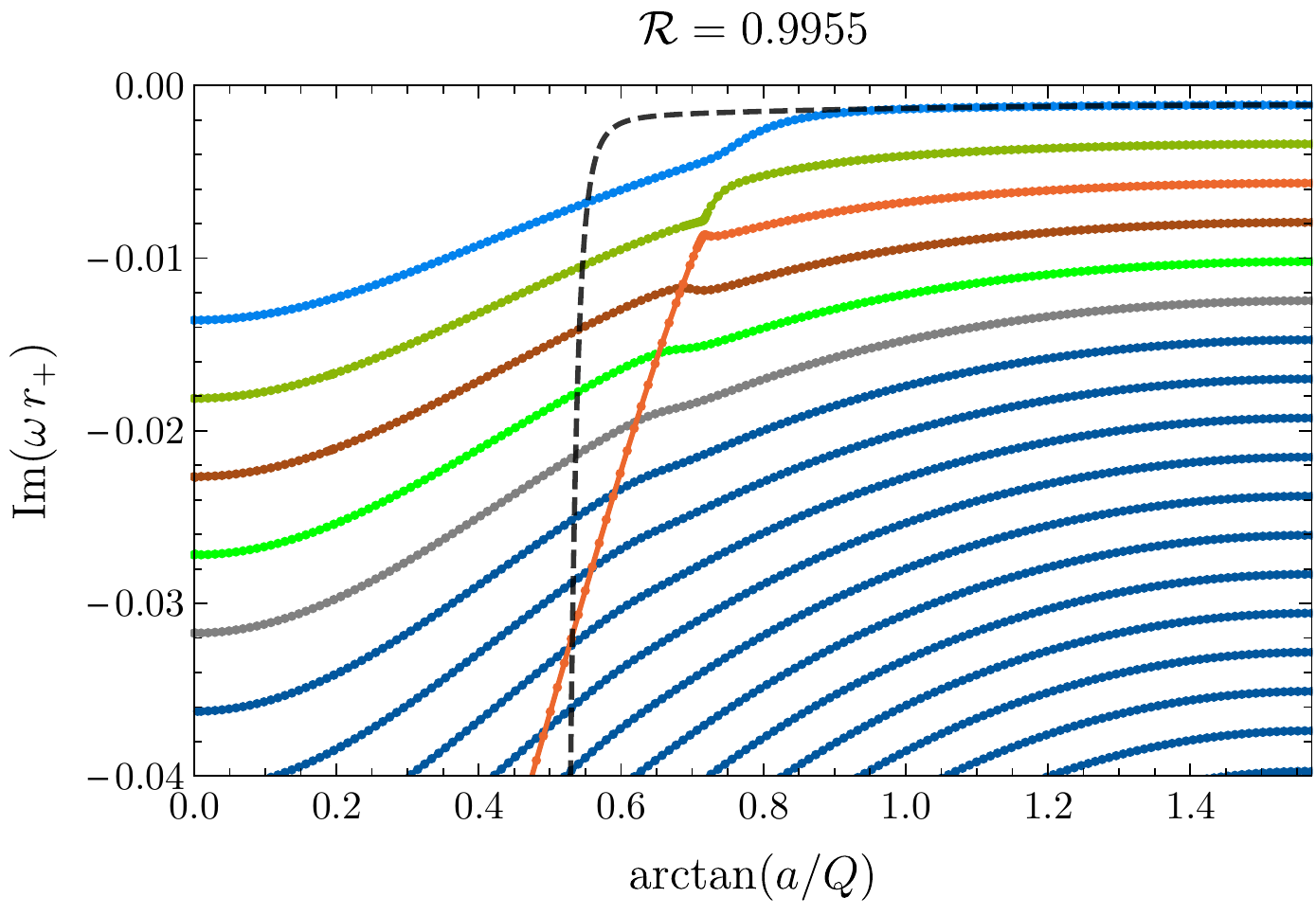}
\hspace{0.0cm}
\includegraphics[width=.49\textwidth]{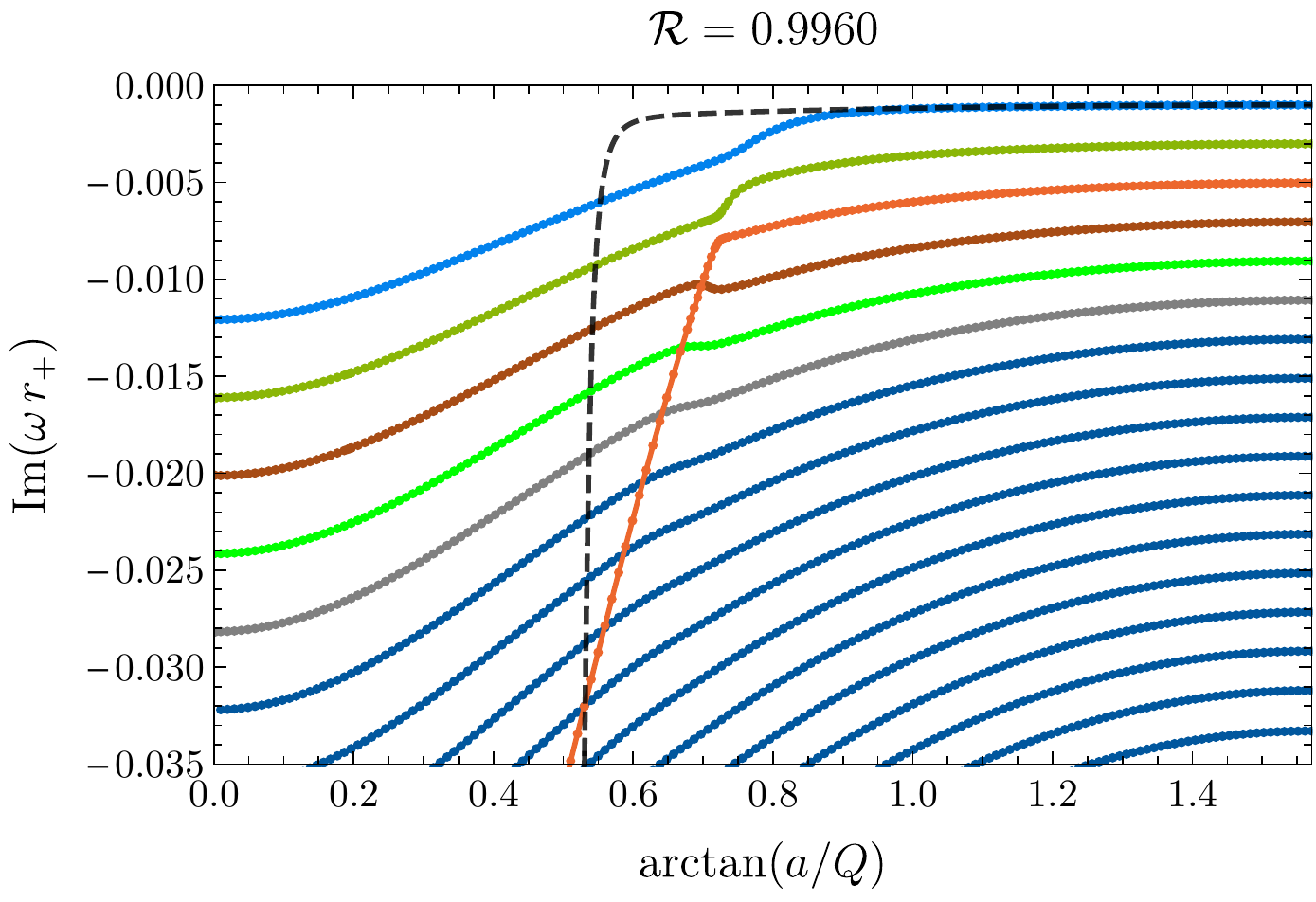}
\vskip 0.2cm
\includegraphics[width=.49\textwidth]{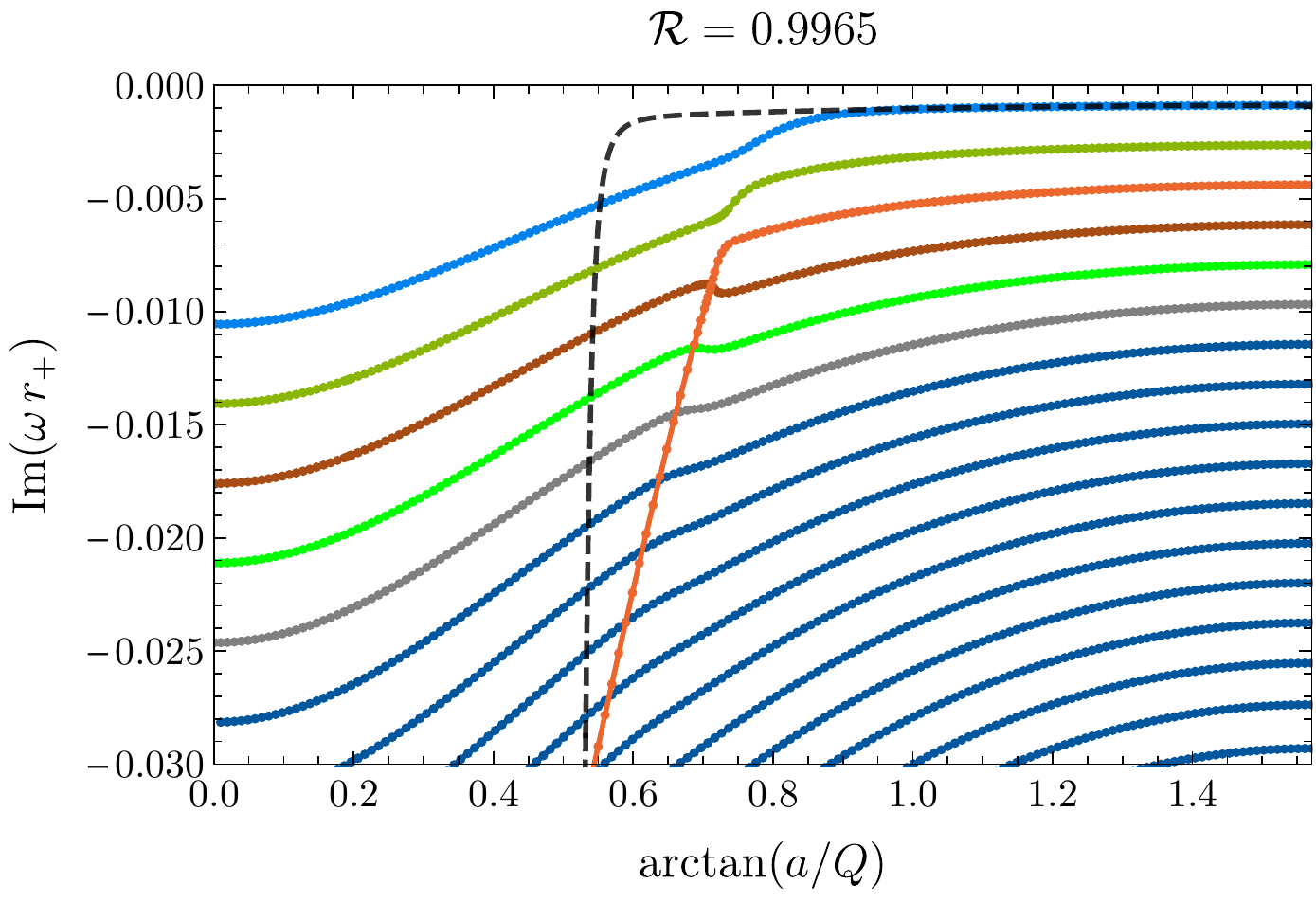}
\hspace{0.0cm}
\includegraphics[width=.49\textwidth]{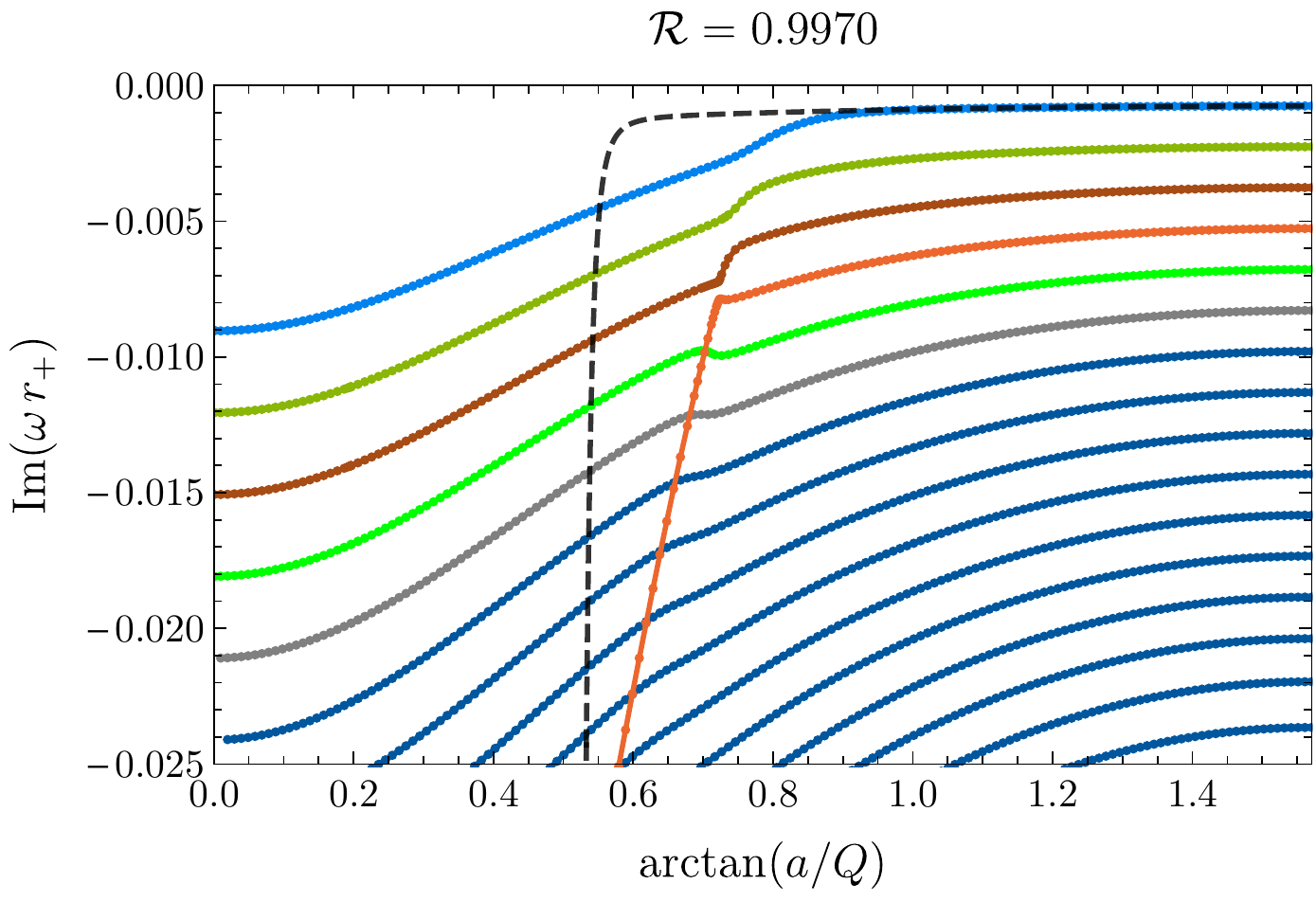}
\caption{Imaginary part of the frequency as a function of $\Theta=\arctan\left( a/Q\right)$ for the $n=0$ PS (orange disks) and the  $n=0,1,2,3, 4,\cdots, 16$ NH (blue, dark-green, brown, green and, for $n\geq 4$, dark-blue diamonds) families of QNM  with  $m=\ell=2$ for a KN family with  $\mathcal{R}=0.9940$,  $\mathcal{R}=0.9950$,  $\mathcal{R}=0.9955$ and $\mathcal{R}=0.9960$, $\mathcal{R}=0.9965$ and $\mathcal{R}=0.9970$ (following the lexicographic order). We also display the WKB result $\tilde{\omega}_{\hbox{\tiny WKB}}$ for $n=0$ (dashed black line). (This series of plots continues in Fig.~\ref{Fig:WKB-NH-Im_near_ext_2} for larger $\mathcal{R}$).}
\label{Fig:WKB-NH-Im_near_ext_1}
\end{figure}

\begin{figure}[t]
\centering
\includegraphics[width=.49\textwidth]{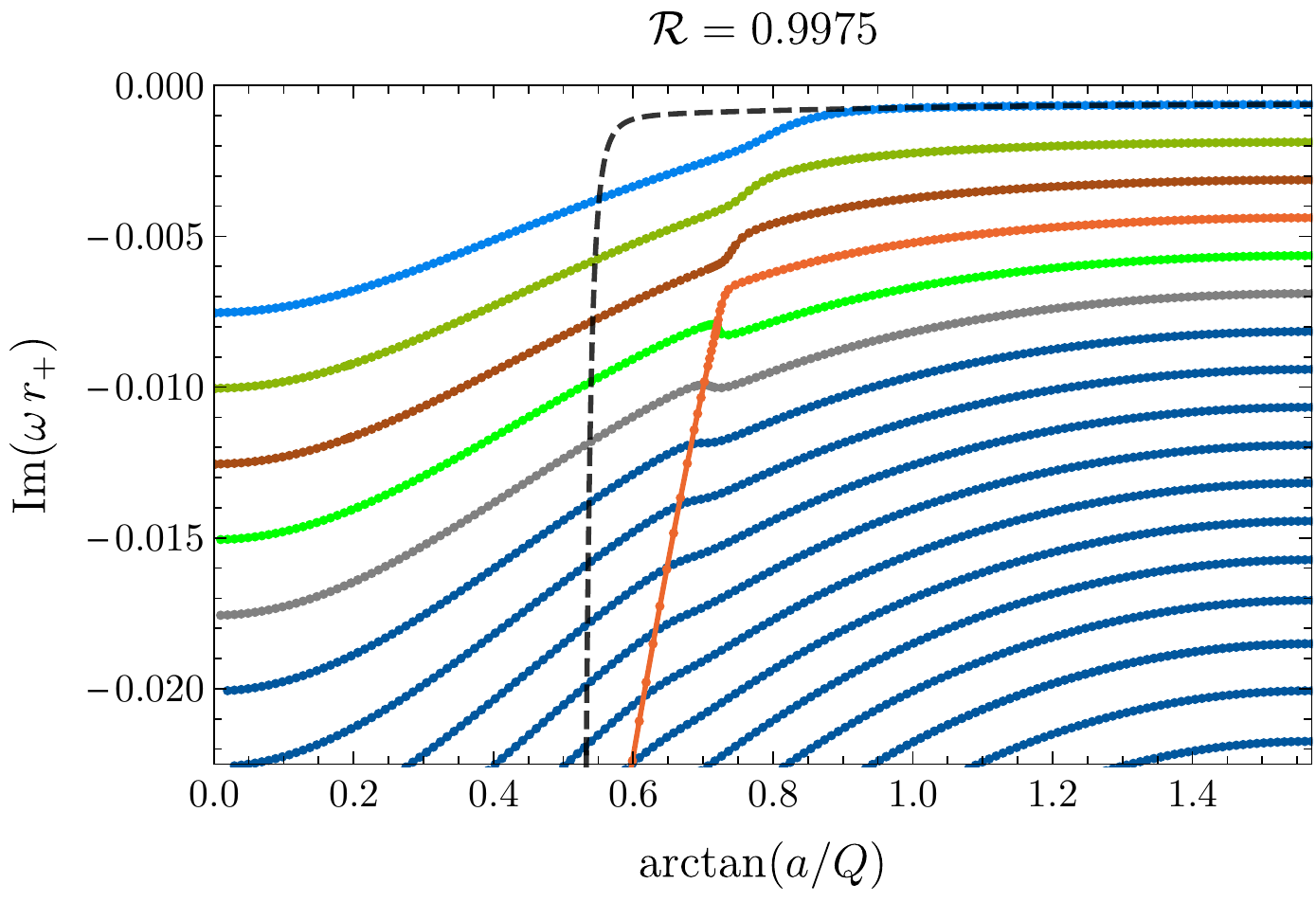}
\hspace{0.0cm}
\includegraphics[width=.49\textwidth]{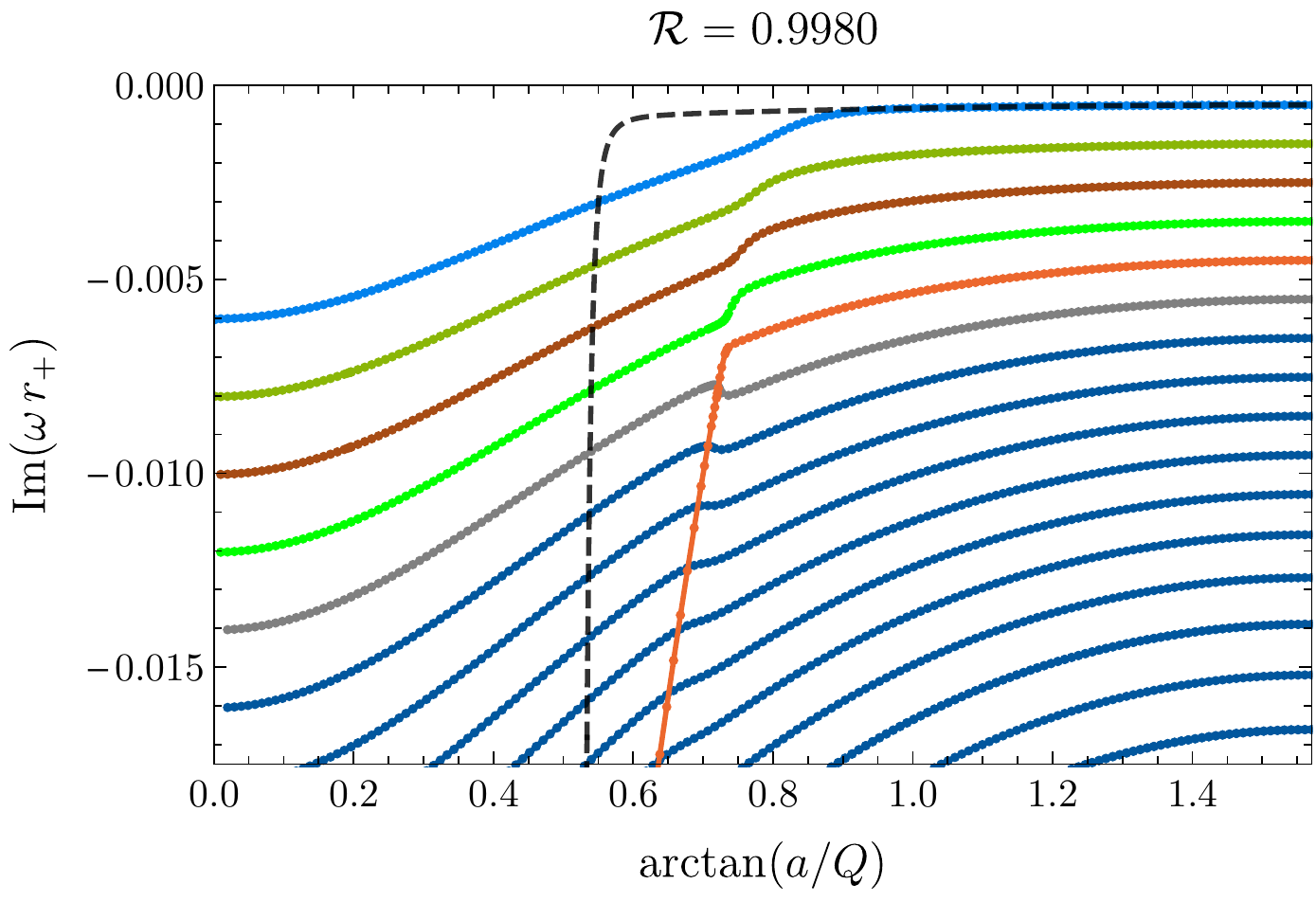}
\vskip 0.2cm
\includegraphics[width=.49\textwidth]{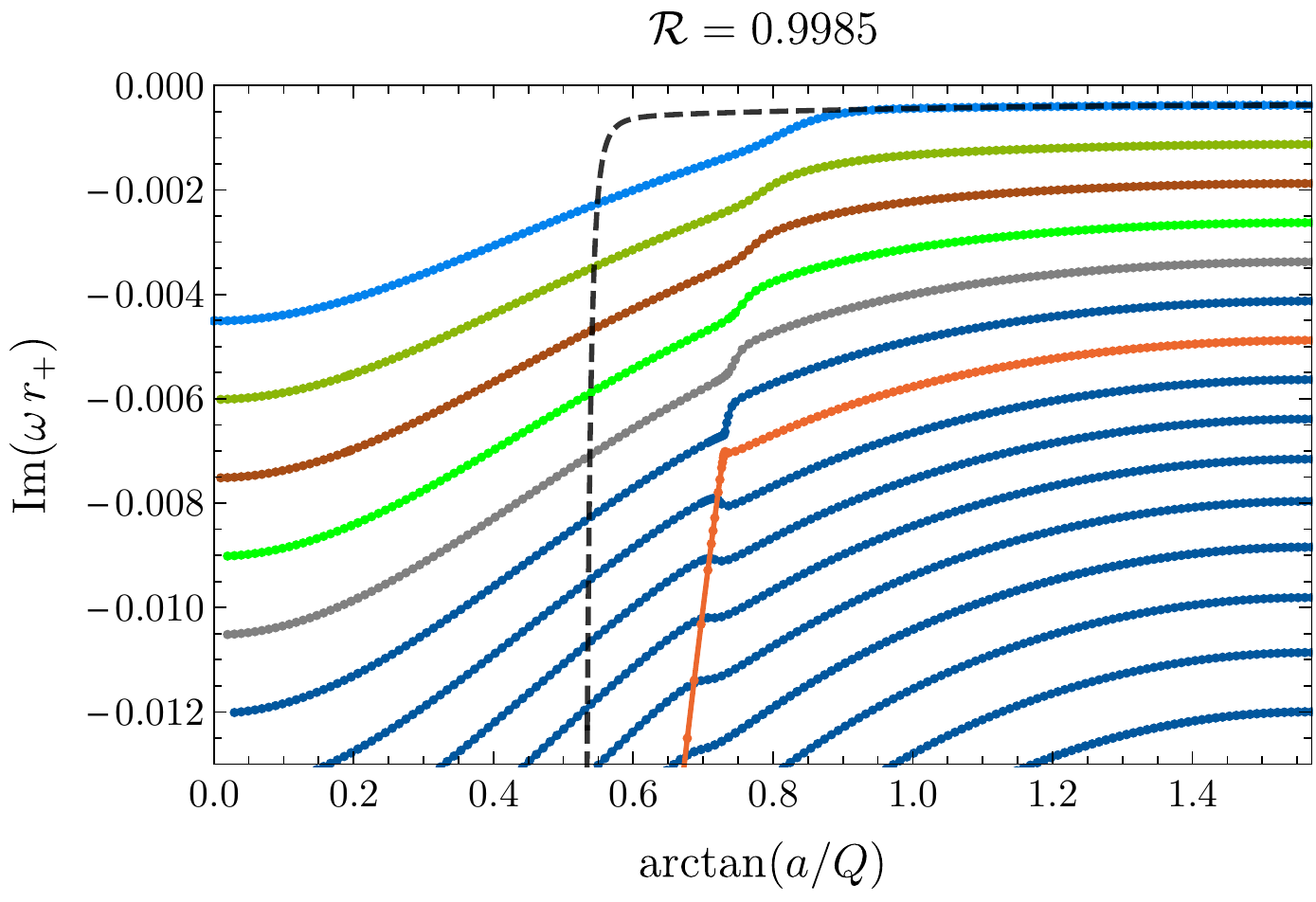}
\hspace{0.0cm}
\includegraphics[width=.49\textwidth]{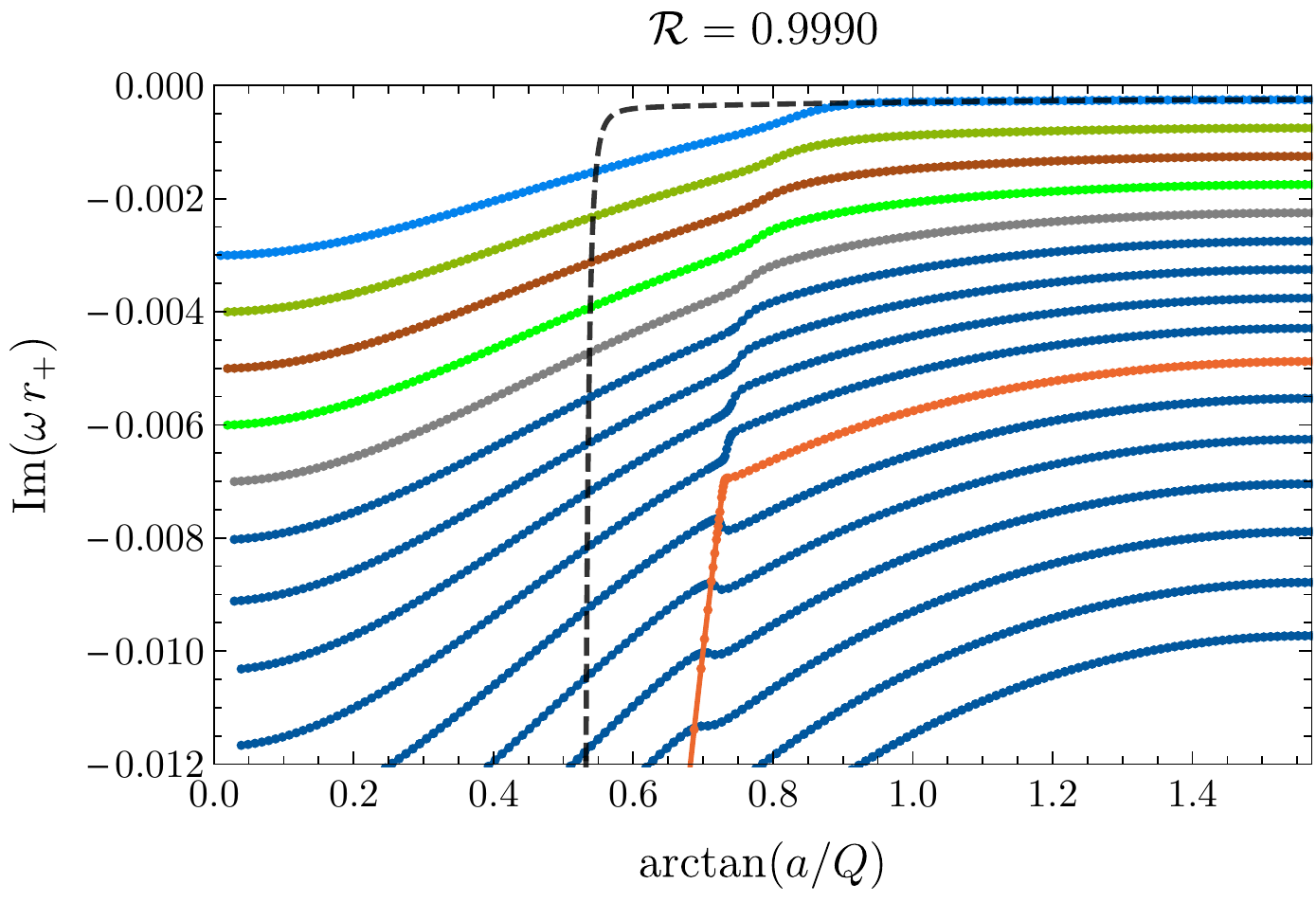}
\caption{Continuation of Fig.~\ref{Fig:WKB-NH-Im_near_ext_1}, this time for KN families with $\mathcal{R}=0.9975$, $\mathcal{R}=0.9980$, $\mathcal{R}=0.9985$ and $\mathcal{R}=0.9990$ (following the lexicographic order).}
\label{Fig:WKB-NH-Im_near_ext_2}
\end{figure}  

So far we have focused our attention in the parameter region $\mathcal{R}\leq 0.993$. But we might also ask what happens when we approach extremality ($\mathcal{R}=1$) even further. This question is addressed in Figs.~\ref{Fig:WKB-NH-Im_near_ext_1}-\ref{Fig:WKB-NH-Im_near_ext_2} (where the latter plots are to be seen as a continuation of the former) which describes what happens for $ 0.993 <\mathcal{R}\leq 0.999$, namely for $\mathcal{R}=0.9940, 0.9950, 0.9955, 0.9960,$ $0.9965, 0.9970$ (Fig.~\ref{Fig:WKB-NH-Im_near_ext_1}) and $\mathcal{R}= 0.9975, 0.9980, 0.9985, 0.9990$ (Fig.~\ref{Fig:WKB-NH-Im_near_ext_2}); note that unlike in the previous figures here we are increasing $\mathcal{R}$ as we move along the lexicographic order.
We see that further eigenvalue repulsions happen. Indeed, when moving from $\mathcal{R}=0.9950$ (top-right panel of Fig.~\ref{Fig:WKB-NH-Im_near_ext_1}) to $\mathcal{R}=0.9955$ (middle-left panel), one notes an eigenvalue repulsion between the $n=0$ PS family (orange curve) and the $n=1$ NH family (dark-green curve). Then, when moving from $\mathcal{R}=0.9965$ (bottom-left panel of Fig.~\ref{Fig:WKB-NH-Im_near_ext_1}) to $\mathcal{R}=0.9970$ (bottom-right panel), one observes an eigenvalue repulsion between the $n=0$ PS family (orange curve) and the $n=2$ NH family (brown curve). Continuing our analysis now in Fig.~\ref{Fig:WKB-NH-Im_near_ext_2},  when moving from $\mathcal{R}=0.9975$ (top-left panel of Fig.~\ref{Fig:WKB-NH-Im_near_ext_2}) to $\mathcal{R}=0.9980$ (top-right panel), we see a further eigenvalue repulsion this time between the $n=0$ PS family (orange curve) and the $n=3$ NH family (green curve). Finally, we see clear evidence that a series of further eigenvalue repulsions keep happening at an increasingly higher rate (in the sense that small increments $\mathcal{R}$ produce more repulsions) as we further approach $\mathcal{R}=1$. Indeed, in the bottom panel of Fig.~\ref{Fig:WKB-NH-Im_near_ext_2} we find that at $\mathcal{R}=0.9985$, for large $\Theta$, the $n=0$ PS curve is now below the $n=5$ NH curve and then, at $\mathcal{R}=0.9990$, for large $\Theta$, the $n=0$ PS curve is now below the $n=8$ NH curve. This overwhelmingly suggests that as $\mathcal{R}\to 1$, there is a (possibly infinite) cascade of eigenvalue repulsions where, for large $\Theta$, the $n=0$ PS curve gets below the $n$-th overtone NH curve for an increasingly higher value of $n$ (possibly with $n\to\infty$).

\begin{figure}
\centering
\includegraphics[width=.48\linewidth]{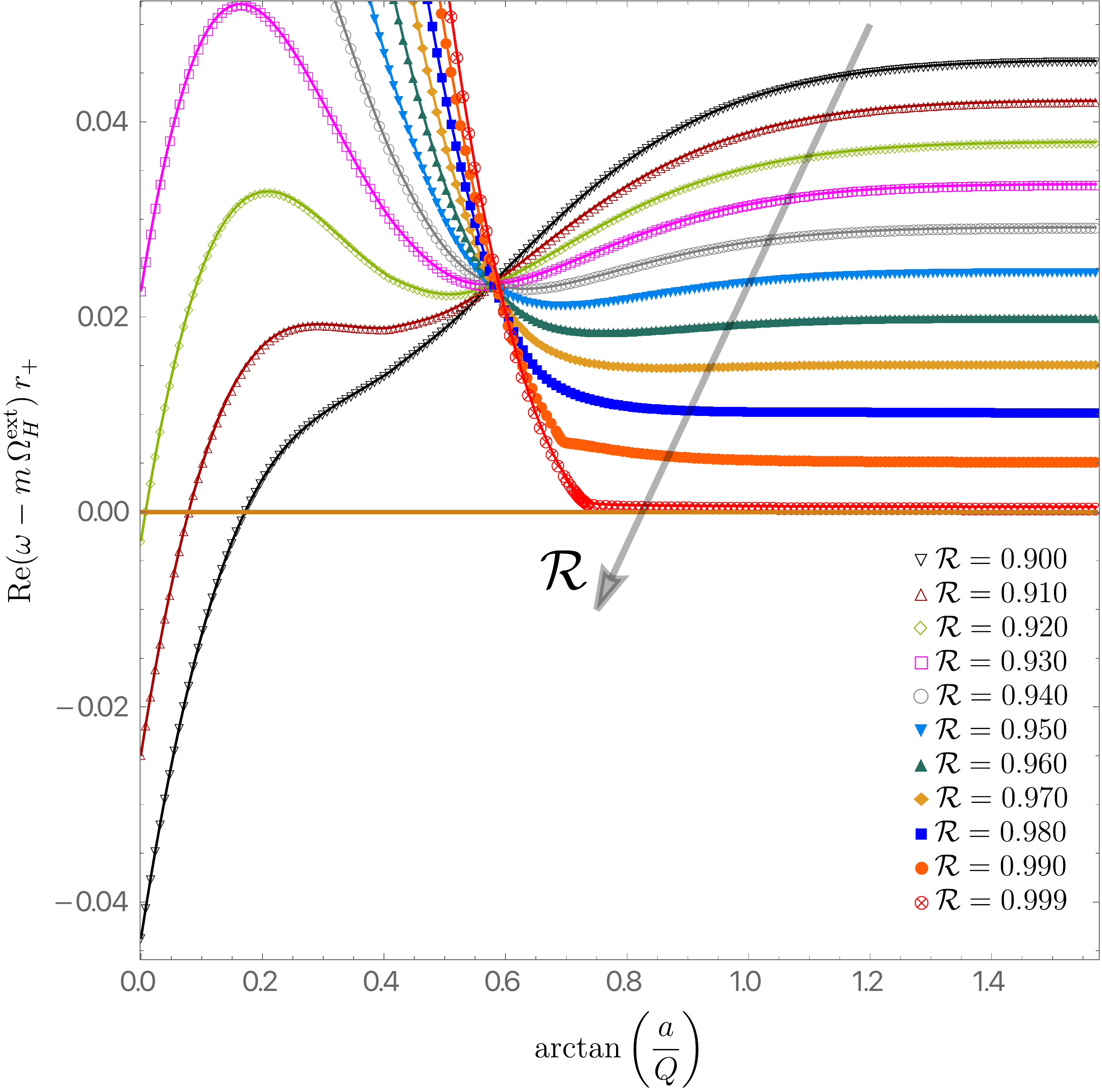}
\hspace{0.1cm}
\includegraphics[width=.48\linewidth]{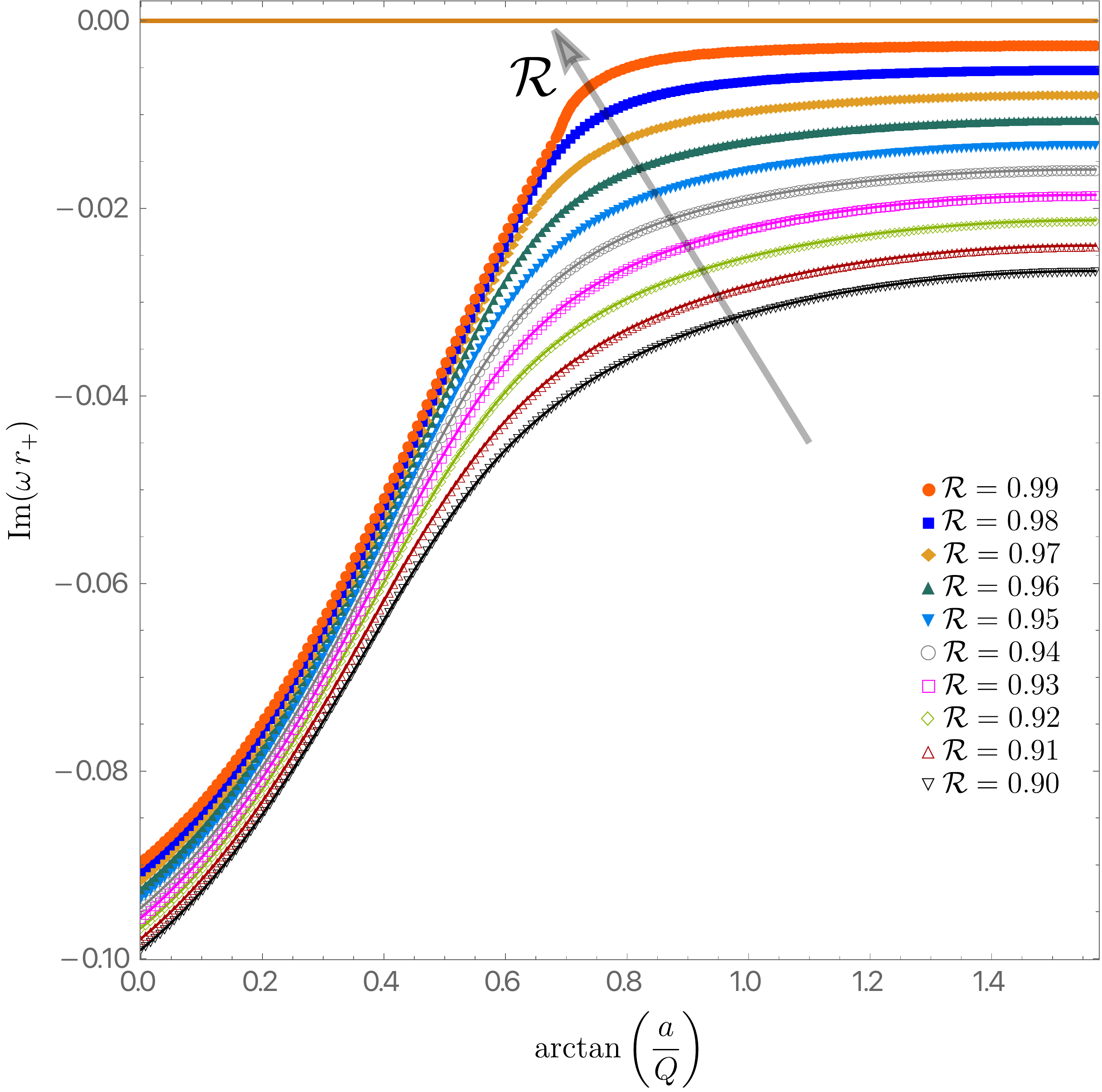}
\vskip 0.1cm
\includegraphics[width=.48\linewidth]{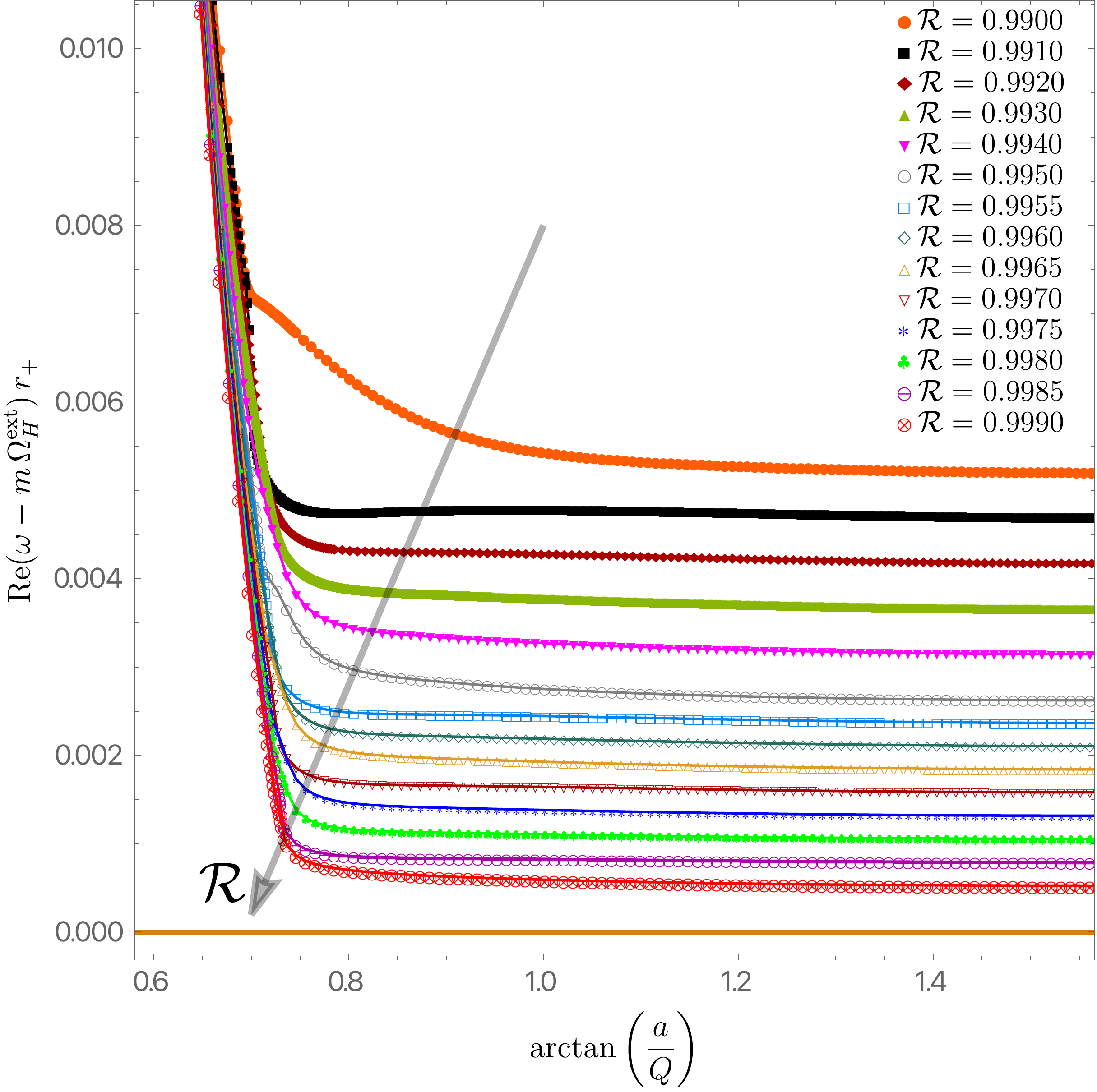}
\hspace{0.1cm}
\includegraphics[width=.48\linewidth]{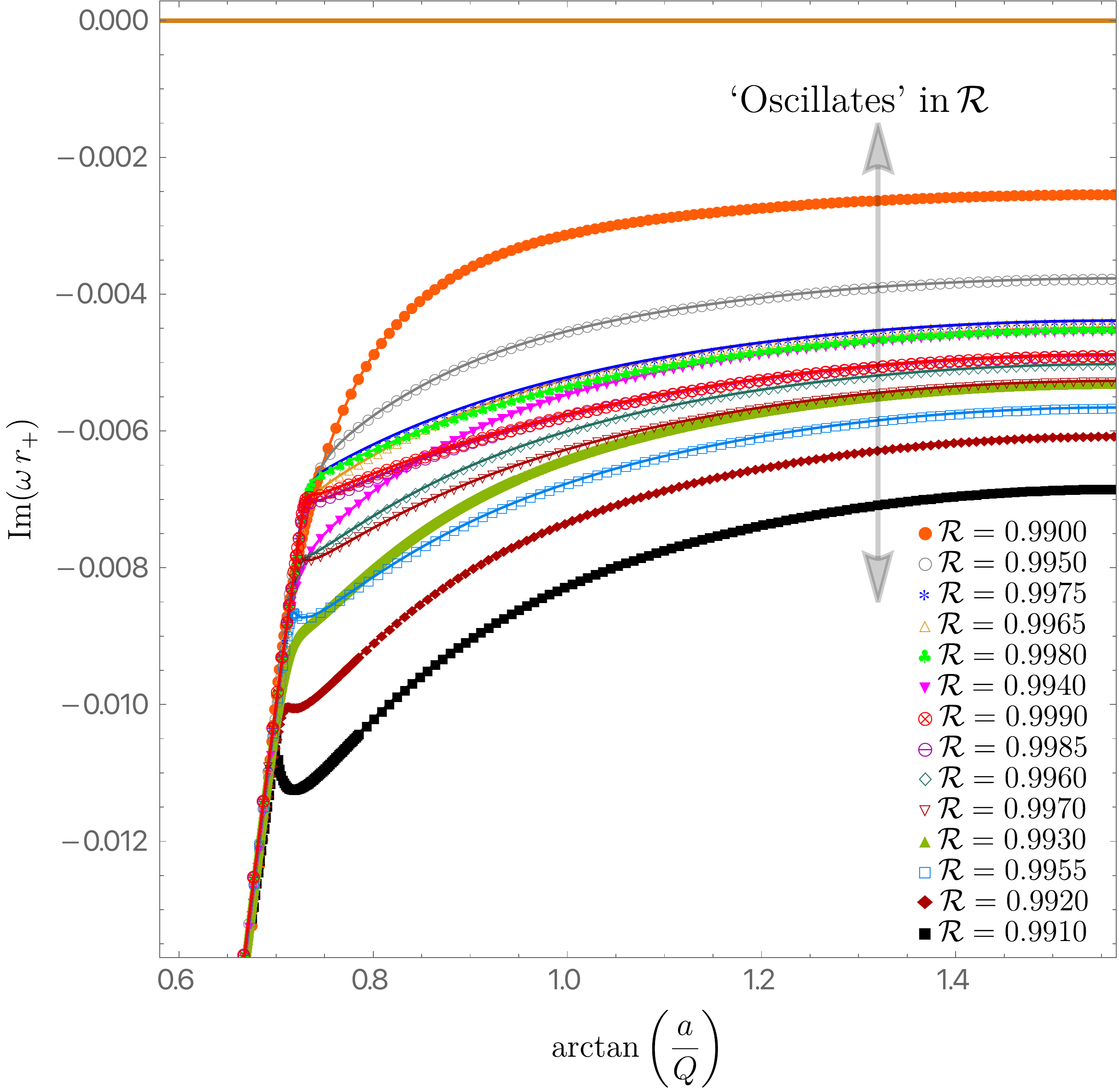}
\caption{Information about the real (left) and imaginary (right) parts of the photon sphere (PS) frequency as a function of $\Theta$ for several fixed values of $\mathcal{R}$. Namely, in the top panels ones has curves for $0.90\leq \mathcal{R}\leq 0.999$ (top-left) or for $0.90\leq \mathcal{R}\leq 0.99$ (top-right); see exact values of $\mathcal{R}$ and associated color code in the legends of the plots. On the other hand, in the bottom panels we focus our attention closer to extremality, in the region  $0.990\leq \mathcal{R}\leq 0.999$.}
\label{Fig:Star}
\end{figure} 

Altogether, from the analyses of Figs.~\ref{Fig:WKBn0-NHn0-Im}-\ref{Fig:WKB-NH-Im_near_ext_2} there is a fundamental property that emerges and that should be further highlighted and discussed. The orange $n=0$ photon sphere mode (as we identify it in the RN limit and trace forward for higher values of $\Theta$) seems to be trying  to reach $\mathrm{Re}\, \tilde{\omega}=m\tilde{\Omega}_H^{\rm ext}$ and $\mathrm{Im}\,\tilde{\omega}=0$ at extremality and for sufficiently large $\Theta$ --- note that the scale of the vertical axis is changing between Figs~\ref{Fig:WKBn0-NHn0-Im}-\ref{Fig:WKB-NH-Im_near_ext_2}. This would be in line with the WKB analysis (and its NH limit) which predicts that PS modes should indeed have this behaviour as $\mathcal{R}\to 1$ and for $\Theta\geq \Theta_\star\gtrsim \Theta_{\star}^{\footnotesize \rm WKB}>\pi/6$ (see the dashed WKB line in previous figures), where $\Theta_\star$ was introduced in the second to last paragraph of section~\ref{sec:EigenvalueRepulsionsA} (recall also  footnote~\ref{foot:ThetaStar}). But the second KN family of QNMs $-$ namely the NH family $-$ already sits at $\mathrm{Re} \, \tilde{\omega}=m\tilde{\Omega}_H^{\rm ext}$ and $\mathrm{Im} \, \tilde{\omega}=0$ at extremality (and for any value of $0 \leq \Theta \leq\pi/2$, not only for $\Theta_\star \leq \Theta \leq\pi/2$). So, making contact with the three possible scenarios enumerated in the end of section~\ref{sec:EigenvalueRepulsionsA}, our numerical data seems to be suggesting that in the KN QNM spectra there are {\it no} eigenvalue crossings in isolated {\it non}-extremal points of the parameter space. Instead, we are seeing evidence that two families of eigenvalues (the PS and NH modes as we identify them in the RN limit and their overtones) can meet and {\it terminate} along a continuous portion of the KN extremal boundary (very much like several QNM overtones already meet and terminate at the extremal endpoint of the RN and Kerr black holes). To gather more evidence in favour of this scenario, it is thus worthy to analyse in more detail how the PS modes (attempt to) force their pathway towards $\tilde{\omega}=m\tilde{\Omega}_H^{\rm ext}$ at extremality. For that  we collect some of the data of Figs.~\ref{Fig:WKBn0-NHn0-Im}-\ref{Fig:WKB-NH-Im_near_ext_2} in a single plot where we show the evolution of the PS mode frequency as a function of $\Theta$ for several fixed values of $\mathcal{R}$. This is done in Fig.~\ref{Fig:Star}.
In  the top panels we show the evolution of the PS family when we are not too close to extremality, namely for $0.90\leq \mathcal{R}\leq 0.99$ ($\mathcal{R}=0.90,0.91,0.92,0.93,0.94,0.95,0.96,0.97,0.98,0.99$); the same monotonic behaviour is observed in the evolution for smaller values of $\mathcal{R}$ but we do not show it here (see later Fig.~\ref{Fig:KN3d} with the full phase space). On the other hand, in the bottom panels we plot the same quantities but this time for families of KN black holes that are even closer to extremality in the region $0.990\leq \mathcal{R}\leq 0.999$, namely we plot several curves with constant 
$\mathcal{R}=0.990,0.991,0.992,0.993,0.994,0.995,0.9955,0.996,0.9965,0.997, 0.9975$, $0.998, 0.9985, 0.999$ (see legends in the plots).
In the right panels of Fig.~\ref{Fig:Star} we  plot the imaginary part of the frequency, $\mathrm{Im}\,\tilde{\omega}$, as a function of $\Theta$. On the other hand, in the left panels, instead of simply plotting the real part of the frequency, we plot $\mathrm{Re}\,\tilde{\omega}-m\tilde{\Omega}_H^{\rm ext}$ where $\tilde{\Omega}_H^{\rm ext}$ is the extremal value ($\mathcal{R}=1$) of the frequency for a given $\Theta$ as given by \eqref{KN:thermo}-\eqref{PolarParametrization}: $\tilde{\Omega}_H^{\hbox{\footnotesize ext}} =\alpha_{\hbox{\footnotesize ext}}/(1+\alpha_{\hbox{\footnotesize ext}}^2)=\sin\Theta/(1 + \sin^2\Theta)$. This quantity has the advantage that it vanishes when $\mathrm{Re}\,\tilde{\omega}\to m\tilde{\Omega}_H^{\rm ext}$ at extremality.

From the top panels of Fig.~\ref{Fig:Star} we see that the PS frequency indeed attempts to approach $\mathrm{Re}\,\tilde{\omega}= m\tilde{\Omega}_H^{\rm ext}$ and $\mathrm{Im}\,\tilde{\omega}=0$ for $\Theta>\Theta_\star$ as we keep decreasing the distance to extremity (i.e. as we approach $\mathcal{R}=1$ from below), where $\Theta_\star$ is roughly around $0.8$  (doing a rough extrapolation of the $\mathcal{R}=0.999$ curve all the way up to $\mathrm{Im}\, \tilde{\omega}=0$). But around $\mathcal{R}\sim 0.99$, the system `realizes' that the PS modes are dangerously approaching the NH modes and eigenvalues repulsions (reported in Figs.~\ref{Fig:WKBn0-NHn0-Im}-\ref{Fig:WKB-NH-Im_near_ext_2}) kick in. 
In more detail, for $\mathcal{R}>0.9$, the bottom-left panel of Fig.~\ref{Fig:Star} shows that, interestingly, the real part of the PS frequencies still keeps monotonically approaching $m\tilde{\Omega}_H^{\rm ext}$ as $\mathcal{R}$ increases from 0.990 to 0.999 following a pattern that seems to be blind to any worries with level repulsion. But, perhaps to avoid eigenvalue crossing in the complex plane as $\mathcal{R}\to 1$, the $\mathrm{Im}\,\tilde{\omega}$ (see bottom-right panel)  reacts and starts `oscillating' in $\mathcal{R}$, i.e. for $\mathcal{R}\gtrsim 0.990$ one finds that $|\mathrm{Im}\,\tilde{\omega}|$ no longer decreases monotonically towards zero with increasing $\mathcal{R}$. Instead, for a small increment of $\mathcal{R}$ (and fixed $\Theta$), sometimes  $|\mathrm{Im}\,\tilde{\omega}|$ decreases and other times it increases in such a way that in practice (for the values of $\mathcal{R}\leq 0.999$ that we computed),  for $\Theta>\Theta_\star$, it stays in-between the top orange disk curve (with $\mathcal{R}=0.990$) and the bottom black square curve (with $\mathcal{R}=0.991$).
This is eigenvalue repulsion in action at its best. 

What happens if we approach even further extremality, i.e. if we plunge into the region $0.999<\mathcal{R}\leq 1$? We have not attempted to explore this region. To put into context, in the Kerr limit, the line of constant  $\mathcal{R}=a/r_+=0.999$ already corresponds, in mass units, to $a/M\simeq 0.999999$. So we have not attempted to increase this value even further since it becomes very costly computationally. Furthermore, whether the PS modes do or do not reach $\mathrm{Im}\, \omega\to 0$ and $\mathrm{Re}\, \omega\to m\Omega_H^{\hbox{\footnotesize{ext}}}$ at extremality for $\Theta\geq \Theta_\star\simeq 0.881$ is a lesser question, since even if it does, it does so with an $|\mathrm{Im}\,\tilde\omega|$ that is always larger than (or equal to) the $|\mathrm{Im}\,\tilde\omega|$ of the NH modes (i.e. they approach at smaller rate) and thus they are not the dominant modes in this region of the parameter space. It should however be noticed that, in the QNM spectra of Kerr, \cite{Yang:2013uba} report the existence of modes whose $|\mathrm{Im}\,\tilde\omega|$ has `damped oscillations' in $\mathcal{R}$ and approaches zero at extremality (see the discussion of figure 3 of \cite{Yang:2013uba}). Moreover, similar damped oscillations of $\mathrm{Im}\,\tilde\omega$ as extremality is approached was observed in the (charged) scalar field QNM spectra of Reissner-Nordstr\"om-de Sitter \cite{Dias:2018ufh} (see discussion of figures 9-12 of \cite{Dias:2018ufh}; especially the latter). 
So we cannot exclude the possibility (also very much suggested by the WKB analysis of section~\ref{sec:PSwkb}) that the `oscillations' observed in the bottom-right plot of Fig.~\ref{Fig:Star} persist all the way to extremality with an exponential decay envelope (sufficiently close to extremality) such that the curves ultimately hit $\mathrm{Im}\, \omega\to 0$ and $\mathrm{Re}\, \omega\to m\Omega_H^{\hbox{\footnotesize{ext}}}$ at extremality ($\mathcal{R}=1$) for $\Theta>\Theta_\star$. It is very tempting to claim it is the case and we believe it is for the above reasons. If so, the PS and NH modes meet and terminate, with $\mathrm{Im}\, \omega\to 0$ and $\mathrm{Re}\, \omega\to m\Omega_H^{\hbox{\footnotesize{ext}}}$, at extremality $\mathcal{R}=1$ for $\Theta>\Theta_\star$ (this is the third possibility enumerated in the end of our discussion of section~\ref{sec:EigenvalueRepulsionsA}).
A similar behaviour was observed for the gravito-electromagnetic modes of KN \cite{Dias:2021yju,Dias:2022oqm}.

\begin{figure}[t]
\centering
\includegraphics[width=.49\textwidth]{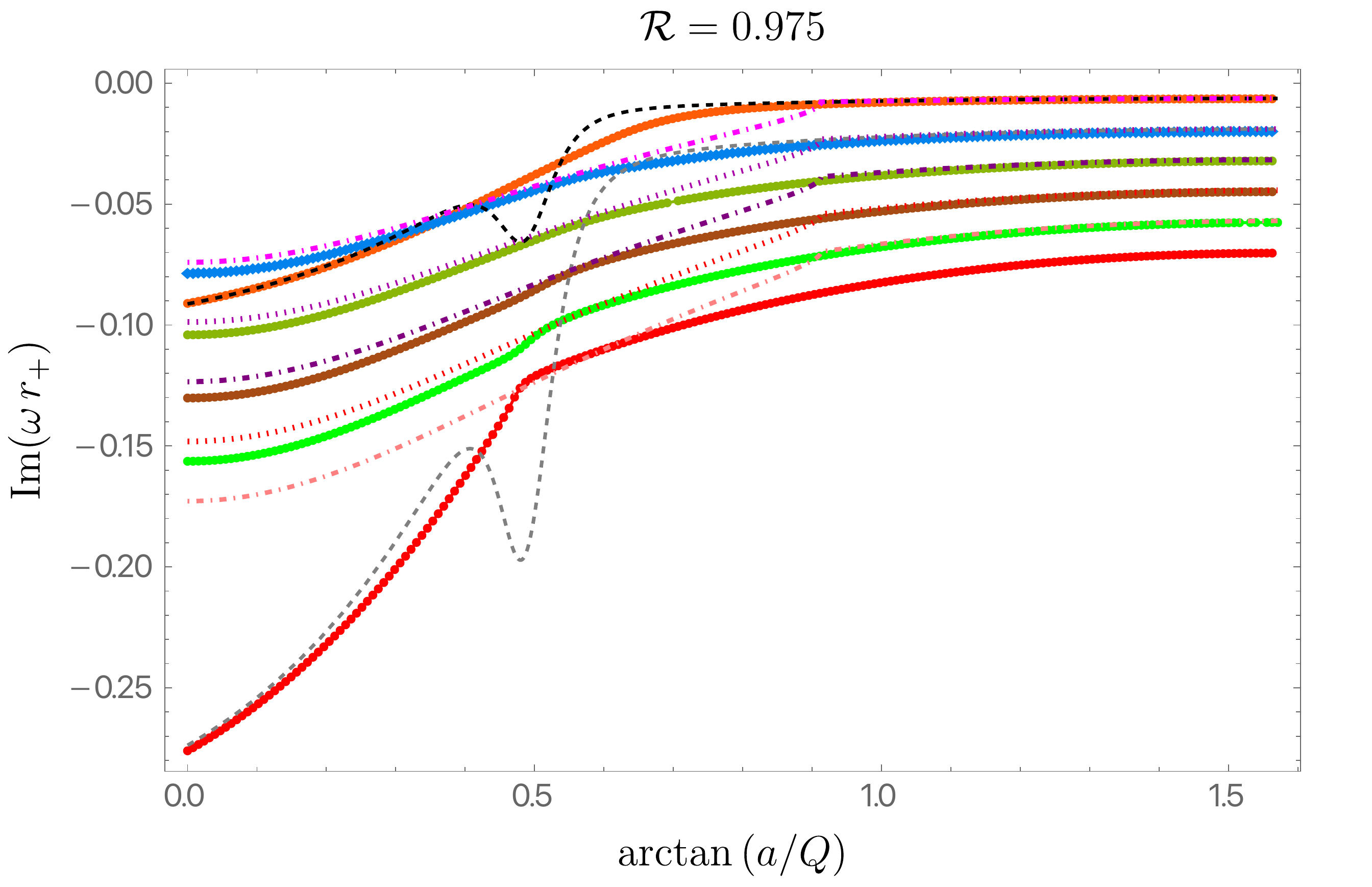}
\hspace{0.0cm}
\includegraphics[width=.49\textwidth]{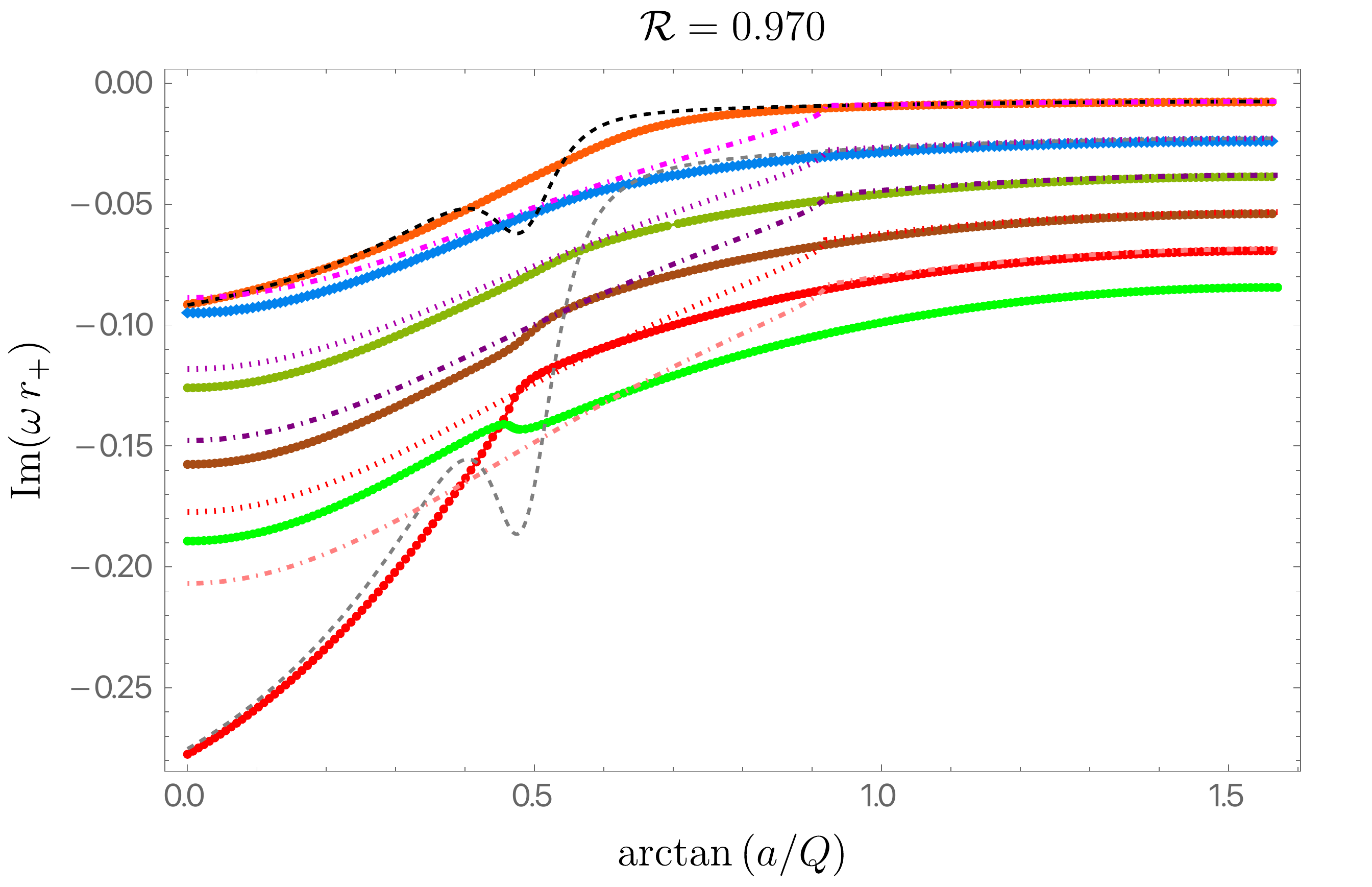}
\vskip 0.2cm
\includegraphics[width=.49\textwidth]{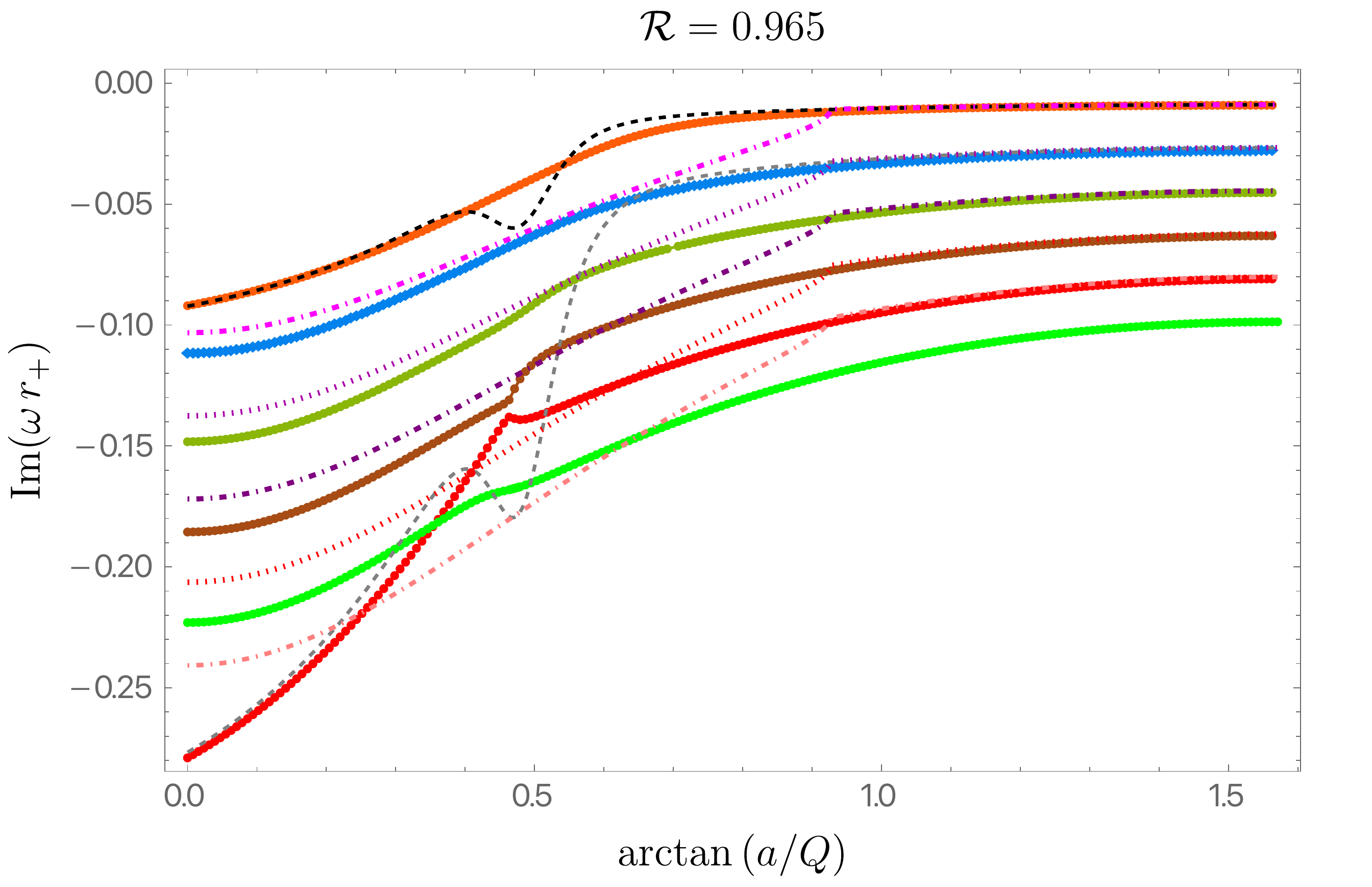}
\hspace{0.0cm}
\includegraphics[width=.49\textwidth]{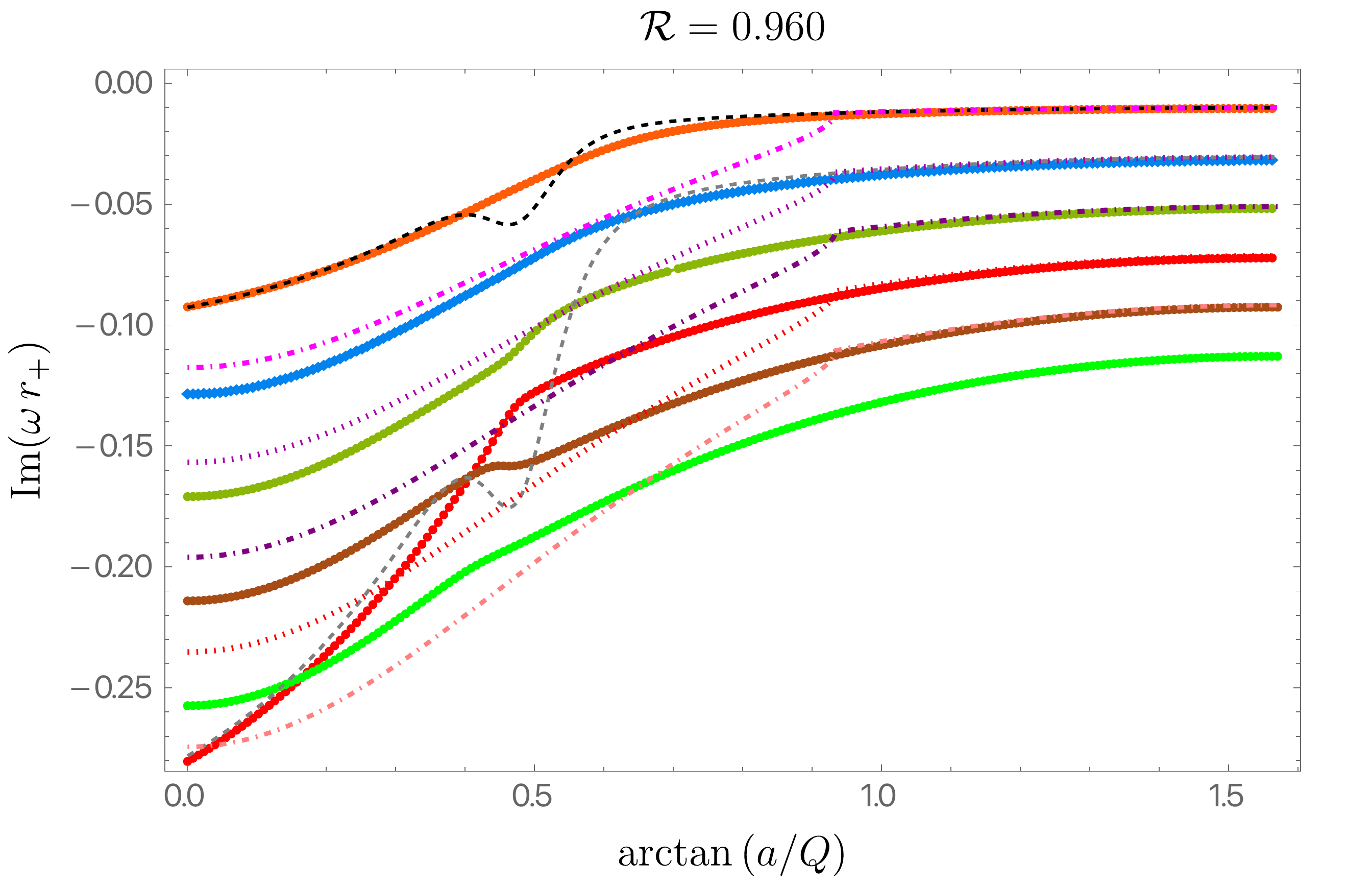}
\vskip 0.2cm
\includegraphics[width=.49\textwidth]{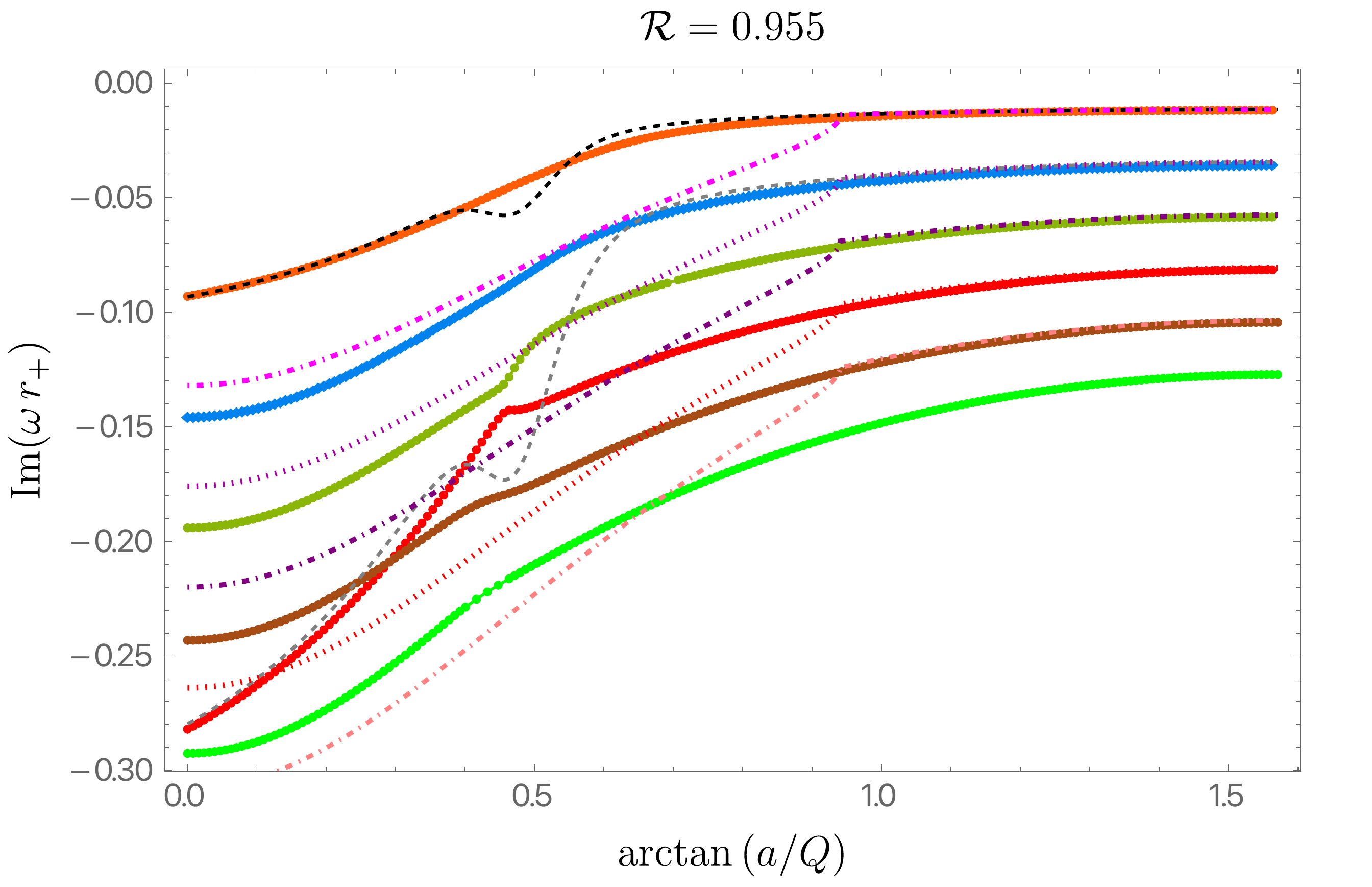}
\hspace{0.0cm}
\includegraphics[width=.49\textwidth]{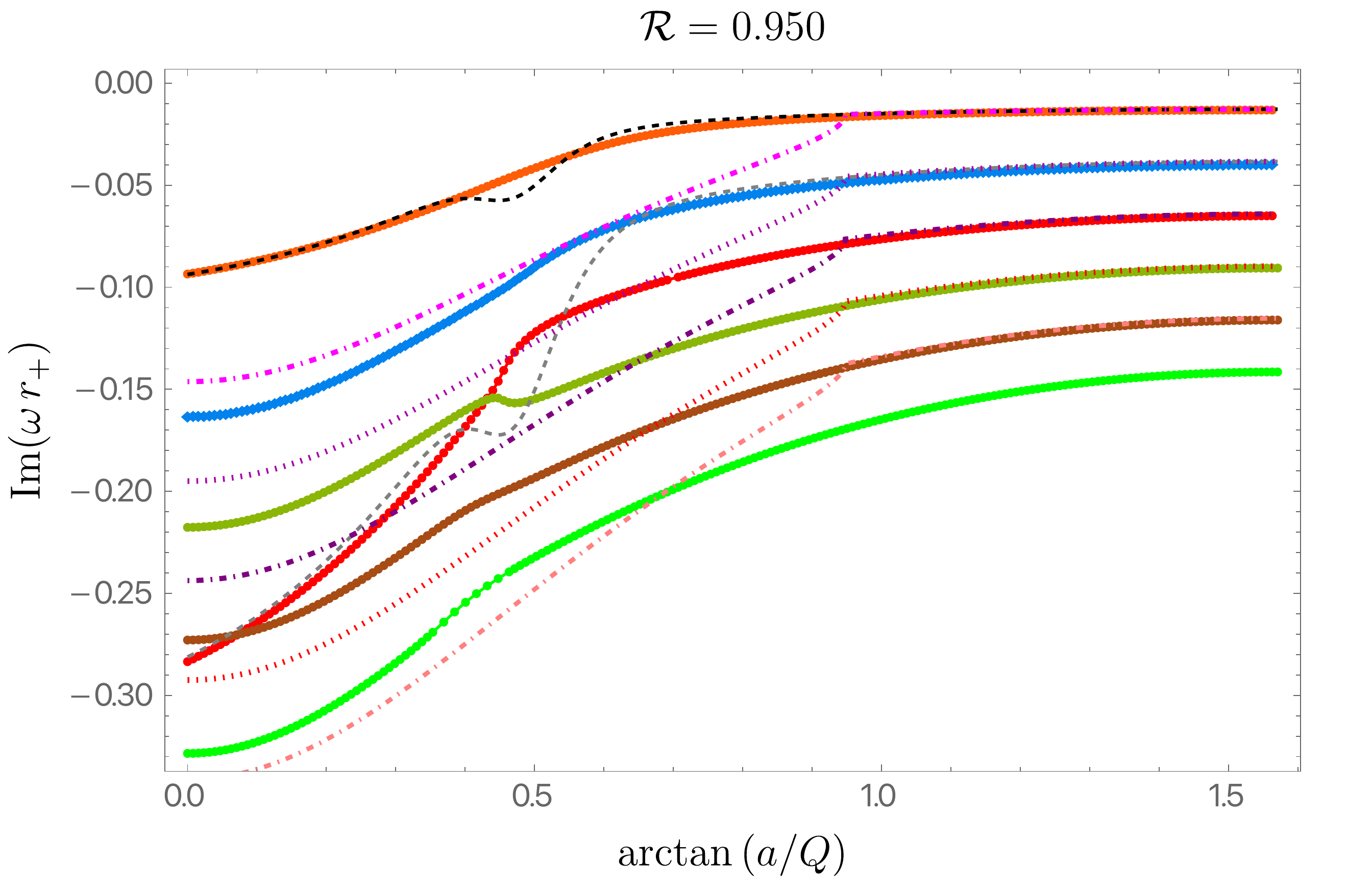}
\caption{Imaginary part of the frequency as a function of $\Theta=\arctan\left( a/Q\right)$ for the $n=0$ PS (orange disks),  $n=1$ PS (dark-red disks; not present in Figs.~\ref{Fig:WKBn0-NHn0-Im}$-$\ref{Fig:WKBn0-NHn4-Im})  and the $n=0,1,2,3$ NH (blue, dark-green, brown and green diamonds) families of QNMs with  $m=\ell=2$ for a KN family with $\mathcal{R}=0.975$, $\mathcal{R}=0.970$, $\mathcal{R}=0.965$, $\mathcal{R}=0.960$,  $\mathcal{R}=0.955$ and $\mathcal{R}=0.950$ (following the lexicographic order). We also display the WKB result $\tilde{\omega}_{\hbox{\tiny WKB}}$ for $n=0,1$ (dashed black and dashed gray lines, respectively) and the near-extremal frequency $\tilde{\omega}_{\hbox{\tiny MAE}}$ for $p=0,1,2,3$ (dot-dashed magenta, dotted dark magenta, dot-dashed purple, dotted pink, dot-dashed pink  lines, respectively). (This series of plots continues in Fig.~\ref{Fig:WKBn1-NHn4-ImB} for smaller $\mathcal{R}$).}
\label{Fig:WKBn1-NHn4-Im}
\end{figure}  

\begin{figure}[t]
\centering
\includegraphics[width=.49\textwidth]{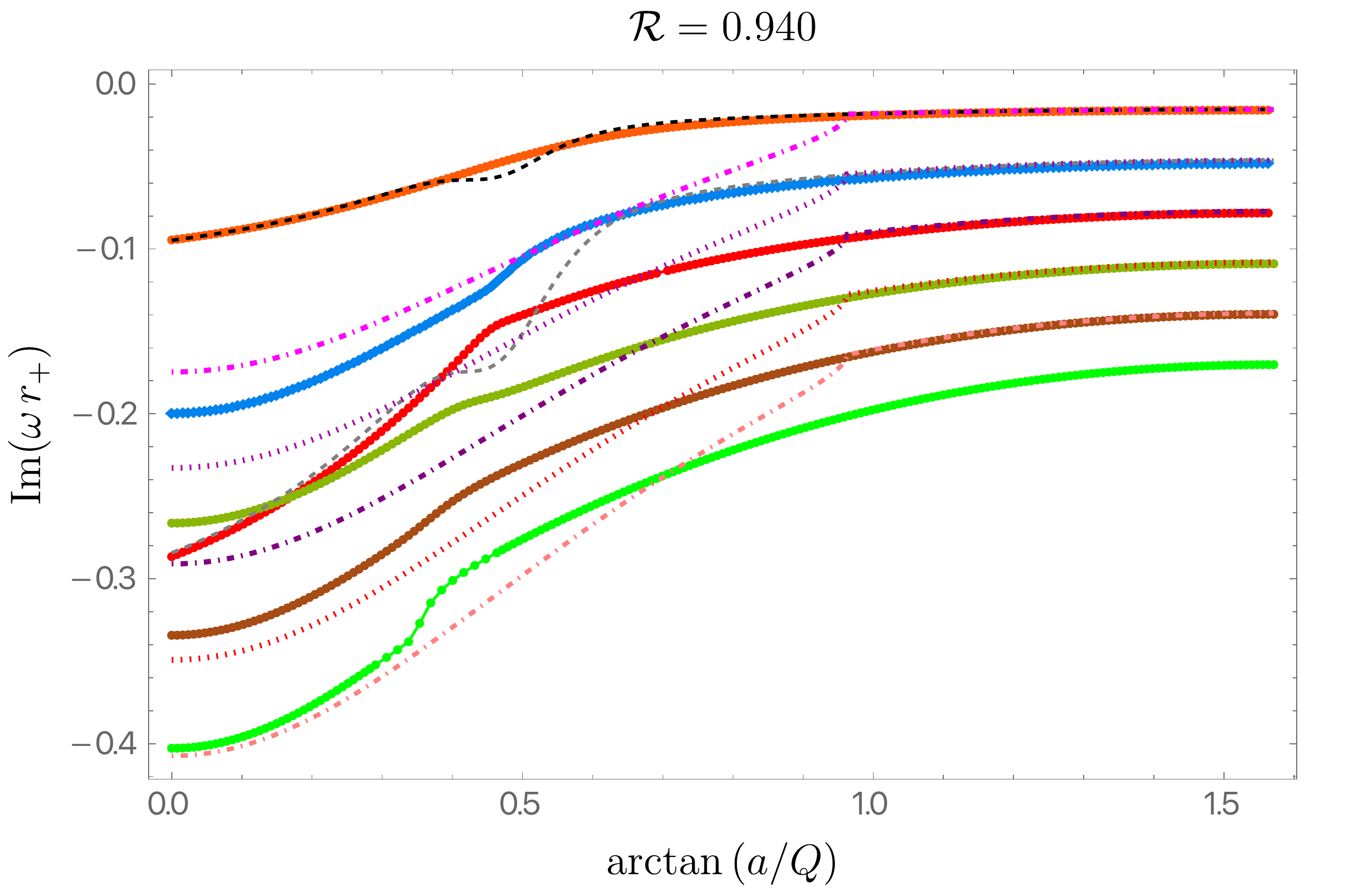}
\hspace{0.0cm}
\includegraphics[width=.49\textwidth]{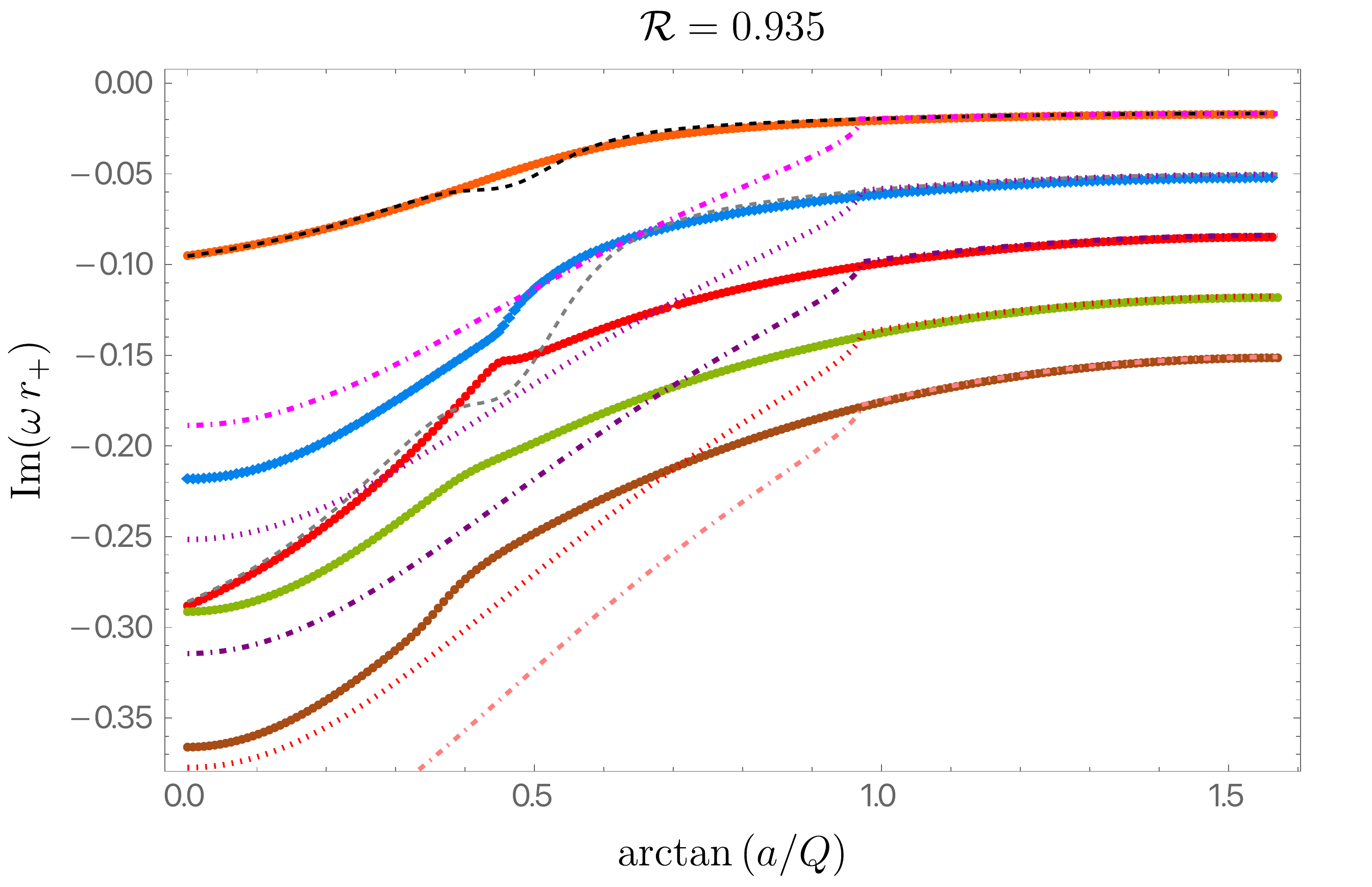}
\vskip 0.2cm
\includegraphics[width=.49\textwidth]{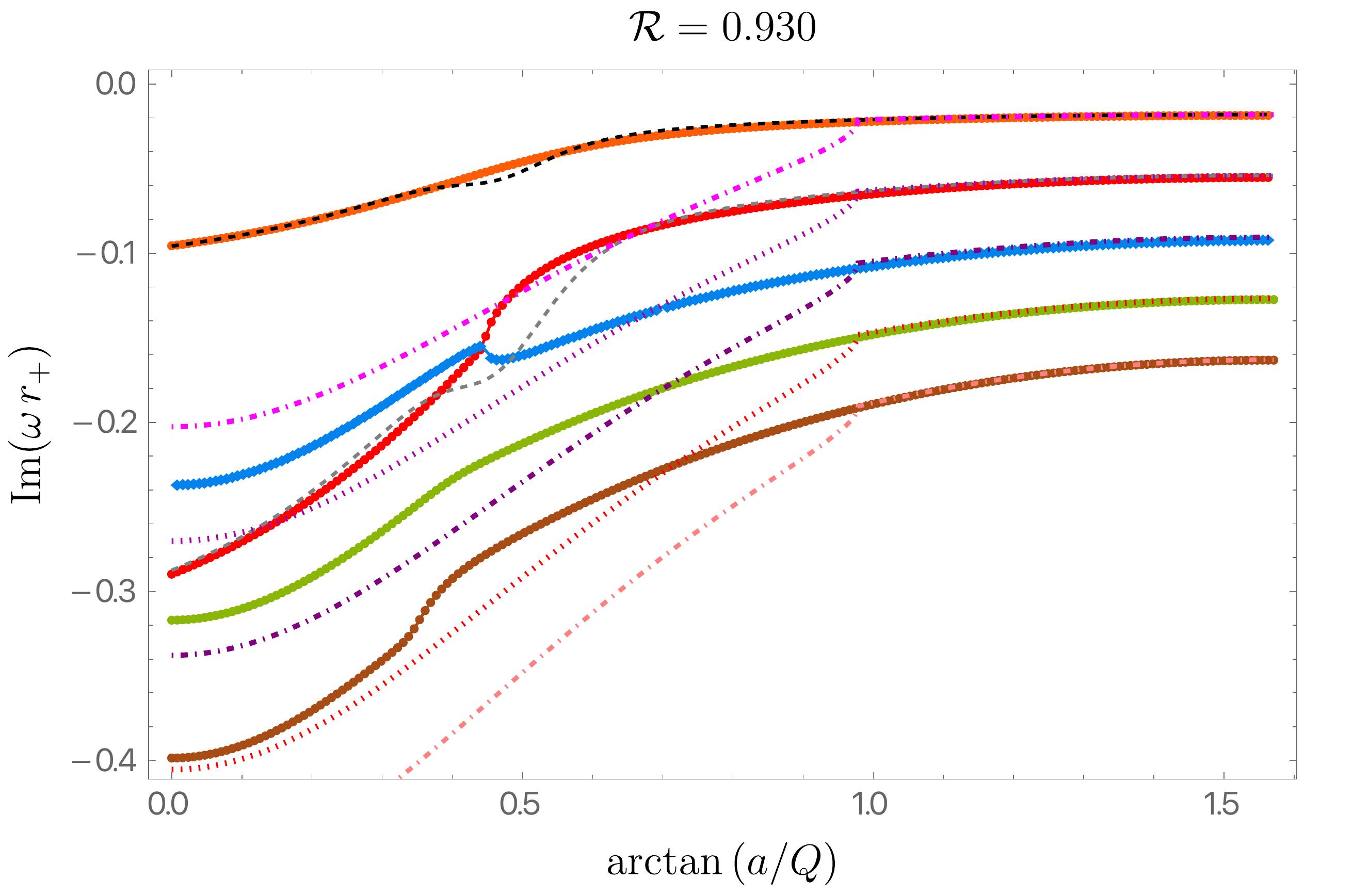}
\hspace{0.0cm}
\includegraphics[width=.49\textwidth]{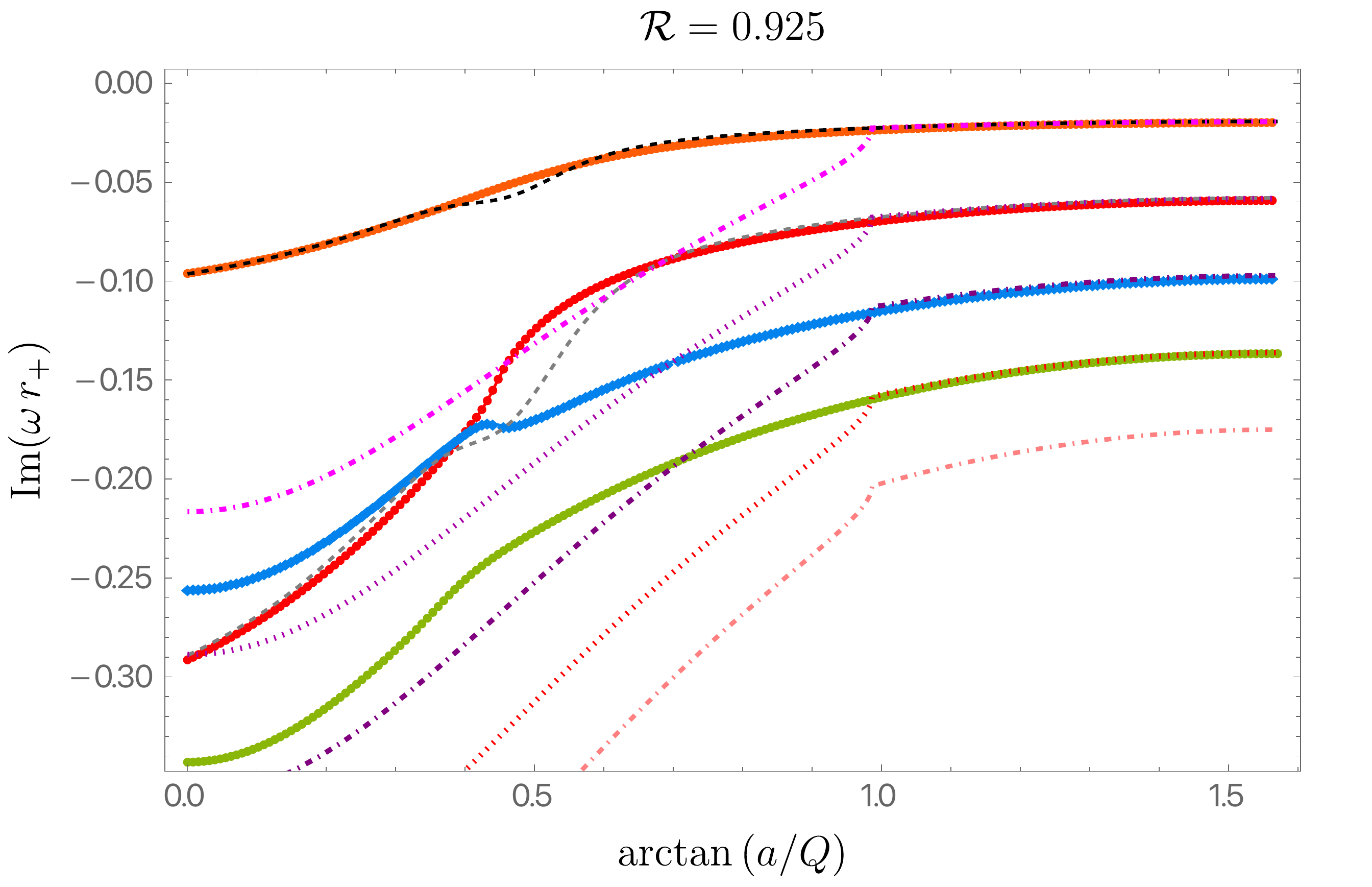}
\vskip 0.2cm
\includegraphics[width=.49\textwidth]{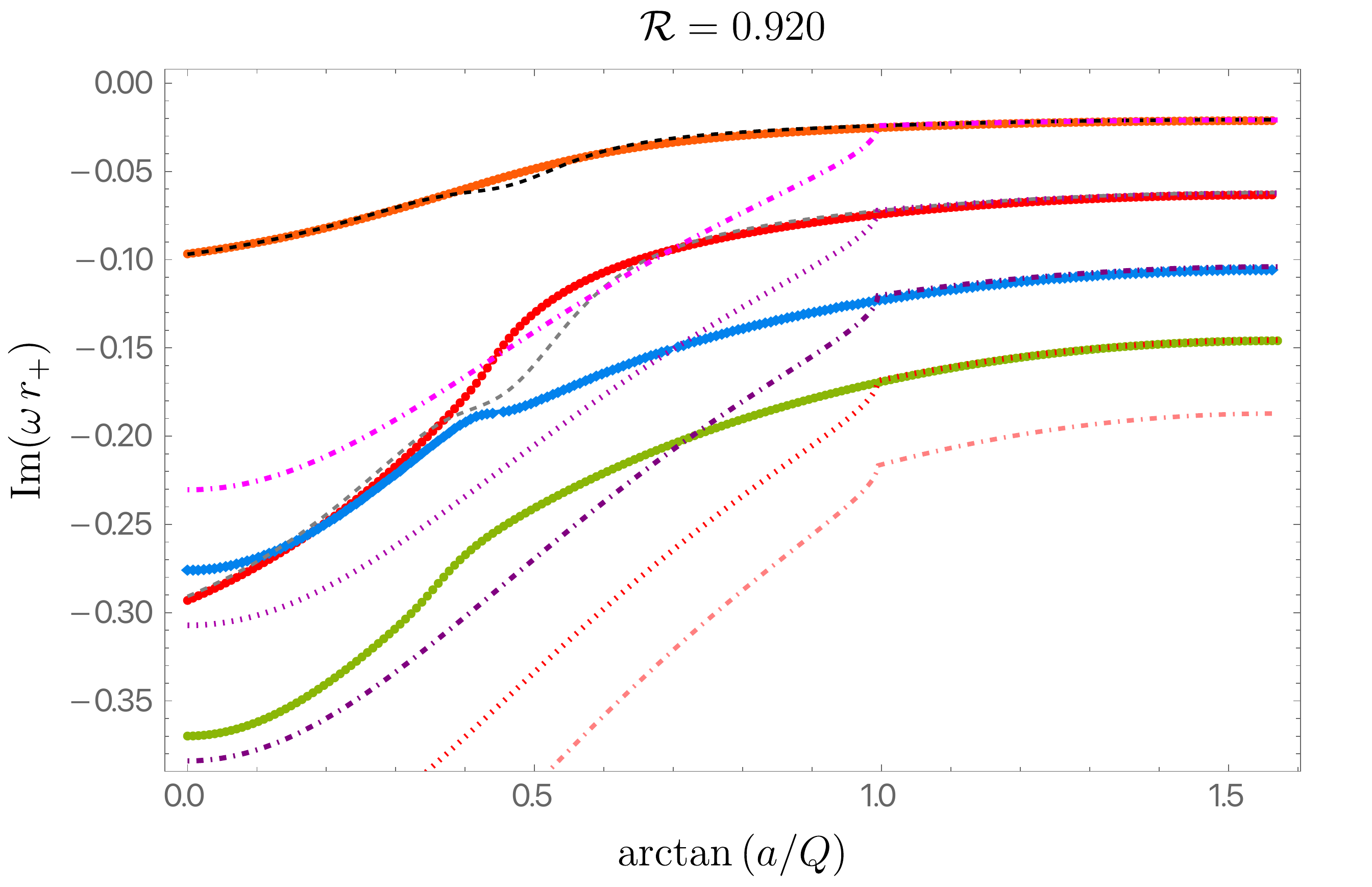}
\hspace{0.0cm}
\includegraphics[width=.49\textwidth]{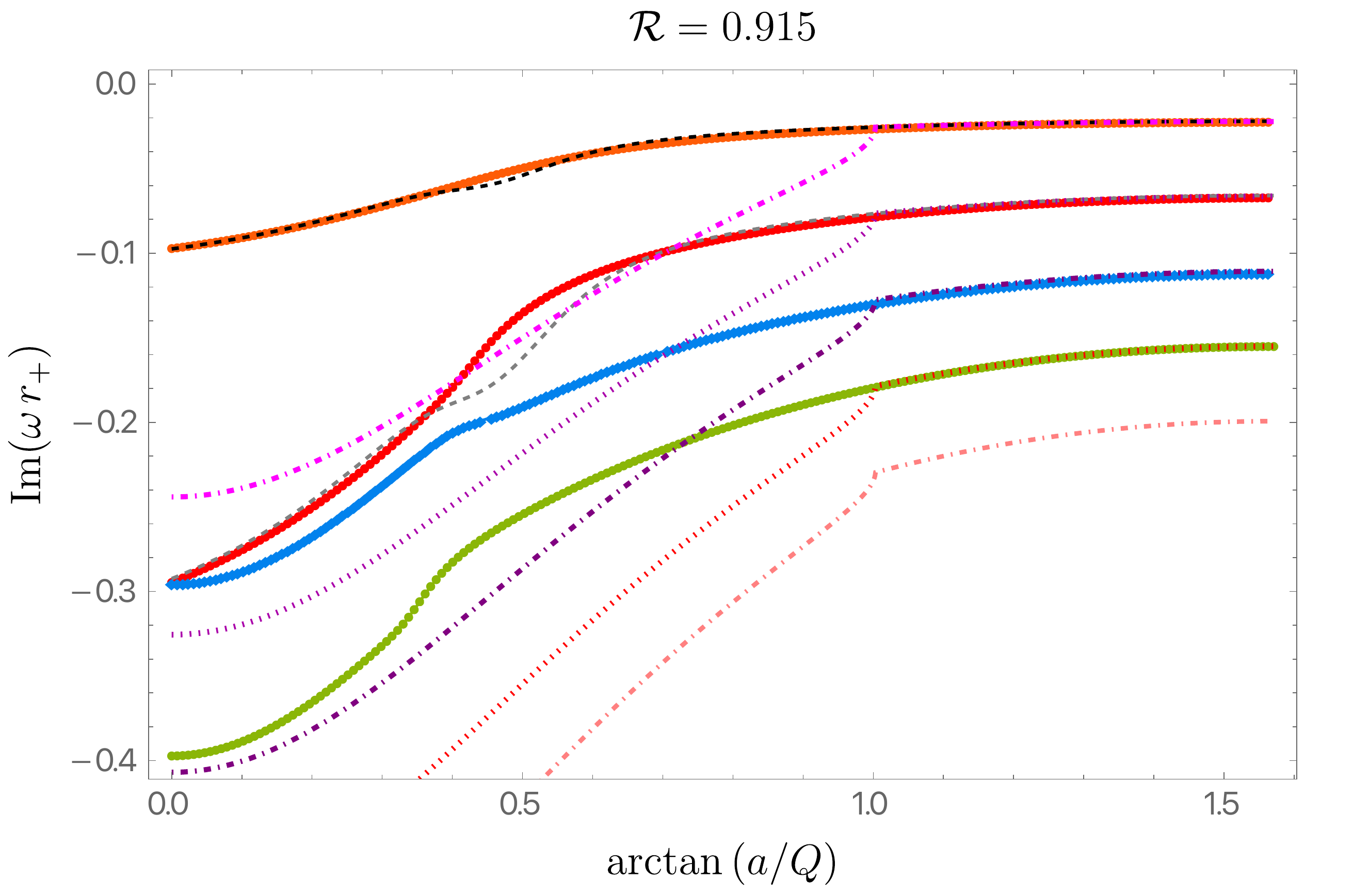}
\caption{This series of plots continues the one started in  Fig.~\ref{Fig:WKBn1-NHn4-Im} but this time for $\mathcal{R}=0.940$, $\mathcal{R}=0.935$, $\mathcal{R}=0.930$, $\mathcal{R}=0.925$,  $\mathcal{R}=0.920$ and $\mathcal{R}=0.915$ (following the lexicographic order).}
\label{Fig:WKBn1-NHn4-ImB}
\end{figure}  

For completeness, we may now ask whether the eigenvalue repulsions observed in  Figs.~\ref{Fig:WKBn0-NHn0-Im}-\ref{Fig:Star} between the $n=0$  PS (orange disks) and NH families also extend to the higher PS overtones, namely for $n=1$ (or higher). The answer is yes, and in fact the eigenvalue repulsions in the $n=1$ PS family are already very visible further away from extremality. In detail, we address this question in Figs.~\ref{Fig:WKBn1-NHn4-Im}$-$\ref {Fig:WKBn1-NHn4-ImB}, where the latter plots are to be seen as a continuation of the former. In particular, Fig.~\ref{Fig:WKBn1-NHn4-Im} displays the cases $\mathcal{R}=0.975$, $\mathcal{R}=0.970$, $\mathcal{R}=0.965$, $\mathcal{R}=0.960$,  $\mathcal{R}=0.955$ and $\mathcal{R}=0.950$  (following the lexicographic order) and then Fig.~\ref{Fig:WKBn1-NHn4-ImB} continues this series of plots to even smaller values of $\mathcal{R}$ for the values $\mathcal{R}=0.940$, $\mathcal{R}=0.935$, $\mathcal{R}=0.930$, $\mathcal{R}=0.925$,  $\mathcal{R}=0.920$ and $\mathcal{R}=0.915$.\footnote{\label{footColorCode} Recall that the color code was introduced in Fig.~\ref{Fig:RN} for the RN case ($\Theta=0$). Then we {\it locked} this color code and, at fixed $\mathcal{R}$, we increase $\Theta$ to follow each QNM family from its RN limit ($\Theta=0$) until its Kerr limit ($\Theta=\pi/2$). Further recall that the nomenclature PS/NH refers to the nature of the modes in the RN limit.} In these figures we always plot the imaginary part of the frequency as a function of $\Theta$ for the $m=\ell=2$ modes, and we display the $n=0$ (orange disks) and $n=1$ (red disks) PS families together with the $n=0,1,2,3$ NH (blue, dark-green, brown and green diamonds) families of QNMs. Moreover, we also plot the higher-order WKB frequency $\tilde{\omega}_{\hbox{\tiny WKB}}$ $-$ as given by~\eqref{WKBfrequency} $-$ with $p=0$ (dashed black line) and with $p=1$ (dashed gray line),  and the near-extremal frequency $\tilde{\omega}_{\hbox{\tiny MAE}} $ $-$ as given by~\eqref{NEfreq} $-$ for $p=0,1,2,3$ (dot-dashed magenta, dotted dark magenta, dot-dashed purple, dotted pink, dot-dashed pink  lines, respectively). We conclude that a series of eigenvalue repulsions (with characteristic cusp formation followed by breakup/merge of distinct branches), similar to those described in Fig.~\ref{Fig:WKBn0-NHn0-Im}, do occur, namely between:
\begin{description}
    \item $\bullet$ The $n=1$ PS red curve and $n=3$ NH green curve ($\small 0.975 \gtrsim \mathcal{R}\gtrsim 0.970$ in Fig.~\ref{Fig:WKBn1-NHn4-Im}).
    
    \item $\bullet$ The $n=1$ PS red curve and $n=2$ NH brown curve ($ 0.965 \gtrsim \mathcal{R}\gtrsim 0.960$ in Fig.~\ref{Fig:WKBn1-NHn4-Im}).
    
    \item $\bullet$ The $n=1$ PS red curve and $n=1$ NH dark-green curve ($ 0.955 \gtrsim \mathcal{R}\gtrsim 0.950$ in Fig.~\ref{Fig:WKBn1-NHn4-Im}).
    
    \item $\bullet$ The $n=1$ PS red curve and $n=0$ NH blue curve ($ 0.935 \gtrsim \mathcal{R}\gtrsim 0.930$ in Fig.~\ref{Fig:WKBn1-NHn4-ImB}).
\end{description}
This shows that the eigenvalue repulsion phenomena observed in Figs.~\ref{Fig:WKBn0-NHn0-Im}$-$\ref{Fig:WKBn0-NHn0-Re} for the $n=0$  PS  and $n=0$ NH families is not unique. Instead, it is a common feature for other overtones.

We started our discussion of the KN QNM spectra with the RN limit ($\Theta=0$; see Fig.~\ref{Fig:RN}). It is thus enlightening to terminate our journey with a discussion of the `opposite' Kerr limit ($\Theta=\pi/2$).
Therefore, in Fig.~\ref{Fig:Kerr} we display the $m=\ell=2$ QNM spectra for the Kerr black hole. 
\begin{figure}
\centering
\includegraphics[width=.66\linewidth]{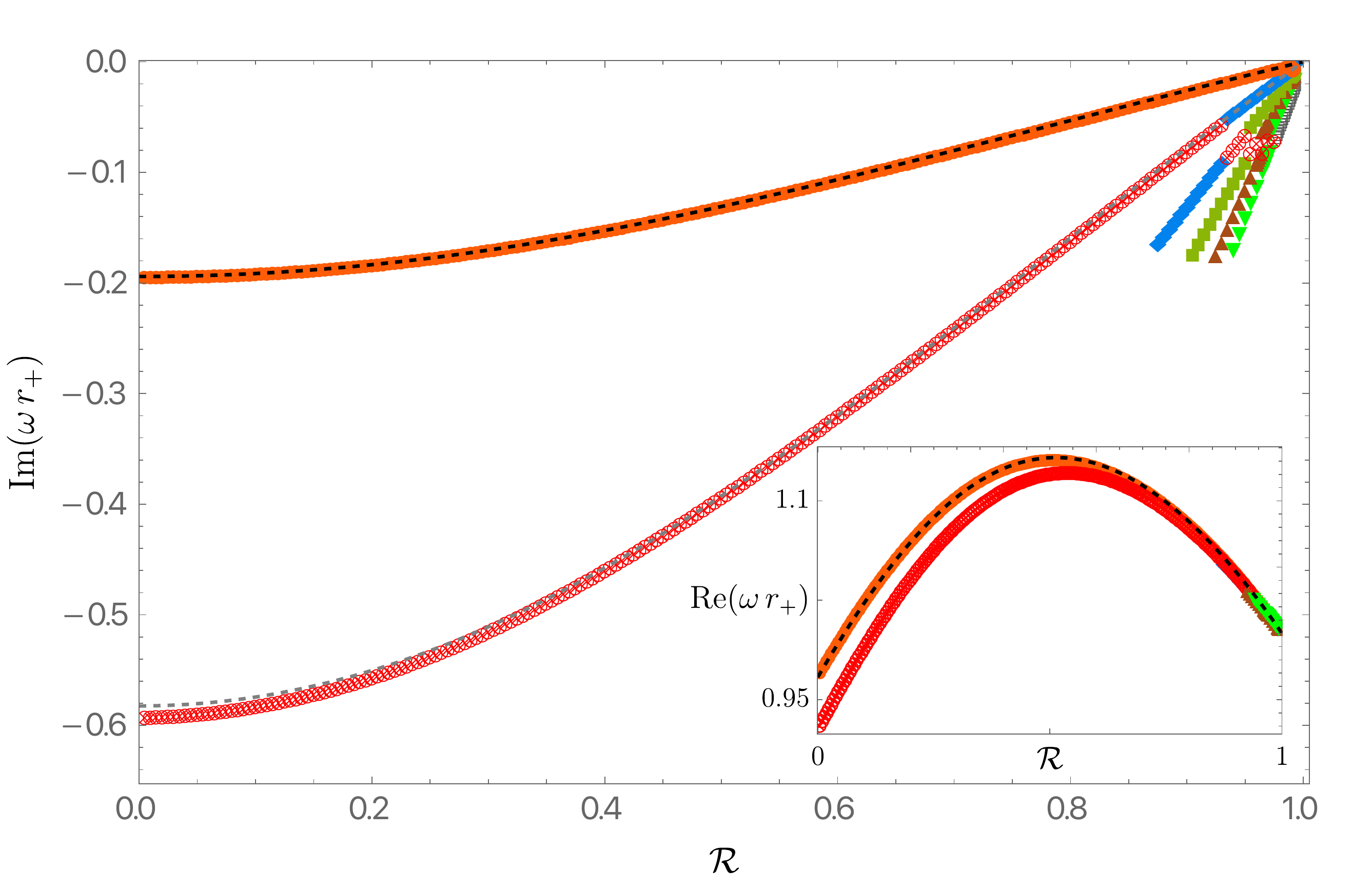}
\includegraphics[width=.66\linewidth]{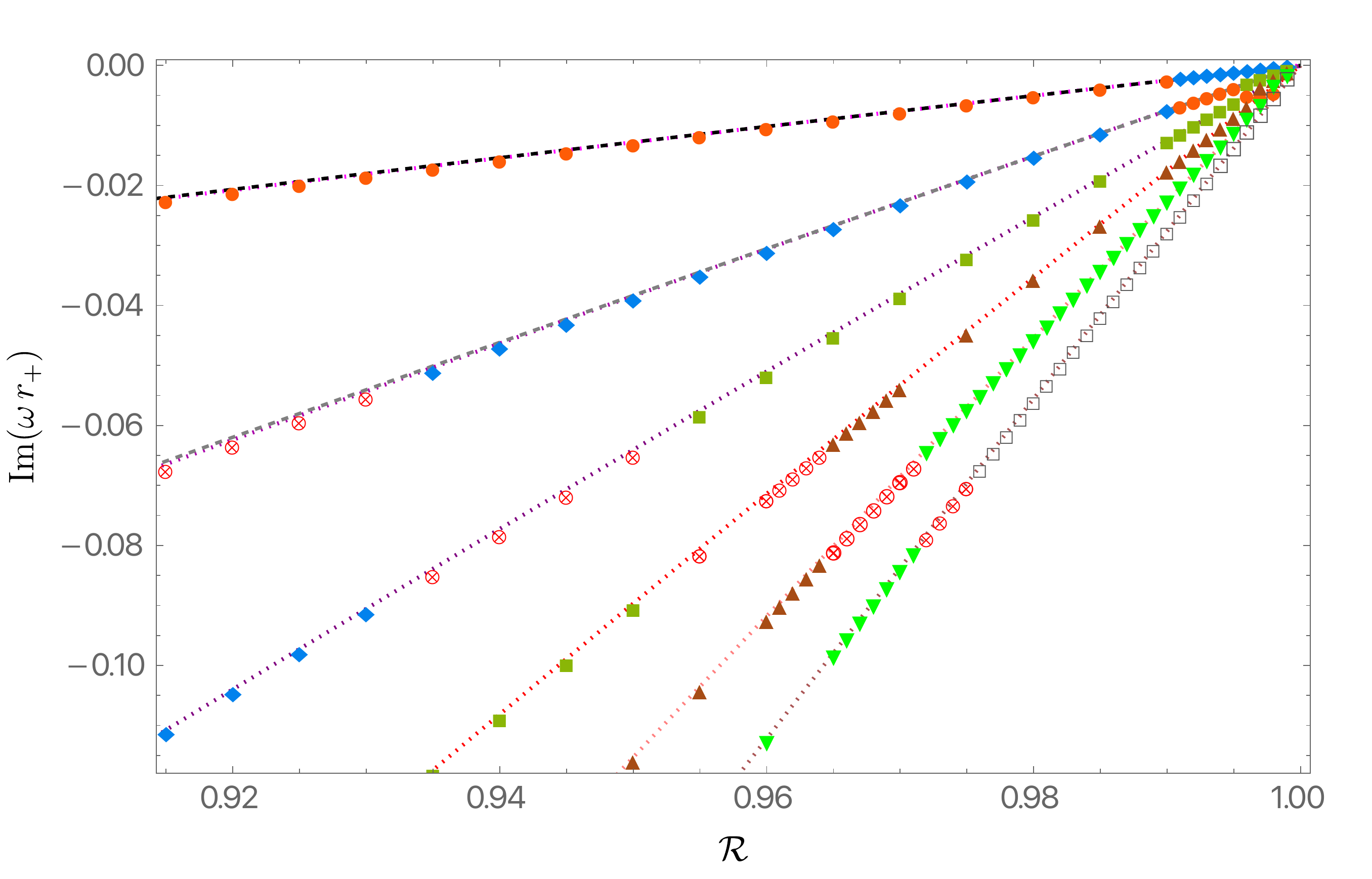}
\caption{The first few overtone PS-NH families of QNMs with  $m=\ell=2$ for the Kerr black hole (\ie the KN black hole with $\Theta=\pi/2$ and thus $a/r_+=\mathcal{R}$). 
The orange~$\hbox{\Large $\bullet$}$ and red~$\hbox{\small $\otimes$}$ are points that connect smoothly, when we decrease $\Theta$ from $\Theta=\pi/2$ down to $\Theta=0$, to the $n=0$ and $n=1$ PS family of the RN black hole, while the blue~$\hbox{\large $\blackdiamond$}$, dark-green~${\hbox{\tiny $\blacksquare$}}$, brown~${\hbox{\small $\blacktriangle$}}$, green $\hbox{\Large $\blacktriangledown$}$ and gray $\hbox{\Large $\square$}$ points connect smoothly to the NH families of the RN black hole with $n=0,1,2,3,4$, respectively.
 {\bf Top panel:} the main plot displays the imaginary part of the dimensionless frequency as a function of $\mathcal{R}$. On the other hand, the inset plot displays the real part of the 
 frequency. The black (gray) dashed line that is almost on the top of the orange~$\hbox{\Large $\bullet$}$ and red~$\hbox{\small $\otimes$}$ numerical points are the analytical WKB approximation $\tilde{\omega}_{\hbox{\tiny WKB}}$ given by \eqref{WKBfrequency} with $p=0,1$.
 {\bf Bottom panel:} Zoom of the left panel in the near-extremal region (\ie around $\mathcal{R}\sim 1$) where all the overtone NH families approach $\hbox{Im}\,\tilde{\omega} \to 0$ as $\mathcal{R}\to 1$. This time we also show, as dotted magenta/purple lines, the near-extremal approximation 
$\tilde{\omega}_{\hbox{\tiny MAE}}$ as can be read from \eqref{NEfreq} for $p=0,1,2,3,4,5$. We see that the latter approximate the `PS-NH' frequencies very well when we are close to extremality (\ie as $\mathcal{R}\to 1$), as expected. (The counterpart to this figure for the RN case is displayed in Fig.~\ref{Fig:RN}).}
\label{Fig:Kerr}
\end{figure} 
In the top panel, we show the two families with the lowest $|{\rm Im}\,\tilde{\omega}|$ for the full range $0\leq \mathcal{R}\leq 1$. These are families that, except in a small neighbourhood around extremality ($\mathcal{R}=1$), are smoothly connected to the $n=0$ and $n=1$ PS families in the RN limit.
However, for the same reasons discussed above for KN, in the special Kerr case ($\Theta = \pi/2$) it is still true that it is more appropriate to denote the modes of Fig.~\ref{Fig:Kerr} as the $n=0$ PS-NH family (orange~$\hbox{\Large $\bullet$}$ curve) and an $n=1$ PS-NH family (red~$\hbox{\small $\otimes$}$ curve).  Indeed, on one hand, for the full range of $\mathcal{R}$ these families are very well approximated by the high order WKB approximation $\tilde{\omega}_{\hbox{\tiny WKB}}$, namely by \eqref{WKBfrequency} with $p=0$ (black dashed line on top of orange~$\hbox{\Large $\bullet$}$ curve) and with $p=1$ (gray dashed line almost on top of the red~$\hbox{\small $\otimes$}$ curve). This seems to support the idea that these are photon sphere modes. On the other hand, this classification is challenged by what happens near extremality ($\mathcal{R}\to 1$). Indeed, note that in the bottom panel of Fig.~\ref{Fig:Kerr} we also display the magenta dot-dashed lines that are the analytical MAE approximation $\tilde{\omega}_{\hbox{\tiny MAE}}$ given by \eqref{NEfreq} with $p=0$ and $p=1$ (and with $p=2,3,4,5$). We see that, sufficiently close to extremality, this MAE approximation is also on top of the orange $\hbox{\Large $\bullet$}$ and the red~$\hbox{\small $\otimes$}$ curves of the top panel. So, from this perspective, we could instead say that the orange~$\hbox{\Large $\bullet$}$ and red~$\hbox{\small $\otimes$}$ curves (as best identified in the top panel) are NH families. Altogether, since near-extremality of Kerr (or of KN for large $\Theta$, as shown previously) the orange~$\hbox{\Large $\bullet$}$ and red~$\hbox{\small $\otimes$}$ curves are simultaneously well described by $\tilde{\omega}_{\hbox{\tiny WKB}}$ and $\tilde{\omega}_{\hbox{\tiny MAE}}$, it is more appropriate (as advocated previously) to denote these modes as a single `PS-NH' family of QNMs with several overtones $n=0,1,\cdots$, {\it all} of which approach (the higher overtones with higher slope) $\mathrm{Im}\,\tilde{\omega}=0$ and $\mathrm{Re}\,\tilde{\omega}=\tilde{\Omega}_H^{\rm \tiny ext}$ at extremality.
 In the top panel of Fig.~\ref{Fig:Kerr}, in addition to the first two overtone families with $n=0,1$, we also present the next four overtones with $n=2,3,4,5$ but, in these cases, we just show the curves near extremality for $\mathcal{R}\gtrsim 0.85$. These extra four overtone curves are also simultaneously well described by $\tilde{\omega}_{\hbox{\tiny WKB}}$ and $\tilde{\omega}_{\hbox{\tiny MAE}}$ (with $p\geq 2$). The $n=0$ and $n=1$ `PS-NH' curves in Fig.~\ref{Fig:Kerr} agree with the frequencies of the $\ell=m=2$ scalar field QNMs first computed in figures 3 and 4 of \cite{Glampedakis:2003dn} and also reproduced previously in figures 7 and 8 of \cite{Berti:2003jh} (the higher overtones in Fig.~\ref{Fig:Kerr} also agree with \cite{Berti:2003jh}; recall that we convert our units $\omega r_+$ into $\omega M$ using~\eqref{Mr+}).

 This observation that in the Kerr black hole the PS and NH families lose their individual identity and merge into a single `PS-NH' family and its overtones is better illustrated if we focus our attention in the near-extremal region. We do this in the bottom panel of Fig.~\ref{Fig:Kerr}, by zooming into the region $\mathcal{R}\gtrsim 0.91$ of the top panel. In this plot we identify the first 6 overtone families of the top panel using the color code convention of the RN QNMs described in footnote~\ref{footColorCode}. Looking into the details that are not clear in the top panel, we first notice that the $n=0$ overtone (top curve) is essentially filled with orange~$\hbox{\Large $\bullet$}$ except for $\mathcal{R}\geq 0.991$ where it is instead continued with blue~$\hbox{\large $\blackdiamond$}$. So this Kerr $n=0$ PS-NH curve $-$ that is simultaneously well described by $\tilde{\omega}_{\hbox{\tiny WKB}}$ (black dashed line) and $\tilde{\omega}_{\hbox{\tiny MAE}}$ (magenta dotted line) $-$ can be seen has being formed out of solutions (points) that connect smoothly to either the PS and NH families in the RN limit ($\Theta=0$). Similarly, the $n=1$ PS-NH overtone curve (second from top) is filled with red~$\hbox{\small $\otimes$}$ (which are $n=1$ PS modes in RN), blue~$\hbox{\large $\blackdiamond$}$ (which are $n=0$ NH modes in RN), orange~$\hbox{\Large $\bullet$}$ (that are $n=0$ PS modes when traced back to RN) and finally a few dark-green~${\hbox{\tiny $\blacksquare$}}$ (that are $n=1$ NH modes in RN) points. Comparing these first two curves on the bottom panel, we identify the trade-off that occurs at $ 0.990 \lesssim \mathcal{R}\lesssim 0.991$: see the few blue~$\hbox{\large $\blackdiamond$}$ that are on the $n=0$ `PS-NH' curve (that is otherwise dominated by the orange~$\hbox{\Large $\bullet$}$) and the few orange~$\hbox{\Large $\bullet$}$ points in the $n=1$ `PS-NH' curve. This is nothing but the eigenvalue repulsion already documented in detail (for KN including the $\Theta= \pi/2$ case) in Fig.~\ref{Fig:WKBn0-NHn0-Im}. Indeed, the transitions between the orange~$\hbox{\Large $\bullet$}$ and the blue~$\hbox{\large $\blackdiamond$}$ branches in Fig.~\ref{Fig:Kerr} map  to the cusp formation and consequent mergers of `old left/right' branches of Fig.~\ref{Fig:WKBn0-NHn0-Im}.
 
 Moreover, as we move along the sequence of curves that describe the $n=1,2,3,4,5$ `PS-NH' families, we also clearly identify the eigenvalue repulsions between the red~$\hbox{\small $\otimes$}$ modes, \ie the $n=1$ PS modes in the RN limit, and the $n=0,1,2,3,4$ NH modes in the RN limit described, respectively, by the 
 blue~$\hbox{\large $\blackdiamond$}$, dark-green~${\hbox{\tiny $\blacksquare$}}$, brown~${\hbox{\small $\blacktriangle$}}$, green $\hbox{\Large $\blacktriangledown$}$ and gray $\hbox{\Large $\square$}$ points. This series of eigenvalue repulsions were already identified and studied in detail in the discussions of Figs.~\ref{Fig:WKBn1-NHn4-Im}$-$\ref{Fig:WKBn1-NHn4-ImB}. For example, the trade-off between the red~$\hbox{\small $\otimes$}$ and blue~$\hbox{\large $\blackdiamond$}$ around $\mathcal{R}\sim 0.93$ when we move from the $n=1$ to the $n=2$ `PS-NH' curves is in a one-to-one correspondence with the cusp/merger observed in the transition between the plots of $\mathcal{R}=0.0935$ and $\mathcal{R}=0.0930$ of Fig.~\ref{Fig:WKBn1-NHn4-ImB}. As another example, the trade-off between the red~$\hbox{\small $\otimes$}$ and dark-green~${\hbox{\tiny $\blacksquare$}}$ around $\mathcal{R}\sim 0.95$ when we move from the $n=2$ to the $n=3$ `PS-NH' curves is in a one-to-one correspondence with the cusp/merger observed in the transition between the plots of $\mathcal{R}=0.0955$ and $\mathcal{R}=0.0950$ of Fig.~\ref{Fig:WKBn1-NHn4-Im}. And the trade-off between the red~$\hbox{\small $\otimes$}$ and brown~${\hbox{\small $\blacktriangle$}}$ around $\mathcal{R}\sim 0.965$ when we move from the $n=3$ to the $n=4$ `PS-NH' curves is in a one-to-one correspondence with the cusp/merger observed in the transition between the plots of $\mathcal{R}=0.0965$ and $\mathcal{R}=0.0960$ of Fig.~\ref{Fig:WKBn1-NHn4-Im}.\footnote{The following observations might further help interpreting the bottom panel of Fig.~\ref{Fig:Kerr}. In this plot, let us fix our attention at constant $\mathcal{R}=0.92$: we identify the orange~$\hbox{\Large $\bullet$}$, red~$\hbox{\small $\otimes$}$ and blue~$\hbox{\large $\blackdiamond$}$ that are observed at $\Theta=\pi/2$ (Kerr limit) in the bottom-left plot of Fig.~\ref{Fig:WKBn1-NHn4-ImB}. As another example, consider now $\mathcal{R}=0.94$ in Fig.~\ref{Fig:Kerr}: we identify the orange~$\hbox{\Large $\bullet$}$, blue~$\hbox{\large $\blackdiamond$}$, red~$\hbox{\small $\otimes$}$ and dark-green~${\hbox{\tiny $\blacksquare$}}$ also seen at $\Theta=\pi/2$ in the top-left plot of Fig~\ref{Fig:WKBn1-NHn4-ImB}. As a final example, consider this time $\mathcal{R}=0.97$ in Fig.~\ref{Fig:Kerr}: we identify the orange~$\hbox{\Large $\bullet$}$, blue~$\hbox{\large $\blackdiamond$}$, dark-green~${\hbox{\tiny $\blacksquare$}}$, brown~${\hbox{\small $\blacktriangle$}}$, red~$\hbox{\small $\otimes$}$ and green $\hbox{\Large $\blacktriangledown$}$ that are observed at $\Theta=\pi/2$ in the top-right plot of Fig~\ref{Fig:WKBn1-NHn4-Im}. This exercise further helps understanding how several overtones of what we call  PS and NH families of the RN QNM spectra ($\Theta=0$; see Fig.~\ref{Fig:RN}) entangle between each other as $\Theta$ increases to collectively generate, in an intricate combination, the single `PS-NH' QNMs (and its overtones) of the Kerr black hole when one reaches $\Theta=\pi/2$.   
 
 }

An important feature that emerges from Fig.~\ref{Fig:Kerr} is that one could say the families denoted as $n=0$ and $n=1$ PS modes in RN limit are completely `swallowed' by near-horizon modes in the extremal Kerr limit, and we simply have the entangled `PS-NH' curves (describing several overtones) shown in the bottom panel of Fig.~\ref{Fig:Kerr}. Another property of the Kerr QNM spectra worth mentioning in the context of the eigenvalue repulsion discussion of section~\ref{sec:EigenvalueRepulsionsA} is the fact that the several overtone PS-NH frequencies do meet and terminate with $\mathrm{Im}\, \omega\to 0$ and $\mathrm{Re}\, \omega\to m\Omega_H^{\hbox{\footnotesize{ext}}}$ at the extremal Kerr point $\mathcal{R}=1=a/r_+$. So we clearly can have different modes meeting and terminating at the boundary of the Kerr parameter space.

As a conclusion or reflection to settle ideas after the above detailed discussions, it is perhaps enlightening to again observe the $m=\ell=2$ QNM spectra of the $\Theta=0$ RN black hole (Fig.~\ref{Fig:RN}) and its counterpart in the $\Theta=\pi/2$ Kerr black hole (Fig.~\ref{Fig:Kerr}), keeping in mind the color code nomenclature fixed in footnote~\ref{footColorCode}. In the RN case, the PS families and the NH families extend all the way to extremality while preserving their individual identity. However, eigenvalue repulsions entangle the PS and NH modes as we turn on angular momentum, so that by the time we reach the Kerr case we instead have the `PS-NH' family of modes and their radial overtones.

\section{QNM spectra of scalar field perturbations in Kerr-Newman} \label{sec:QNMspectra}

In the previous section we focused our attention on the region of KN parameter space relevant to eigenvalue repulsions (typically the near-extremal region for large $\Theta$). For completeness, we now discuss the QNM spectra for the whole $(\mathcal{R}, \Theta)$ parameter space of KN, for both $m = \ell = 2$ and, this time, also $m = \ell = 0$ scalar field modes.
Recall that the $m = \ell = 2$ QNM spectra is important because it should and does capture properties (\eg eigenvalue repulsions) that are in common with the gravito-electromagnetic perturbations of KN 
\cite{Dias:2015wqa,Carullo:2021oxn,Dias:2021yju,Dias:2022oqm}. On the other hand, the $m = \ell = 0$ QNM spectra is relevant because these angular quantum numbers match the spin 0 of the scalar field.

As discussed previously, we choose to parametrize the KN family using the `polar' quantities $\{\mathcal{R},\Theta\}$, such that the dimensionless rotation $\alpha$ and charge $\tilde{Q}$ are given by \eqref{PolarParametrization}, \ie $\alpha= \mathcal{R}\, \sin \Theta$ and $\tilde{Q}=\mathcal{R}\, \cos \Theta$. Thus,  $\Theta$ ranges from the Reissner-Nordstr\"om solution ($\Theta = 0$) to the Kerr solution ($\Theta=\pi/2$). On the other hand, $\mathcal{R}$ is an off-extremality measure, such that extremality of KN (and hence RN and Kerr) is at $\mathcal{R}=1$, while $\mathcal{R}=0$ describes the Schwarzschild BH.

\subsection{The QNM spectra of $m = \ell = 2$ modes \label{sec:m=l=2}}

\begin{figure}[th]
\centering
\includegraphics[width=.49\textwidth]{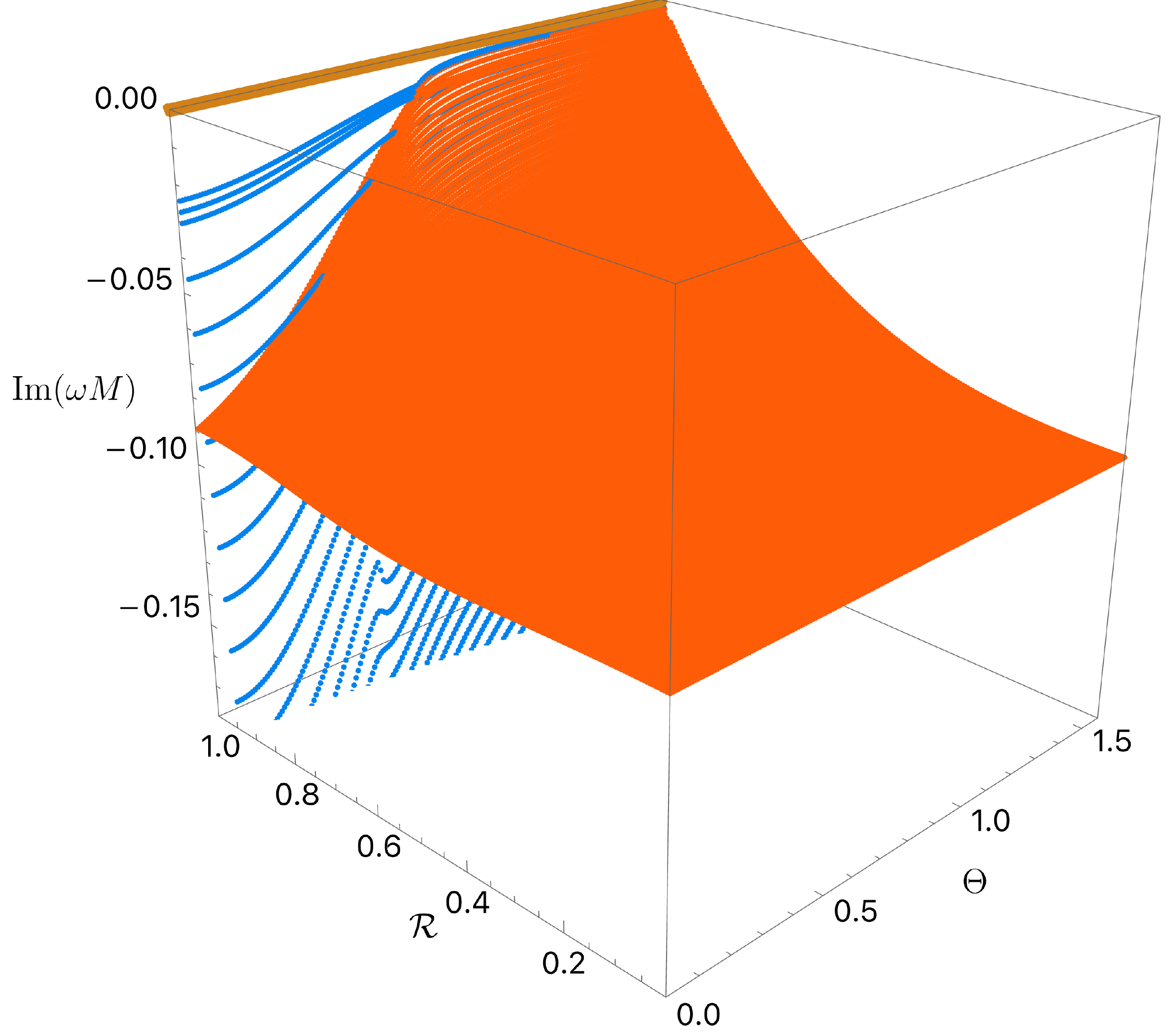}
\hspace{0.0cm}
\includegraphics[width=.49\textwidth]{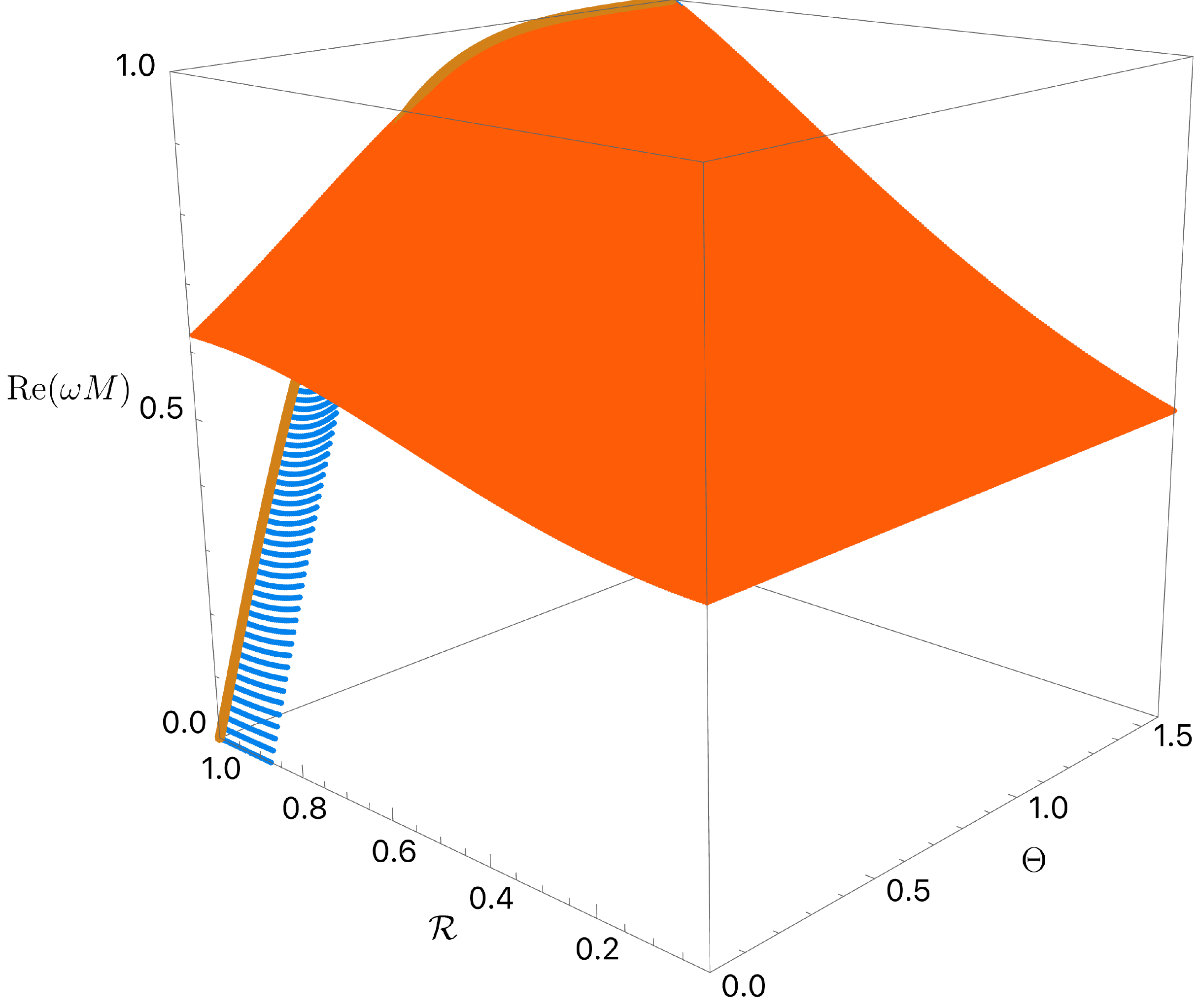}
\caption{Dominant $m=\ell=2$ QNMs of the KN black hole parametrized by the `polar' parameters $(\mathcal{R},\Theta)$. The orange (blue) surface describes the QNM family that reduces to the $n=0$ PS (NH) modes in the  RN limit.
The light brown curve at $\mathcal{R}=1$ has $\hbox{Re}(\omega M)=m\Omega_H$ and $\hbox{Im}\, \omega =0$.
 {\bf Left panel:} Imaginary part of the dimensionless frequency (in mass units) as a function of  $\{\mathcal{R},\Theta\}$.  {\bf Right panel:} Real part of the dimensionless frequency as a function of  $\{\mathcal{R},\Theta\}$. In both panels we only display the NH modes for $0.875\leq \mathcal{R}\leq 1$ (since they plunge very deeply to very negative values of $\hbox{Im}\, (\omega M)$ for smaller values of $\mathcal{R}$).}
\label{Fig:KN3d}
\end{figure}

\begin{figure}[th]
\centering
\includegraphics[width=.49\textwidth]{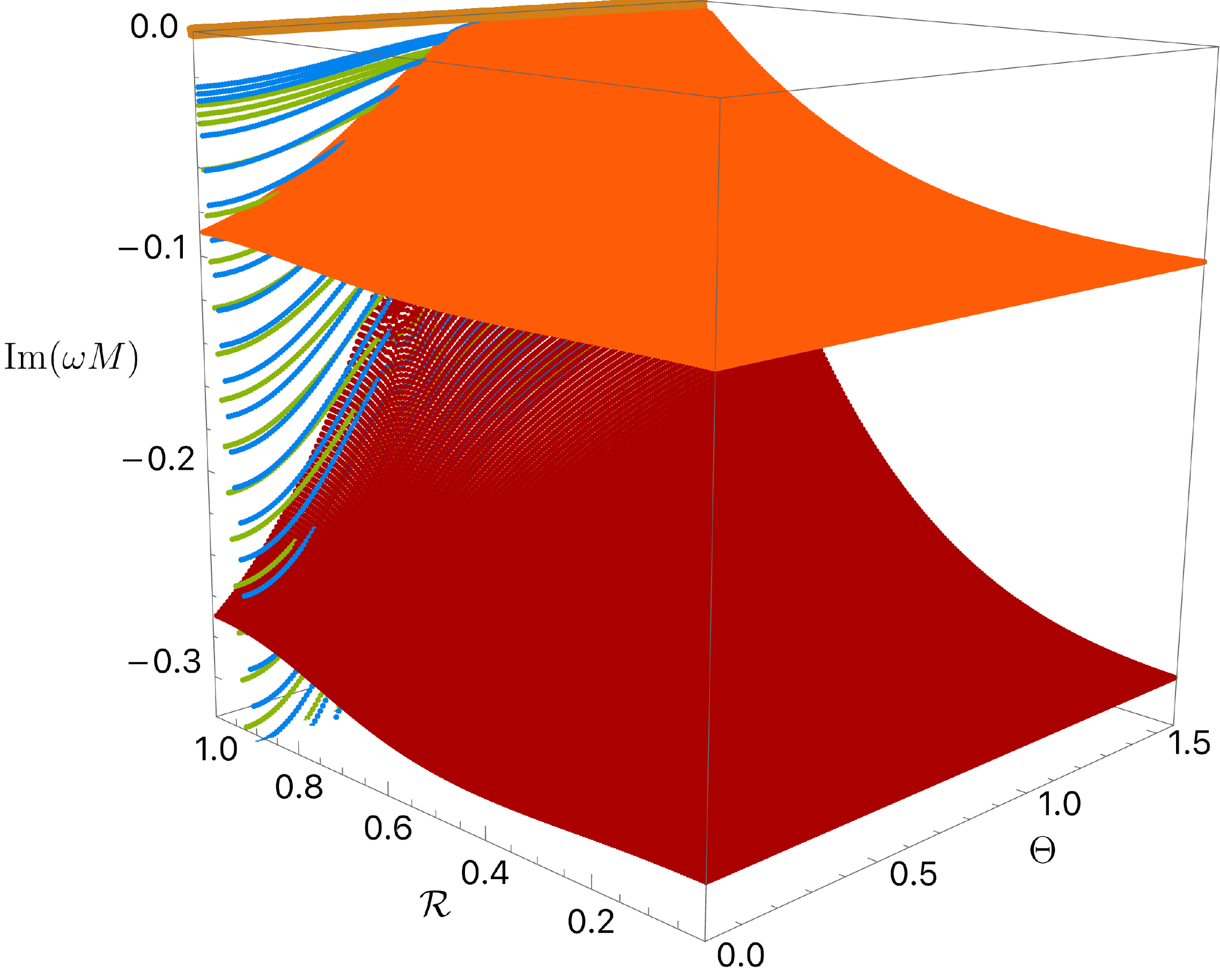}
\hspace{0.0cm}
\includegraphics[width=.49\textwidth]{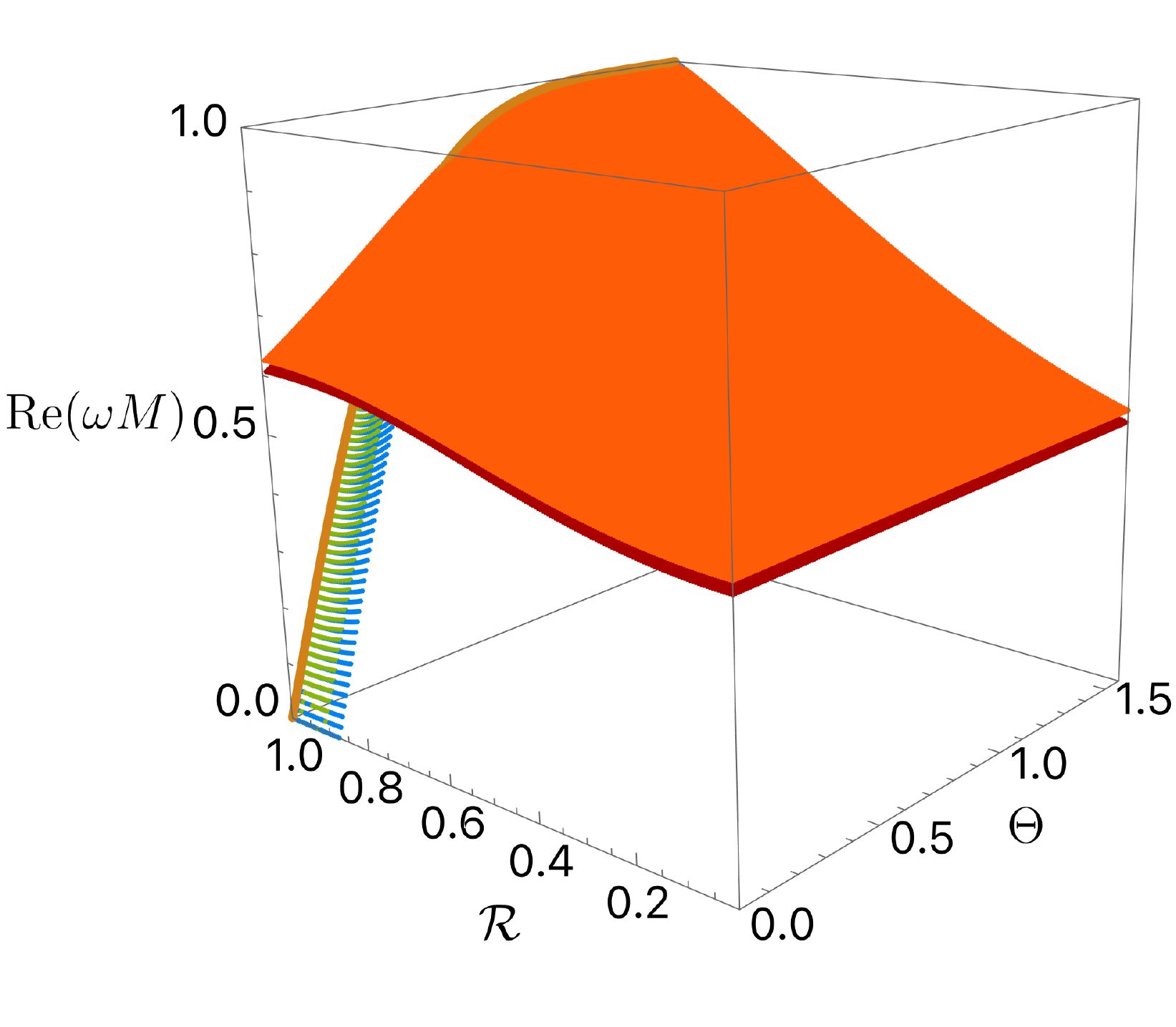}
\caption{In addition to the two surfaces (orange and blue) already presented in Fig.~\ref{Fig:KN3d}, we also display the red family of modes that connect to the $n=1$ PS family of modes in the RN limit, and the dark-green surface which describes the modes that connect to the $n=1$ NH family in the RN limit.}
\label{Fig:KN3d-4}
\end{figure}
   
In Fig.~\ref{Fig:KN3d} we consider the full parameter space ($\mathcal{R}\in [0,1]$ and $\Theta \in [0,\pi/2]$)  of the KN black hole and we plot the two $\ell=m=2$ modes that can dominate the spectra (\ie that can have the smallest $|{\rm Im}(\omega M)|$) in some region of the phase space.

In Fig.~\ref{Fig:KN3d} and all figures of this section we choose to provide the dimensionless frequency in units of the mass ($\omega M$), because this is the standard unit in astrophysical studies.\footnote{The intricate properties of the spectra due to the eigenvalue repulsions are however best seen if we use units of $r_+$. This is the reason we used this unit in previous figures. To convert $\omega r_+$ into $\omega M$ we use~\eqref{Mr+}.} 
Still following the color convention of footnote~\ref{footColorCode}, the orange surface is the family of modes that connects to the $n=0$ PS family of modes in the RN limit, and is typically well approximated by the high order WKB approximation $\tilde{\omega}_{\hbox{\tiny WKB}}$, namely by \eqref{WKBfrequency} with $p=0$. On the other hand, the blue surface describes the modes that connect to the $n=0$ NH family in the RN limit, and is typically well approximated by the MAE frequency $\tilde{\omega}_{\hbox{\tiny MAE}}$ given by \eqref{NEfreq} with $p=0$. 
The light brown curve at $\mathcal{R}=1$ in Fig.~\ref{Fig:KN3d} has $\hbox{Im}\, \omega =0$ and $\hbox{Re}\,\omega =m\Omega_H$. As already concluded from the analysis of Figs.~\ref{Fig:WKBn0-NHn0-Im} and~\ref{Fig:WKBn0-NHn4-Im}, Fig.~\ref{Fig:KN3d} clearly shows that the orange family is the dominant mode (\ie it has smaller $|\hbox{Im}(\omega M)|$) in most of the parameter space except in a small neighbourhood around extremality $\mathcal{R}\sim 1$ where the blue family dominates.
Note that the $\Theta=0$ plane of Fig.~\ref{Fig:KN3d} is the RN spectrum shown in Fig.~\ref{Fig:RN}, and the $\Theta=\pi/2$ plane of Fig.~\ref{Fig:KN3d} is the Kerr spectrum of Fig.~\ref{Fig:Kerr} (after converting units $\omega r_+ \to \omega M$ and keeping only the orange and blue modes of those earlier figures). Finally, several planes at constant $\mathcal{R}$ of Fig.~\ref{Fig:KN3d} can be found in previous figures, \eg in Figs.~\ref{Fig:WKBn0-NHn0-Im}$-$\ref{Fig:WKBn0-NHn0-Re}.

For completeness, in~Fig.~\ref{Fig:KN3d-4} we again display the two QNM families of Fig.~\ref{Fig:KN3d}, but this time we also add the next two subdominant modes. Namely, the red family of modes that connects to the $n=1$ PS family of modes in the RN limit and that is typically well approximated by the high order WKB approximation $\tilde{\omega}_{\hbox{\tiny WKB}}$ with $p=1$, and the dark-green surface which describes the modes that connect to the $n=1$ NH family in the RN limit and which is typically well approximated by the MAE frequency $\tilde{\omega}_{\hbox{\tiny MAE}}$  with $p=1$.\footnote{Again, the planes $\Theta=0$ and $\Theta=\pi/2$ in~Fig.~\ref{Fig:KN3d-4} are given by the RN and Kerr plots displayed in Fig.~\ref{Fig:RN} and Fig.~\ref{Fig:Kerr}, respectively (after unit conversion), and several planes at constant $\mathcal{R}$ of Fig.~\ref{Fig:KN3d-4} can be found in previous figures, namely in Figs.~\ref{Fig:WKBn1-NHn4-Im}$-$\ref{Fig:WKBn1-NHn4-ImB}.}

Concluding, the scalar field $m = \ell = 2$ QNM spectra is qualitatively similar to that of gravito-electromagnetic perturbations of KN~\cite{Dias:2021yju,Dias:2022oqm}. For example, this is evident when we compare the scalar field Fig.~\ref{Fig:KN3d-4} with the gravito-electromagnetic Fig.~15 of \cite{Dias:2022oqm}. The intricate properties associated to eigenvalue repulsions are also similar for the two sectors of perturbations. However, since the scalar field perturbations are described by a single pair of radial/angular ODEs (instead of a coupled pair of PDEs), and using the polar parametrization of the KN parameter space introduced in the present manuscript, we were able to explore fine-tuned details of the eigenvalue repulsions that were not so easy to extract in the gravito-electromagnetic case of 
\cite{Dias:2021yju,Dias:2022oqm}.

A comparison between our findings and those of \cite{Zimmerman:2015trm}, where the scalar QNM spectra of KN was also analysed and the existence of two families of QNMs was extensively discussed, is in order.
Overall, our study is complementary to the one of \cite{Zimmerman:2015trm} (see also \cite{Yang:2012pj,Yang:2013uba}), but it offers a fresh perspective of the RN/KN/Kerr QNM spectra, identifies and studies the features of eigenvalue repulsions, and helps to clarify the Kerr QNM spectra from the perspective of the RN QNM spectra. 
For each pair  $\{\ell,m\}$ of quantum numbers, Refs.~\cite{Yang:2012pj,Yang:2013uba,Zimmerman:2015trm} 
define the quantity $\mu=\frac{m}{\ell+1/2}$. Furthermore, these references analyse the properties of the Schr\"oedinger potential of the system in the eikonal limit, namely its maximum and whether it is located outside the event horizon. From this analysis, \cite{Yang:2012pj,Yang:2013uba,Zimmerman:2015trm}
conclude that there a separatrix curve $\mu=\mu_c(\tilde{a})\gtrsim 0.74$ (where the lower bound holds in the Kerr limit) such that  the qualitative behaviour of the QNM spectra is  distinct depending on whether $\{\ell,m\}$ are such that $\mu\lesssim \mu_c(\tilde{a})$ or $\mu\gtrsim \mu_c(\tilde{a})$.
In appendix \ref{app:bf_bound}, following a similar analysis in higher-dimensional Kerr-de Sitter (Myers-Perry-de Sitter) \cite{Davey:2022vyx}, we provide a complementary first-principles analysis that (also) identifies this separatrix curve $\mu_c(\tilde{a})$ (and agrees with the one found in \cite{Yang:2012pj,Yang:2013uba,Zimmerman:2015trm}). In our analysis we use the fact that the near-horizon geometry of the extremal KN black hole (NHEKN)
is a spacetime similar to $AdS_2 \times S^2$ (remaining a solution of the Einstein equation). This NHEKN geometry has an $SL(2,R)\times U(1)$ isometry group, where the $U(1)$ is inherited from the axisymmetry of the KN solution and the $SL(2, R)$ extends the KN time-translation symmetry. The Klein-Gordon equation in this NHEKN geometry (which can be equivalently obtained taking the near-horizon limit of the Klein-Gordon equation of the extremal KN) naturally reduces to the equation for a scalar field in AdS$_2$ with an effective mass that depends on $\{\ell,m\}$. In AdS$_2$, the scalar field mass must be higher than the 2-dimensional Breithl\"ohner-Freedman (BF) bound for the solutions to have finite energy \cite{Breitenlohner:1982jf,mezincescu_stability_1985}. This AdS$_2$ BF bound turns out to define the separatrix boundary $\mu=\mu_c(\tilde{a})$ $-$ see in particular Table~\ref{table:bf_crit} and \eqref{eqn:mu_crit_eikonal} $-$
that is essentially the same 1-parameter curve that was found in \cite{Yang:2012pj,Yang:2013uba,Zimmerman:2015trm} by looking at the location of the maxima of the eikonal Schr\"oedinger potential of the KN Klein-Gordon problem. These two criteria are in close numerical agreement despite the fact that $\mu_c(\tilde{a})$ is strictly-speaking only valid in the eikonal limit (as discussed further in Appendix~\ref{app:bf_bound}).

As stated above, \cite{Yang:2012pj,Yang:2013uba,Zimmerman:2015trm} found that the qualitative behaviour of the QNM spectra is significantly different depending whether $\{\ell,m\}$ are such that $\mu\lesssim \mu_c(\tilde{a})$ or $\mu\gtrsim \mu_c(\tilde{a})$ which is closely related with whether the near-horizon quantity $\delta^2$ defined in \eqref{def:delta}, i.e. the argument of the square root in~\eqref{NEfreq}, is negative or positive.
Our results confirm this is the case. In particular, for the $\ell=m=2$ modes ($\mu=0.8$) we are discussing in this section, one is in the regime $\mu\lesssim \mu_c(\tilde{a})$ for $\Theta<\Theta_\star$ and $\mu\gtrsim \mu_c(\tilde{a})$ for $\Theta\geq \Theta_\star$ (where recall that $\Theta_\star\simeq 0.881$). For $\mu\gtrsim \mu_c(\tilde{a})$, \cite{Yang:2012pj,Yang:2013uba,Zimmerman:2015trm} find that the QNM spectra has only zero-damped modes (ZDMs), i.e. modes whose imaginary part of the frequency approaches zero at extremality ($\mathcal{R}\to 1$). Essentially, this is consistent with our analysis for $\Theta\geq \Theta_\star$ if, very close to extremality, the PS modes indeed approach $\tilde{\omega}=m\tilde{\Omega}_H^{\mathrm{ext}}$  (please see detailed discussion of the bottom panels of Fig.~\ref{Fig:Star} that we do not repeat here) as do the NH modes. We would however emphasize that for $\Theta\geq \Theta_\star$ (i.e. $\mu\gtrsim \mu_c(\tilde{a})$) there are not one but two  families of modes that have a distinct origin when we trace them back to the RN case (as we discuss next).
On the other hand, for smaller $\Theta<\Theta_\star$, in agreement with \cite{Zimmerman:2015trm}, we find that the system has (in the nomenclature of \cite{Yang:2012pj,Yang:2013uba,Zimmerman:2015trm}) both ZDMs and damped modes (DMs) that approach a finite $\mathrm{Im}\,\tilde{\omega}$ at extremality) as summarized in the top panel of Fig.~\ref{Fig:Star} and in Figs.~\ref{Fig:WKBn0-NHn0-Im}-\ref{Fig:WKB-NH-Im_near_ext_2}. We emphasize that these DMs are the PS modes that become ZDMs for $\Theta>\Theta_\star$.
This and  the differences between the two regimes $\mu\lesssim \mu_c(\tilde{a})$ and $\mu\gtrsim \mu_c(\tilde{a})$ are clearly observed in Figs.~\ref{Fig:KN3d}-\ref{Fig:KN3d-4}: in short, the NH surfaces always end in the extremal brown line, while the PS modes do not do so for $\Theta<\Theta_\star$.

One should also provide an important clarification that helps bridge our findings with those of \cite{Yang:2012pj,Yang:2013uba,Zimmerman:2015trm} and avoid an apparent inconsistency. In regimes $\mu\lesssim \mu_c(\tilde{a})$ where there are both ZDMs and DMs, \cite{Yang:2012pj,Yang:2013uba,Zimmerman:2015trm} state that far away from extremality the system only has a DM family which exists all the way till extremality. But, at a critical value of $\mathcal{R}$, these references state that the DM family `{\it bifurcates}' into two branches: one continues to describe DMs till extremality while the other only exists above this critical $\mathcal{R}$ and describes ZDMs. We find that this is because \cite{Yang:2012pj,Yang:2013uba,Zimmerman:2015trm} effectively considered modes that have not only fixed $\{\ell,m\}$ but also {\it fixed} overtone $p$. On the other hand, by construction, in our analysis we always fix $\{\ell,m\}$ (but {\it not} the overtone $p$) and, starting from the PS and NH modes at the RN limit ($\Theta=0$), we follow these modes as $\Theta$ grows till reaching the Kerr limit ($\Theta=\pi/2$). This is a continuous process where no bifurcations are observed (although we have eigenvalue repulsions with the associated cusp formations and break/trade-off/merge of branches). That is to say, at each fixed $\Theta$ we always have two modes, below and above any possible critical $\mathcal{R}$. However, the fact that we do not observe bifurcations in our analysis is not inconsistent with the analysis of \cite{Yang:2012pj,Yang:2013uba,Zimmerman:2015trm}. The key observation here is that the overtone number $p$ of the continuous curves/surfaces we follow effectively changes as we march in $\Theta$ when eigenvalue repulsions kick in. For example (among many others described previously), the overtone of the orange PS and blue NH curves in Fig.~\ref{Fig:WKBn0-NHn0-Im} change when making the transition $\mathcal{R}=0.991 \to \mathcal{R}=0.990$ (middle panels). If we insist on fixing our attention on modes with fixed overtone, we would interpret some results as being a `bifurcation' in an $\mathrm{Im}\,\omega$ vs $\Theta$ plot and this was the approach followed in \cite{Yang:2012pj,Yang:2013uba,Zimmerman:2015trm}. We do not do so and thus, instead of a bifurcation, we do observe two surfaces intersecting or crossing in an $\mathrm{Im}\,\omega$ vs $\{\Theta,\mathcal{R}\}$ plot (or two curves crossing in a $\mathrm{Im}\,\omega$ vs $\Theta$ plot); 
note that it is only $\mathrm{Im}\,\omega$, but not $\mathrm{Re}\,\omega$, that  coincides along the crossing.

\subsection{The QNM spectra of $m = \ell = 0$ modes}\label{sec:ml0Perturbations}

The focus of this paper so far has been the $m = \ell = 2$ QNMs due to their close analogy with the gravito-electromagnetic perturbations computed in~\cite{Dias:2021yju,Dias:2022oqm,Dias:2015wqa}. However, for completeness, in this section we consider QNMs with $m = \ell = 0$ since these quantum numbers match the spin of the scalar field perturbations. Moreover, these modes are important as they dominate the $m = \ell = 2$ modes (\ie they have lower $|{\rm Im}(\omega M)|$) in certain regions of the parameter space. 

\begin{figure}[t]
\centering
\includegraphics[width=0.49\textwidth]{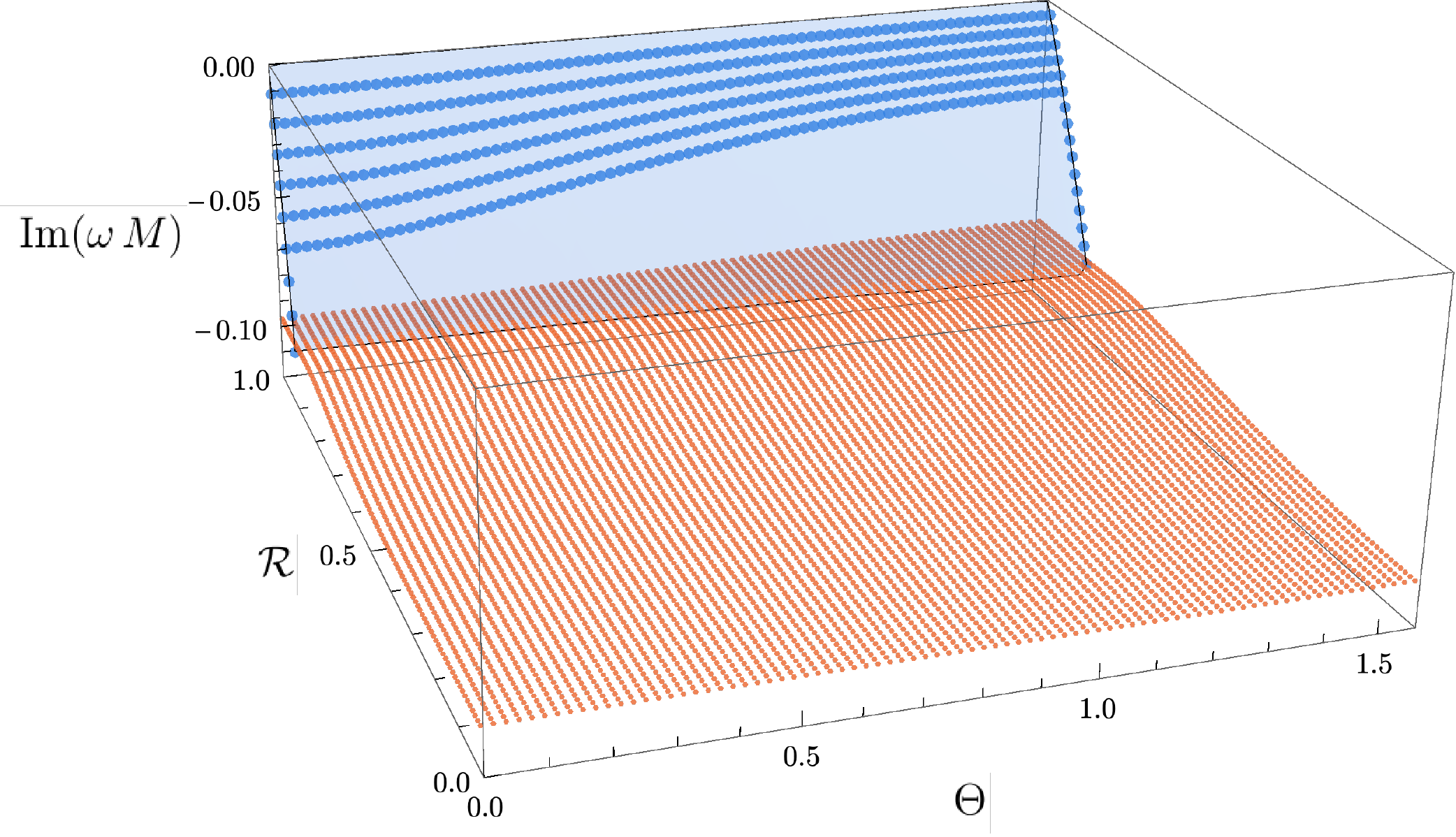}
\includegraphics[width=0.49\textwidth]{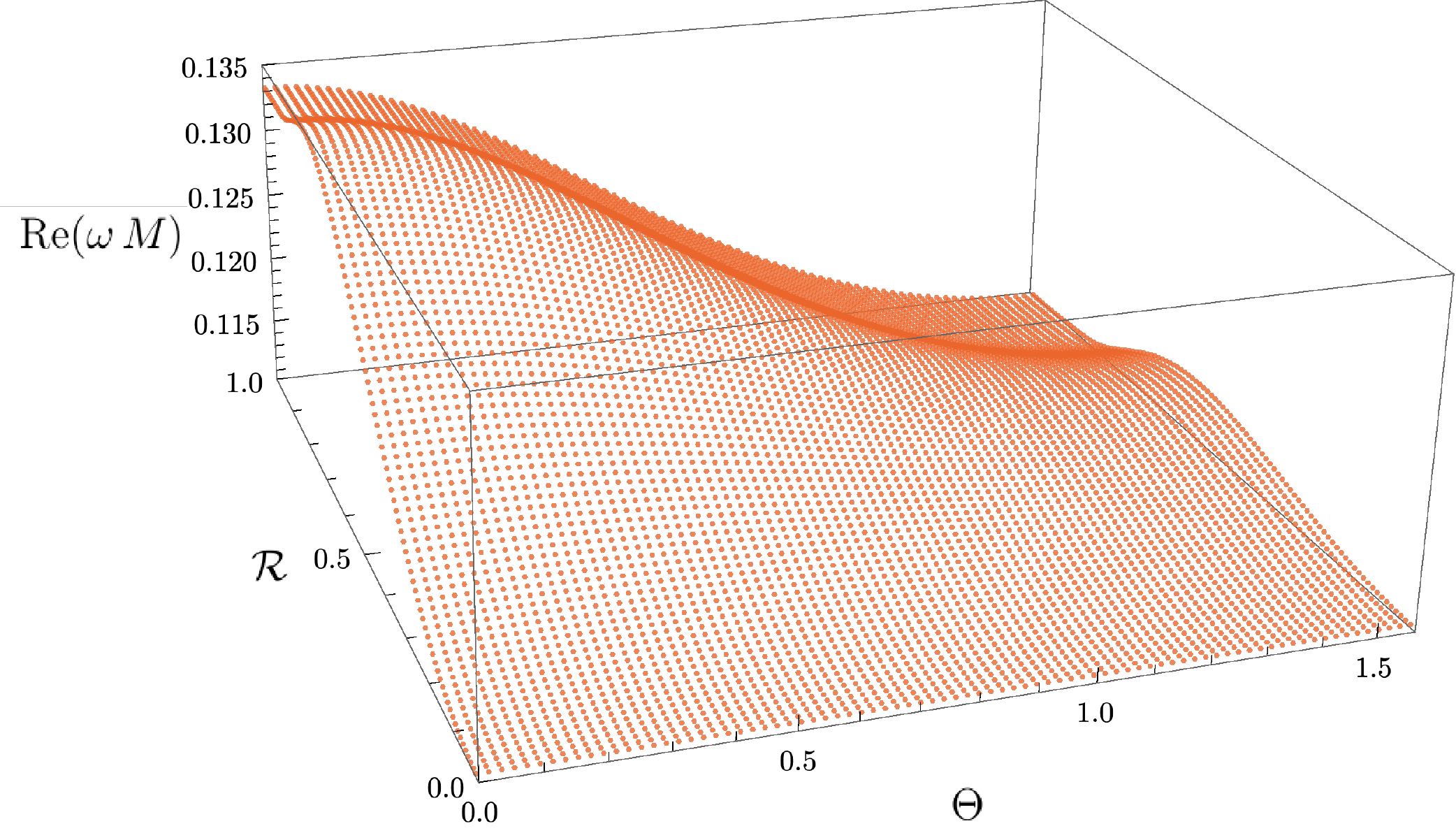}
\caption{The frequency of the slowest-decaying PS (orange disk) and NH (blue disk) QNMs with $m = \ell = 0$ for the whole parameter space $\{\mathcal{R}, \Theta\}$ of KN. \textbf{Left panel:} The imaginary parts of the frequency. With the exception of the Kerr and RN limits, only the NH modes down to $\mathcal{R} = 0.94$ were found, due to computational limitations. The blue surface represents the extrapolation of the points found until the intersection with the PS modes. \textbf{Right panel:} The real part of the PS frequencies. The NH frequencies always have zero real part.}
\label{Fig:NH-PS-full-m0}
\end{figure}

Perturbations with $m = 0$ have an enhanced $t \to -t$ symmetry, so QNM frequencies form pairs $\{\omega, -\omega^{*}\}$.  When $m = 0$, the WKB approximation $\tilde{\omega}_{\hbox{\tiny WKB}}$ in~\eqref{WKBfrequency} is not valid (for obvious reasons since it is an expansion at large $m$), however the matched asymptotic expansion $\tilde{\omega}_{\hbox{\tiny MAE}}$~\eqref{NEfreq} remains a very good approximation provided we are close to extremality, and simplifies as follows. When $m = 0$, the angular equation~\eqref{KG:ang} at extremality can be solved exactly, with the angular eigenvalue $\lambda_{\rm ext}^{(m = 0)} = \ell(\ell+1)$, where $\ell = 0, 1, 2, \dots$ denotes the number of zeros of the angular eigenfunction, as usual. The MAE frequency~\eqref{NEfreq} simplifies to
\begin{equation}\label{NEfreqm0}
    \tilde{\omega}_{\hbox{\tiny MAE}}^{(m = 0)} = - \frac{i(\ell +p + 1)}{2(1+ \alpha^{2})}\sigma + \mathcal{O}(\sigma^{2}), \qquad\quad p = 0, 1, 2, \dots.
\end{equation}
This frequency is purely imaginary, for all $\Theta = \arctan(a/Q)$, in contrast to the $m = l = 2$ case. Since the PS frequencies remain complex (as we will discuss later), this hints at a major difference between the $m = \ell = 2$ and  $m = \ell = 0$ QNM spectra: in the latter there are no eigenvalue repulsions. Consequently, for $m = \ell = 0$ we can always unambiguously identify the photon sphere (PS) and near-horizon (NH) modes in the full parameter space of KN (not only in the RN limit).
These properties can be seen in Fig.~\ref{Fig:NH-PS-full-m0} where we display
the spectrum of KN QNMs with $m = \ell = 0$ for all $(\mathcal{R}, \Theta)$. We have found the full spectrum of the PS modes (orange disks), however the NH modes (blue disks) have proven to be much more difficult to find numerically, and we have only computed those down to $\mathcal{R} = 0.94$, except in the RN ($\Theta = 0$) and Kerr ($\Theta = \pi/2$) limits. For intermediate values of $\Theta$, the blue surface in the left panel of Fig.~\ref{Fig:NH-PS-full-m0} represents the extrapolation of the NH data till the point they intersect with the PS modes and become subdominant. 

\begin{figure}[t]
\centering
\includegraphics[width=0.49\textwidth]{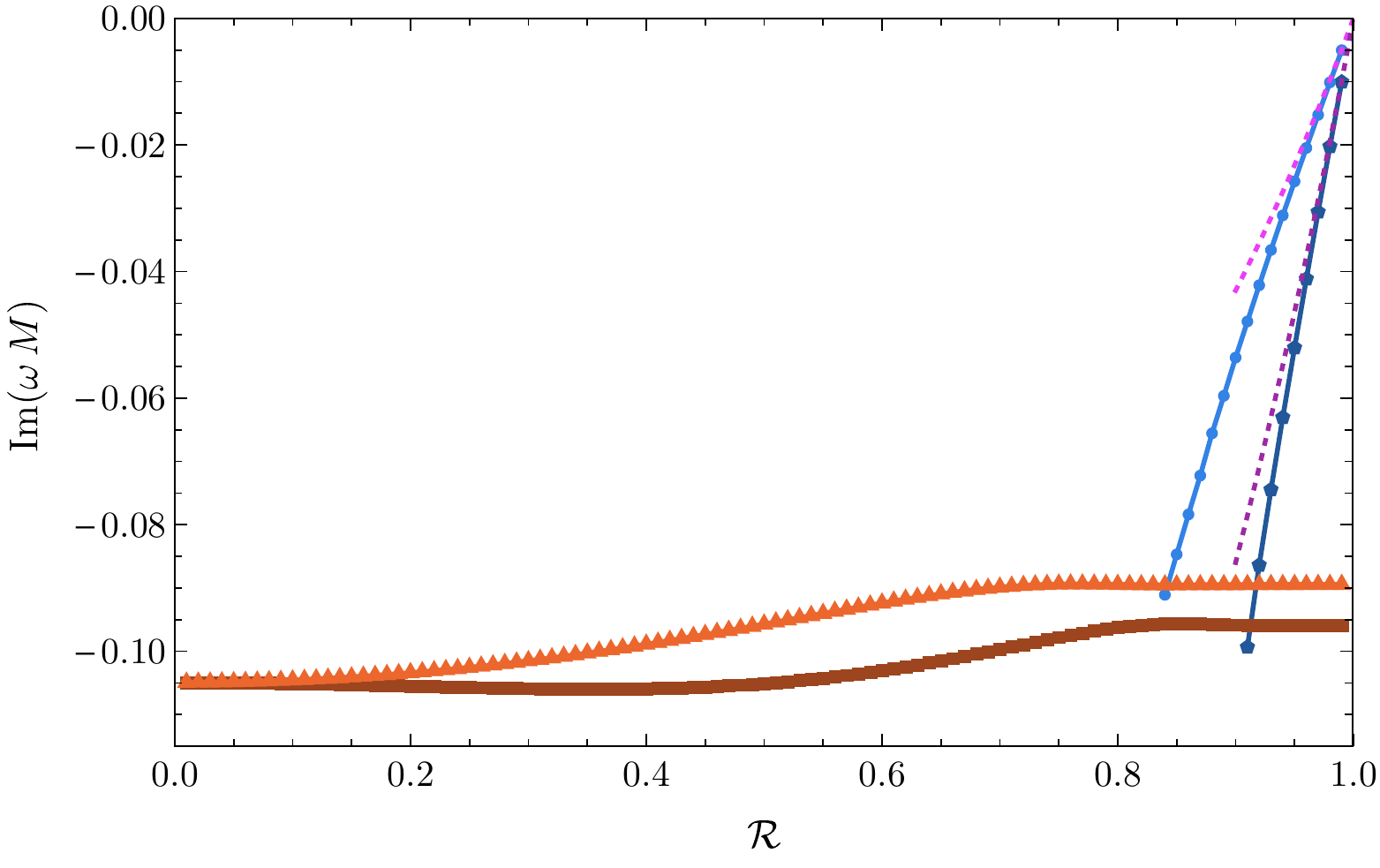}\hfill
\includegraphics[width=0.48\textwidth]{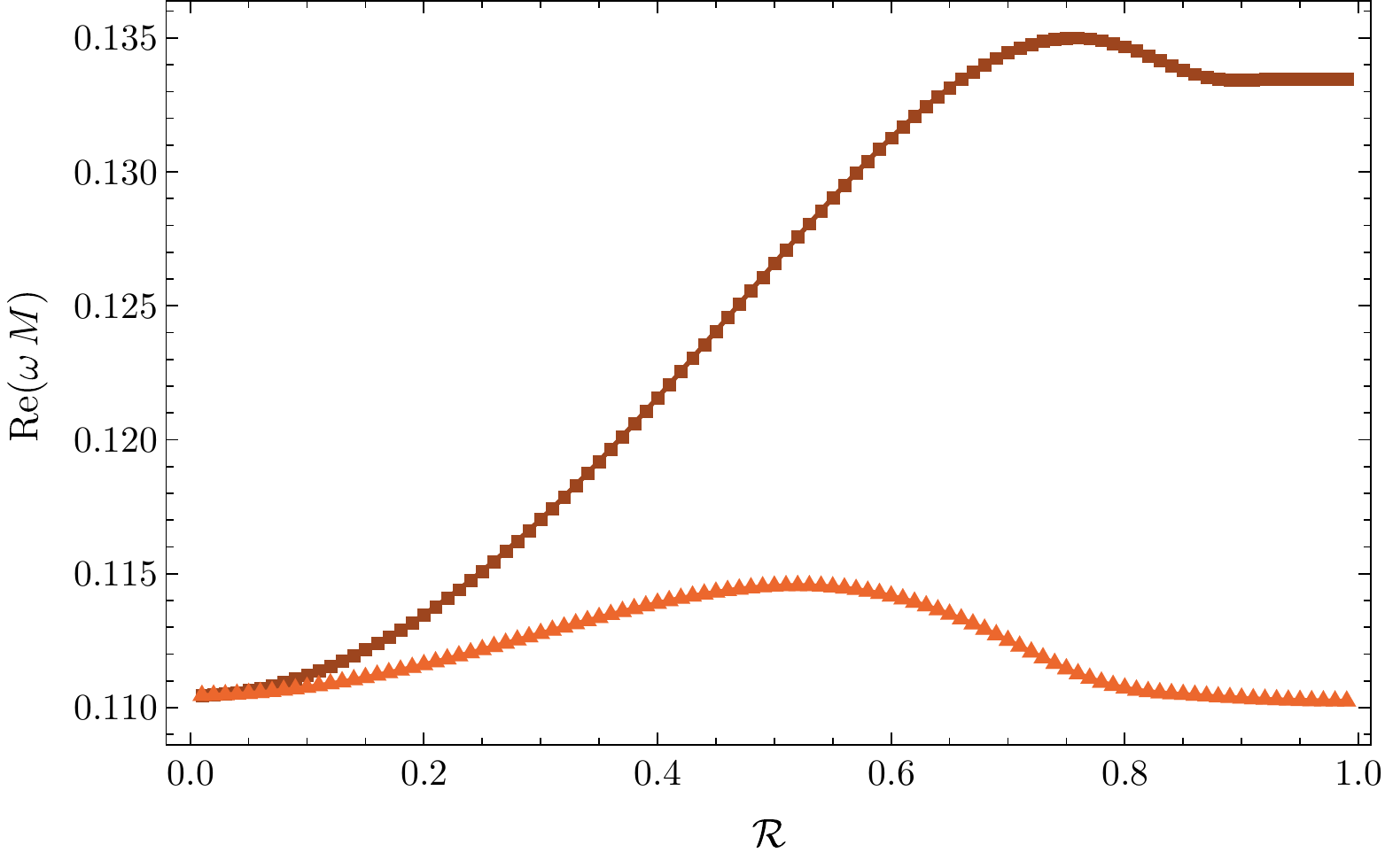}
\caption{The frequency of the slowest-decaying PS and NH QNMs in the RN ($\Theta = 0$) and Kerr ($\Theta = \pi/2$) limits. For Kerr (RN), the PS modes are orange triangles (brown squares) and the NH modes are light blue disks (dark blue pentagons).
 \textbf{Left panel:} The imaginary parts of the frequency. The dashed magenta and dotted purple lines emerging from $\mathcal{R}=1$ describe the matched asymptotic expansion $\tilde{\omega}_{\hbox{\tiny MAE}}^{(m = l = 0)}$~\eqref{NEfreqm0} for $p=0$ and $p=1$, respectively. \textbf{Right panel:} The real part of the PS frequencies for Kerr and RN. The NH modes always have zero real part.}
\label{Fig:NH-PS-full-m0KerrRN}
\end{figure}

For clarity, the QNM spectra in the special RN ($\Theta = 0$) and Kerr ($\Theta = \pi/2$) limits are displayed in Fig.~\ref{Fig:NH-PS-full-m0KerrRN}, as well as the matched asymptotic expansion frequency approximation~\eqref{NEfreqm0}. For Kerr (RN), the PS modes are orange triangles (brown squares) and the NH modes are light blue disks (dark blue pentagons), with the matched asymptotic expansion $\tilde{\omega}_{\hbox{\tiny MAE}}^{(m = \ell = 0)}$ of~\eqref{NEfreqm0} with $p=0$ represented by a dashed magenta (dotted purple) line. Note that the QNM spectra only depends weakly on $\Theta$, most likely because these are axisymmetric perturbations, and therefore are not significantly affected by the angular momentum of KN. In the RN case, our PS frequencies (brown squares) in Fig.~\ref{Fig:NH-PS-full-m0KerrRN} agree with those first computed in \cite{Kokkotas:1988fm,Leaver:1990zz,Andersson:1993,Onozawa:1995vu,Andersson:1996xw}. 
On the other hand, to the best of our knowledge, the RN NH QNM spectrum (dark blue pentagons) is first computed exactly (within numerical accuracy) in Fig.~\ref{Fig:NH-PS-full-m0KerrRN} (their existence is predicted in \cite{Zimmerman:2015trm}; see however the discussion in the next paragraph). 
In the Kerr case, our PS frequencies (orange triangles) in Fig.~\ref{Fig:NH-PS-full-m0KerrRN} agree with those first computed in \cite{Seidel:1989bp,Berti:2003jh} (see  figure 6 of \cite{Berti:2003jh}). On the other hand, as far as we know, the Kerr NH spectrum (light blue disks)  is first computed exactly (within numerical accuracy) in Fig.~\ref{Fig:NH-PS-full-m0KerrRN} (their existence follows from the analysis of \cite{Leaver:1985ax,Yang:2012pj,Yang:2013uba}).

In agreement with the matched asymptotic expansion $\tilde{\omega}_{\hbox{\tiny MAE}}^{(m = 0)}$ in~\eqref{NEfreqm0}, the real part of the NH modes is zero. 
In contrast, the PS modes are always complex with $|\operatorname{Re}(\omega M)| > 0.11$ across the full parameter space. 
Therefore, there is no single point in the KN parameter space where the PS and NH frequencies can coincide (not even at extremality), strongly suggesting that the associated phenomenon of eigenvalue repulsion should also be absent.
Indeed, unlike the $m=\ell=2$ case, for the $m=\ell=0$ modes we find no evidence of the presence of eigenvalue repulsions in the spectra. In particular, the PS family never develops cusps or intricate features that could be a sign of eigenvalue repulsions: the imaginary part of the frequency always increases monotonically with  $\mathcal{R}$ and $\Theta$ (this is the reason we found no need to push our numerics further to find more NH modes than those in Fig.~\ref{Fig:NH-PS-full-m0}). 

Given the lack of eigenvalue repulsions, we can unambiguously identify NH modes throughout the KN parameter space, by tracing them to the extremal limit $\mathcal{R} \to 1$ where we can compare them to the matched asymptotic expansion~\eqref{NEfreqm0} (see the left panel of Fig.~\ref{Fig:NH-PS-full-m0KerrRN}), unlike the $m = \ell = 2$ modes. A similar phenomenon occurs in other black hole spacetimes, such as higher-dimensional Kerr-dS (Myers-Perry-dS), where eigenvalue repulsions between the PS and NH modes occur when $m \ne 0$, but not for $m = 0$~\cite{Davey:2022vyx}.

\begin{figure}[t]
\centering
\includegraphics[width=0.49\textwidth]{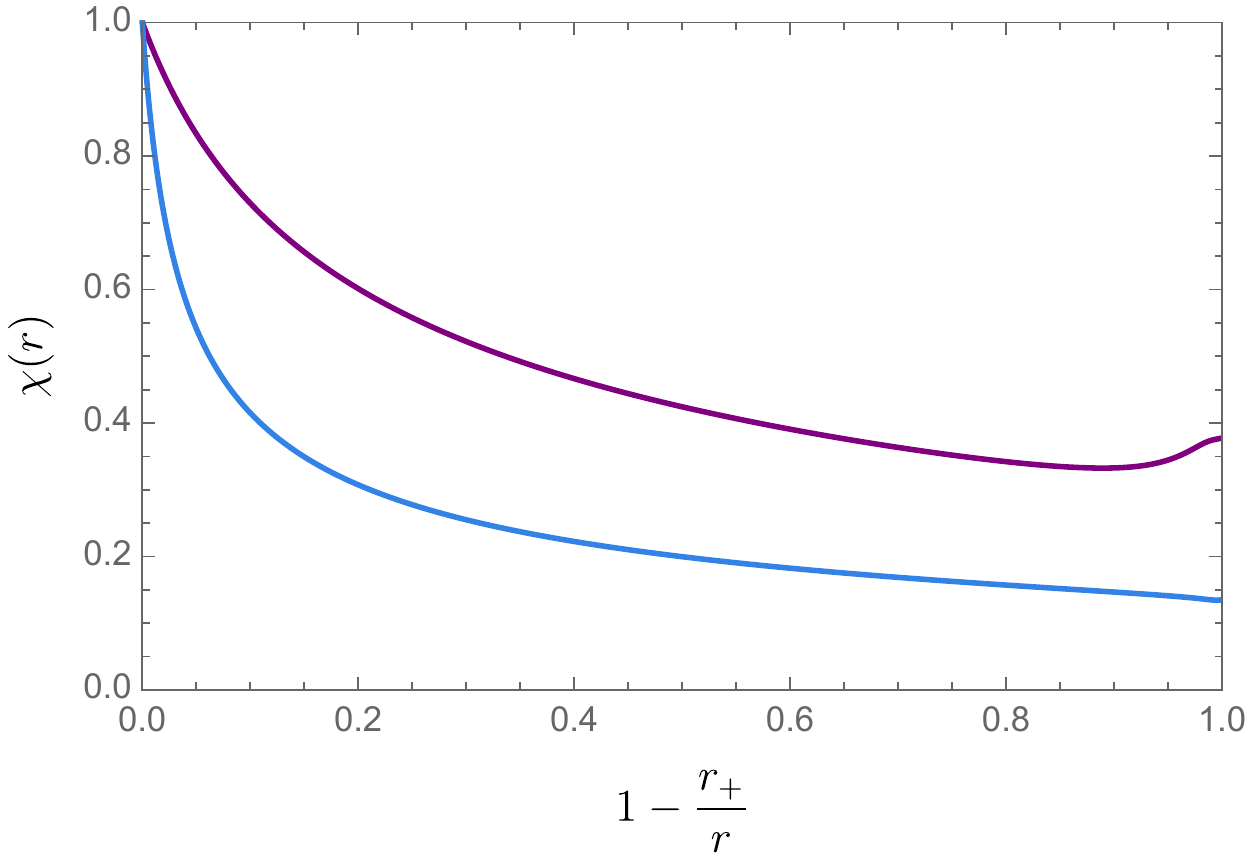}
\includegraphics[width=0.49\textwidth]{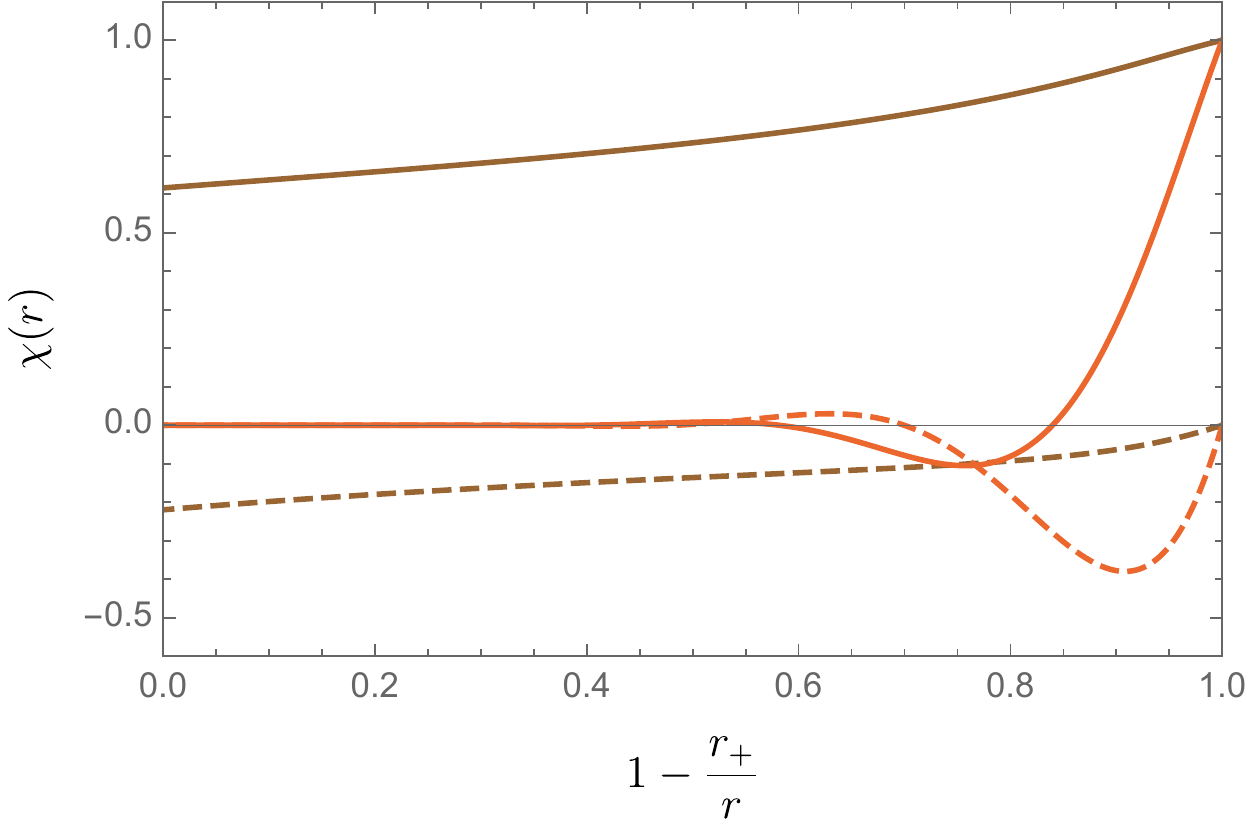}
\caption{Typical eigenfunctions $\chi(r)$ of KN QNMs with $m = \ell = 0$. $\chi(r)$ is the redefined radial function as in~\eqref{eigenfunction_redef}. \textbf{Left panel:} NH eigenfunctions for $\mathcal{R} = 0.95$ (purple) and $\mathcal{R} = 0.99$ (blue) at $\Theta = \pi/4$. The dominant NH eigenfunctions are purely real, peaked at the horizon radius $r = r_{+}$, with no zeros. \textbf{Right panel:} Real part (solid lines) and imaginary part (dashed lines) of the PS eigenfunctions for $\mathcal{R} = 0.8$ (brown) and $\mathcal{R} = 0.99$ (orange). The PS eigenfunctions are complex with non-zero support at large $r$, and the number of zeros depends on the distance to extremality.}
\label{fig:eigenfunctions_ml0}
\end{figure}

The two QNM families can also be distinguished by their eigenfunctions. In Fig.~\ref{fig:eigenfunctions_ml0} we plot the eigenfunctions $\chi(r)$ as defined by~\eqref{eigenfunction_redef} for the dominant NH modes (left panel) and PS modes (right panel) at a representative value of $\Theta = \pi/4$.
The eigenfunctions of the NH modes are purely real (blue and purple curves for $\mathcal{R} = 0.95$ and $0.99$, respectively, at $\Theta = \pi/4$), peaking at the horizon radius $r = r_{+}$, and becoming increasingly peaked as we approach extremality $\mathcal{R} \to 1$. Furthermore, the eigenfunctions of the dominant NH modes do not have any zeros. 
In contrast with the NH modes, the PS modes (orange and brown curves for $\mathcal{R} = 0.8$ and $0.99$, respectively, for $\Theta = \pi/4$ in right panel of Fig.~\ref{fig:eigenfunctions_ml0}) have both real and imaginary parts. Far away from extremality the PS eigenfunction is nearly flat, with no zeroes, but becomes oscillatory, with the number of zero crossings increasing the closer we approach extremality.\footnote{On the other hand, when $m = \ell = 2$, the NH modes acquire an imaginary part, but otherwise the NH and PS eigenfunctions are qualitatively similar to those in Fig.~\ref{fig:eigenfunctions_ml0} in the Kerr and RN limits. However, due to the eigenvalue repulsions, they can smoothly transition from one type to another as we vary $\Theta$, such that the NH-type eigenfunctions in the RN limit become PS-type in the Kerr limit, and vice versa. Essentially, in the regions of the parameter space where eigenvalue repulsions occur, we cannot use the eigenfunctions to distinguish the nature of the modes.}

Compared to the $m = \ell = 2$ modes, the PS family of $m = \ell = 0$ modes is always subdominant, \ie with larger $|\operatorname{Im} \omega|$ than the corresponding PS QNM at any point in the KN parameter space. On the other hand, at least near extremality, the NH modes with $m = \ell = 0$ dominate modes with $m = \ell = 2$. However, the difference in imaginary part between the two shrinks as $\mathcal{R}$ decreases.

As in the previous subsection, it is appropriate to make contact between our findings and those of \cite{Yang:2012pj,Yang:2013uba,Zimmerman:2015trm}. 
In terms of the quantity $\mu_c(\tilde{a})$ introduced in the discussion below Fig.~\ref{Fig:KN3d-4} of subsection~\ref{sec:m=l=2}
(and originally in \cite{Yang:2012pj,Yang:2013uba,Zimmerman:2015trm}), 
 the $\ell=m=0$ modes satisfy the condition $\mu<\mu_c(\tilde{a})$ in the whole range of $\{\mathcal{R},\Theta\}$. 
 Refs. \cite{Yang:2012pj,Yang:2013uba,Zimmerman:2015trm} do not present results for the scalar $\ell=m=0$ Kerr or RN modes displayed in Fig.~\ref{Fig:NH-PS-full-m0KerrRN}, but do discuss in detail other cases (including with spin $s=1,2$ in the Dudley-Finley approximation) in the same class $\mu<\mu_c(\tilde{a})$. 
As explained in the last paragraph of subsection~\ref{sec:m=l=2}, Refs.~\cite{Yang:2012pj,Yang:2013uba,Zimmerman:2015trm} fix not only $\{\ell,m\}$ but also the overtone $p$ when describing/interpreting their results (unlike in our analysis). Consequently, \cite{Yang:2012pj,Yang:2013uba,Zimmerman:2015trm} describe cases like the one in Figs.~\ref{Fig:NH-PS-full-m0}-\ref{Fig:NH-PS-full-m0KerrRN} as a `bifurcation' happening in the  $\mathrm{Im}\,\omega$ plot along a $\mathcal{R}(\Theta)$ line (or at a point $\mathcal{R}_c$ in the Kerr or RN case) close to extremality, while in our analysis (where we do not constrain the overtone to be fixed) we see that there is simply a `crossing' of two surfaces (or curves in the RN/Kerr limits) that intersect along the curve $\mathcal{R}(\Theta)$ (or along a point in RN/Kerr); note that it is only $\mathrm{Im}\,\omega$, but not $\mathrm{Re}\,\omega$, that  coincides along the ``crossing". Thus the two analyses are consistent and not contradictory once we identify the conditions underlying the selected choice of language. The compatibility of our findings and those of \cite{Yang:2012pj,Yang:2013uba,Zimmerman:2015trm}  is further confirmed when we analyse the results (for a gravitational mode with $\ell=2$ and $m=1$ that fits in the $\mu<\mu_c(\tilde{a})$ class) of figure 8 of \cite{Yang:2013uba}, which reproduces figure 3.b of \cite{Leaver:1985ax} (the latter was the first instance where the simultaneous existence of zero-damped and damped modes was observed). 

\smallskip

\subsection*{Acknowledgments}
The authors acknowledge the use of the IRIDIS High Performance Computing Facility, and associated support services at the University of Southampton, in the completion of this work.
 O.~C.~D. acknowledges financial support from the STFC ``Particle Physics Grants Panel (PPGP) 2018" Grant No.~ST/T000775/1.  J.~E.~S. has been partially supported by STFC consolidated grants ST/P000681/1, ST/T000694/1. The research leading to these results has received funding from the European Research Council under the European Community's Seventh Framework Programme (FP7/2007-2013) / ERC grant agreement no. [247252].

\appendix


\section{Near-horizon geometry and Breitenl\"ohner-Freedman bound} \label{app:bf_bound}

In the main text we have studied the QNM spectra of the $\ell=m=2$ and $\ell=m=0$ modes. These are representative elements of the two distinct classes or sectors of QNM that KN can have. As discussed in \cite{Yang:2012pj,Yang:2013uba,Zimmerman:2015trm}, these two classes of QNM can be (within a very good approximation) identified analysing the  sign of the quantity $\delta^2$ $-$ defined in~\eqref{def:delta} $-$ that appears naturally in the near-horizon analysis of modes. Concretely, for the QNM sector with $\delta^{2} < 0$ (e.g. the $\ell=m=0$ modes), the QNM spectra has two independent and clearly distinct families of modes $-$ PS and NH modes $-$ in the entire parameter space of KN. On the other hand, the QNM sector with $\delta^{2} > 0$ (e.g. the $\ell=m=2$ modes) is such that sufficiently close to extremality and for high $\Theta$ one can state that there is a single PS-NH family of QNMs and its overtones (although for small $\Theta$ one might still say that the PS and NH families are present and clearly distinguishable).
Thus, the condition $\delta^2(m, \ell, \Theta) = 0$, which depends on $\{m, \ell, \Theta\}$, provides a good approximation for the boundary that separates the two QNM sectors. In certain points of the main text, we used the equivalent notion of a critical value $\Theta_c(m, \ell)$ at which $\delta^2 = 0$. Furthermore, in the literature other critical quantities such as $\mu_c$, $\mathcal{F}_0$ and $\mathcal{J}$ (all defined later) are used to describe this  boundary~\cite{Yang:2012pj,Yang:2013uba,Hod_2012,Zhao_2015,Zimmerman:2015trm}. In this appendix, we attempt to clarify the relationship between the plethora of critical quantities, and provide a complementary first-principles identification of this boundary in terms of the effective AdS\textsubscript{2} Breitenl\"ohner-Freedman (BF) bound \cite{Breitenlohner:1982jf,mezincescu_stability_1985} on the near-horizon extremal Kerr-Newman (NHEKN) geometry.

To find the near-horizon geometry, we take the Kerr-Newman metric~\eqref{KNsoln} at extremality, and zoom into the horizon by first making the coordinate and gauge transformations
\begin{align}\label{eqn:nh_transformations}
 & r \to r_{+} + (r_{+}^{2}+a^{2}) \epsilon \, \rho, \qquad t \to \frac{\tau}{\epsilon}, \qquad \phi \to \psi + \Omega_H^{\rm ext} \frac{\tau}{\epsilon}, \nonumber \\
 & A \to \mathcal{A} + \frac{r_+ Q}{r_+^2+a^2} \frac{d\tau}{\epsilon},
\end{align}
where $\Omega_H^{\rm ext} = \left.\frac{a}{r_+^2+a^2}\right|_{\rm ext}$.
The near-horizon limit is then given by the limit $\epsilon \to 0$. The $\phi$-transformation ensures that the coordinates $(\tau, \rho, \theta, \psi)$ co-rotate with the horizon, and the gauge transformation of $A$ is required for the near-horizon limit to yield a finite gauge field. Using the fact that at extremality $\Delta$ has a double root, $\Delta|_{r_+} \sim (r-r_{+})^{2}$, the resulting near-horizon (NHEKN) geometry is
\begin{align}\label{eqn:nh_geometry}
  ds^{2} &= \Sigma(r_+, \theta) \left( - \rho^{2} d\tau^{2} + \frac{d\rho^{2}}{\rho^{2}}\right) + \frac{(r_{+}^{2}+a^{2})^{2} \sin(\theta)^{2}}{\Sigma(r_+, \theta)} \left( d\psi + 2 r_{+} \Omega_H \rho d\tau\right)^{2} + \Sigma(r_+, \theta) d\theta^{2},\nonumber \\
  \mathcal{A} &= Q \left[ \rho d\tau - \frac{r_+}{\Sigma(r_+, \theta)} \left( 2 r_+ \rho d\tau + a \sin^2 \theta d\psi \right) \right],
\end{align}
where $\Sigma(r_+, \theta) \equiv r_{+}^{2} + a^{2} \cos^2\theta$. This is still a solution of 4-dimensional Einstein-Maxwell theory (with zero cosmological constant). The $\tau-\rho$ part of this metric describes the 2-dimensional anti-de Sitter ($AdS_{2}$) spacetime, a solution of the vacuum Einstein equations with a negative cosmological constant, with Ricci scalar $R_{(2)} = - 2/\Sigma(r_{+}, \theta)$, and hence AdS\textsubscript{2} radius $L_{\rm AdS}^{2} = \Sigma(r_{+}, \theta)$. The geometry is much like $\textrm{AdS}_{2} \times S^{2}$, being exactly $\textrm{AdS}_{2} \times S^{2}$ in the RN limit $a = 0$, or e.g. when $\theta = 0$ or $\theta = \pi$. The isometry group $SL(2, R) \times U(1)$ of this near-horizon or \emph{throat} geometry was first described in~\cite{Bardeen:1999px}, and has been studied extensively in the context of the \emph{Kerr/CFT correspondence}~\cite{Guica2009}, which has since been extended to include Kerr-Newman. See~\cite{Compre2012} for a review of NHEKN and its CFT interpretation.

We now want to find the Klein-Gordon equation on NHEKN. This can be done in two equivalent ways, either by directly computing the Klein-Gordon equation on the near-horizon geometry~\eqref{eqn:nh_geometry}, or by taking the near-horizon limits~\eqref{eqn:nh_transformations} of the Klein-Gordon equation~\eqref{KG:radial}-\eqref{KG:ang} on the full KN geometry. We take the latter approach. First, note that in order to preserve the form of the Fourier ansatz $e^{-i \omega t + i m \phi}$, all finite frequencies must approach the superradiant frequency in the near-horizon limit, $\omega \to m \Omega_H + \epsilon \, \hat{\omega}$. Applying this, and the near-horizon transformations~\eqref{eqn:nh_transformations} to~\eqref{KG:radial}-\eqref{KG:ang}, we get the radial and angular ODEs for the Klein-Gordon equation on NHEKN,
\begin{subequations}
\begin{align}\label{eqn:nh_eqn_radial}
 & \frac{d}{d\rho} \left( \rho^{2} \frac{dR}{d\rho} \right) + \left[ (2 r_+^2 + a^2) m^2 \Omega_H^2 + \frac{1}{\rho^2} \left( \hat{\omega} + 2 m r_+ \rho \, \Omega_H \right)^2 - \lambda \right] R(\rho) = 0, \\\label{eqn:nh_eqn_angular}
 &  \frac{d}{dx} \left( (1-x^{2}) \frac{dS}{dx}\right) + \left[ \frac{m^2}{x^2 - 1} + \left(a\, m\,  \Omega_H\,x\right)^2 + \lambda \right] S(x) = 0,
\end{align}
\end{subequations}
where recall that $x=\cos\theta$ and $\lambda$ is the separation constant of the problem.
Note that the angular equation~\eqref{eqn:nh_eqn_angular} is still the oblate spheroidal harmonic equation, but it does not contain the eigenfrequency $\hat{\omega}$ since in the near-horizon limit (and at leading order) the original frequency $\omega$ has been replaced by the superradiant frequency $\omega \to m \Omega_H$. In the limits $m = 0$ or $a = 0$ the angular  Klein-Gordon equation can be solved exactly, since the eigenvalues are simply those of the spherical harmonics $\lambda = \ell(\ell+1)$, but otherwise we can solve the NHEKN angular equation~\eqref{eqn:nh_eqn_angular} numerically, using the same redefinition~\eqref{eigenfunction_redef_angular} as the angular equation~\eqref{KG:ang} on the full KN geometry.

What is the interpretation of the radial part~\eqref{eqn:nh_eqn_radial} of the NHEKN Klein-Gordon equation? Using the standard Fourier decomposition $\Phi = e^{- i \hat{\omega} \tau + i m \psi} R(\rho)$, we can rewrite the near-horizon radial equation~\eqref{eqn:nh_eqn_radial} as a Klein-Gordon equation for a  massive charged scalar field on AdS\textsubscript{2},
\begin{equation}
    \left(\hat{\nabla} - i q_{\rm eff} A_{\rm eff}\right)^2 \Phi = L_{\rm AdS}^{-2} \big[ \lambda - m^2 (a^2 + 2 r_+^2) \Omega_H^2 \big] \, \Phi, \qquad q_{\rm eff} \equiv -2 m r_{+} \Omega_H
\end{equation}
where $\hat{\nabla} - i q_{\rm eff} A_{\rm eff}$ is the gauge covariant derivative of pure $AdS_2$ with a homogeneous electric field:
\begin{equation}
    ds_{\rm AdS_2}^2 = L_{\rm AdS}^2 \left( - \rho^2 d\tau^2 + \frac{d\rho^2}{\rho^2}\right), \qquad\qquad A_{\rm eff} = -\rho \, d\tau.
\end{equation}
In other words, starting with a massless and uncharged scalar field and taking the near-horizon limit~\eqref{eqn:nh_transformations}, the scalar field acquires an effective mass and charge from the perspective of the near-horizon geometry, where the charge arises as a consequence of the horizon rotation $\Omega_H$. 

In pure AdS\textsubscript{2}, it is well known that massive scalar field perturbations are normalisable even if their squared mass $\xi^{2}$ is negative, provided it is above the Breitenl\"ohner-Freedman (BF) bound: $\xi^{2} L_{\rm AdS}^2 \ge - \frac{1}{4}$~\cite{Breitenlohner:1982jf, mezincescu_stability_1985}. On the other hand, if the mass is below the BF bound, the scalar field perturbation is not stable. A similar argument applies to the NHEKN geometry. The asymptotic behaviour of a solution of the near-horizon radial equation~\eqref{eqn:nh_eqn_radial} is $R|_{\rho \to\infty} \sim \rho^{\Delta_{\pm}}$, where $\Delta_{\pm}$ are the 2-dimensional conformal scaling dimensions
\begin{align}\label{eqn:M_bf_definition}
  \Delta_{\pm} = \frac{1}{2} \left(-1 \pm \sqrt{ 1 + 4 \xi_{\rm eff}^{2} L_{\rm AdS}^2}\right), \qquad \xi_{\rm eff}^{2} L_{\rm AdS}^2 &\equiv \lambda -  (6 r_{+}^{2} + a^2)m^{2} \Omega_H^{2} \\
  &= \lambda - m^{2}\, \frac{\sin^{2}\Theta \left(6 + \sin^{2}\Theta\right)}{\left(1 + \sin^{2}\Theta\right)^2}.
\end{align}
These solutions do not oscillate at large $\rho$  (i.e. they are normalisable, with finite energy) provided that
\begin{equation}\label{eqn:bf_bound}
  \xi_{\rm eff}^2 L_{\rm AdS}^2 \ge - \frac{1}{4}.
\end{equation}
This defines the effective BF bound for NHEKN\footnote{Note that an instability of the near-horizon geometry does not \emph{necessarily} imply an instability of the full extremal KN geometry, see~\cite{durkeePerturbationsNearhorizonGeometries2011,Hollands:2012sf} for a detailed discussion about this point regarding Kerr.}. Note that while the effective AdS\textsubscript{2} length scale $L_{\rm AdS}$ is a function of $\theta$ (and consequently so is the mass $\xi_{\rm eff}$), the physically relevant quantity is the dimensionless mass $\xi_{\rm eff} L_{\rm AdS}$ which is independent of $\theta$.

The quantity $\delta^2 =(6 r_+^2 +a^2)m^2 \Omega_H^2 -\frac{1}{4}-\lambda$ was defined in~\eqref{def:delta} in the context of the matched asymptotic expansion of the NH modes $\tilde{\omega}_{\hbox{\tiny MAE}}$ of~\eqref{NEfreq}. As discussed at the end of section~\ref{sec:NHanalytics}, the sign of $\delta^{2}$ indicates two distinct types of behaviour of the matched asymptotic expansion of the NH modes, since it determines whether $\delta$ gives a real or imaginary contribution to $\tilde {\omega}_{\hbox{\tiny MAE}}$. By studying the numerical QNM spectrum, it has also been observed that the sign of $\delta^2$ approximately corresponds to the phase boundary of the families of QNMs (\ie whether there are one or two distinct families): see  \cite{Yang:2012pj,Yang:2013uba,Zimmerman:2015trm} and the discussion in section~\ref{sec:m=l=2}. 

Complementing the analysis of \cite{Yang:2012pj,Yang:2013uba,Zimmerman:2015trm} we add  that from the perspective of the NHEKN geometry, we see that $\delta^2$ is related to the effective near-horizon mass by
\begin{equation}\label{eqn:bf_delta}
  1 + 4 \xi_{\rm eff}^{2} L_{\rm AdS}^2 = -4\delta^{2}.
\end{equation}
Therefore, we conclude that the sign of $\delta^{2}$ effectively indicates whether the effective AdS\textsubscript{2} BF bound of the near-horizon geometry is violated or not. Namely, one has $\delta^2 \le 0$ when the BF bound~\eqref{eqn:bf_bound} is respected and $\delta^2 > 0$ when the BF bound is violated. In the RN limit ($\Theta = 0$), the BF bound is always respected. However, as we move to the Kerr limit ($\Theta = \pi/2$), perturbations with certain combinations of the angular quantum numbers $\{m, \ell\}$ violate the BF bound. Concretely, in Table~\ref{table:bf_crit} we display the critical value $\Theta_\star$ of $\Theta = \arctan(a/Q)$ above which the BF bound~\eqref{eqn:bf_bound} is violated (and thus $\delta^2>0$), while the values of $\{m, \ell\}$ which always respect the BF bound, for all $\Theta$, are indicated by a dash. In the former case, there are two families of QNMs (the NH and PS modes) for $0\leq \Theta<\Theta_\star$ and a single `NH-PS' family for $\Theta_\star<\Theta\leq \pi/2$, while in the latter case there are two clearly distinct families of PS and NH modes for any $\Theta$. For example, the $m = \ell = 2$ entry is $\Theta_{\star} \sim 0.881 $ as discussed in section~\ref{sec:NHanalytics}: for $0\leq \Theta<0.881$ there are two distinct families (PS and NH) of QNMs but for $0.881<\Theta\leq \pi/2$ there is a single `NH-PS' family (and its overtones). Our results are consistent with previous discussions in the Kerr limit; e.g. compare our Table~\ref{table:bf_crit} with Fig.~1 in~\cite{Yang:2012pj}, which computed the phase boundary in Kerr numerically, by searching the QNM spectrum for each pair of quantum numbers $\{m, \ell\}$.

\begin{table}[]
\centering
\begin{tabular}{cccccccccc}
\toprule
& \multicolumn{9}{c}{azimuthal number $m$} \\ \cmidrule{2-10}
$\ell$   & 2    & 3    & 4    & 5    & 6    & 7    & 8    & 9    & 10   \\\midrule
2  & 0.88 &      &      &      &      &      &      &      &      \\
3  & -    & 0.71 &      &      &      &      &      &      &      \\
4  & -    & -    & 0.66 &      &      &      &      &      &      \\
5  & -    & -    & -    & 0.62 &      &      &      &      &      \\
6  & -    & -    & -    & 1.00 & 0.60 &      &      &      &      \\
7  & -    & -    & -    & -    & 0.88 & 0.59 &      &      &      \\
8  & -    & -    & -    & -    & -    & 0.80 & 0.59 &      &      \\
9  & -    & -    & -    & -    & -    & -    & 0.75 & 0.58 &      \\
10 & -    & -    & -    & -    & -    & -    & 1.10 & 0.71 & 0.57 \\\bottomrule
\end{tabular}
\caption{The critical value $\Theta_\star$ of $\Theta = \arctan(a/Q)$ above which the effective AdS\textsubscript{2} BF bound~\eqref{eqn:bf_bound} in KN is violated, for a range of values of the angular quantum numbers $\{m, \ell\}$. Dashed entries indicate that the BF bound is respected for all $\Theta$. Perturbations with $\ell < 2$ respect the BF bound for all $\Theta$ and so are not displayed.\label{table:bf_crit}}
\end{table}

In Kerr, at extremality, one can transform the radial part of the Klein-Gordon equation to a Schr\"odinger-type equation with a real potential (under the assumption that the frequencies approach the superradiant bound $\omega \to m \Omega_H$), and determine the location of the peak of the effective potential~\cite{Yang:2012pj}. A critical quantity $\mathcal{F}_0$ was defined such that the effective potential has a peak outside the horizon when $\mathcal{F}_0^2 > 0$. $\mathcal{F}_0^2$ differs from the Kerr limit of $\delta^2$ by $1/4$. In~\cite{Zimmerman:2015trm} it was argued that the generalisation of $\mathcal{F}_0$ to Kerr-Newman is a quantity $\mathcal{J}^2$, which again differs from $\delta^2$ by a quarter
\begin{equation}
    \mathcal{J}^2 = \delta^2 + 1/4.
\end{equation}
By comparison with~\eqref{eqn:bf_delta} we see that $\mathcal{J}$ is essentially the near-horizon effective mass, $\mathcal{J}^2 = - \xi_{\rm eff}^2 L_{\rm AdS}^2$. Hence the criteria $\mathcal{J}^2 = 0$ is the point at which the near-horizon mass $\xi_{\rm eff}$ vanishes, while the condition $\delta^2 = 0$ is when the effective mass $\xi_{\rm eff}^2$ decreases even further and violates the BF bound. Thus our BF bound analysis explains the physical relevance of the two  quantities $\delta^2$ and $\mathcal{J}^2$ introduced in \cite{Yang:2012pj,Yang:2013uba,Zimmerman:2015trm} and why they differ by $1/4$.

In the eikonal limit, we can parameterize the phase boundary or separatrix curve between the two QNM behaviours by just two parameters $\{\mu, \Theta\}$, where $\mu \equiv \frac{m}{\ell + 1/2}$ is called the \emph{inclination parameter}~ \cite{Yang:2012pj,Yang:2013uba,Zimmerman:2015trm}. By considering the location of the peak of the WKB potential\footnote{Note that this is different to the WKB expansion derived in section~\ref{sec:PSwkbHighOrders}, as we allow the possibility that $m \ne \ell$ here.} in the eikonal limit $(\ell + \frac{1}{2}) \gg 1$, a criterion for a peak to exist outside the event horizon was given in~\cite{Zimmerman:2015trm} (see also \cite{Hod_2012,Zhao_2015}):
\begin{equation}\label{eqn:mu_crit_eikonal}
  \mu^{2} \le \mu_{c}^{2}, \qquad\qquad \mu_{c}^{2} \equiv \frac{1}{2} \left( 3 + \frac{12 - \sqrt{136 + 56 (a/M)^{2} + (a/M)^{4}}}{(a/M)^{2}} \right),
\end{equation}
and it was also shown that the exact numerical phase boundary is well-described by $\mu = \mu_c$. The inclination parameter $\mu$ is bounded by $0 \le \mu^2 \le 1$. Setting $\mu_c = 1$, corresponding to equatorial modes $|m| = \ell$, in the eikonal prediction~\eqref{eqn:mu_crit_eikonal} yields the criteria that $a/M = 1/2$, or $\Theta = \pi/6$. As expected, this is the eikonal ($m = \ell$) critical value $\Theta_\star^{\rm eik} = \pi/6$ at which the PS modes have vanishing imaginary part in the extremal limit, as discussed in section~\ref{sec:EigenvalueRepulsionsA}.
The criteria $\mu^2 \le \mu_c^2$ that one gets from considering the location of the peak of the WKB potential can also be found by taking the eikonal limit of the BF bound criteria~\eqref{eqn:bf_bound} in the eikonal limit, as we now show. In~\cite{Zimmerman:2015trm, Yang:2012he}, in addition to the eikonal approximation $(\ell + \frac{1}{2}) \gg 1$, a further approximation was made for the angular eigenvalues of~\eqref{eqn:nh_eqn_angular}:
\begin{equation}
  \lambda \sim \left( \ell + \frac{1}{2} \right)^{2} + \frac{1}{2} \left[ -1 + \frac{m^{2}}{\left( \ell  + 1/2 \right)^{2}} \right] m^{2} a^{2} \Omega_H^{2}.
\end{equation}
Inserting this into the BF bound criteria~\eqref{eqn:bf_bound}, and solving for $\mu$, we find that the leading-order eikonal condition for the BF bound to be respected is equivalent to the eikonal condition~\eqref{eqn:mu_crit_eikonal} above. The criteria for the vanishing of the near-horizon mass $\xi_{\rm eff}^2 L_{\rm AdS}^2$ (or equivalently the vanishing of $\mathcal{J}^2$) also yields the same $\mu_c$~\eqref{eqn:mu_crit_eikonal} in the eikonal limit, since the factor of $\frac{1}{4}$ is sub-leading.

Strictly speaking, the eikonal condition $\mu^{2} \le \mu_{c}^{2}$ in~\eqref{eqn:mu_crit_eikonal} is only valid in the eikonal limit $(\ell + \frac{1}{2}) \gg 1$, whereas the BF bound criteria~\eqref{eqn:bf_bound} has no such restriction (both are derived in the near-extremal limit). However, in practice the two criteria are in very close agreement even for small $\ell$. For example, even as low as $m = \ell = 2$, the eikonal prediction is $\Theta|_{\tiny\mu=\mu_c} = 0.876$ for the $m = l = 2$ modes, versus $\Theta_\star = 0.881$ for the BF bound criteria --- see Table~\ref{table:bf_crit}. Given that these are both only considered to be approximate criteria we use them interchangeably in the main text.

The computation of the near-horizon geometry in this appendix generalises naturally to other spacetimes, even those which are not asymptotically flat. For example, in~\cite{Davey:2022vyx}, a violation of the near-horizon BF bound was found to be an important indicator of potential Strong Cosmic Censorship violations in Myers-Perry de Sitter, and these are intimately related to the fact that for sufficiently large $|m| = \ell$ there is only a single family of modes (just like in Kerr and KN).

\bibliography{eigenRepulsionKN_scalar}{}
\bibliographystyle{apsrev}

\end{document}